%% file: ms.tex
\documentclass[pdflatex,sn-mathphys]{sn-jnl}

\usepackage[utf8]{inputenc}
\usepackage{caption}
\usepackage{subcaption}
\usepackage[dvipsnames,table,xcdraw]{xcolor}
\usepackage{amssymb}
\usepackage{amsmath}
\usepackage{amsthm}
\usepackage{multirow}
\usepackage{lscape}
\usepackage{longtable}
\usepackage{array}
\usepackage{adjustbox}
\usepackage{soul,color}
\usepackage{caption} 
\usepackage{framed}
\usepackage{cjhebrew}
\usepackage{lmodern}

\newcommand{\RR}{\ensuremath{\mathbb{R}}}

\newcommand{\EE}{\ensuremath{\mathbb{E}}}

\DeclareMathOperator*{\argmin}{arg\,min}

\DeclareUnicodeCharacter{1E24}{\d{H}}
\DeclareUnicodeCharacter{1E25}{\d{h}}

\usepackage{mathscinet}

\jyear{2024}%

\theoremstyle{thmstyleone}%

\theoremstyle{thmstyletwo}%

\theoremstyle{thmstylethree}%

\begin{document}

\title[Dating ancient manuscripts using \textsuperscript{14}C and AI]{Dating ancient manuscripts using radiocarbon and AI-based writing style analysis}

\author*[1]{\fnm{Mladen} \sur{Popovi\'c}}\email{m.popovic@rug.nl}

\author[1,2]{\fnm{Maruf A.} \sur{Dhali}}\email{m.a.dhali@rug.nl}

\author[2]{\fnm{Lambert} \sur{Schomaker}}\email{l.r.b.schomaker@rug.nl}

\author[3]{\fnm{Johannes} \sur{van der Plicht}}\email{j.van.der.plicht@rug.nl}

\author[4]{\fnm{Kaare} \sur{Lund Rasmussen}}\email{klr@sdu.dk}

\author[5]{\fnm{Jacopo} \sur{La Nasa}}\email{jacopo.lanasa@unipi.it}

\author[5]{\fnm{Ilaria} \sur{Degano}}\email{ilaria.degano@unipi.it}

\author[5]{\fnm{Maria} \sur{Perla Colombini}}\email{maria.perla.colombini@unipi.it}

\author[6]{\fnm{Eibert} \sur{Tigchelaar}}\email{eibert.tigchelaar@kuleuven.be}

\affil*[1]{\orgdiv{Qumran Institute}, \orgname{University of Groningen}, \orgaddress{\country{The Netherlands}}}

\affil[2]{\orgdiv{Artificial Intelligence}, \orgname{University of Groningen}, \orgaddress{\country{The Netherlands}}}

\affil[3]{\orgdiv{Center for Isotope Research}, \orgname{University of Groningen}, \orgaddress{\country{The Netherlands}}}

\affil[4]{\orgdiv{Department of Physics, Chemistry, and Pharmacy}, \orgname{University of Southern Denmark}, \orgaddress{\country{Denmark}}}

\affil[5]{\orgdiv{Department of Chemistry and Industrial Chemistry}, \orgname{University of Pisa}, \orgaddress{\country{Italy}}}

\affil[6]{\orgdiv{Faculty of Theology and Religious Studies}, \orgname{KU Leuven}, \orgaddress{\country{Belgium}}}

\abstract{Determining the chronology of ancient handwritten manuscripts is essential for reconstructing the evolution of ideas. For the Dead Sea Scrolls, this is particularly important. However, there is an almost complete lack of date-bearing manuscripts evenly distributed across the timeline and written in similar scripts available for palaeographic comparison. Here, we present Enoch, a state-of-the-art AI-based date-prediction model, trained on the basis of new \textsuperscript{14}C dated samples of the scrolls. Enoch uses established handwriting-style descriptors and applies Bayesian ridge regression. The challenge of this study is that the number of radiocarbon-dated manuscripts is small, while current machine learning requires an abundance of training data. We show that by using combined angular and allographic writing style feature vectors and applying Bayesian ridge regression, Enoch could predict the \textsuperscript{14}C-based dates from style, supported by leave-one-out validation, with varied MAEs of 27.9 to 30.7 years relative to the \textsuperscript{14}C dating. Enoch was then used to estimate the dates of 135 unseen manuscripts, revealing that 79\% of the samples were considered ‘realistic’ upon palaeographic post-hoc evaluation. We present a new chronology of the scrolls. The \textsuperscript{14}C ranges and Enoch’s style-based predictions are often older than the traditionally assumed palaeographic estimates. In the range of 300–50 BCE, Enoch’s date prediction provides an improved granularity. The study is in line with current developments in multimodal machine-learning techniques, and the methods can be used for date prediction in other partially-dated manuscript collections. This research shows how Enoch’s quantitative, probability-based approach can be a tool for palaeographers and historians, re-dating ancient Jewish key texts and contributing to current debates on Jewish and Christian origins.}

\keywords{Palaeography, Artificial Intelligence, Radiocarbon Dating, Dead Sea Scrolls}
\maketitle
\input{tex-sections/00-main}
\input{tex-sections/01-radiocarbon}
\input{tex-sections/02-ai}
\input{tex-sections/03-05-results}
\input{tex-sections/06-09-discussion}

\clearpage
\newpage

\begin{appendices}

\clearpage
\newpage
\input{tex-sections/A-palaeography}

\clearpage
\newpage
\input{tex-sections/B-radiocarbon}

\clearpage
\newpage
\input{tex-sections/C-OxCal-plots}

\clearpage
\newpage
\input{tex-sections/D-palaeography-C14}

\clearpage
\newpage
\input{tex-sections/E-ai}

\clearpage
\newpage
\input{tex-sections/F-deep-learning}

\clearpage
\newpage
\input{tex-sections/G-test-135}

\clearpage
\newpage
\input{tex-sections/H-combineplots}

\clearpage
\newpage
\input{tex-sections/I-imagelist}

\clearpage
\newpage
\input{tex-sections/J-sample-locations}

\clearpage
\newpage
\input{tex-sections/K-datasheetrun}

\clearpage
\newpage
\input{tex-sections/L-worksheet-comparison}

\end{appendices}

\end{document}

%% file: tex-sections/00-main.tex
One of the main aims of palaeography—the study of ancient handwriting—is the dating of manuscripts on the basis of their handwriting~\cite{Nongbri2019, Orsini2018-edit, OrsiniClarysse2012}. Determining the chronology of ancient handwritten manuscripts is essential for reconstructing the evolution of ideas. This is particularly important for the Dead Sea Scrolls from ancient Judaea. These contain the oldest manuscripts of the Hebrew Bible and many previously unknown ancient Jewish texts,  mostly written in Aramaic/Hebrew script. The discovery of the scrolls in the 1940s–1960s fundamentally transformed our knowledge of Jewish and Christian origins~\cite{Brooke2018-kl}. 

Aramaic/Hebrew script in Judaea evolved from the imperial Aramaic script of the fifth and fourth centuries BCE to the Jewish square script in the first and second centuries CE. For the centuries in between, palaeographers have constructed a model of successive developmental stages, each characterized by distinct features of the script. This model was used to date manuscripts and thus affected the study of religious, cultural, and historical developments~\cite{Tigchelaar2020, Puech2017, Cross2003, Avigad1965}. However, these palaeographic distinctions are not reliably grounded (see Appendix~\ref{appen:A}). 

For palaeographic comparison, one requires enough date-bearing manuscripts that are evenly distributed across the timeline and written in similar script. Yet, only some of the very oldest, fourth century BCE, and the very youngest, first and second century CE, manuscripts have calendar dates (see Section~\ref{appen:A1:fewdate} in Appendix~\ref{appen:A}). To compensate for this scarcity for the centuries in between, palaeographers have turned to inscriptions on other surfaces which would be historically datable~\cite{Puech2017,Cross2003,yardeni2000-mq}), but these, too, have no absolute dates (see Section~\ref{appen:A2.1:notabs} in Appendix~\ref{appen:A}). Historical hypotheses of a slow development of the Aramaic/Hebrew script in the third century BCE and the emergence and rapid development of a national script around the mid-second century BCE~\cite{Puech2017, Cross2003,Yardeni1990} remain unsubstantiated. Neither inscriptions nor historical hypotheses enable us to reliably date the Dead Sea Scrolls (see Section~\ref{appen:A2.2:unsub} in Appendix~\ref{appen:A}). 

In this study, we bridge the palaeographic gap between the fourth century BCE and the second century CE, and advance palaeography in general, by combining new radiocarbon (\textsuperscript{14}C) dates with state-of-the-art artificial intelligence(AI)-based writing style analysis. A straightforward approach is to use machine-learning algorithms that are able to learn from a small set of labeled, i.e., dated, examples. This requirement is in conflict with the need for labeled data in current supervised deep learning, typically a thousand examples per class~\cite{Krizhevsky2017}. An abundance of data points is needed to warrant the stable estimation of, e.g., a neural-network model with millions of coefficients, in order to minimize the risk of arriving at a seemingly ‘good’ model~\cite{Vapnik2000}. A simple example is a linear function that has two coefficients and consequently needs a minimum of two data points to be determined. It may be appreciated that there exists a fundamental problem if the number of manuscript-date reference points is in the order of two dozen, while a computational model requires hundreds of thousands of coefficients or more. We address this challenge by applying methods that can (a) operate under sparse data conditions, (b) are explainable, and (c) do not require (pre)training from an extraneous, alien image collection. So, while it is tempting to use modern methods of deep learning, as we have done before~\cite{He2019,He2020,He2021,Zhang2022,Ameryan2023}, we will present several arguments for not using such approaches for the proposed style-based date prediction on a very small data set, i.e., at the current stage of Dead Sea Scrolls research on handwriting-style based manuscript dating (Appendix~\ref{appen:J:transferdeeplearningillustration}).

Since a general, large, representative, and labeled data set is not available for the period of the scrolls, we apply dedicated pattern recognition and machine-learning models, only using the relevant scrolls data for training a date-prediction model. Given the importance of the topic, it is expected that the use of pretrained deep transfer learning on the basis of extraneous material would elicit valid concerns among palaeographers about the relation between the scrolls’ target data and training data from a (very) different origin and period. Like the Ithaca approach~\cite{assael2022restoring}, a deep neural network making chronological attributions of ancient Greek inscriptions based on the totality of textual content, we focus on predicting chronological development, but unlike Ithaca, we use shape-style information from handwritten manuscript images instead.

We present Enoch, a machine-learning-based date-prediction model using established handwriting-style descriptors and applying Bayesian ridge regression. Enoch, named after the ancient Jewish science hero, was trained on the basis of new \textsuperscript{14}C-dated samples of the scrolls, providing reliable, absolute time markers that can bridge the palaeographic gap. Because of possible castor oil contamination issues with previous \textsuperscript{14}C datings of the scrolls~\cite{Bonani1992,Jull1995}, which would give a misleading \textsuperscript{14}C age that was ‘younger’ than the true age of the samples, new \textsuperscript{14}C dating was necessary for this study~\cite{rasmussen2001effects, rasmussen2003reply, rasmussen2009effects}. 

Enoch integrates multiple dating methods, using both physical, material-based evidence from \textsuperscript{14}C dating and geometric, writing style-based evidence from AI methods. With a new set of \textsuperscript{14}C-dated scrolls for temporal reference, the corresponding handwritten style features in those tested manuscripts are used for date estimation for undated manuscripts from the collection by applying machine-learning-based writing style analysis. Subsequently, interpolation of writing style features over time allows Enoch to make estimates for samples that do not have a \textsuperscript{14}C date and are only available as a digital image. Thus, Enoch offers date predictions as probability-based options that can aid palaeographers and historians in their decision-making and contribute to historical debates. 
 

%% file: tex-sections/01-radiocarbon.tex
\section{Radiocarbon dating}\label{sec:c14}

We performed \textsuperscript{14}C dating on 30 undated manuscripts from 4 sites, spanning an estimated 5 centuries: 25 from the Qumran caves, 1 from Masada, 2 from the Murabba\lasp at caves, and 2 from the Naḥal Ḥever caves. Twenty-eight manuscripts were made of animal skin or parchment, and 2 of papyrus.

Samples were selected because of their script and presumed period, the manuscripts having a sufficient number of characters for Enoch to be trained, and also on the basis of practical and conservational considerations (see Section~\ref{appen:B1:selec} in Appendix~\ref{appen:B}). One date-bearing document, Mur19, was used as a validation test for \textsuperscript{14}C, but did not go into the training of Enoch because of its cursive script.  

The scrolls are extremely delicate material. As in the previous attempts made at dating the scrolls~\cite{Bonani1992,Jull1995}, we, too, had to adjust the standard chemical AAA (Acid-Alkali-Acid) pretreatment (see Section~\ref{appen:B2:pretreat} in Appendix~\ref{appen:B}). Also, many fragments are contaminated with castor oil, which scholars in the 1950s used to improve the readability of the scrolls’ text~\cite{rasmussen2001effects, rasmussen2003reply, rasmussen2009effects}. This study is the first to apply, prior to \textsuperscript{14}C dating, a chemical treatment specifically designed for removing fatty materials, employing solvent extraction (see Sections~\ref{appen:B2:pretreat} and~\ref{appen:B7.1:Soxhlet} in Appendix~\ref{appen:B}). Further specialized analytical chemistry methods were applied before and after the sample pretreatment to demonstrate that the total amount of lipid materials is below a threshold that does not significantly skew the \textsuperscript{14}C date (see Sections~\ref{appen:B7.2:ramanOManalysis}–\ref{appen:B7.5:HPLCMSresults} in Appendix~\ref{appen:B}). 

The samples were dated by two Accelerator Mass Spectrometry (AMS) machines (see Section~\ref{appen:B3:AMSmeas} in Appendix~\ref{appen:B}). 

%% file: tex-sections/02-ai.tex
\section{Integration of multiple dating methods}\label{sec:ai}
We used 24 manuscripts from the \textsuperscript{14}C samples with accepted dates as labeled data for the primary training set for Enoch (see Sections~\ref{appen:B4:amsdating}–\ref{appen:B6:reject} in Appendix~\ref{appen:B} and Section~\ref{appen:C2:combipalaeoradiocarbon} in Appendix~\ref{appen:C-new}). For the data labels, we used OxCal v4.4.2~\cite{Oxcal, Oxcal2} to obtain the raw data points for the probability distributions. This is because the \textsuperscript{14}C results are not single dates, as with date-bearing documents, but represent date ranges with probability distributions. The \textsuperscript{14}C data input for training Enoch consists of the probability distributions of accepted 2-sigma (2$\sigma$) calibrated ranges (see Section~\ref{appen:E6:datepredictionmodel} in Appendix~\ref{appen:C}).

In addition to the primary training set, we created different combinations of training data to perform comparative analyses and further check the robustness of the model. These combinations include the tentative addition or omission of 4Q52, some previously tested \textsuperscript{14}C samples~\cite{Bonani1992, Jull1995}, date-bearing documents from the fifth–fourth centuries BCE and the second century CE (see Tables~\ref{old-c14-list} and \ref{tab:internally-dated} in Appendix~\ref{appen:D} for complete lists of manuscripts), the Maresha ostracon from 176 BCE (see Section~\ref{appen:A1:fewdate} in Appendix~\ref{appen:A}), and leave-one-out of the training data points.

\subsection{Deep neural networks for detection of handwritten ink-trace patterns}\label{deepneuralnetworksmainarticle}
The physical 24 \textsuperscript{14}C-dated manuscripts are visually available on many individual images of the IAA’s Leon Levy Dead Sea Scrolls Digital Library collection~\cite{dssllweb}. We also use images from Brill Publishers~\cite{lim1995dead}, especially in cases where the manuscript is unavailable in the IAA collection. For this study, these images underwent multiple preprocessing measures to become suitable for pattern recognition-based techniques. It should be noted that the images are extremely difficult to work with (some examples can be seen in Figure~\ref{fig:4Q319proccess} in Appendix~\ref{appen:C} and Figure~\ref{fig:howto-4q544} in Appendix~\ref{appen:H}; see also Section~\ref{appen:H2:dataquality} in Appendix~\ref{appen:H}). We are not dealing with digitally encoded text but with pixel images of highly degraded manuscripts as input.

We utilize multispectral band images of each fragment and employ an in-house image fusion technique~\cite{dhali2019binet} to generate a three-channel image. The resulting image representation enhances ink-vs-background contrast and therefore facilitates the effective separation of ink from backgrounds, commonly called binarization. For this purpose, we employ BiNet~\cite{dhali2019binet}, an artificial neural network based on an encoder-decoder U-net architecture designed to binarize the diverse range of scroll images. The resulting binarized images consist solely of black foreground pixels (ink) against a white background, ensuring that subsequent analyses focus exclusively on the handwritten patterns while minimizing inadvertent matches due to material-texture attributes. We further correct the rotation of the binarized images and divide them into multiple parts to maintain a balanced distribution of handwritten characters within each new image. No extraneous image material was used to train this binarization method.

Thus, we obtained a data set of 75 images from the 24 \textsuperscript{14}C-dated manuscripts. We used 62 of these images to train Enoch. The remaining 13 images, chosen deliberately and randomly, were passed as unseen test data to cross-validate the robustness and reliability of Enoch’s performance. The prediction of these 13 images by Enoch gives an 85.14\% overlap to the original \textsuperscript{14}C probability distributions (see Table~\ref{tab:trainingc14-in-testing} in Appendix~\ref{appen:D}). The image samples typically contain 150–200 characters, which has been shown to be sufficient for the comparable task of writer identification~\cite{Brink2008}. 

\subsection{Extracting features for style attribution}
It should be noted that in this context, ‘style’ is not related to textual content or wording. In fact, for characterizing the handwritten shapes, small shapes along the ink trace are used, largely uncoupled from the textual content, because we want to avoid spurious matches or date predictions on the basis of textual content. Once the training images were available, we could perform feature extraction techniques to translate handwriting patterns into feature vectors. The feature vectors relate directly to the shape-based evidence of the ink traces in the manuscripts and have a solid basis in writer identification~\cite{Bulacu2003, Schomaker2004, Bulacu2007} and document dating~\cite{Dhali2020, He2016}. We extract features from both the allographic and textural levels of characters~\cite{PlosOne}. An overview of machine-learning methods can be found in~\cite{Sommerschield2023}.

The first, allographic, method uses a self-organized character map obtained using a Kohonen neural network. As an example, this allographic codebook feature allows for a 93\% ($\pm ~ \sigma=2.3$\%) accuracy classification of the scripts ‘Hasmonaean’ vs. ‘Herodian’, using PCA, on 590 labeled manuscripts, results averaged over 32 random odd/even splits for training/testing \cite{monknet}. The second, textural, method uses statistical pattern recognition on angular information. The ‘hinge’ method for estimating the curvature distribution has been used extensively in writer verification and dating studies~\cite{Bulacu2007, Adam2018, Dhali2020}. Whereas the allographic feature addresses stylistic elements at the character level, the ‘hinge’ method concerns a micro-level feature directly related to the original writing activity that yielded the curvature of the ink trace. Therefore, we make a weighted combination of textural and allographic features to obtain an adjoined feature vector for each manuscript image. Such a feature vector constitutes the input data to Enoch.

\subsection{Bayesian ridge regression}
Due to the limited size of the data set, we cannot employ high-parametric models like period-specific temporal codebooks~\cite{He2016}. Instead, we utilize conditional modeling with Bayesian ridge regression~\cite{Hoerl2000}. This approach applies Bayesian inference to estimate model parameters for date prediction. By placing a prior distribution on the parameters and updating it with observed data using Bayes’ rule, we obtain the posterior distribution of the parameters and predicted dates. The Bayesian approach is chosen because our target output data represents probability curves for \textsuperscript{14}C dates (i.e., a vector) containing the accepted 2$\sigma$ calibrated ‘OxCal’ ranges. This probabilistic approach enables us to incorporate all available information while maintaining interpretability. Moreover, instead of producing a single number for the estimated date of a sample, it provides a comprehensive posterior distribution that allows us to assess the uncertainty associated with the estimated dates. Additionally, Enoch can utilize the Bayesian approach to provide error margins for predictions on unseen data.

\subsection{Testing Enoch\label{sec:recipe}}
Once Enoch was trained, we performed the validation by leave-one-out tests to check its performance. At this point, we took the calibrated style-based date estimation method of Enoch and applied it to a collection of 135 unseen manuscripts from the Dead Sea Scrolls as test data (see Table~\ref{tab:allTest135} in Appendix~\ref{appen:D}). 

We use two types of data-balancing techniques to compensate for the imbalanced distribution of the training data over different periods. One balancing technique involves data augmentation using random elastic morphing~\cite{bulacu2009morph} to create a balanced training data distribution. The second balancing is done on the output date predictions. This post-data-balancing uses accumulated training probabilities and training data point counts with 5\%, 10\%, and 20\% threshold values to avoid under-sampled time regions. 

The general recipe for Enoch’s analysis of manuscript images is presented in Table~\ref{tab:recipe}. Before applying this to scrolls manuscripts, we tried out a known mediaeval, dated benchmark data set of charters, MPS~\cite{He2016}, with success~\cite{Koopmans2023}.

\begin{table}[!ht]
\centering
\caption{Style-based date prediction recipe for Enoch\label{tab:recipe}}
    \begin{framed} \footnotesize
        \begin{enumerate}
            \item {\bf Select and crop} the relevant manuscript images based on scholarly identification criteria;
            \item In the images, perform a separation of the ink trace from the material background texture by using a deep-learning-based {\bf U-net variant} for
            {\bf multispectral image-intensity binarization}~\cite{dhali2019binet};
            \item For each manuscript, compute {\bf two shape descriptors}: a histogram of allographic fraglet occurrence and a histogram of angular co-occurrence along the ink-trace edges~\cite{Bulacu2007,Schomaker2004};
            \item {\bf Adjoin} the two {\bf feature vectors}, properly weighted, to a single handwriting-style vector~\cite{bulacu2009morph};
            \item In order to decorrelate the features, avoid collinearity, and minimize the necessary number of parameters in the next stage, perform a strong  {\bf dimensionality reduction (PCA, 20 dimensions)}.
            \item Take the {\bf \textsuperscript{14}C-dated manuscript-image samples}
            for {\bf training} Enoch as a style-based {\bf Bayesian ridge-regression model} with a scalar date estimate as the target output. In this training, augment the image data set by using random elastic morphing to obtain
            a sufficient and balanced number of examples per
            \textsuperscript{14}C-dated reference. This step is an essential, new contribution that allows a merger of \textsuperscript{14}C-based and style-based information in the date estimation. For validating Enoch, use the {\bf leave-one-out} approach: each sample that is under evaluation does not occur in the training data;
            \item {\bf Harvesting}: estimate style-based dates for {\bf undated manuscripts}.
        \end{enumerate}
    \end{framed}
\end{table}

%% file: tex-sections/03-05-results.tex
\section{\texorpdfstring{\textsuperscript{14}C dates and palaeographic estimates}{}}
The AMS results yielded 26 accepted \textsuperscript{14}C dates (see Sections~\ref{appen:B4:amsdating}–\ref{appen:B6:reject} in Appendix~\ref{appen:B}), which are shown in Table~\ref{tab:summarized-c14} (Appendix~\ref{appen:B}). The historical date preserved in Mur19 is consistent with the calibrated age range obtained by \textsuperscript{14}C (see Section~\ref{appen:B4:amsdating} in Appendix~\ref{appen:B}). Overall, we improved and extended the existing series of \textsuperscript{14}C-dated Dead Sea Scrolls~\cite{Bonani1992,Jull1995}.

Figure~\ref{fig:AIvsC14PAL} shows the comparison between the 2$\sigma$ calibrated ranges and traditional palaeographic estimates (in blue and red). This demonstrates that 17 of the 26 sampled manuscripts have whole or partial overlap, and 9 out of 26 samples yield calibrated ages that do not overlap with previous palaeographic estimates (see Appendix~\ref{appen:C-new}). 

Overall, the \textsuperscript{14}C results indicate older date ranges for individual manuscripts as well as for the emergence of the so-called ‘Hasmonaean’ and ‘Herodian’ scripts. Only two manuscripts have date ranges that go in the direction of a younger possible range. The \textsuperscript{14}C results for most manuscripts confirm the basic distinction between Hasmonaean-type manuscripts that are older, and Herodian-style manuscripts that are younger, and also between so-called ‘Archaic’ and Hasmonaean-type manuscripts. 

However, the \textsuperscript{14}C date ranges for manuscripts that are traditionally considered Hasmonaean and Herodian are quite differently distributed across the timeline. As can be seen in Figure~\ref{fig:AIvsC14PAL} (in blue), Hasmonaean-type manuscripts are all grouped together in a narrower part of the timeline but Herodian-type manuscripts are more spread out across the timeline, extending from the second century CE all the way back to the second century BCE (see Sections~\ref{appen:B8.1:overlap}–\ref{appen:B8.3:conclcomp} in Appendix~\ref{appen:B}). 

Sample 4Q114 is one of the most significant findings of the \textsuperscript{14}C results. The manuscript preserves Daniel 8–11, which scholars date on literary-historical grounds to the 160s BCE~\cite{SchmidSchroeter,zenger9}. The accepted 2$\sigma$ calibrated peak for 4Q114, 230–160 BCE, overlaps with the period in which the final part of the biblical book of Daniel was presumably authored (see Section~\ref{appen:B8.2:nooverlap} in Appendix~\ref{appen:C-new}). 

\section{Validation of Enoch}\label{main-valid-enoch}
\begin{figure}[!ht]
    \centering
    \includegraphics[width=1.1\textwidth]{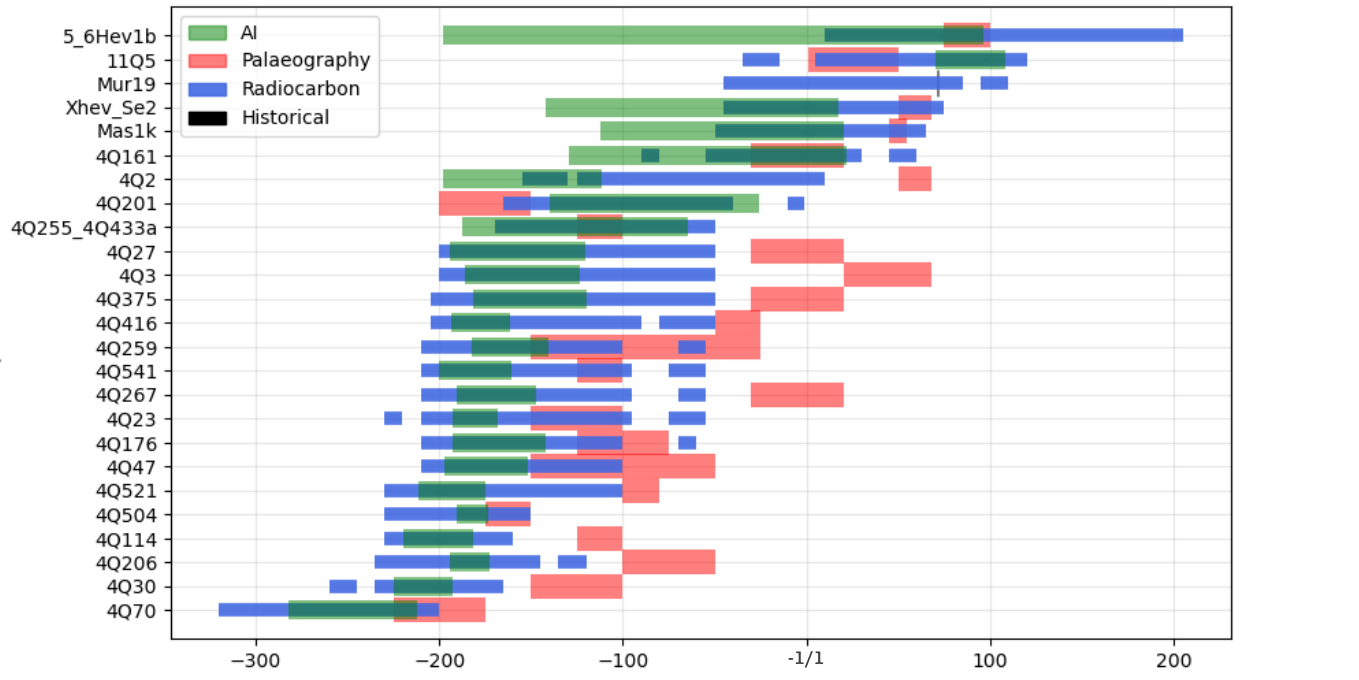}
    \captionsetup{format=hang}
    \caption{Overview of date estimations by three information sources and a calendar date: (accepted) 2$\sigma$ calibrated ranges \textsuperscript{14}C (blue), Enoch (green), palaeography (red), and historical (black). The vertical axis contains the manuscript numbers, and the horizontal axis contains dates: BCE in negative and CE in positive.}
    \label{fig:AIvsC14PAL}
\end{figure}

Figure~\ref{fig:AIvsC14PAL} (in green) also shows the results of cross-validation and leave-one-out tests for training Enoch. The choice for the bandwidths (2$\sigma$ date ranges for \textsuperscript{14}C, 1$\sigma$ uncertainties of the ridge regression for style-based predictions) is based on the intrinsic reliability of the two information sources. \textsuperscript{14}C date ranges are evidently superior to style-based predictions. 

Enoch’s style-based predictions largely follow the \textsuperscript{14}C results, even though the validation samples (rows) are in no way present in the training data. In the range 300–50 BCE, Enoch’s estimates provide a more fine-grained distribution than the \textsuperscript{14}C results. For samples 5/6Hev1b, Mas1k, and XHev/Se2, the style-based estimate is earlier and more uncertain. However, 11Q5 shows that in this late date range, a fairly certain style-based date estimate above 100 CE can also be achieved. This may go against historical reconstructions according to which the scrolls were hidden in the Qumran caves before the summer of 68 CE~\cite{Popovic2012}. Yet, we did not impose here a chronological limit on the model, because of the \textsuperscript{14}C result for 11Q5, and in order to examine the possibility of style continuation after 70 CE. 

Regarding the differences between the \textsuperscript{14}C date ranges and Enoch’s script style-based estimates, the mean absolute error (MAE) is $30.7$ years. The MAE drops to $27.9$ years when minor peaks (less than $4\%$ in all cases except for $5.2\%$ in 4Q2 and $9.4\%$ in 4Q416) are ignored (see Figure~\ref{fig:AIvsC14PAL-wo-minorpeaks} in Appendix~\ref{appen:E-plots}). In manuscript dating, MAE is commonly used~\cite{HamidMAE2019} for evaluation of a regression method. The difference with rms error is limited~\cite{HodsonMAE202}. With the chosen 2$\sigma$ (\textsuperscript{14}C) and 1$\sigma$ (AI) bandwidths, the error for the leftmost margin is $6.4$ years while for the rightmost margin it equals $-38.4$ years, indicating that Enoch’s style-based estimate range ends earlier than the \textsuperscript{14}C range. For each sample, the date ranges of the two information sources have partial to full overlap with an average of $88.8\%$. For Ithaca~\cite{assael2022restoring}, AI and epigraphy were used as two information sources to predict dates for ancient Greek inscriptions. Their prediction provides an average distance of 29.3 years from the target dating brackets, with a median distance of 3 years based on the totality of texts. We also aim for date prediction tasks, but, unlike Ithaca, we utilize three information sources: \textsuperscript{14}C, shape-based writing style analysis (AI), and palaeography.

Figure~\ref{fig:AIvsC14PAL} shows the general result that, on average, \textsuperscript{14}C date ranges and Enoch’s predictions indicate older dates than palaeography. Only 4Q201 and 11Q5 have older palaeographic date estimates, although there is an overlap with the \textsuperscript{14}C results (see Section~\ref{appen:B8.1:overlap} in Appendix~\ref{appen:C-new}). 

\section{Harvest of Enoch’s date predictions for previously undated manuscripts} \label{harvest-enoch}

\begin{table}[!ht]
\centering
\caption{Expert validation of Enoch’s date predictions}\label{tab:undated}
\resizebox{0.75\textwidth}{!}{%
\begin{tabular}{@{}|l|lrr|r|@{}}
\toprule
\textbf{Prediction is:}               & \multicolumn{1}{l|}{\textbf{Subcategory}} & \multicolumn{2}{c|}{\textbf{Manuscript count}} & \multicolumn{1}{c|}{\textbf{Percentage}} \\ \midrule
\textbf{Realistic}                    & \multicolumn{3}{r|}{107}                                                                    & 79.26\%                                  \\ \midrule
\multirow{3}{*}{\textbf{Unrealistic}} & \multicolumn{1}{l|}{Indecisive}           & \multicolumn{1}{r|}{4}   & \multirow{3}{*}{28} & \multirow{3}{*}{20.74\%}                 \\ \cmidrule(lr){2-3}
                                      & \multicolumn{1}{l|}{Too old}              & \multicolumn{1}{r|}{13}  &                     &                                          \\ \cmidrule(lr){2-3}
                                      & \multicolumn{1}{l|}{Too young}            & \multicolumn{1}{r|}{11}   &                     &                                          \\ \midrule
\textbf{Total manuscripts}            & \multicolumn{3}{r|}{135}                                                                   & 100.00\%                                 \\ \bottomrule
\end{tabular}}
\end{table}

Table~\ref{tab:undated} summarizes Enoch’s date predictions for 135 previously undated manuscripts. Expert palaeographers evaluated the style-based date predictions, condensing the prediction into two main categories: $realistic$ and $unrealistic$, the latter subdivided into $too$ $old$ and $too$ $young$ (see Appendix~\ref{appen:H}). 

As can be seen in Table~\ref{tab:undated}, $107$ ($79\%$) of the undated manuscripts were dated realistically, according to the palaeographers. Enoch’s date prediction task is not a $50/50$, binary decision task but regressive, with many possible years in the interval 300 BCE–200 CE. Assuming a coarseness of 25 years, as in the MPS project~\cite{He2016}, the date range would consist of 20 bins, with a $5\%$ prior-probability hit rate. Therefore, a success rate of $79\%$ is unlikely to be accidental. For $21\%$of the manuscripts, the palaeographers judged Enoch’s date predictions to be unrealistic. Enoch’s 28 unrealistic predictions were divided between too old ($46\%$) and too young ($39\%$). 

\begin{figure}[!ht]
    \centering
    \includegraphics[width=\textwidth]{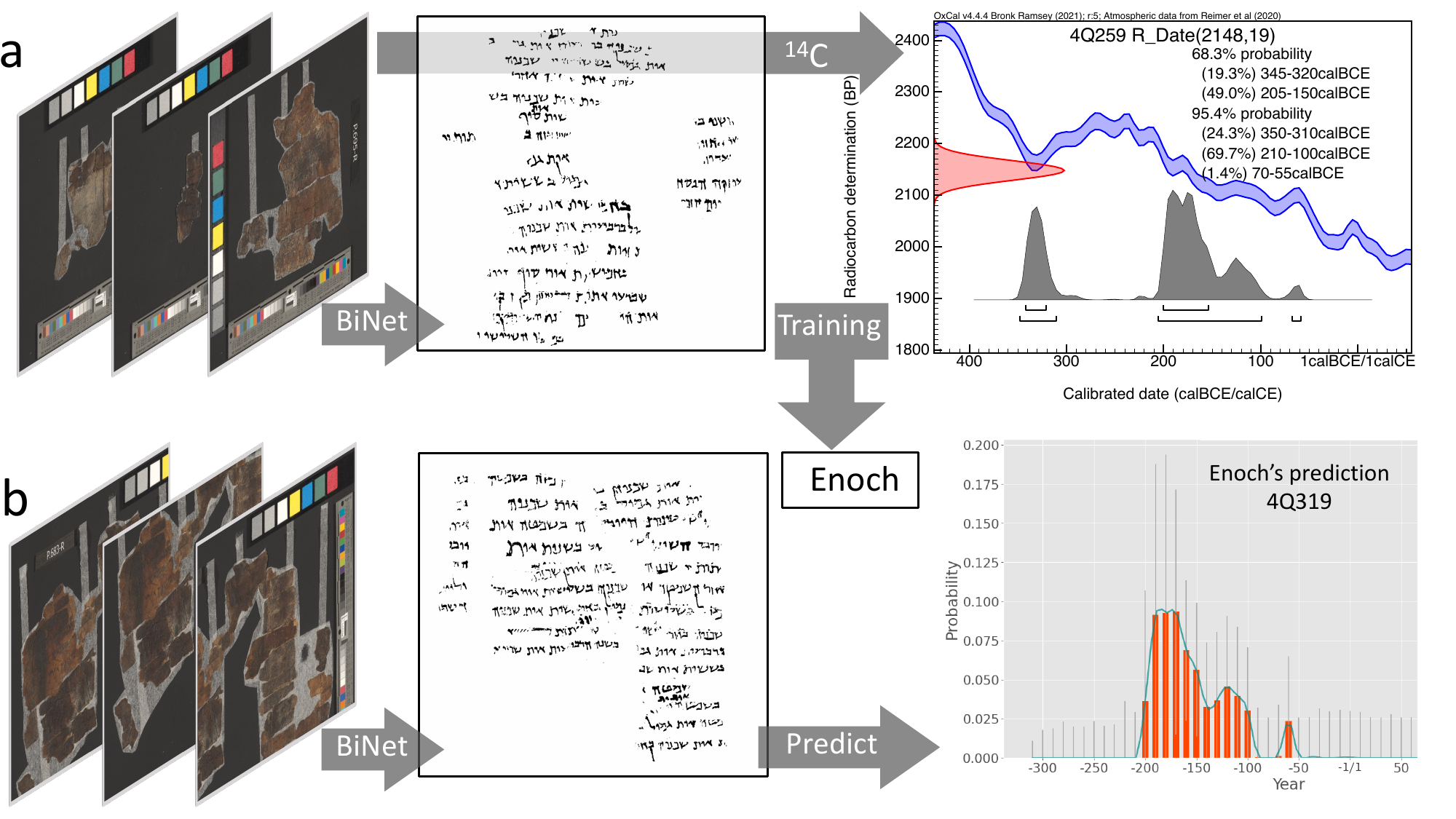}
    \caption{\textbf{a}, from full spectrum colour image to binarized image to \textsuperscript{14}C plot for 4Q259 that went into the training of Enoch. \textbf{b}, from full spectrum colour image to binarized image to Enoch’s date prediction plot for 4Q319 (see also~Fig. \ref{fig:4Q319proccess}). Red bars represent the probability of each date bin. The blue curve shows the smoothed distribution. Grey spikes indicate the local uncertainty of the estimate.}
    \label{fig:4Q319result}
\end{figure}

Samples 4Q259 and 4Q319 show that Enoch can accurately find the same date estimate for the same writing styles. The accepted 2$\sigma$ calibrated range of 4Q259 was used to train Enoch. Images of 4Q319 were part of a test set already in 2021. 4Q259 contains text that is part of the so-called Rule of the Community. 4Q319 contains a calendrical text. Because of perceived generic differences, 4Q319 received a separate classification number but is materially actually part of the same manuscript as 4Q259~\cite{Hempel2020}. At the time of the test, 6 July 2021, this identity was not known to the AI experts. Figure~\ref{fig:4Q319result} shows that Enoch was able to give a date prediction estimate for 4Q319 that matches the accepted 2$\sigma$ calibrated range of 4Q259 (see Section~\ref{appen:6July2021Test} in Appendix~\ref{appen:H}).

\begin{figure}[!ht]
    \centering
    \includegraphics[width=\textwidth]{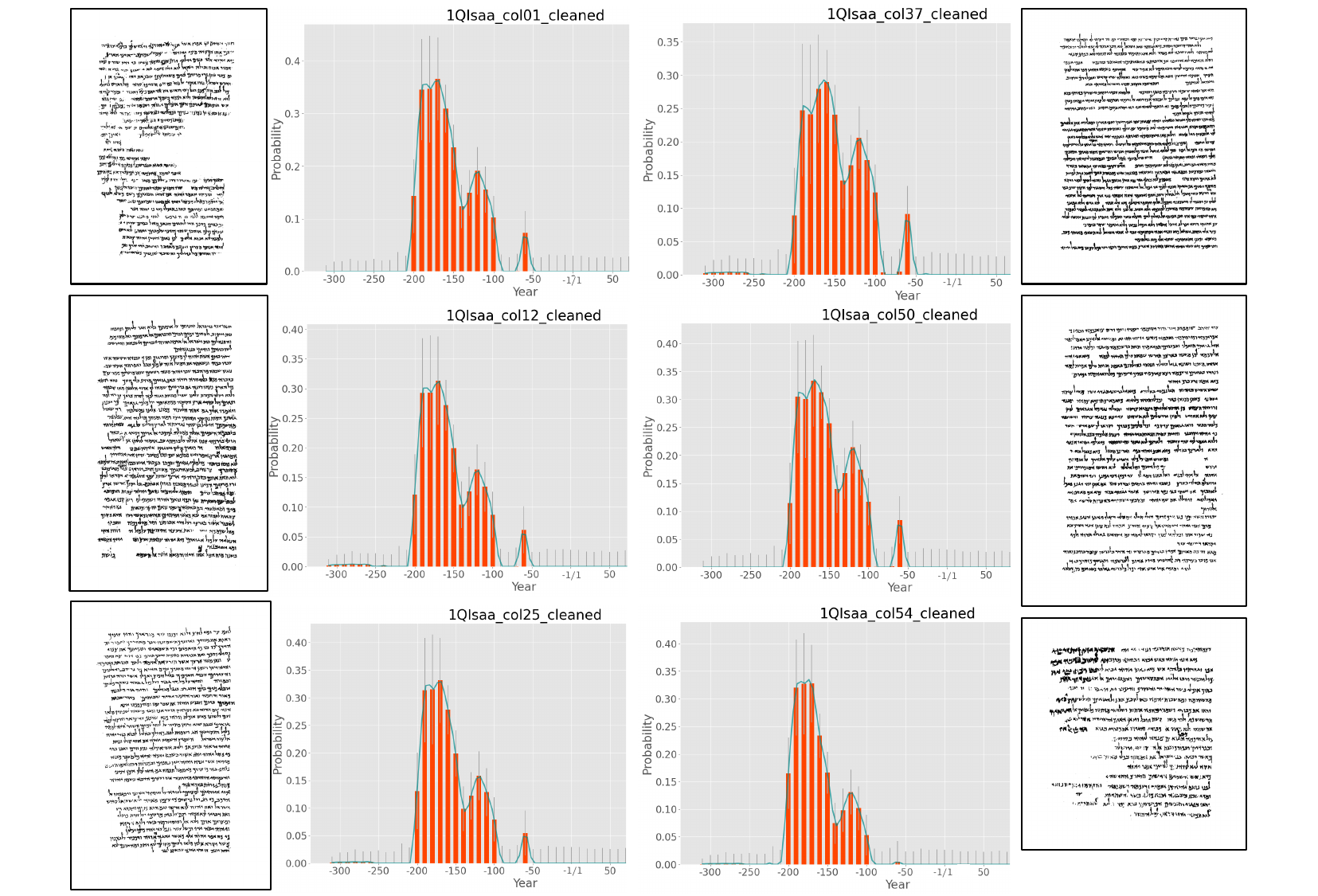}
    \caption{Enoch’s date prediction plots for 6 of the 54 columns from the two halves of 1QIsa\textsuperscript{a} (the left 3 columns are from the first half of the manuscript, the right 3 columns are from the second half of the manuscript).}
    \label{fig:1qisaaresult}
\end{figure}

Previously, we demonstrated that two scribes were at work in the Great Isaiah Scroll~\cite{PlosOne}. Now, Enoch shows that there is no temporal difference between the two halves of the manuscript as if one part were written significantly later than the other. On the contrary, both scribes are estimated to have worked on their respective part of the scroll of 1QIsa\textsuperscript{a} in the same period. Figure~\ref{fig:1qisaaresult} shows that Enoch dates the two halves consistently between 180–100 BCE.

%% file: tex-sections/06-09-discussion.tex
\section{Discussion and conclusions}\label{sec:discussion}

\subsection{Aramaic/Hebrew script development in ancient Judaea}
This study in style-based date prediction using the Enoch approach is a first step. The \textsuperscript{14}C data generated in this study in combination with machine-learning-based writing style analysis enabled us to examine Aramaic/Hebrew script in individual manuscripts with an empirically based precision that was not possible before. We combined palaeography, AI, and \textsuperscript{14}C to create a date-prediction model that leads to a new chronology of the scrolls during the third century BCE until the second century CE. We give four novel insights into Aramaic/Hebrew script development during this period and the date of individual manuscripts. 

First, \textsuperscript{14}C date ranges and Enoch’s style-based estimates are overall older than previous palaeographic estimates. These older dates for the scrolls are realistic. Hasmonaean-type manuscripts have accepted 2$\sigma$ calibrated ranges that allow for older dates in the first half of the second century BCE, and sometimes slightly earlier, instead of only circa 150–50 BCE. There are no compelling palaeographic or historical reasons that preclude these older dates as reliable time markers for the ‘Hasmonaean’ script. This also applies to the accepted 2$\sigma$ calibrated range for 4Q70 and its ‘Archaic’ script.  

Second, ‘Herodian’ script emerged earlier than previously thought. This suggests that the ‘Hasmonaean’ and ‘Herodian’ scripts were not transitioning from the mid-first century BCE onward, but that they existed next to each other at a considerably earlier date. 

Third, this novel approach of palaeography leads to a new chronology of the scrolls that impacts our understanding of the history of ancient Judaea and the people behind the scrolls. Hypotheses about whether the movement behind the scrolls originated in the second or first century BCE will need to be reconsidered in light of Enoch’s second-century BCE date predictions for Hasmonaean-type manuscripts such as 1QS and 4Q163 (see Appendix~\ref{appen:H}), bearing texts that are regarded typical for the movement. Scholars often assume that the rise and expansion of the Hasmonaean kingdom from the mid-second century BCE onward caused a rise in literacy and gave a push to scribal and intellectual culture. Yet, the results of this study attest to the copying of multiple literary manuscripts before this period. One example is 4Q109, a copy of the biblical book of Ecclesiastes, a book which scholars tentatively date to the end of the third century BCE~\cite{SchmidSchroeter}, for which Enoch gives a third-century BCE date prediction (see Appendix~\ref{appen:H}), close to Archaic-type manuscripts such as 4Q52 and 4Q70---copies of the biblical books of Samuel and Jeremiah. 

Fourth, this study’s \textsuperscript{14}C result for 4Q114 and Enoch’s date prediction for 4Q109 now establish these to be the first known fragments of a biblical book from the time of their presumed authors~\cite{SchmidSchroeter}. Also, Enoch’s integration of multiple dating methods yields a strongly improved value of sources of evidence and allows for a mutual confirmation of evidence from the two sources---physical (material) and geometric (shape-based).

The results of this study thus dismantle unsubstantiated historical suppositions and chronological limitations, and call into question the validity of the default model’s relative typology. This relative typology can only be maintained with restrictions. The spread of the Hasmonaean-type manuscripts over the timeline does not affect the default relative typology in a major way, but the older, second-century BCE date ranges of the Herodian-type manuscripts challenge the relative typology. More research is needed to solve this issue.

\subsection{The Enoch approach to dating ancient manuscripts}\label{discussion-Enochapproach}
To our knowledge, Enoch is the first complete machine-learning-based model that employs raw image inputs to deliver probabilistic date predictions for handwritten manuscripts utilizing the entire probability distribution from \textsuperscript{14}C output, and that is completed by palaeographic input while ensuring transparency and interpretability through its explainable design. Palaeographers and historians may now use Enoch’s quantitative, probability-based approach to palaeography as a tool to examine date predictions. The probability-based options can help in decision-making and to explicate qualitative palaeographic reasoning. Also, the methods underpinning Enoch can be used for date prediction in other partially-dated manuscript collections. 

It could be argued that the style-based predictions are influenced by the \textsuperscript{14}C-based training of the model. However, the leave-one-out validation results indicate that unseen samples obtain their interpolated position on the time axis based on the detected handwriting style in the images. The placement of an unseen sample on the time axis is not fundamentally constrained. Any date in the time range of 300 BCE to 200 CE could have been reached, looking at all style-based dates empirically covered by the model. 

In this study, we have avoided using palaeographic estimates as target values for machine learning because our goal is to provide physical (material) and geometric (shape-based) evidences for manuscript dating. While the use of palaeographic estimates as target values for machine learning is technically possible, we consider it too risky, given the existing uncertainties and lack of consensus associated with the precise dating of individual manuscripts. 

It becomes apparent that a broader time axis, with a sufficient number of samples at the tails---both at the BCE and CE ends---will allow for a larger time range of predictions. It would be very valuable if new manuscript samples could be added to the current collection. The consequences of each newly added manuscript sample to the Dead Sea Scrolls \textsuperscript{14}C reference collection can now easily be computed using the Enoch approach.

Enoch’s $79$\% success rate in date prediction is potentially interesting in view of the fully undated status of the manuscripts before the analysis was performed. Moreover, the images for the test data were not treated with the same care as those for the training of Enoch. All the training images underwent rotation and alignment correction, followed by a clean arrangement of smaller fragments within each manuscript, to obtain accurate feature extractions for the style periods represented by those manuscripts. If the same image preparation treatment were applied to every single image of the test data, it is to be expected that the percentage of realistic date predictions would exceed $79$\%. The $28$ ($21$\%) manuscripts that received an unrealistic date prediction in the current test may be due to image quality issues (see Figure~\ref{fig:11q7issues} in Appendix~\ref{appen:C}). The results of the test samples are likely to be better if more accurate manual cropping and rotation correction had been performed, similar to what has been done to the training samples. 

In its manuscript analysis, Enoch differs from traditional palaeographic approaches. Enoch emphasizes shared characteristics and similarity matching between trained and test manuscripts, whereas traditional palaeography focuses on subtle differences that are assumed to be indicative for style development. Combining dissimilarity matching and adaptive reinforcement learning can uncover hidden patterns. This interdisciplinary fusion may enrich our understanding of textual content, material properties, and historical context, leading to enhanced interpretations of the past. This remains a task for the future. New \textsuperscript{14}C evidence or, with new discoveries, a whole range of date-bearing manuscripts can be added to Enoch’s training data for further refinement and precision, continuously improving accuracy. Although the limited data were insufficient for a full deployment of deep-learning in the prediction task (see Appendix~\ref{appen:J:transferdeeplearningillustration}), future research needs to address the problems of sparse labeling and high dimensionality. It is to be expected that new solutions will appear here, because these problems are encountered in many application domains. If palaeographers are willing to accept the use of ‘black box’, pre-trained deep-learning models that are based on completely extraneous large image and photograph collections, future research may be directed at adapting the output of such models to the vectorial regression-based date-prediction task that is proposed in the current article.

\section{Online content}\label{sec:online}
All data, code, and test film associated with this article are publicly available on Zenodo with the following DOIs:\\
Data and prediction plots (v3): \href{https://doi.org/10.5281/zenodo.10998958}{https://doi.org/10.5281/zenodo.10998958}\\
Code and feature files (v5): \href{https://doi.org/10.5281/zenodo.11371749}{https://doi.org/10.5281/zenodo.11371749}\\
Film (see details in Appendix~\ref{appen:H}): \href{https://doi.org/10.5281/zenodo.8167946}{https://doi.org/10.5281/zenodo.8167946}

\section{Supplementary information}
This article has ten supplementary materials:
\begin{itemize}
    \item Appendix~\ref{appen:A}: The dating problem of the Dead Sea Scrolls
    \item Appendix~\ref{appen:B}: Radiocarbon dating of the Dead Sea Scrolls
    \item Appendix~\ref{appen:K}: \textsuperscript{14}C determinations and calibrated date plots
    \item Appendix~\ref{appen:C-new}: Palaeography and radiocarbon dating of the Dead Sea Scrolls
    \item Appendix~\ref{appen:C}: Artificial intelligence (AI) in dating the scrolls
    \item Appendix~\ref{appen:J:transferdeeplearningillustration}: On the use of pre-trained deep learning methods for image-based dating
    \item Appendix~\ref{appen:H}: Enoch’s date predictions for 135 previously undated manuscripts
    \item Appendix~\ref{appen:E-plots}: Comparative plots for different information sources
    \item Appendix~\ref{appen:D}: List of images for different tests
    \item Appendix~\ref{appen:F-sample}: Radiocarbon sample information
    \item Appendix~\ref{appen:G}: Data-sheet radiocarbon runs    
    \item Appendix~\ref{appen:I}: Worksheet of comparative data for 2$\sigma$ \textsuperscript{14}C  dates and traditional palaeographic estimates
\end{itemize}

\section{Acknowledgments}
The authors thank P. Shor, J. Uziel, T. Bitler, H. Libman, B. Riestra, O. Rosengarten, and S. Halevi at the Dead Sea Scrolls Unit of the Israel Antiquities Authority (IAA) and E. Boaretto (advisor to the IAA from the Weizmann Institute of Science, Jerusalem) for providing physical samples and multispectral images of the scrolls---courtesy of the Leon Levy Dead Sea Scrolls Digital Library; Brill Publishers for the Dead Sea Scrolls images from the Brill Collection; A. Aerts-Bijma and D. Paul for handling and measuring the \textsuperscript{14}C samples at the Center for Isotope Research (Groningen); S. Legnaioli for the Raman analyses performed at the CNR-ICCOM (Pisa); A. Krauss and T. van der Werff for their contributions to developing and testing Enoch; L. Bouma for cleaning images; D. Longacre, G. Hayes, A.W. Aksu, H. van der Schoor, C. van der Veer, and M. van Dijk for their contributions to preparing images for training Enoch; M.W. Dee for advising on and inspecting the code and data acquisition process from OxCal to the Enoch model at the Center for Isotope Research (Groningen). This project has received funding by the European Research Council under the European Union’s Horizon 2020 research and innovation programme under grant agreement no. 640497 (HandsandBible). M.P. and E.T. were also supported by NWO, Netherlands Organisation for Scientific Research and FWO, the Research Foundation - Flanders (SV-15-29).

\section*{Declarations}
Please check the Instructions for Authors of the journal to which you are submitting to see if you need to complete this section. If yes, your manuscript must contain the following sections under the heading `Declarations':

\begin{itemize}
\item Funding:\\
The work has been supported by an ERC Starting Grant of the European Research Council (EU Horizon 2020): \textit{The Hands that Wrote the Bible: Digital Palaeography and Scribal Culture of the Dead Sea Scrolls} (HandsandBible \# 640497).
\item The authors have no conflict of interest/Competing interests
\item Availability of data, materials, and code: see Section~\ref{sec:online} 
\item Authors' contributions: all the authors contributed equally to the article. 
\end{itemize}

%% file: tex-sections/A-palaeography.tex
\section{The dating problem of the Dead Sea Scrolls}\label{appen:A}

There is broad agreement in scholarship about the long-term lines of development of Aramaic and Hebrew script in Judaea from the fourth century BCE until the second century CE as an evolution from imperial Aramaic chancery script of the fourth century BCE to what became the dominant Jewish square script in the first and second centuries CE. However, when we zoom into the specifics of the centuries in between, the finer typological and chronological distinctions—misleadingly connected with historical-political eras—are not reliably grounded in the data; rather, they rely on so-called absolute pegs that are not absolute at all and on unsubstantiated suppositions about historical processes that would have influenced palaeographic developments.

The main problem is that there is a palaeographic gap between the third century BCE and the second century CE. There is a lack of absolute dates across the time period of the scrolls.

\subsection{Too few date-bearing manuscripts to compare with}\label{appen:A1:fewdate}
Palaeographic comparison of undated and dated manuscripts with a similar script is not possible. Only few date-bearing manuscripts have survived and those are at the outer limits of the date range. The oldest, from fourth-century BCE Wadi Daliyeh~\cite{Gropp2001-dy} and fourth-century BCE Bactria~\cite{Naveh2012-vv}, have script comparable to only one or two manuscripts, 4Q52 and 4Q70 (see also Sections~\ref{appen:A2.2:unsub} in Appendix~\ref{appen:A} and~\ref{appen:B8.1:overlap} in Appendix~\ref{appen:B}), but not the vast majority of the scrolls. The manuscripts from fifth-century BCE Elephantine are even further away in time~\cite{PortenYardeni19861999}. 

The youngest, from first- and second-century CE Murabba\lasp at~\cite{yardeni2000-mq, Benoit1961} and Naḥal Ḥever~\cite{yardeni2000-mq, CottonYardeni1997}, are mostly in cursive script and cannot be used to compare and date the vast majority of the Hebrew and Aramaic scrolls written in more formal scripts. Those dated manuscripts include about 30 documentary texts, mainly from Murabba\lasp at and Naḥal Ḥever. From the same period are 15 undated but datable letters, mostly in cursive script, to and from Simon bar Kokhba, the leader of the revolt against the Romans in 132–135 CE. Dated documents written in formal or bookhand script are limited to a farming contract from Murabba\lasp at (Mur24) and three leases of land from Naḥal Ḥever (5/6Ḥev 44, 45, 46), from 133 and 134 CE. 

Only one dated ostracon, from 176 BCE from Maresha~\cite{eshel1996aramaic} is known from the crucial period between the third century BCE and the first century CE. Another ostracon, from Khirbet el-Qom, is partially dated, and could date from 277, 241, or 217/6 BCE\cite{geraty1975khirbet}. Yet, these can hardly be used for dating formal hands, and cannot even serve as an indicative time marker to tie in manuscripts with a semicursive handwriting.

\subsection{Weak workarounds}\label{appen:A2:weak}
The way taken by Cross and others around the lack of date-bearing documents in formal, semiformal, or semicursive script from the third century BCE until the first century CE does not solve the problem. The relative development and absolute chronology of the scrolls’ palaeography was determined by taking recourse to a combination of a. supposed absolute pegs and b. unsubstantiated palaeographic and historical suppositions:

\subsubsection{Not so absolute time markers}\label{appen:A2.1:notabs}
Cross~\cite{Cross2003} claimed that his model was pegged by a series of absolute datings, in scores if not hundreds of documents inscribed on a variety of materials, especially from the late first century BCE and first century CE. Puech~\cite{Puech2017} provided additional pegs, specifically for the less formal Hasmonaean hands. Cross and Puech relied on inscriptions on other surfaces such as stone and metal, but here too there are no absolute dates, not even for the most important pegs, such as the Benei Ḥezir tomb and the Jason’s tomb inscriptions. Avigad~\cite{Avigad1965} acknowledged this, but his caution seems to have been forgotten. 

A telling example is the estimated date of the Benei Ḥezir tomb inscription in Jerusalem’s Kidron Valley (CIIP 137~\cite{eck2010corpus}), which, according to Cross, had been dated securely, on the basis of archaeological and historical evidence, to the end of the first century BCE. Based on architectural typology of the Hellenistic-style façade and Josephus’s description of the Maccabees’ family tomb in Modi\lasp in, Avigad~\cite{avigad1954} initially suggested to date the tomb to the mid-second century BCE. He then estimated the inscription, which lists eight priests from two generations who had been interred in the tomb, to have been made on the façade one or two generations after the construction of the tomb, in the first half of the first century BCE. Later, he dated the inscription palaeographically to the second half of the first century BCE, or to the Herodian period, and on that basis redated the tomb to the end of the Hasmonaean period~\cite{Avigad1965}. The precise length of time between the construction of the façade and the writing of the inscription (how many years are one to two generations?) is a conjecture. 

After the 2000–2001 exploration of the Benei Ḥezir and Zechariah tombs, Barag~\cite{Barag2003} put forward new data and interpretations which would indicate that the tomb dated to the period of flourishing in Jerusalem between ca. 132/1 and 63 BCE, most likely in the first century BCE. For example, it features the new type of tombs typical of the Hasmonaean period, which became common in the first century BCE. In the same direction point correspondences with Nabataean tomb architecture (undated but supposed to go back to the first century BCE), which, Barag argued, likely inspired the Benei Ḥezir tomb. As for the inscription, which he conjectured to be 50–100 years younger than the construction of the tomb, he compared its writing to that of the bronze coins of the 25th year of Alexander Jannaeus (79/8) BCE, and posited that the script of the Benei Ḥezir inscription would seem to be slightly later, from the late Hasmonaean or early Herodian period.

Without mentioning the Benei Ḥezir inscription, Naveh~\cite{Naveh1986} had identified the script on the Alexander Jannaeus coinage as ‘vulgar semiformal’ and saw its closest parallels to the letters found on ossuaries. Cross~\cite{Baillet1962} had described this style as a “crude simplified derivative” of the formal Herodian hand. Naveh’s aligning of the letters of the coins with those of the ossuaries might suggest that this type of Herodian hand was already anticipated by the Jannaeus coins. Naveh therefore referred to the palaeographical significance of these coins. One should note, however, that neither Naveh nor Barag carefully analysed the letters of the coins.     

Summarizing, all scholars associate the Benei Ḥezir tomb with the Maccabaees or the Hasmonaean period (either early or late), and date its inscription to the first century BCE. Yet, Cross’s claim that a late first-century BCE date is secure and an absolute peg, cannot be sustained. The date estimates of the tomb and its inscriptions are not only based on architectural typology, but also on the palaeographic typology. None of the evidence argues against a mid-first century BCE or even earlier date of the inscription. 

This is one example to demonstrate that inscriptions in Hebrew and Aramaic on other surfaces, such as stone and metal, cannot fill the void of absolute dating pegs between the third century BCE until the first century CE. In addition to the Benei Ḥezir burial inscription, this applies also, for example, to the so-called Queen Helena inscription (CIIP 123~\cite{eck2010corpus}) and Uzziah plaque (CIIP 602~\cite{eck2010corpus}) from the first century CE. Strictly speaking, these are not absolutely dated. The same applies to the Jason’s tomb inscriptions (CIIP 392-397~\cite{eck2010corpus}), the date of which is not fixed either. Puech~\cite{Puech2001,Puech1983} had initially argued on the basis of his reconstruction of the historical background of the inscriptions that the main one in Aramaic (CIIP 392) must be dated to 82/1 BCE, but more recently he stated that the Aramaic inscription dates palaeographically to about the middle of the first century BCE or slightly earlier \cite{Puech2017}. Yet, Yardeni dated the inscription shortly before the destruction of the tomb by an earthquake in 31 BCE \cite{eck2010corpus}. 

Another example are the hundreds of ossuary inscriptions, which Cross~\cite{Cross1955} said to virtually all belong to the Herodian era. A post-20/15 BCE date for the ossuaries may be archaeologically correct~\cite{Magness2005}, but the political and historical framing to the Herodian period does not limit the emergence of the script exhibited on the ossuaries to that period. The question when the so-called Herodian script came into being is decided somewhat arbitrarily. Cross took 30 BCE, Milik and Baillet 50 BCE~\cite{Tigchelaar2020}. Avigad also took 50 BCE or slightly earlier. Furthermore, Avigad already acknowledged that scrolls referred to as ‘Herodian’ may easily be earlier than this period~\cite{Avigad1965}. In other words, even for the ‘Herodian’ script, just as for the ‘Hasmonaean’ script (see below), the emergence is difficult to establish. In terms of typological development, we have to reckon with the possibility of longer, broader time frames for both scripts.

\subsubsection{Unsubstantiated palaeographic and historical premises}\label{appen:A2.2:unsub}

Even if one were to accept Cross’s recourse to a series of absolute datings, these would support mainly late first-century BCE and first-century CE comparisons. They do not help to establish the beginnings of the ‘Hasmonaean’ script. Lacking dated material from the third and second centuries BCE, Cross had to take further recourse to two premises to attempt to establish the upper range of the oldest scripts, ‘Archaic’ and ‘Hasmonaean’, from the scrolls, and to limit the earliest dating of the scrolls mainly to the second century BCE, with only a few exceptions for older ‘Archaic’ manuscripts such as 4Q52 and 4Q70. 

In addition to a lack of time markers, two palaeographic and historical premises by Cross, Yardeni and others stand out: a slow development of the Aramaic/Hebrew script in the early Hellenistic period (third century BCE); and the emergence of a national script as a watershed around 200–150 BCE. 

The presumed slow development of the Aramaic/Hebrew script in the early Hellenistic period is not supported by any dated evidence of that period. The assumption was in part based on a few undated cursive Aramaic papyri from Egypt containing Greek names (hence assumed to be from the third century BCE), but the later discovery of the dated Wadi Daliyeh papyri showed that there were different lines of development, some having taken place much earlier~\cite{Cross1974, Yardeni1990}, thus challenging the premise of the slow development, and reducing the importance of those Hellenistic Egyptian Aramaic papyri for establishing the evolution of the Aramaic/Hebrew script. For Judaea, Cross~\cite{Cross1955} also referred in passing to a conservative palaeography for the copying of sacred texts, but without further explanation or supporting evidence. 

Cross initially dated 4Q52 (4QSam\textsuperscript{b}) to “the last quarter of the third century B.C.”~\cite{Cross1955}, “no doubt late in the century”~\cite{Cross1961}, but after the discovery of the Wadi Daliyeh manuscripts, simply to “ca. 250 B.C.”~\cite{Cross1974} or “the mid-third century BCE”~\cite{Cross1998, Cross2003}. He seems to have been reluctant to date 4Q52 and also 4Q70 (4QJer\textsuperscript{a}) earlier, and therefore assumed a very slow evolution of the script, so as not to have a large time gap with the manuscripts written in what he called the “early Hasmonaean” script and which he dated to ca. 150 BCE. 

Yardeni, too, regarded 4Q52 and 4Q70 as examples of a transitional stage from the fourth and third century BCE Aramaic script in the direction of the ‘Hasmonaean’ script~\cite{Yardeni1990}. Her conclusion that these two manuscripts could therefore be dated to the late third or early second century BCE seems to be based rather on the supposed proximity to this national script than on the correspondences with the earlier Aramaic scripts. 

However, the palaeographic principle is to date an undated manuscript by comparing its script to that of dated writings with a similar script. This means that the oldest manuscript of the scrolls, 4Q52, must be compared to the Aramaic evidence from Wadi Daliyeh from the fourth century BCE. 4Q52 should then be chronologically closer to those manuscripts, especially WDSP 1 (335 BCE).

The hypothesis of the emergence of a national script around 200–150 BCE and the supposition that the ‘Hasmonaean’ script was a development of the Hasmonaean period after 150 BCE are not supported by any dated evidence but based on historical assumptions, given in passing, about a “nationalistic expansion and resurgent Orientalism”~\cite{Cross2003} after the death of the Seleucid king Antiochus IV Epiphanes (164 BCE). These unfounded assumptions were then imposed as an interpretative framework on the manuscript evidence. But, given the absence of dated material from the third and second century BCE, there are no historical, typological or other palaeographic reasons for limiting the rise of the script which Cross called ‘Hasmonaean’ to the mid-second century BCE. 

This means that manuscripts written in ‘Hasmonaean’ script may date already from the early second century or from the third century BCE. This older dating is also realistic when manuscripts written in so-called ‘Archaic’ script, such as 4Q52 or 4Q70, can be dated earlier in the third century BCE or, for 4Q52, even perhaps in the late fourth century BCE. Furthermore, this older dating can be independently supported by the \textsuperscript{14}C dating results in this study (see Section~\ref{appen:B8:comparison} in Appendix~\ref{appen:C-new}).

\subsection{The way out of the gap}\label{appen:A3:wayout}
Summarizing, the dating problem of the Dead Sea Scrolls, due to the absence of calendar dates, is further confounded by the fact that there are no other date-bearing manuscripts in similar script available for palaeographic comparison. This lack of date-bearing documents cannot be overcome by using inscriptions on other surfaces instead because these, too, have no absolute dates. Also, datable inscriptions mainly date from the first century BCE and first century CE and thus cannot shed light on script developments in the third and second centuries BCE. Historical premises and assumptions remain unsubstantiated and devoid of factual support, and they fail to support a chronological framework for the palaeography and the manuscript evidence. These assumptions cannot determine or sufficiently constrain the dates connected to the writing of the scrolls. 

Therefore, \textsuperscript{14}C dates derived from manuscript samples are needed as absolute time markers to lead the way out of the palaeographic gap. In the absence of an abundance of date-bearing manuscripts written in similar script available for palaeographic comparison, \textsuperscript{14}C dating, a scientific measurement (“yardstick of time”), provides more reliable time markers, and in combination with our style-based date-prediction model Enoch even more precise time markers.

%% file: tex-sections/B-radiocarbon.tex
\section{Radiocarbon dating of the Dead Sea Scrolls}\label{appen:B}

Two series of Dead Sea Scrolls were radiocarbon dated in the 1990s, in Zurich and in Tucson, Arizona~\cite{bonani1991radiocarbon,Bonani1992,Jull1995}. In addition, three samples were submitted to Oxford but in all three cases the chemistry is recorded as having “failed,” i.e., no sample to measure; probably the samples completely dissolved during the pretreatment phase (communication from R. Hedges, Research Laboratory for Archaeology, Oxford, 7 January 2005). 

Although scrolls were radiocarbon dated in the 1990s, new radiocarbon dating was necessary because of castor oil contamination issues with these previous dates. Furthermore, since then, radiocarbon dating methods and procedures have improved significantly in terms of better calibration, higher precision obtained by more modern methods and instruments, and also more effective cleaning procedures for dealing with contaminated samples. 

In this study, we have taken the following analytical steps for the samples:

\begin{enumerate}
    \item They were precleaned by a Soxhlet procedure in Odense (see Sections~\ref{appen:B2:pretreat} and~\ref{appen:B7.1:Soxhlet});
    \item Subsequently, they were further pretreated by standard methods in Groningen (see Section~\ref{appen:B2:pretreat});
    \item The cleaned samples were dated by Accelerator Mass Spectrometry (AMS) in Groningen (see Sections~\ref{appen:B3:AMSmeas}–\ref{appen:B6:reject});
    \item During the study, the residual lipids in the extracts of all 30 samples after the Soxhlet cleaning were analysed, and 17 samples have been further investigated by specialized analytical chemistry methods in Pisa regarding the nature of the contamination (see Sections~\ref{appen:B7.2:ramanOManalysis}–\ref{appen:B7.5:HPLCMSresults}).
\end{enumerate}


\subsection{Selection of Samples}\label{appen:B1:selec}
\begin{figure}[ht]
    \centering
    \includegraphics[width=1.2\textwidth]{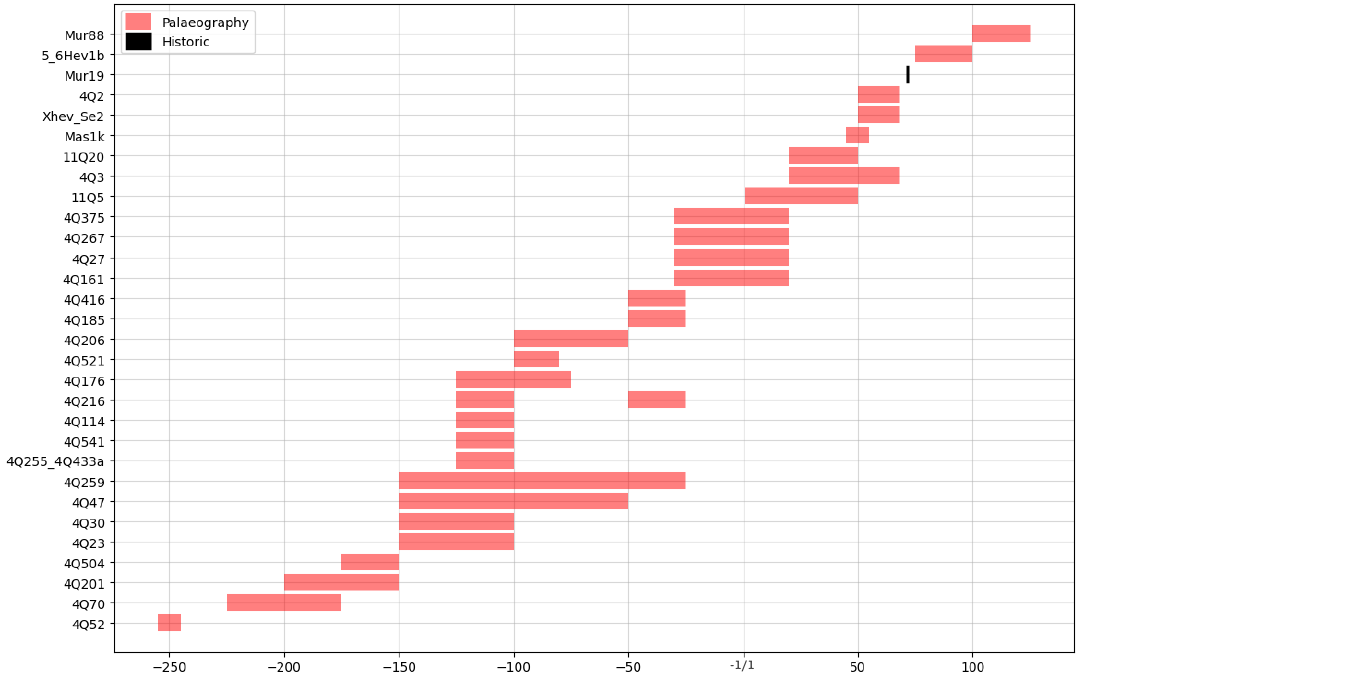}
    \caption{The selection of 30 manuscript samples according to their traditional palaeographic date estimates.}
    \label{fig:pal-chronology}
\end{figure}

The 30 samples we received from the Israel Antiquities Authority (IAA) were selected on the basis of script and presumed period so as to obtain reliable time markers in the palaeographic gap between the fourth century BCE and the second century CE. We made this selection at the start of the project on the basis of the default model in the field (see Appendix~\ref{appen:A}). The dates associated with the manuscripts according to this traditional model provided balanced coverage of the timeline under investigation (as can be seen in Figure~\ref{fig:pal-chronology}). Also, because the \textsuperscript{14}C dates are needed to go into the date-prediction model, we selected manuscripts that contain a sufficient number of characters in their extant material, 150–200~\cite{Brink2008}. The manuscript identity and presumed palaeographic periods of the samples were not known to the staff of the laboratories in Groningen, Odense, and Pisa at the time of the measurement. One of the 30 samples, from a date-bearing document (Mur19), was added as a control text. Its identity and date were also unknown to the laboratories at the time of measurement. Furthermore, in consultation with the IAA, the final selection of samples was determined also on the basis of practical and conservational considerations regarding specific manuscript remains. The IAA provided general indications concerning where the physical samples were taken from (see Appendix~\ref{appen:F-sample}). In our sample set, we have 28 manuscripts of animal skin, and 2 of papyrus (4Q255/4Q433a and Mur19).

From the first century CE onward, a clear distinction appears in the manuscript evidence between the square bookhand script and the standard cursive style~\cite{yardeni2000-mq}, but such a distinction is less pronounced in the manuscript evidence of earlier periods. This also applies to the distinctions made between formal, semiformal, and semicursive styles. Across the continuum of the chronological range covered by the scrolls, exemplary specimens for some styles are lacking~\cite{Cross2003}. Often manuscripts exhibit a mixture of these presumed styles~\cite{Popovic2023,Tigchelaar2020,https://doi.org/10.25592/uhhfdm.739}. Therefore, our sampled manuscripts cover all three categories and their mixtures. The cursive style has been excluded from our sampling, except for Mur19 which was used as a validation test for \textsuperscript{14}C. 


\subsection{Soxhlet Treatment and AAA Pretreatment}\label{appen:B2:pretreat}

Castor oil was used in the 1950s by the original team of scholars reconstructing and editing the Dead Sea Scrolls to clean the manuscripts and to improve readability of the text. But the castor oil needs to be removed, because it would give a misleading \textsuperscript{14}C age that was “younger” than the true age of the sample. Later testing showed that not all castor oil will be removed even by the standard AAA (Acid-Alkali-Acid) protocol, let alone by the reduced form of the standard protocol used in the 1990s in Zurich and Tucson \cite{rasmussen2001effects, rasmussen2003reply, rasmussen2009effects}. 

Before the actual start of the project, we received 2 test samples from the IAA which were relatively large (tens of milligrams). These were materials without context but of scrolls origin according to the IAA. Both samples were subjected to the standard AAA treatment, but the material immediately started dissolving before our eyes during the first acid step. This meant we could not apply the standard method; also considering the test samples were much larger than the identified manuscript samples we were to receive.

In our project the first step was to pre-clean the samples by a liquid extraction with suitable solvents, performed inside a Soxhlet apparatus to remove the castor oil contamination. The Soxhlet treatment was carried out in Odense, initially in three but subsequently in four extraction steps. The latter was done for redundancy, and not because of suspicion that the three-step procedure was not sufficient within the given dating uncertainty. The fourth step furthermore guaranteed that even more potential contaminants were removed. Castor oil is a plant product, which consists of several triglycerides and free fatty acids. The Soxhlet treatment is designed to remove lipid material to a high extent, and the analyses done in Pisa by HPLC-MS and Py-GC/MS are performed to demonstrate that the amount of the remaining lipid material, including fatty acids, is below a threshold which does not significantly skew the radiocarbon date (see section B.7).

The scrolls are extremely delicate material. Their fragility is an issue for the chemical cleaning of samples for radiocarbon dating. In the 1990s, the Zurich and Tucson laboratories had to adjust or stop the AAA pretreatment because the samples were dissolving \cite{Bonani1992, Jull1995}. Most of these samples were much larger than the samples in the current study. With much smaller sample materials, we also had to adjust the standard chemical pretreatment. 

Following the Soxhlet treatment in Odense, the samples were further prepared for dating in Groningen. The pretreatment was adapted to Acid only in a “soft” form: 0.5–1\% HCl, refrigerator temperature (ca. 4°C) and only for 10 minutes. Next, we dried the sample in an oven at a temperature of 80°C overnight. Using diluted HCl and skipping the Alkali step is necessary because of the delicate nature of the samples. This is justified because of the conditions the scrolls were kept in. No significant amounts of foreign materials that could cause errors larger than the measurement uncertainties were observed. Our procedure is proven correct because the sample with a known historical date (Mur19) was \textsuperscript{14}C dated correctly. Combined with the Soxhlet treatment, this is the optimum treatment for this delicate material, and generally effective. 

The scrolls were stored in caves in the Judaean desert in the absence of humic acids and constant groundwater. In particular the humic acids constitute a problem for many other archaeological excavations worldwide and they are the main reason that necessitates the alkaline bath in the standard pretreatment protocol (the second A in AAA). The environment in the caves can be characterized as limestone, gypsum and marls — none of which has the potential to inflict alkali-soluble compounds onto the parchments. Similarly, bat guano and excretions from other small animals who have possibly found their way into the caves over the centuries are unlikely to contain humic acids, and therefore their deposits are likely to be dissolvable in either the more polar solvents of the Soxhlet treatment (i.e., the ethanol) or in the acidic bath of the pretreatment in the radiocarbon laboratory. And even further, the pyrolysis-gas-chromatography measurements did not reveal any compounds unaccounted for (see Section~\ref{appen:B7.4:PyGCMSresults}); that includes the alkanes that can be considered markers for bat guano \cite{Queffelec2018}.


\subsection{AMS Measurements}\label{appen:B3:AMSmeas}

After cleaning, the samples were combusted into CO\textsubscript{2} gas. For the GrA dates, the gas is subsequently reduced to graphite using H\textsubscript{2}. Subsequently, the \textsuperscript{14}C content was measured in this graphite. This method was also applied by the GrM machine for routine dating. However, this machine also has the option to measure the \textsuperscript{14}C content in CO\textsubscript{2}, skipping the graphite production step. This is very useful for small samples, as is the case for many scroll samples. Therefore, for scroll samples measured by the Micadas, the gas source was used. For more details on measurement procedures see~\cite{Dee2019}. 

For the 30 samples in this study, there is a grand total of 131 individual AMS runs. This total number includes duplicate samples and multiple runs. In most cases a solid date can be calculated for the separate runs done for a particular scroll, based on averaging. The numbers reported reflect the measurements by AMS. In addition, there are aspects of sample integrity and pretreatment which are hard or even impossible to quantify. We have rejected 10 AMS runs for technical reasons, resulting in a final number of 121 accepted runs.

The \textsuperscript{14}C content in the sample is measured by AMS. The original AMS was a 2.5 MV Tandetron accelerator \cite{van2000status}. It was decommissioned in 2017, and replaced by a Micadas system \cite{synal2007micadas}. This took place during the project, so that both machines have been used to date the scroll samples. This allows for internal intercomparison (see Table~\ref{tab:summarized-c14}). The Tandetron dates have laboratory code GrA; for the Micadas, this is GrM.


\subsection{AMS Dating Results}\label{appen:B4:amsdating}

Radiocarbon dates are reported by convention in BP, using a defined halflife and reference radioactivity for \textsuperscript{14}C, and a correction for isotopic fractionation using the stable isotope \textsuperscript{13}C \cite{mook1999reporting}. The BP dates are converted to calendar dates, using the IntCal20 calibration curve (\cite{reimer2020intcal20}) and OxCal software \cite{Oxcal}. The calibration results in a non-Gaussian probability distribution of calendar dates. This distribution is given in 1$\sigma$ (68.3\% confidence) and 2$\sigma$ (95.4\% confidence) date ranges.

For the 30 samples, 27 yielded accepted dates; only 3 samples yielded inconsistent results and had to be technically rejected (4Q216, 11Q20, and Mur88; see Section~\ref{appen:B6:reject}). Also, it appeared that the sample received for 4Q185 could not be ascertained as belonging to that particular manuscript. This sample is not used in our analysis (see Section~\ref{appen:B5:4Q185}).

The resulting \textsuperscript{14}C dates for the 26 samples are shown in Table~\ref{tab:summarized-c14}. Each individual \textsuperscript{14}C sample receives a unique laboratory number. As the table shows, each scroll is dated at least twice. In addition, many measurement batches were repeated (thus yielding two dates per graphite sample). The resulting \textsuperscript{14}C age shown is the averaged number for all accepted runs. Overall, the logistics is complex. For example, the sample 4Q114 (4QDaniel\textsuperscript{c}) has been dated in 7 runs. Two samples were received from the IAA. Graphite was prepared from all material of the first sample, and it was dated by the GrA machine. There were 3 runs from the same graphite (to increase the \textsuperscript{14}C statistics), so all have the same GrA number; the 3 runs are triplicates and can be taken together as 1 GrA date. An additional second sample was received later. From this sample we dated 4 subsamples in 4 runs by the GrM machine. Hence there are 4 GrM numbers.

The resulting BP dates are very precise, with 1$\sigma$ uncertainties of only 15–28 years. For the full results of all runs with more details (in particular Carbon yield and $\delta$\textsuperscript{13}C value), see Appendix~\ref{appen:G}.

Table~\ref{tab:summarized-c14} shows the summarized results of 26 accepted \textsuperscript{14}C dates: laboratory code, sample identification, \textsuperscript{14}C age (BP), its sigma (BP), and calibrated dates (both 1$\sigma$ and 2$\sigma$ ranges). The OxCal plots can be seen in Appendix~\ref{appen:K}. 

Although the most recent calibration curve, IntCal20, has a resolution of 1 calendar year that does not mean 1-year resolution is significant. The measurement precision for the \textsuperscript{14}C dates is, at best, 15 \textsuperscript{14}C years, and often a few decades. Moreover, OxCal can be calculated for 1 year, but the default resolution of OxCal is 5 years without any interpolation. However, if the resolution is set to less than 5 years, the curve will be interpolated by a cubic function. A cubic function is a polynomial function of degree 3, which, in the case of OxCal, performs interpolation of two different data points to obtain intermediate points. This is a mathematical formulation and not a calibration of 1-year resolution. Hence, we do not take a 1-year interpolated resolution but present the raw 5-year resolution data from OxCal. For more details, we refer to https://c14.arch.ox.ac.uk/oxcalhelp/hlp\_analysis\_inform.html. 

Furthermore, for the time range relevant for the scrolls our calibrated results are often bimodal, especially for 2$\sigma$ distributions which we use for our further analyses for firmer grounding of our date-prediction model. The calibrated results from the 1990s were also often bimodal~\cite{bonani1991radiocarbon,Bonani1992,Jull1995}. This bimodality is an effect of the calibration curve not being linear, showing peaks and other irregularities caused by variations in the cosmic ray flux which produces \textsuperscript{14}C in the earth’s atmosphere~\cite{vanderPlicht2022}. 

Table~\ref{tab:result-4q185} shows the valid and acceptable radiocarbon date of the sample received for 4Q185 but the date cannot be used (see Section~\ref{appen:B5:4Q185}).

Samples of the 3 scrolls 4Q216, 11Q20 and Mur88 did not produce acceptable \textsuperscript{14}C dates; these are summarized in Table~\ref{tab:rejected-c14} (see Section~\ref{appen:B6:reject}). 

\include{tables/TableX}

As was done in the 1990s~\cite{Bonani1992,Jull1995}, we also tested our procedure by dating a date-bearing manuscript, Mur19. The text of Mur19 refers to “year 6 of Masada”, which is now understood as a reference from the first Jewish revolt against Rome to 71/72 CE \cite{Benoit1961,koffmahn1963dating,yadin1965excavation,goodblatt1999dating,eshel2003documents,Eshel2005,Wise2015}. The 2$\sigma$ calibrated range is 45 BCE–85 CE (91.5\%), 95–110 CE (3.9\%). The \textsuperscript{14}C date is clearly consistent with the historical date, 71/72 CE.

\begin{figure}
    \centering
    \includegraphics[width=\textwidth]{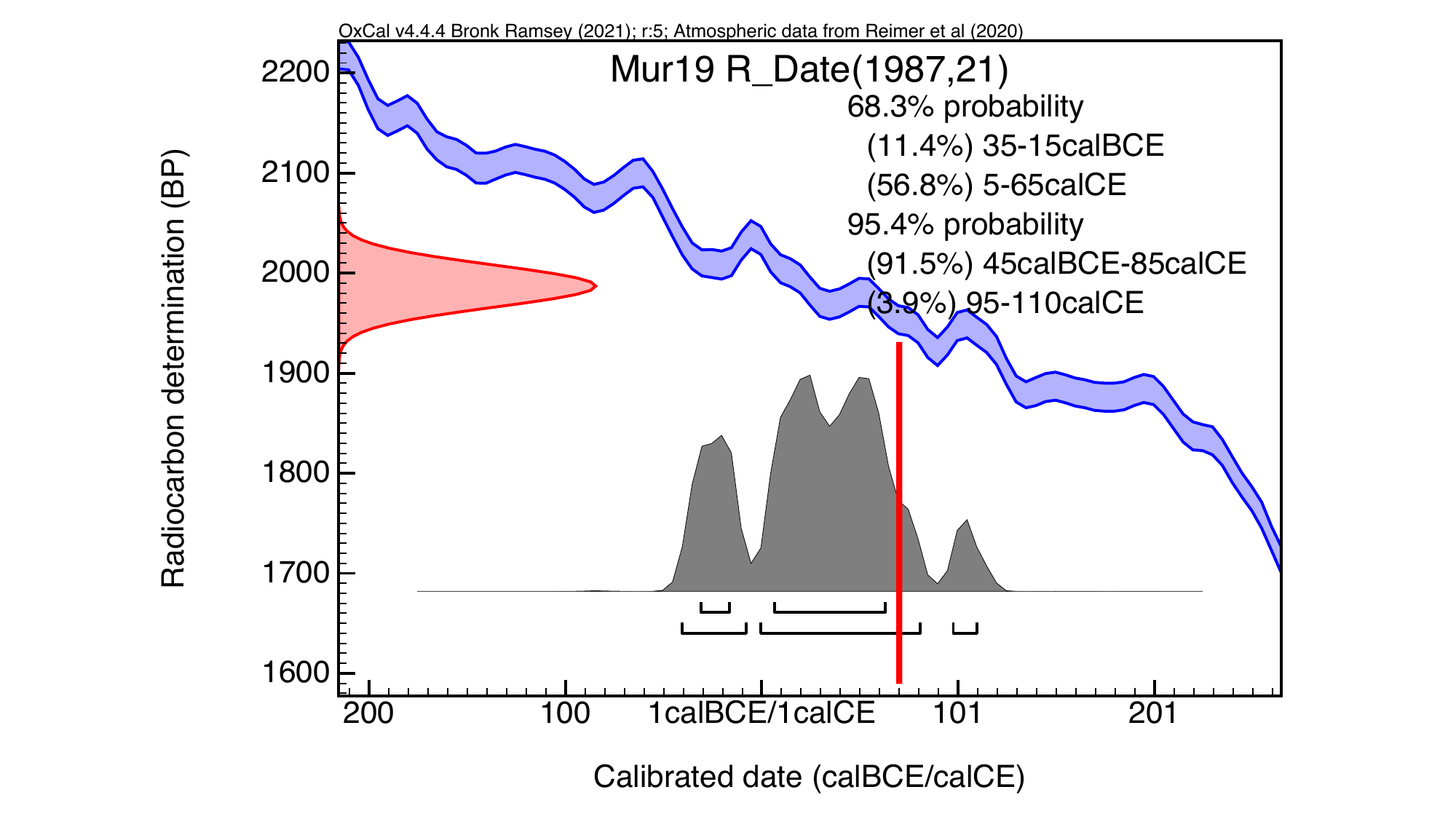}
    \caption{OxCal plot for Mur19 with \textcolor{red}{\textbf{red}} vertical line indicating the calendar date 71/72 CE.}
    \label{fig:mur19-oxcal}
\end{figure}

\subsection{Result not to be used for palaeography: 4Q185}\label{appen:B5:4Q185}
From a radiocarbon point of view, the dating of the sample is a valid and acceptable result. However, because the sample fragment cannot be attributed to a larger manuscript, the date cannot be used for our palaeographic analysis. 

For 4Q185 (4QSapiential Work), we had requested Plate 801 fragment 1. Because Plate 801 fragment 1 was sewn and encapsulated for exhibition, the IAA sent sample Plate 801 fragment 3 instead. Unfortunately, it is very uncertain that this sampled fragment is part of manuscript 4Q185. From a palaeographic perspective, identification with 4Q185 is doubtful. E.g., the letter \textit{ayin} is different from other occurrences in the manuscript (see also~\cite{Pajunen2011}). For that reason, the measurement results cannot be used for our palaeographic purposes. 

\input{tables/TableY }


\subsection{Technically rejected results: 4Q216, 11Q20, and Mur88}\label{appen:B6:reject}
The various AMS runs for scrolls 4Q216, Mur88, and 11Q20 resulted in internally inconsistent results. No valid \textsuperscript{14}C date could be deduced. Therefore, the results are rejected for technical reasons. 

For all three scrolls, different samples were received from the IAA in subsequent batches. The first samples were measured by the GrA machine, the subsequent samples were later during the project measured by the GrM machine.

For 4Q216 (4QJub\textsuperscript{a}), the first sample was measured for graphite (GrA-69799). For the second sample, two gas samples were measured (GrM-10675, 10676). The GrA and GrM measurements do not provide mutually consistent dates. In other words, both samples received from the IAA do not give consistent results. In addition, the measurements yield \textsuperscript{14}C dates which are impossibly old. We conclude that the sample material may not be homogeneous. 

For 11Q20 (11QTemple\textsuperscript{b}), the first sample was measured for graphite in triplicate (GrA-69800). For the second sample, two different parts of the scroll sample were taken, and two gas samples measured for each (GrM-10681, 10682, 18827, 18828). The 3 GrA measurements are internally consistent, the same for the 4 GrM results. However, GrA and GrM do not provide mutually consistent dates. Also here, both samples received from the IAA do not give consistent results. 
We conclude that the sample material may not be homogeneous.

For Mur88 (MurXII), the first sample was measured for graphite in triplicate (GrA-69806). For the second sample, two different parts of the scroll sample were taken, and two gas samples measured for each of them (GrM-10663, 10664, 18829, 18830). The resulting GrA and GrM measurements yield three different \textsuperscript{14}C dates. Also here, the sample material may not be homogeneous.

For the full results of these runs with more details (in particular Carbon yield and $\delta$\textsuperscript{13}C value) see Appendix~\ref{appen:G}.

\input{tables/TableZ}


\subsection{Analytical Chemistry}\label{appen:B7:analych}

\subsubsection{Soxhlet extraction}\label{appen:B7.1:Soxhlet}

Upon arrival of the samples in Odense, they were photographed, if this was not already done in Groningen. Detailing what was said in section B.2, the chemical cleaning procedure developed to remove later added contamination such as, e.g., castor oil, was the following. Three Soxhlet apparatuses were operated in parallel, with three samples mounted simultaneously one in each chamber. The Soxhlet apparatuses had different volumes: the first one operated with 100 mL of solvent, the second with 70 mL and the third with 50 mL of solvent. All solvents were of the highest quality available (LC-grade for Liquid Chromatography).

The cleaning procedure was initiated by running the whole set of solvents with no sample mounted, intended to clean the apparatus, the stainless-steel cage and glass utensils. Then a sample was placed in the stainless-steel cage mounted in a Soxhlet apparatus chamber. The first solvent was added to the lower flask. The first solvent was LC-grade ethanol LiChrosolv (1.11727.2500 from Merck). This was operated for one hour corresponding to ca. 50 turnovers of the solvent over the sample. The second solvent was LC-grade \textit{n}-hexane LiChrosolv (1.03701.2500 from Merck), which was operated for four hours, corresponding to ca. 240 turnovers of the solvent over the sample. The third solvent applied was LC-grade ethanol LiChrosolv (1.11727.2500 from Merck), operated for one hour, corresponding to ca. 50 turnovers. After each step in the cleaning procedure samples of 8 mL of each of the solvent were transferred to pre-cleaned glass vials. That is, three samples of 8 mL of ethanol, hexane, and ethanol were procured after each step in the cleaning procedure. They were placed in a heating apparatus operating at 80°C, which evaporated the solvents in the glass vials to dryness, after which the glass vials were sealed with a lid. The condensate was later to be re-dissolved and analyzed by HPLC-MS in Pisa (see section 7.2). After cleaning, the samples were removed from the stainless-steel cages and brought to dryness for one night at 60°C at zero humidity in a Memmert HCP 108 Climate chamber. Following this, the samples were weighed, packed, and shipped to Groningen, there to undergo pretreatment and dating following \textsuperscript{14}C protocols.        

This three-step Soxhlet protocol, which was developed by~\cite{rasmussen2009effects}, was applied to the first batch of 10 samples (4Q52, 4Q114, 4Q161, 4Q176, 4Q185, 4Q216, 4Q504, 11Q20, Mur88, 5/6Hev1b) which were analyzed in the project. Following the chromatographic-mass spectrometric analyses in Pisa of this first set of solvents, it was decided that a fourth cleaning step should be added to the procedure for the remaining 20 samples. This was done for redundancy, and not because of proof or suspicion that the three-step procedure was not sufficient within the given dating uncertainty. The fourth step was added to further ensure that castor oil and many other contaminants were removed even in the worst case scenario. The fourth Soxhlet step was performed using a 30:70 mixture of dichloromethane:hexane, both of LC-grade purity (dicholoromethane CHROMASOLV 34856 by Sigma-Aldrich, and \textit{n}-hexane as described above), operated for one hour, corresponding to ca. 60 turnovers of the solvent over the sample.

\subsubsection{Raman spectroscopy, optical microscopy, Py-GC/MS, and HPLC-MS analysis}\label{appen:B7.2:ramanOManalysis}
The study of the materials constituting the scrolls was performed in Pisa using a multi-analytical approach based on chromatographic and spectroscopic analytical techniques. The use of these complementary approaches allowed us to characterize both the original materials of the parchments and to evaluate the possible occurrence of modern materials used for consolidating/restoring the scrolls. These results were used to define the best cleaning strategy to remove from the scrolls the modern materials that could affect the dating, and to evaluate the efficiency of the purification steps. In detail:

\begin{itemize}
    \item \textit{Raman spectroscopy and optical microscopy} (OM) were used as non-invasive and non-destructive methods to evaluate the general appearance of the parchments and to characterize the possible occurrence of inorganic materials.
    
    \item \textit{Analytical pyrolysis coupled with gas chromatography and mass spectrometry} (Py-GC/MS) analyses were performed on small (ca. 100 µg) sub-samples of the samples before these went into cleaning treatment by Soxhlet and AAA to characterize the organic material constituting the scrolls and to evaluate the possible presence of modern synthetic materials used as consolidating materials. This technique represents one of the best methods to obtain a complete picture of the organic materials in a sample \cite{degano2018recent}. Pyrolysis consists of a thermal decomposition of organic materials in absence of oxygen. This process leads to the formation of low molecular weight species that can be separated by gas chromatography and identified by mass spectrometry. This analytical approach allows to obtain specific molecular markers that can be used to identify the source of organic materials. 

    \item \textit{Liquid chromatography coupled with mass spectrometry} (HPLC-MS) was applied to evaluate the content of lipid materials present in Soxhlet extracts from parchments during the cleaning steps. This is among the best approaches for the separation and characterization of complex mixtures of lipid materials, such as castor oil. The use of mass spectrometry as detection system allows to obtain information on the glyceride chemical structure \cite{la2021liquid}. This information cannot be achieved using more conventional analytical approaches such as GC/MS. Moreover, this method allows to detect very low amounts of analytes. 

\end{itemize}

\subsubsection{Results of the optical microscopy and Raman spectroscopy analyses performed on 17 samples}\label{appen:B7.3:ramanOMresults}

The microscopy observations and micro-Raman analyses were performed on samples 4Q2, 4Q3, 4Q27, 4Q30, 4Q114, 4Q201/4Q338, 4Q206, 4Q216, 4Q259, 4Q267, 4Q375, 4Q416, 4Q521, 4Q541, Mas1k, Mur19, XHev/Se2. All these samples were characterized by similar appearance, except for sample Mur19 that showed a different morphology, suggesting the use of a different material as a writing support. 

\begin{figure}[ht]
    \centering
    \includegraphics[width=\textwidth]{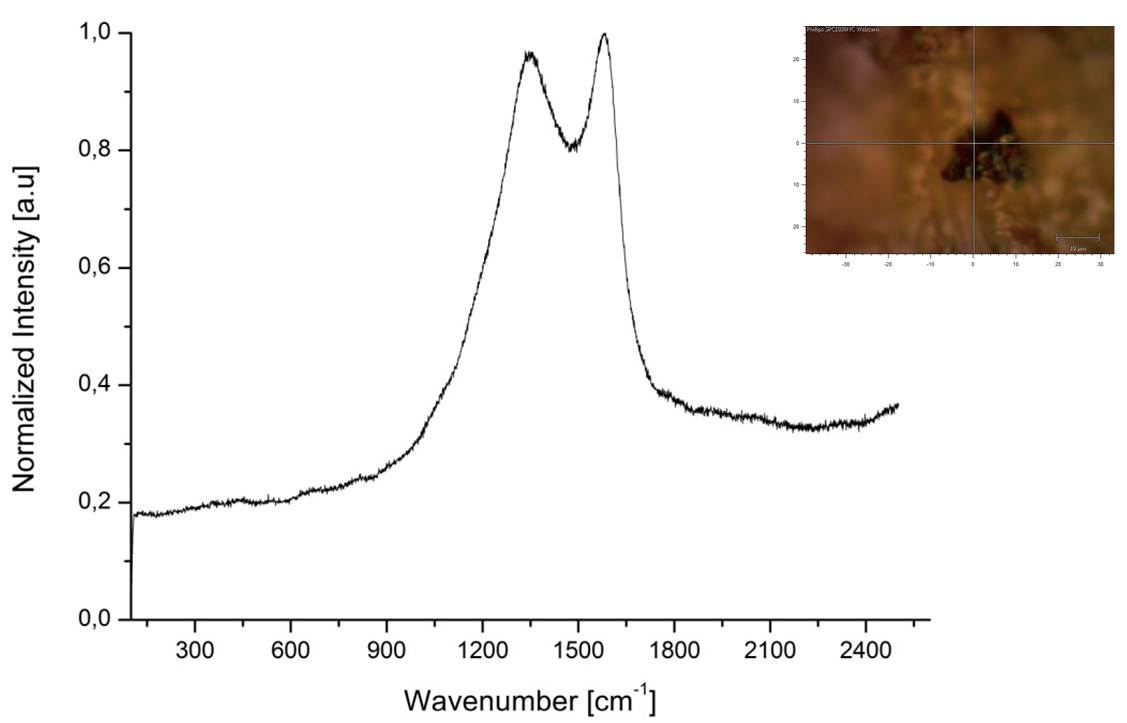}
    \caption{Raman spectrum obtained for one spot in sample 4Q216}
    \label{fig:b7-raman}
\end{figure}

Several samples featured microscopic black spots with diameters in the range of 10-200 µm, except for sample 4Q114 that was characterized by one black spot of approximately 600 µm. Raman spectroscopy was applied to investigate the chemical composition of the spots. For several samples, the Raman spectra featured the typical peaks at 1350 and 1580 cm${}^{-1}$ corresponding to the Raman wavenumbers typical of C-C of amorphous carbon (signals not detected in the background). For example, Figure \ref{fig:b7-raman} reports the spectrum obtained from one spot on the sample 4Q216, and Table~\ref{tab:B7XX} presents the OM photographs along with a description of the observed surface and summarizes the relevant information obtained by Raman spectroscopy.

The biggest black spot from the scroll 4Q114 was sampled separately, and radiocarbon dated to 2390±60 BP (GrM-13256). The size of the black spot was ca. 600 µm in diameter, with an observed thickness of ca. 50 µm, translating into a calculated mass of ca. 14 µg. Thus, with a sample mass of 6.2 mg for the sample radiocarbon dated for 4Q114, the contamination mass fraction from the black spot would be ca. 0.2\% and thus, the effect of such contamination is negligible, whatever its age.

\input{tables/TableB7-XX}

\subsubsection{Results of the Py-GC/MS analysis performed on 17 samples}\label{appen:B7.4:PyGCMSresults}

Py-GC/MS was used in order to evaluate the possible presence of synthetic materials used as consolidating materials on the scrolls and to characterize the original parchment material: the 17 samples (4Q2, 4Q3, 4Q27, 4Q30, 4Q114, 4Q201/4Q338, 4Q206, 4Q216, 4Q259, 4Q267, 4Q375, 4Q416, 4Q521, 4Q541, Mas1k, Mur19, XHev/Se2) were directly analyzed without any prior sample pretreatment using a multi-shot pyrolyzer EGA/PY-3030D (Frontier Lab, Japan) coupled with a 6890 N gas chromatography system with a split/splitless injection port, and with a 5973 mass selective single quadrupole mass spectrometer (Agilent Technologies). The complete instrumental conditions are reported in \cite{la2019synthetic}.

\begin{figure}[ht!]
    \centering
    \includegraphics[width=\textwidth]{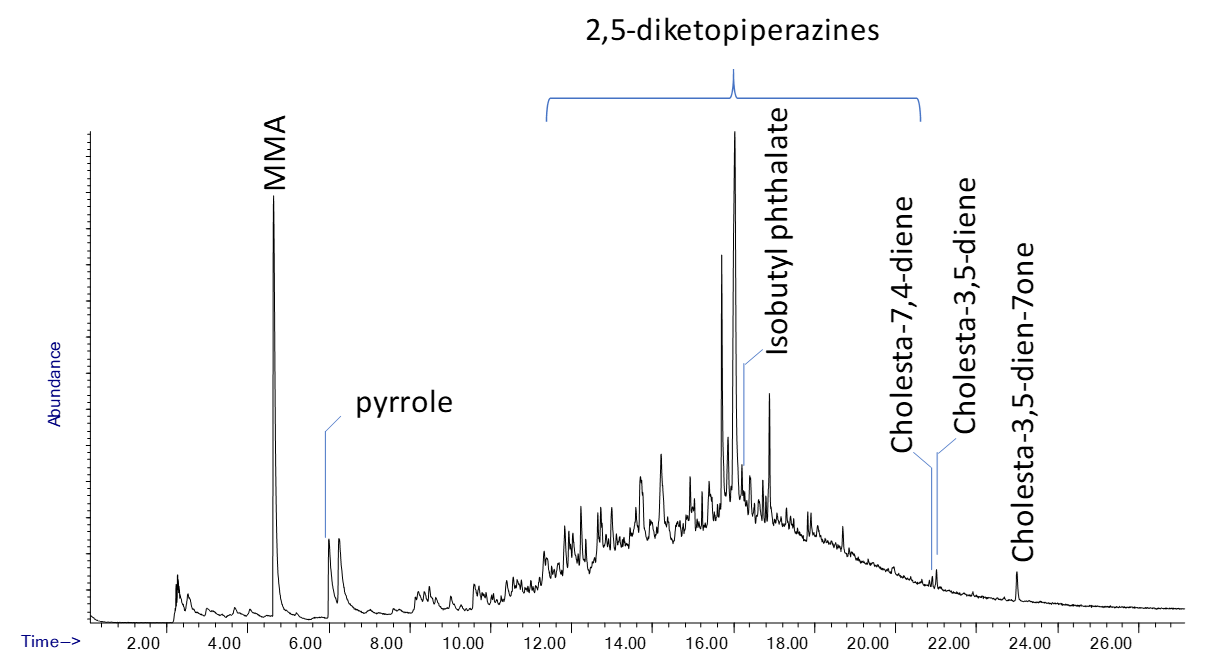}
    \caption{Py-GC-MS chromatogram of sample 4Q521: parchment sample with methyl-methacrylate, an acrylic resin.}
    \label{fig:B7-Py}
\end{figure}

The pyrolytic profile of all these 17 samples featured molecular markers that can be related to the pyrolysis of animal hide or scroll (pyrrole and diketopiperazines), except for sample Mur19 that was instead characterized by the presence of anhydro sugars and levoglucosan, typical of a cellulose-based material \cite{colombini2009organic}. This is consistent with the observation that Mur19 is a papyrus fragment. Figure \ref{fig:B7-Py} reports the chromatogram obtained for the sample from the parchment of 4Q521.

Samples 4Q3 and 4Q206 showed the presence of the markers of polyethylene glycol. The pyrograms of samples 4Q3 and 4Q30 also contain the peaks due to hexadecanonitrile and octadecanonitrile, which are the Py-GC/MS markers characteristic for egg. Samples 4Q521 and Mur19 were characterized by the presence of an acrylic resin. Finally, samples 4Q2, 4Q267, 4Q541, and Mur19 showed the presence of retene: this molecule is a marker characteristic of the combustion of resinous wood and can be indicative of the exposure of the scrolls to a fire in the space where writing took place or could be due to residues related to the illumination with torches. Table \ref{tab:B7-summarize} summarizes the materials detected in the different parchment samples.  

Pyrolysis allowed us to pinpoint the presence of exogenous materials, as consolidation synthetic materials (acrylic resin), or lipids. After disclosing the nature of the contamination, we were able to design the proper cleaning procedures to remove any unwanted consolidant.

The use of a further cleaning step using dichloromethane ensured the total removal of all the synthetic materials, as proven by pyrolysis analyses performed on a subsection of the samples after cleaning and prior to \textsuperscript{14}C dating.

\input{tables/TableB7-X}

\subsubsection{Liquid chromatography-mass spectrometry results of the analysis of residual lipids in the extracts from the 30 samples after cleaning}\label{appen:B7.5:HPLCMSresults}

HPLC-MS was applied to evaluate the presence of lipid materials. The dried extracts were reconstituted in 150 µL of iso-propanol/methanol, 10:90, filtered (PTFE syringe, 0.45 µm pore size) and analyzed. HPLC-ESI-Q-ToF analyses were carried out using a 1200 Infinity HPLC, coupled with a Quadrupole-Time of Flight tandem mass spectrometer 6530 Infinity Q-ToF detector by a Jet Stream ESI interface (Agilent Technologies, USA). The complete instrumental conditions are reported in~\cite{la2013core}.

The analyses were performed on the extracts from the two different sample pretreatments by Soxhlet, i.e., the three-step and the four-step extraction.

\begin{figure}[ht!]
    \centering
    \includegraphics[width=\textwidth]{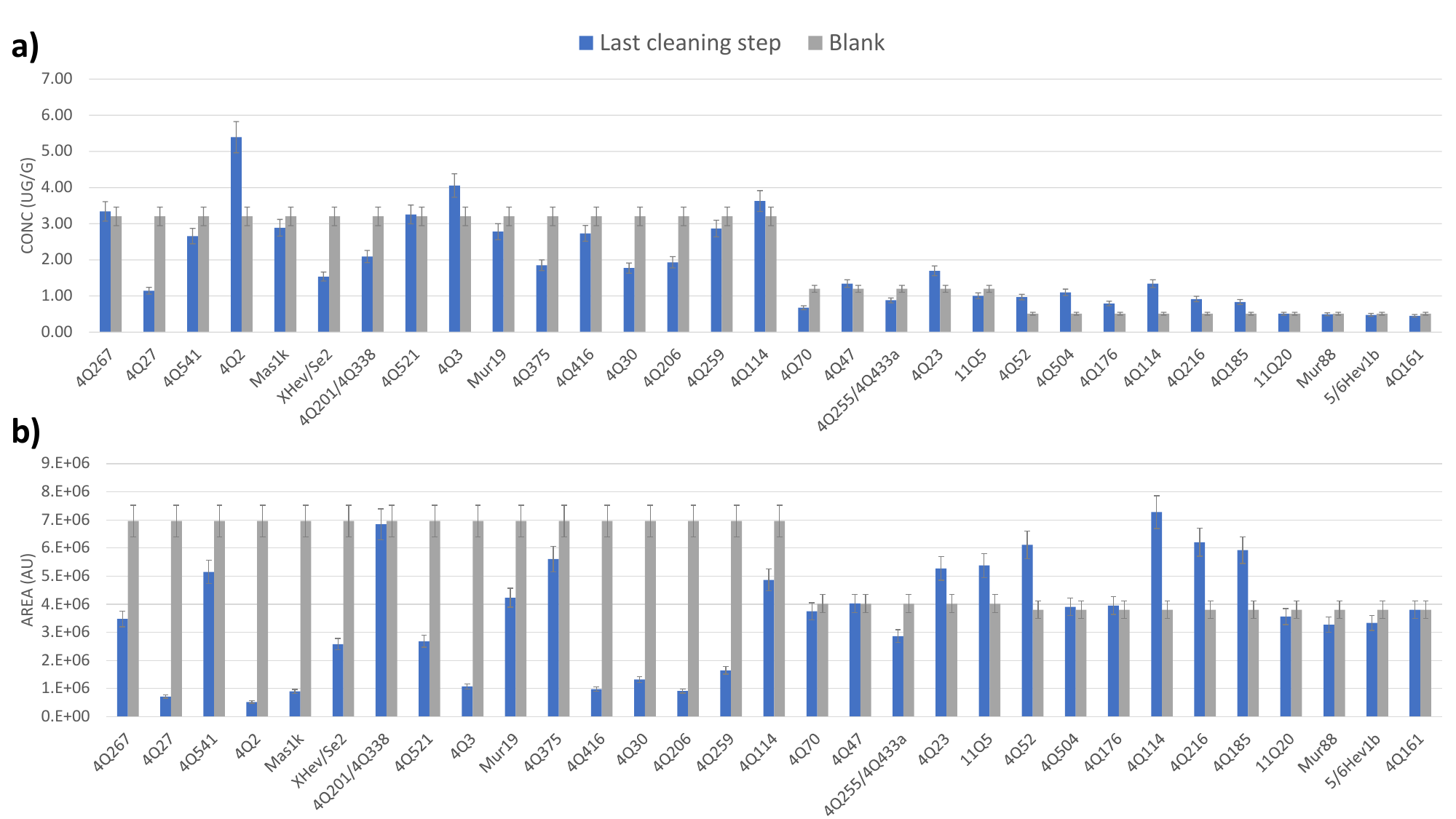}
    \caption{\textit{Top (a)}: Comparison of the free fatty acid concentrations between the blank samples and the final cleaning steps; \textit{Bottom (b)}: comparison of the abundances of TAGs (triacylglycerols) in the last cleaning step with those found in the blank samples (AU: arbitrary unit).}
    \label{fig:B75-ffa-tag}
\end{figure}

The comparison of the results obtained on the extracts with reference blanks allowed us to highlight the effective performances of the cleaning procedures, showing that the glyceride content after the last step was below 7.0 micrograms for both the approaches. The cleaning procedure proved to be effective for removing the lipid materials from the scroll samples, since all the solutions obtained after the last extraction step were characterized by the presence of triglycerides and fatty acids at or below blank level. Figure~\ref{fig:B75-ffa-tag} shows a comparison of all final cleaning steps with the respective blanks for both the fatty acids and the triacylglycerols. 

In particular, the worst case encountered in the entire data set was 7.0 µg of acylglycerols detected in the fourth cleaning step of 4Q3. These triglycerides can originate from the original parchment, or they can originate from later contamination such as castor oil. There is no way to determine the origin; it can also be a mixture of ancient and recent materials. If we, as a worst-case scenario, assume that all the triglycerides detected in 4Q3 were modern contamination, then it would skew a 2000-years old parchment sample with only 12.6 years. 

As stipulated, this is a worst-case scenario depending on all triglycerides to be modern, which is an unlikely assumption because triglycerides are a normal ingredient of animal skin \cite{ghioni2005evidence}. Furthermore, all other samples are well below the 7.0 µg level.

%% file: tables/TableX.tex
\footnotesize
\LTcapwidth=\textwidth
{\tabcolsep=3.3pt
\begin{longtable}{l|l|l|l|l|l}

\caption{Summarized results of 26 accepted \textsuperscript{14}C dates: laboratory code, sample identification, \textsuperscript{14}C age (BP), sigma (BP), calibrated ranges (1$\sigma$ and 2$\sigma$ ranges) in 5-year resolution.}
\label{tab:summarized-c14}\\

\hline\hline 
\multirow{2}{*}{\textbf{lab code}} &
  \multirow{2}{*}{\textbf{scroll}} &
  \multirow{2}{*}{\begin{tabular}[c]{@{}l@{}}\textbf{age}\\ \textbf{(BP)}\end{tabular}} &
  \multirow{2}{*}{\textbf{$\sigma$}} &
  \multirow{2}{*}{\textbf{calibrated ranges (1$\sigma$)}} &
  \multirow{2}{*}{\textbf{calibrated ranges (2$\sigma$)}} \\
 &
   &
   &
   &
   &
   \\ \hline
GrA-68446 &
  P421-Fr004 &
  \multirow{3}{*}{2164} &
  \multirow{3}{*}{16} &
  \multirow{3}{*}{\begin{tabular}[c]{@{}l@{}}345–320, \\ 205–170 BCE\end{tabular}} &
  \multirow{3}{*}{\begin{tabular}[c]{@{}l@{}}355–285, \\ 230–150 BCE \end{tabular}} \\ 
GrA-68447 &
  \begin{tabular}[c]{@{}l@{}}4Q504 \\ (4QDibHam\textsuperscript{a})\end{tabular} &
   &
   &
   &
   \\ \hline
GrA-69793 &
  P206-Fr003 &
  \multirow{3}{*}{2303} &
  \multirow{3}{*}{26} &
  \multirow{3}{*}{405–365 BCE} &
  \multirow{3}{*}{\begin{tabular}[c]{@{}l@{}}410–355, \\ 285–230 BCE\end{tabular}} \\
GrM-10677 &
  4Q52 (4QSam\textsuperscript{b}) &
   &
   &
   &
   \\
GrM-10678 &
   &
   &
   &
   &
   \\ \hline
GrA-69794 &
  P285-Fr002 &
  \multirow{3}{*}{2153} &
  \multirow{3}{*}{19} &
  \multirow{3}{*}{\begin{tabular}[c]{@{}l@{}}345–320, \\ 205–165 BCE\end{tabular}} &
  \multirow{3}{*}{\begin{tabular}[c]{@{}l@{}}355–300, 210–100, \\ 70–60 BCE\end{tabular}} \\
GrM-10679 &
  4Q176 (4QTanh) &
   &
   &
   &
   \\
GrM-10680 &
   &
   &
   &
   &
   \\ \hline
GrA-69795 &
  P224-Fr001 &
  \multirow{5}{*}{2168} &
  \multirow{5}{*}{15} &
  \multirow{5}{*}{\begin{tabular}[c]{@{}l@{}}345–315, \\ 205–175 BCE\end{tabular}} &
  \multirow{5}{*}{\begin{tabular}[c]{@{}l@{}}355–285, \\ 230–160 BCE\end{tabular}} \\
GrM-13252 &
  4Q114 (4QDan\textsuperscript{c}) &
   &
   &
   &
   \\
GrM-13253 &
   &
   &
   &
   &
   \\
GrM-13254 &
   &
   &
   &
   &
   \\
GrM-13255 &
   &
   &
   &
   &
   \\ \hline
GrM-10659 &
  P891-Fr003 &
  \multirow{2}{*}{1940} &
  \multirow{2}{*}{28} &
  \multirow{2}{*}{\begin{tabular}[c]{@{}l@{}}25–45,\\  55–125 CE\end{tabular}} &
  \multirow{2}{*}{10–205 CE} \\
GrM-10660 &
  5/6Hev1b (Ps) &
   &
   &
   &
   \\ \hline
GrA-69810 &
  P585-Fr001 &
  \multirow{3}{*}{2028} &
  \multirow{3}{*}{18} &
  \multirow{3}{*}{45 BCE–10 CE} &
  \multirow{3}{*}{\begin{tabular}[c]{@{}l@{}}90–80 BCE, \\ 55 BCE–30 CE, \\ 45–60 CE\end{tabular}} \\
GrM-10661 &
  4Q161 (4QpIsa\textsuperscript{a}) &
   &
   &
   &
   \\
GrM-10662 &
   &
   &
   &
   &
   \\ \hline
GrM-11151 &
  P1111-Fr010 &
  \multirow{4}{*}{2226} &
  \multirow{4}{*}{17} &
  \multirow{4}{*}{\begin{tabular}[c]{@{}l@{}}365–350, \\ 295–205 BCE\end{tabular}} &
  \multirow{4}{*}{\begin{tabular}[c]{@{}l@{}}375–345, \\ 320–200 BCE\end{tabular}} \\
GrM-11152 &
  4Q70 (4QJer\textsuperscript{a})\footnotemark{} &
   &
   &
   &
   \\
GrM-11170 &
   &
   &
   &
   &
   \\
GrM-11171 &
   &
   &
   &
   &
   \\ \hline
GrM-11153 &
  P1093-Fr005 &
  \multirow{3}{*}{2155} &
  \multirow{3}{*}{19} &
  \multirow{3}{*}{\begin{tabular}[c]{@{}l@{}}345–320, \\ 200–165 BCE\end{tabular}} &
  \multirow{3}{*}{\begin{tabular}[c]{@{}l@{}}355–290, \\ 210–100 BCE\end{tabular}} \\
GrM-11154 &
  4Q47 (4QJosh\textsuperscript{a}) &
   &
   &
   &
   \\
GrM-11172 &
   &
   &
   &
   &
   \\ \hline
GrM-11155 &
  P271-Fr002 &
  \multirow{3}{*}{2152} &
  \multirow{3}{*}{24} &
  \multirow{3}{*}{\begin{tabular}[c]{@{}l@{}}350–315, \\205–150, \\ 130–120 BCE\end{tabular}} &
  \multirow{3}{*}{\begin{tabular}[c]{@{}l@{}}355–285, 230–220, \\ 210–95, 75–55 BCE\end{tabular}} \\
GrM-11156 &
  4Q23 (4QLevNum\textsuperscript{a}) &
   &
   &
   & \\
   &
   &
   &
   &
   &
   \\ \hline
GrM-11166 &
  P177-Fr001 &
  \multirow{6}{*}{2100} &
  \multirow{6}{*}{17} &
  \multirow{6}{*}{\begin{tabular}[c]{@{}l@{}}155–90, \\75–55 BCE\end{tabular}} &
  \multirow{6}{*}{170–50 BCE} \\
GrM-11167 &
  \begin{tabular}[c]{@{}l@{}}4Q255/4Q433a \\ (4QpapS\textsuperscript{a}/4Qpap\\ Hodayot-like Text B)\end{tabular} &
   &
   &
   &
   \\
GrM-11184 &
   &
   &
   &
   &
   \\
GrM-11185 &
   &
   &
   &
   &
   \\ \hline
GrM-11168 &
  P977-Fr004 &
  \multirow{4}{*}{1967} &
  \multirow{4}{*}{18} &
  \multirow{4}{*}{\begin{tabular}[c]{@{}l@{}}20–80, \\ 100–110 CE\end{tabular}} &
  \multirow{4}{*}{\begin{tabular}[c]{@{}l@{}}35–15 BCE, \\ 5–120 CE\end{tabular}} \\
GrM-11169 &
  11Q5 (11QPs\textsuperscript{a}) &
   &
   &
   &
   \\
GrM-11186 &
   &
   &
   &
   &
   \\
GrM-11187 &
   &
   &
   &
   &
   \\ \hline
GrM-14380 &
  P393-Fr005 &
  \multirow{4}{*}{2123} &
  \multirow{4}{*}{21} &
  \multirow{4}{*}{\begin{tabular}[c]{@{}l@{}}175–100, \\ 70–60 BCE\end{tabular}} &
  \multirow{4}{*}{\begin{tabular}[c]{@{}l@{}}340–325, \\ 200–50 BCE\end{tabular}} \\
GrM-14381 &
  4Q3 (4QGen\textsuperscript{c}) &
   &
   &
   &
   \\
GrM-14228 &
   &
   &
   &
   &
   \\
GrM-14229 &
   &
   &
   &
   &
   \\ \hline
GrM-13385 &
  P1081a-Fr002 &
  \multirow{2}{*}{2115} &
  \multirow{2}{*}{26} &
  \multirow{2}{*}{\begin{tabular}[c]{@{}l@{}}175–95, \\ 75–55 BCE\end{tabular} } &
  \multirow{2}{*}{\begin{tabular}[c]{@{}l@{}}340–330, \\ 200–50 BCE\end{tabular}} \\
GrM-13386 &
  4Q27 (4QNum\textsuperscript{b}) &
   &
   &
   &
   \\ \hline
GrM-13387 &
  Px232-Fr001 &
  \multirow{4}{*}{2007} &
  \multirow{4}{*}{18} &
  \multirow{4}{*}{\begin{tabular}[c]{@{}l@{}}45 BCE–25 CE\end{tabular}} &
  \multirow{4}{*}{50 BCE–65 CE} \\
GrM-13388 &
  Mas1k (MasShirShabb) &
   &
   &
   &
   \\
GrM-14175 &
   &
   &
   &
   &
   \\
GrM-14223 &
   &
   &
   &
   &
   \\ \hline
GrM-14382 &
  P386-Fr001 &
  \multirow{4}{*}{2169} &
  \multirow{4}{*}{21} &
  \multirow{4}{*}{\begin{tabular}[c]{@{}l@{}}350–310, \\ 210–170 BCE\end{tabular}} &
  \multirow{4}{*}{\begin{tabular}[c]{@{}l@{}}360–280, 235–145, \\ 135–120 BCE\end{tabular}} \\
GrM-14383 &
  4Q206 (4QEn\textsuperscript{e} ar) &
   &
   &
   &
   \\
GrM-14230 &
   &
   &
   &
   &
   \\
GrM-14241 &
   &
   &
   &
   &
   \\\hline
GrM-14565 &
  P237-Fr007 &
  \multirow{5}{*}{2182} &
  \multirow{5}{*}{18} &
  \multirow{5}{*}{\begin{tabular}[c]{@{}l@{}}355–290 \\ 210–175 BCE\end{tabular}} &
  \multirow{5}{*}{\begin{tabular}[c]{@{}l@{}}360–275, 260–245,  \\ 235–165 BCE\end{tabular}} \\
GrM-14566 &
  \begin{tabular}[c]{@{}l@{}}4Q30 (4QDeut\textsuperscript{c})\end{tabular} &
   &
   &
   &
   \\
GrM-14395 &
   &
   &
   &
   &
   \\
GrM-14242 &
   &
   &
   &
   &
   \\
GrM-14243 &
   &
   &
   &
   &
   \\ \hline
GrM-13389 &
  P904-Fr009 &
  \multirow{4}{*}{2077} &
  \multirow{4}{*}{18} &
  \multirow{4}{*}{110–45 BCE} &
  \multirow{4}{*}{165–40, 10–1 BCE} \\
GrM-13390 &
  \begin{tabular}[c]{@{}l@{}}4Q201/4Q338 \\ (4QEn\textsuperscript{a} ar/\\ 4QGenealogical List)\end{tabular} &
   &
   &
   &
   \\
GrM-14173 &
   &
   &
   &
   &
   \\
GrM-14174 &
   &
   &
   &
   &
   \\ \hline
GrM-14396 &
  P810-Fr011 &
  \multirow{4}{*}{2148} &
  \multirow{4}{*}{19} &
  \multirow{4}{*}{\begin{tabular}[c]{@{}l@{}}345–320, \\ 205–150 BCE\end{tabular}} &
  \multirow{4}{*}{\begin{tabular}[c]{@{}l@{}}350–310, 210–100,\\  70–55 BCE\end{tabular}} \\
GrM-14397 &
  4Q259 (4QS\textsuperscript{e}) &
   &
   &
   &
   \\
GrM-14244 &
   &
   &
   &
   &
   \\
GrM-14245 &
   &
   &
   &
   &
   \\ \hline
GrM-14398 &
  P180-Fr004 &
  \multirow{5}{*}{2130} &
  \multirow{5}{*}{22} &
  \multirow{5}{*}{\begin{tabular}[c]{@{}l@{}}200–100 BCE\end{tabular}} &
  \multirow{5}{*}{\begin{tabular}[c]{@{}l@{}}345–320, 205–90,\\  80–50 BCE\end{tabular}} \\
GrM-14399 &
    \begin{tabular}[c]{@{}l@{}}4Q416 \\(4QInstruction\textsuperscript{b})\end{tabular} &
   &
   &
   &
   \\
GrM-14246 &
   &
   &
   &
   &
   \\
GrM-14359 &
   &
   &
   &
   &
   \\ \hline
GrM-14400 &
  P215-Fr004 &
  \multirow{4}{*}{2059} &
  \multirow{4}{*}{20} &
  \multirow{4}{*}{\begin{tabular}[c]{@{}l@{}}100–70, \\ 60–35, \\ 15 BCE–5 CE\end{tabular}} &
  \multirow{4}{*}{\begin{tabular}[c]{@{}l@{}}155–130 BCE, \\ 125 BCE–10 CE\end{tabular}} \\
GrM-14401 &
  4Q2 (4QGen\textsuperscript{b}) &
   &
   &
   &
   \\
GrM-14360 &
   &
   &
   &
   &
   \\
GrM-14361 &
   &
   &
   &
   &
   \\ \hline
GrM-14567 &
  P122A-Fr001 &
  \multirow{4}{*}{2126} &
  \multirow{4}{*}{23} &
  \multirow{4}{*}{\begin{tabular}[c]{@{}l@{}}195–185, \\180–100, \\ 70–60 BCE\end{tabular}} &
  \multirow{4}{*}{\begin{tabular}[c]{@{}l@{}}345–320, 205–50 BCE\end{tabular}} \\
GrM-14568 &
  \begin{tabular}[c]{@{}l@{}}4Q375 \\(4QapocrMoses\textsuperscript{a})\end{tabular} &
   &
   &
   &
   \\
GrM-14362 &
   &
   &
   &
   &
   \\
GrM-14363 &
   &
   &
   &
   &
   \\ \hline
GrM-13391 &
  P534-Fr002 &
  \multirow{5}{*}{1998} &
  \multirow{5}{*}{20} &
  \multirow{5}{*}{\begin{tabular}[c]{@{}l@{}}40–10 BCE, \\1–30, \\ 40–60 CE\end{tabular}} &
  \multirow{5}{*}{45 BCE–75 CE} \\
GrM-13392 &
  \begin{tabular}[c]{@{}l@{}}XHev/Se2 \\ (XHev/Se Num\textsuperscript{a})\end{tabular} &
   &
   &
   &
   \\
GrM-14224 &
   &
   &
   &
   &
   \\
GrM-14225 &
   &
   &
   &
   &
   \\ \hline
GrM-14569 &
  P147-Fr019 &
  \multirow{5}{*}{2148} &
  \multirow{5}{*}{22} &
  \multirow{5}{*}{\begin{tabular}[c]{@{}l@{}}345–320, \\ 205–150 BCE\end{tabular}} &
  \multirow{5}{*}{\begin{tabular}[c]{@{}l@{}}355–300, 210–95, \\ 75–55 BCE\end{tabular}} \\
GrM-14570 &
    \begin{tabular}[c]{@{}l@{}}4Q541 \\(4QapocrLevi\textsuperscript{b})\end{tabular}&
   &
   &
   &
   \\
GrM-14364 &
   &
   &
   &
   &
   \\
GrM-14365 &
   &
   &
   &
   &
   \\ \hline
GrM-14571 &
  P330-Fr004 &
  \multirow{6}{*}{2159} &
  \multirow{6}{*}{22} &
  \multirow{6}{*}{\begin{tabular}[c]{@{}l@{}}350–315, \\ 205–165 BCE\end{tabular}} &
  \multirow{6}{*}{\begin{tabular}[c]{@{}l@{}}355–285, 230–100 BCE\end{tabular}} \\
GrM-14572 &
  \begin{tabular}[c]{@{}l@{}}4Q521 \\ (4QMessianic \\ Apocalypse)\end{tabular} &
   &
   &
   &
   \\
GrM-14377 &
   &
   &
   &
   &
   \\
GrM-14366 &
   &
   &
   &
   &
   \\ \hline
GrM-13393 &
  P107-Fr010 &
  \multirow{5}{*}{2151} &
  \multirow{5}{*}{21} &
  \multirow{5}{*}{\begin{tabular}[c]{@{}l@{}}345–315, \\ 205–155 BCE\end{tabular}} &
  \multirow{5}{*}{\begin{tabular}[c]{@{}l@{}}355–290, 210–95, \\ 70–55 BCE\end{tabular}} \\
GrM-13394 &
\begin{tabular}[c]{@{}l@{}}4Q267 \\(4QDamascus\textsuperscript{b})\end{tabular} &
   &
   &
   &
   \\
GrM-14226 &
   &
   &
   &
   &
   \\
GrM-14227 &
   &
   &
   &
   &
   \\ \hline
GrM-14573 &
  P879-Fr001 &
  \multirow{4}{*}{1987} &
  \multirow{4}{*}{21} &
  \multirow{4}{*}{\begin{tabular}[c]{@{}l@{}}35–15 BCE, \\ 5–65 CE\end{tabular}} &
  \multirow{4}{*}{\begin{tabular}[c]{@{}l@{}}45 BCE–85 CE, \\ 95–110 CE\end{tabular}} \\
GrM-14574 &
   \begin{tabular}[c]{@{}l@{}}Mur19 pap WrDiv\end{tabular}&
   &
   &
   &
   \\
GrM-14378 &
   &
   &
   &
   &
   \\
GrM-14379 &
   &
   &
   &
   &
   \\ \hline \hline

\end{longtable}}
\normalsize
\footnotetext{This fragment was previously unidentified, but see now for a positive identification~\cite{Tigchelaar2020B}.}

%% file: tables/TableY.tex
{\tabcolsep=3pt
\begin{table}[ht]
\caption{Result not to be used for palaeography: 4Q185, laboratory code, sample identification, 14C age (BP), sigma (BP), calibrated dates (1$\sigma$ and 2$\sigma$ ranges)}
\label{tab:result-4q185}
\resizebox{\textwidth}{!}{%
\begin{tabular}{l|l|l|l|l|l}
\hline \hline
\multirow{2}{*}{\textbf{lab code}} &
  \multirow{2}{*}{\textbf{scroll}} &
  \multirow{2}{*}{\textbf{age (BP)}} &
  \multirow{2}{*}{\textbf{$\sigma$}} &
  \multirow{2}{*}{\textbf{calibrated   date (1$\sigma$)}} &
  \multirow{2}{*}{\textbf{calibrated   date (2$\sigma$)}} \\
          &                           &                       &                     &                             &                                  \\ \hline
GrA-68448 & P801-Fr003                & \multirow{3}{*}{2078} & \multirow{3}{*}{17} & \multirow{3}{*}{107–46 BCE} & \multirow{3}{*}{159–42, 7–5 BCE} \\
GrA-68449 & \begin{tabular}[c]{@{}l@{}}4Q185 \\(4QSapiential Work)\end{tabular} &                       &                     &                             &                                  \\ \hline \hline
\end{tabular}}
\end{table}}

%% file: tables/TableZ.tex
\begin{table}[ht]
\centering
\caption{Technically rejected results: 4Q216, 11Q20, and Mur88, laboratory code, sample identification, 14C age (BP), sigma (BP)}
\label{tab:rejected-c14}
\begin{tabular}{l|l|l|l}
\hline \hline
\textbf{lab code} &
  \textbf{scroll} &
  \textbf{age   (BP)} &
  \textbf{$\sigma$} \\ \hline
GrA-69799 &
  \begin{tabular}[c]{@{}l@{}}P385–Fr011  \\ 4Q216 (4QJub\textsuperscript{a})\end{tabular} &
  2342 &
  51 \\ \hline
\begin{tabular}[c]{@{}l@{}}GrM-10675\\ GrM-10676\end{tabular} &
  \begin{tabular}[c]{@{}l@{}}P385–Fr011\\ 4Q216 (4QJub\textsuperscript{a})\end{tabular} &
  2979 &
  32 \\ \hline
GrA-69800 &
  \begin{tabular}[c]{@{}l@{}}P577-Fr014\\ 11Q20 (11QTemple\textsuperscript{b})\end{tabular} &
  2027 &
  24 \\ \hline
\begin{tabular}[c]{@{}l@{}}GrM-10681   \\ GrM-10682\end{tabular} & \begin{tabular}[c]{@{}l@{}}P577-Fr014\\ 11Q20 (11QTemple\textsuperscript{b})\end{tabular}    & 2183 & 32 \\ \hline
\begin{tabular}[c]{@{}l@{}}GrM-18827   \\ GrM-18828\end{tabular} & \begin{tabular}[c]{@{}l@{}}P577-Fr014   \\ 11Q20 (11QTemple\textsuperscript{b})\end{tabular} & 2202 & 26 \\ \hline
GrA-69806 &
  \begin{tabular}[c]{@{}l@{}}P64-Fr001\\ Mur88 (MurXII)\end{tabular} &
  1950 &
  18 \\ \hline
\begin{tabular}[c]{@{}l@{}}GrM-10663  \\ GrM-10664\end{tabular} &
  \begin{tabular}[c]{@{}l@{}}P64-Fr001\\ Mur88 (MurXII)\end{tabular} &
  1951 &
  30 \\ \hline
\begin{tabular}[c]{@{}l@{}}GrM-18829   \\ GrM-18830\end{tabular} &
  \begin{tabular}[c]{@{}l@{}}P64-Fr001\\ Mur88 (MurXII)\end{tabular} &
  2053 &
  25 \\ \hline \hline
\end{tabular}
\end{table}

%% file: tables/TableB7-XX.tex
{\tabcolsep=3pt
\newcommand{\customstrut}{\rule{0pt}{50pt}\rule[-10.5ex]{0pt}{0pt}}
\begin{longtable}[c]{|l|c|c|m{3cm}|}
\caption{Optical microscope pictures and observations}
\label{tab:B7XX}\\
\hline \hline
\textbf{Sample} &
  \textbf{MO side A} &
  \textbf{MO side B} &
  \textbf{Observation} \\ \hline
\endfirsthead
\endhead
\textbf{4Q3}& 
  \begin{adjustbox}
  {valign=c}\includegraphics[width=0.25\textwidth]{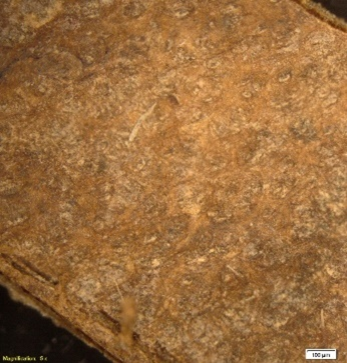}\end{adjustbox} & \customstrut
  \begin{adjustbox}{valign=c}\includegraphics[width=0.25\textwidth]{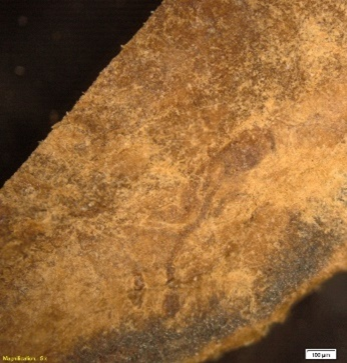}\end{adjustbox} &
  \raggedright Only one detectable black spot. Carbon was identified by Raman analysis. \\ \hline
\textbf{4Q27} &
  \begin{adjustbox}{valign=c}\includegraphics[width=0.25\textwidth]{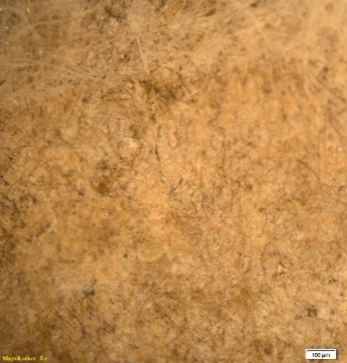}\end{adjustbox} & \customstrut
  \begin{adjustbox}{valign=c}\includegraphics[width=0.25\textwidth]{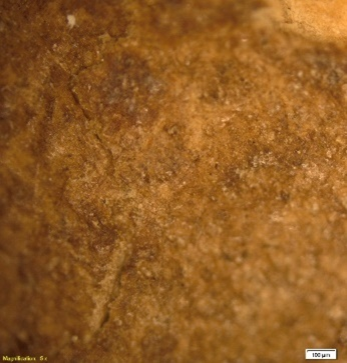}\end{adjustbox} &
  \raggedright Only one detectable black spot. Carbon was identified by Raman analysis. \\ \hline
\textbf{Mas1k} &
  \begin{adjustbox}{valign=c}\includegraphics[width=0.25\textwidth]{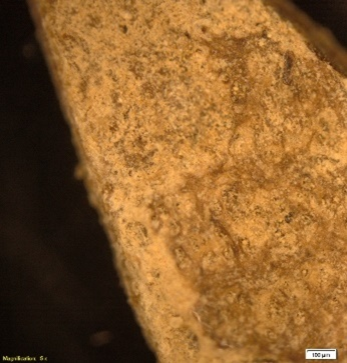}\end{adjustbox} & \customstrut
  \begin{adjustbox}{valign=c}\includegraphics[width=0.25\textwidth]{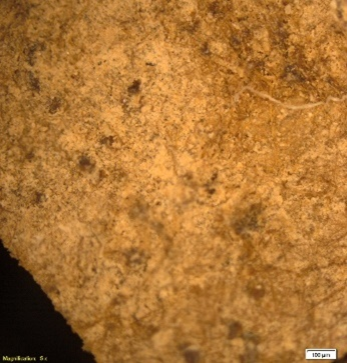}\end{adjustbox} &
  \raggedright The surface of the sample was characterized by high fluorescence and few dark spots. One spot was identified as carbon by Raman analysis. \\ \hline
\textbf{4Q206} &
  \begin{adjustbox}{valign=c}\includegraphics[width=0.25\textwidth]{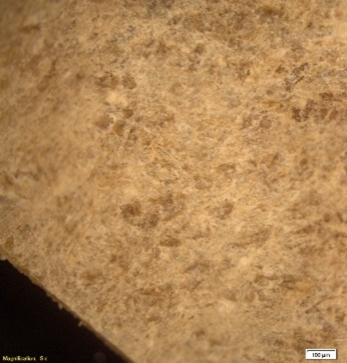}\end{adjustbox} & \customstrut
  \begin{adjustbox}{valign=c}\includegraphics[width=0.25\textwidth]{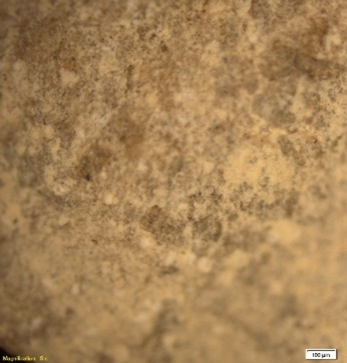}\end{adjustbox} & 
  \raggedright The surface was characterized by the presence of few black spots and few red spots: Raman analysis revealed the presence of carbon and hematite. \\ \hline
\textbf{4Q30} &
  \begin{adjustbox}{valign=c}\includegraphics[width=0.25\textwidth]{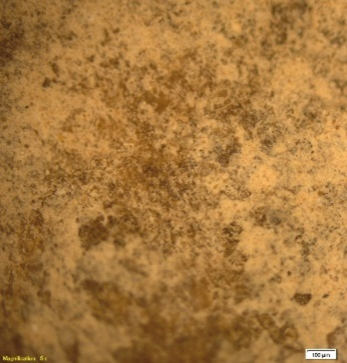}\end{adjustbox} & \customstrut
  \begin{adjustbox}{valign=c}\includegraphics[width=0.25\textwidth]{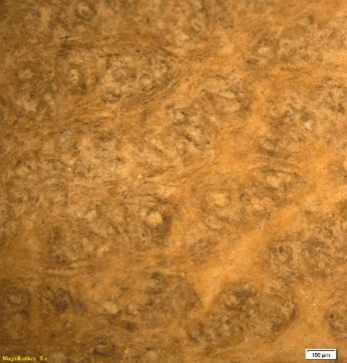}\end{adjustbox} &
  \raggedright The sample was characterized by the presence of few black spots with a size larger than 10 µm. Carbon was identified. \\ \hline
\textbf{\begin{tabular}[c]{@{}l@{}}4Q201/\\ 4Q338\end{tabular}} &
  \begin{adjustbox}{valign=c}\includegraphics[width=0.25\textwidth]{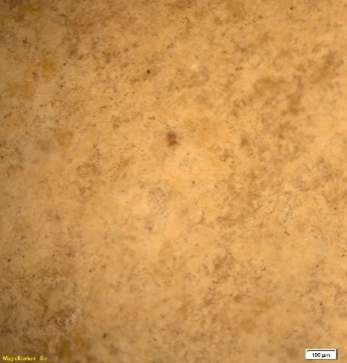}\end{adjustbox} & \customstrut
  \begin{adjustbox}{valign=c}\includegraphics[width=0.25\textwidth]{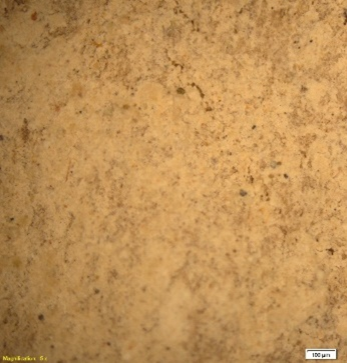}\end{adjustbox} &
  \raggedright The sample was characterized by several black spots. Several red spots were also detected as for sample 386. Due to the high fluorescence of the writing support, Raman spectra evidenced only the presence of carbon. \\ \hline
\textbf{4Q259} &
  \begin{adjustbox}{valign=c}\includegraphics[width=0.25\textwidth]{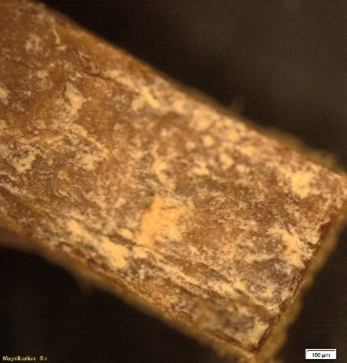}\end{adjustbox} & \customstrut
  \begin{adjustbox}{valign=c}\includegraphics[width=0.25\textwidth]{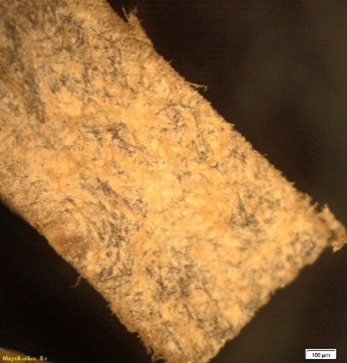}\end{adjustbox} &
  \raggedright Almost clean, no significant spots were detected. \\ \hline
\textbf{4Q416} &
  \begin{adjustbox}{valign=c}\includegraphics[width=0.25\textwidth]{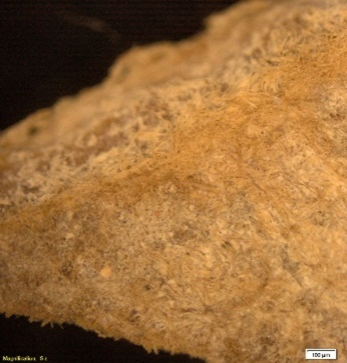}\end{adjustbox} & \customstrut
  \begin{adjustbox}{valign=c}\includegraphics[width=0.25\textwidth]{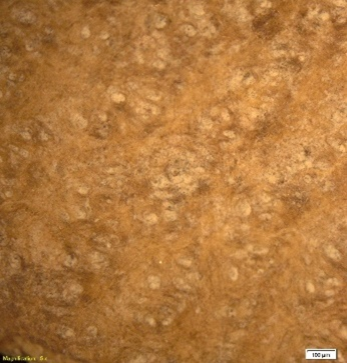}\end{adjustbox} &
  \raggedright The sample was characterized by a rough surface scattering the laser Raman light. The MO observation did not show any significant presence of dark spots. \\ \hline
\textbf{4Q2} &
  \begin{adjustbox}{valign=c}\includegraphics[width=0.25\textwidth]{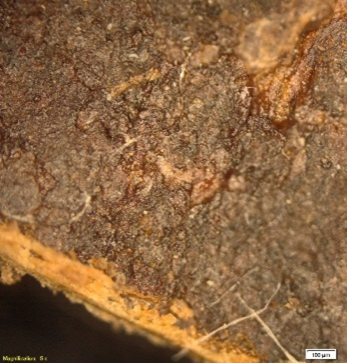}\end{adjustbox} & \customstrut
  \begin{adjustbox}{valign=c}\includegraphics[width=0.25\textwidth]{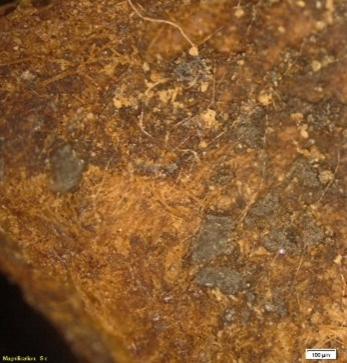}\end{adjustbox} &
  \raggedright The sample was fully covered by a material that did not allow to perform a proper Raman analysis. \\ \hline
\textbf{4Q375} &
  \begin{adjustbox}{valign=c}\includegraphics[width=0.25\textwidth]{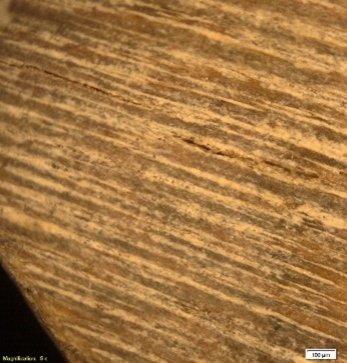}\end{adjustbox} & \customstrut
  \begin{adjustbox}{valign=c}\includegraphics[width=0.25\textwidth]{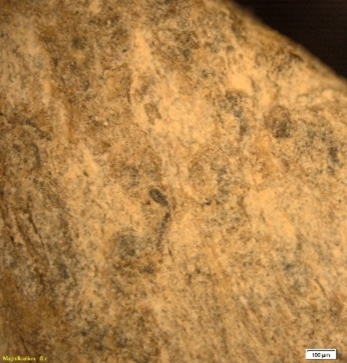}\end{adjustbox} &
  \raggedright The sample was characterized by few black spots with diameter wider than 20 µm. Carbon was identified. \\ \hline
\textbf{\begin{tabular}[c]{@{}l@{}}XHev/\\ Se2\end{tabular}} &
  \begin{adjustbox}{valign=c}\includegraphics[width=0.25\textwidth]{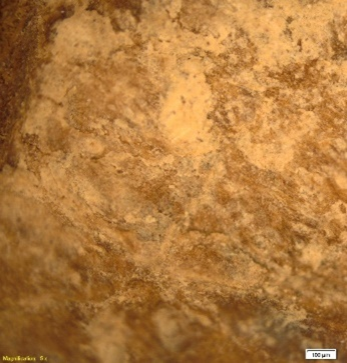}\end{adjustbox} & \customstrut
  \begin{adjustbox}{valign=c}\includegraphics[width=0.25\textwidth]{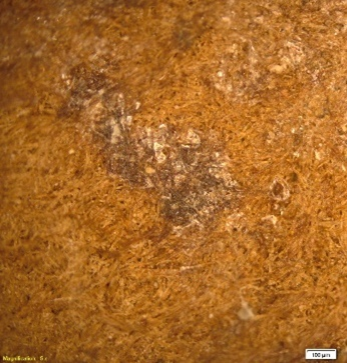}\end{adjustbox} & 
  \raggedright The sample was almost clean. Only two 10 µm in diameter black spots were detected. Carbon was identified. \\ \hline
\textbf{4Q541} &
  \begin{adjustbox}{valign=c}\includegraphics[width=0.25\textwidth]{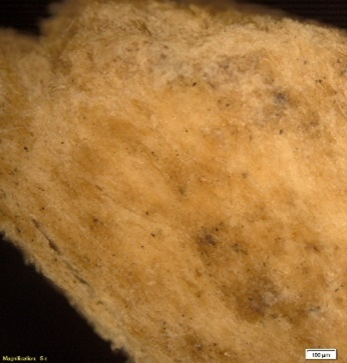}\end{adjustbox} & \customstrut
  \begin{adjustbox}{valign=c}\includegraphics[width=0.25\textwidth]{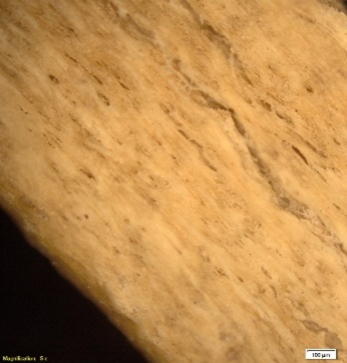}\end{adjustbox} &
  \raggedright The sample was characterized by several black spots. Carbon was identified. \\ \hline
\textbf{4Q521} &
  \begin{adjustbox}{valign=c}\includegraphics[width=0.25\textwidth]{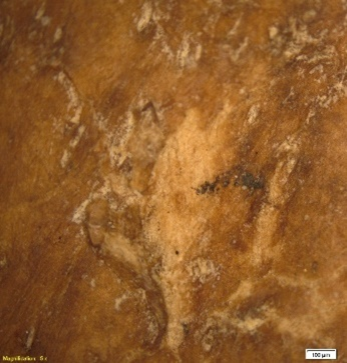}\end{adjustbox} & \customstrut
  \begin{adjustbox}{valign=c}\includegraphics[width=0.25\textwidth]{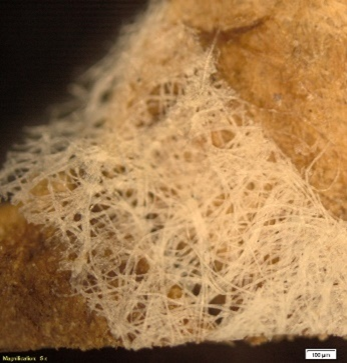}\end{adjustbox} & 
  \raggedright Only one big black spot (100-150 µm) was detected on the surface of the bigger fragment. Carbon was identified. \\ \hline
\textbf{4Q267} &
  \begin{adjustbox}{valign=c}\includegraphics[width=0.25\textwidth]{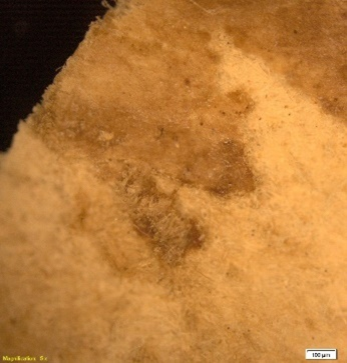}\end{adjustbox} & \customstrut
  \begin{adjustbox}{valign=c}\includegraphics[width=0.25\textwidth]{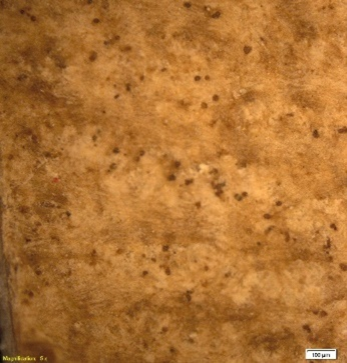}\end{adjustbox} &
  \raggedright The sample was almost clean from dark spots. The size of the identified spots was too small to be investigated by Raman. \\ \hline
\textbf{Mur19} &
  \begin{adjustbox}{valign=c}\includegraphics[width=0.25\textwidth]{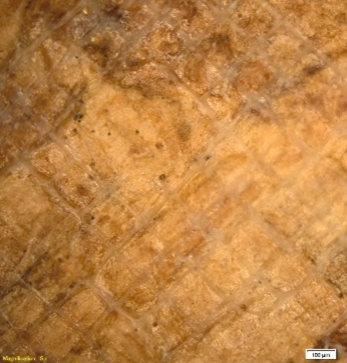}\end{adjustbox} & \customstrut
  \begin{adjustbox}{valign=c}\includegraphics[width=0.25\textwidth]{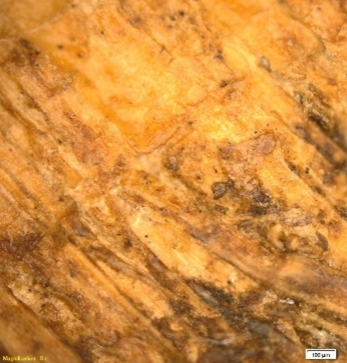}\end{adjustbox} &
  \raggedright The sample was characterized by several black spots and an organic protective. Carbon was identified. \\ \hline
\textbf{4Q216} &
  \begin{adjustbox}{valign=c}\includegraphics[width=0.25\textwidth]{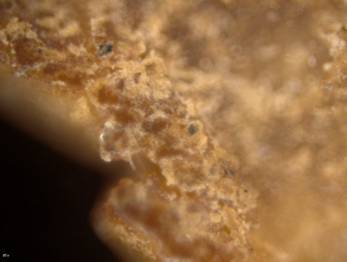}\end{adjustbox} & \customstrut
  \begin{adjustbox}{valign=c}\includegraphics[width=0.25\textwidth]{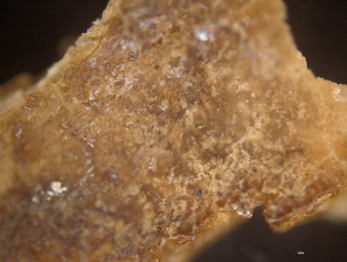}\end{adjustbox} &
  \raggedright The sample was characterized by black spots. Carbon was identified. \\ \hline
\textbf{4Q114} &
  \begin{adjustbox}{valign=c}\includegraphics[width=0.25\textwidth]{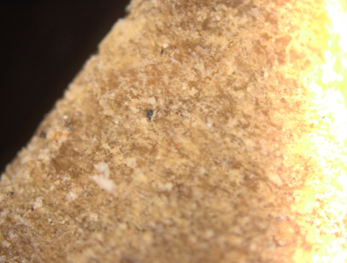}\end{adjustbox} & \customstrut
  \begin{adjustbox}{valign=c}\includegraphics[width=0.25\textwidth]{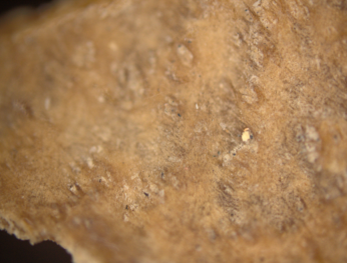}\end{adjustbox} &
  \raggedright The sample was characterized by black spots. Carbon was identified. \\ \hline \hline
\end{longtable}
}

%% file: tables/TableB7-X.tex
\begin{table}[ht!]
\centering
\caption{Summary of the materials detected in the different parchment samples.}
\label{tab:B7-summarize}
\begin{tabular}{l|c}
\hline \hline
\textbf{Samples} & \textbf{Identified organic   materials}               \\ \hline
4Q216            & proteinaceous material, lipid material                \\ \hline
4Q3              & proteinaceous material, lipid material, egg           \\ \hline
4Q27             & proteinaceous material, lipid material                \\ \hline
Mas1k            & proteinaceous material, lipid material                \\ \hline
4Q206            & proteinaceous material, lipid material                \\ \hline
4Q30             & proteinaceous material, lipid material, egg           \\ \hline
4Q201/4Q338      & proteinaceous material, lipid material                \\ \hline
4Q259            & proteinaceous material, lipid material                \\ \hline
4Q416            & proteinaceous material, lipid material                \\ \hline
4Q2              & proteinaceous material, lipid material, retene        \\ \hline
4Q375            & proteinaceous material, lipid material                \\ \hline
XHev/Se2         & proteinaceous material, lipid material                \\ \hline
4Q541            & proteinaceous material, lipid material, retene        \\ \hline
4Q521            & proteinaceous material, lipid material, acrylic resin \\ \hline
4Q267            & proteinaceous material, lipid material, retene        \\ \hline
Mur19            & lignocellulose material, acrylic resin, retene        \\ \hline
4Q114            & proteinaceous material, lipid material                \\ \hline \hline
\end{tabular}
\end{table}

%% file: tex-sections/C-OxCal-plots.tex
C\section{\texorpdfstring{OxCal plots: \textsuperscript{14}C determinations and calibrated date plots}{}}\label{appen:K}
Here, we present the OxCal plots for the 26 accepted samples. No plots were produced for the 3 technically rejected samples (see Section~\ref{appen:B6:reject}), nor for the 1 sample of which the identity could not be ascertained (see Section~\ref{appen:B5:4Q185}).

\begin{longtable}{cc}
    \label{fig:c14-individual-plots}\\
    \includegraphics[width=0.49\textwidth]{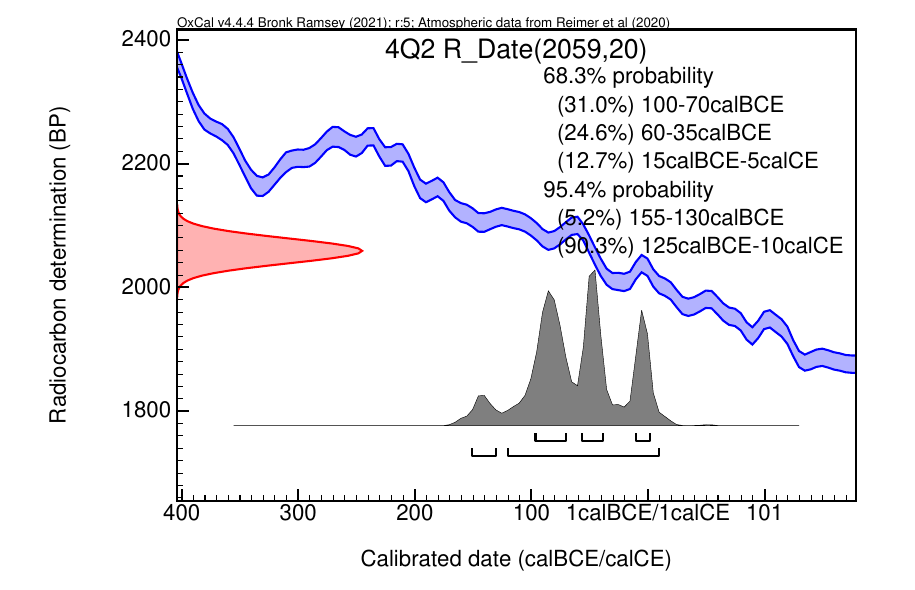} &
    \includegraphics[width=0.49\textwidth]{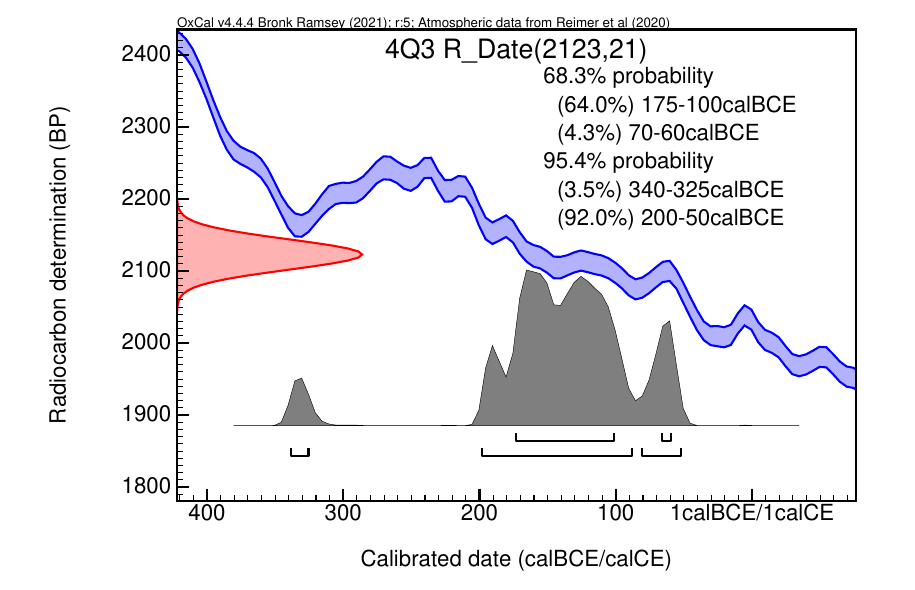} \\
    \includegraphics[width=0.49\textwidth]{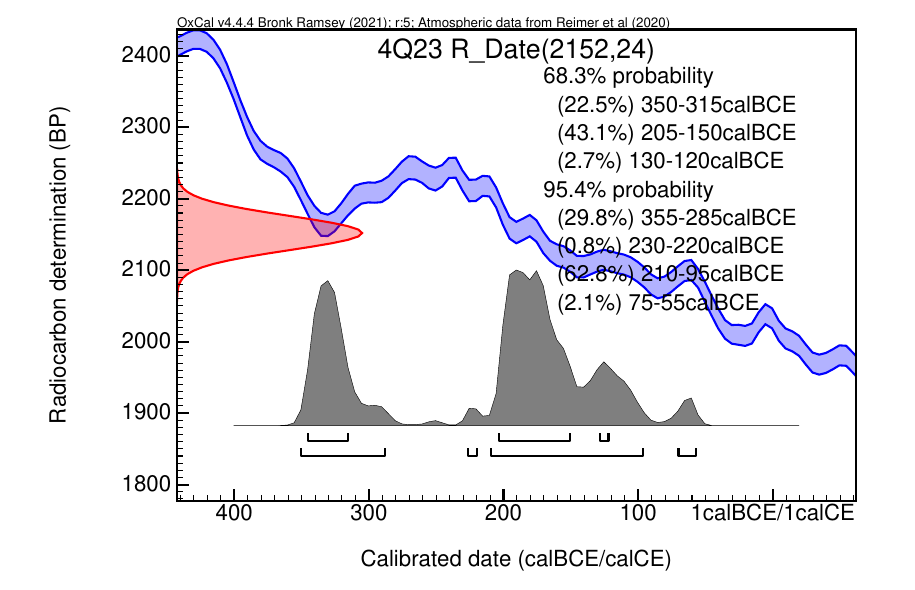} &
    \includegraphics[width=0.49\textwidth]{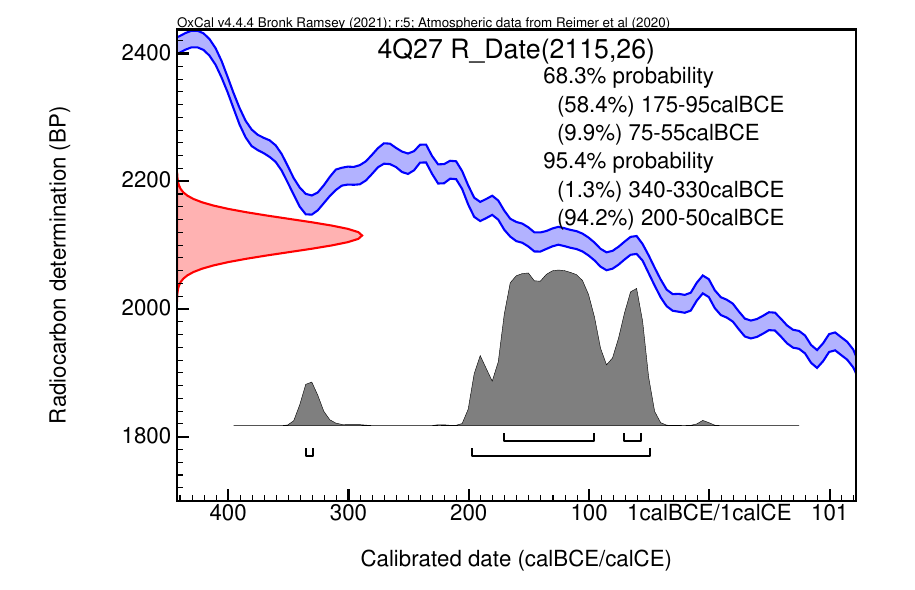} \\
    \includegraphics[width=0.49\textwidth]{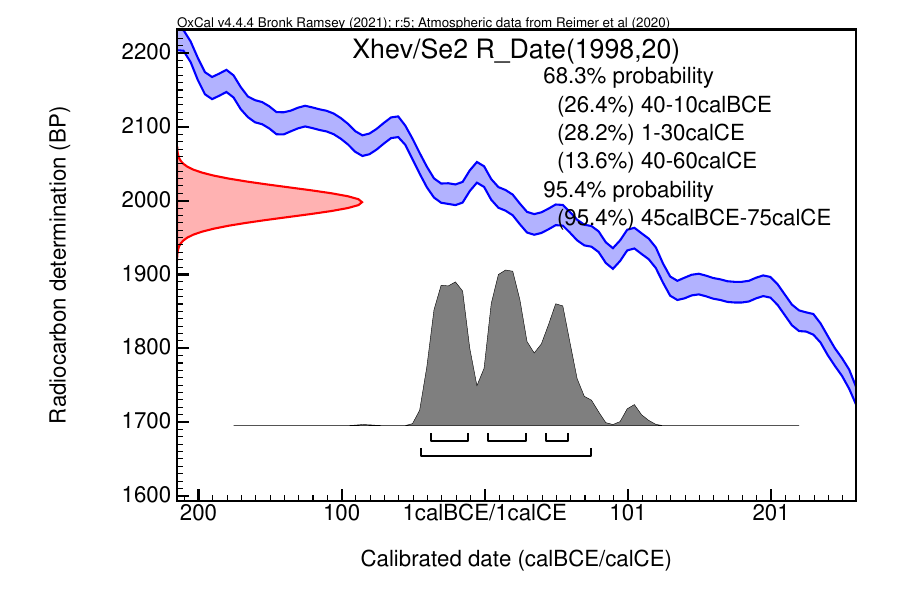} &
    \includegraphics[width=0.49\textwidth]{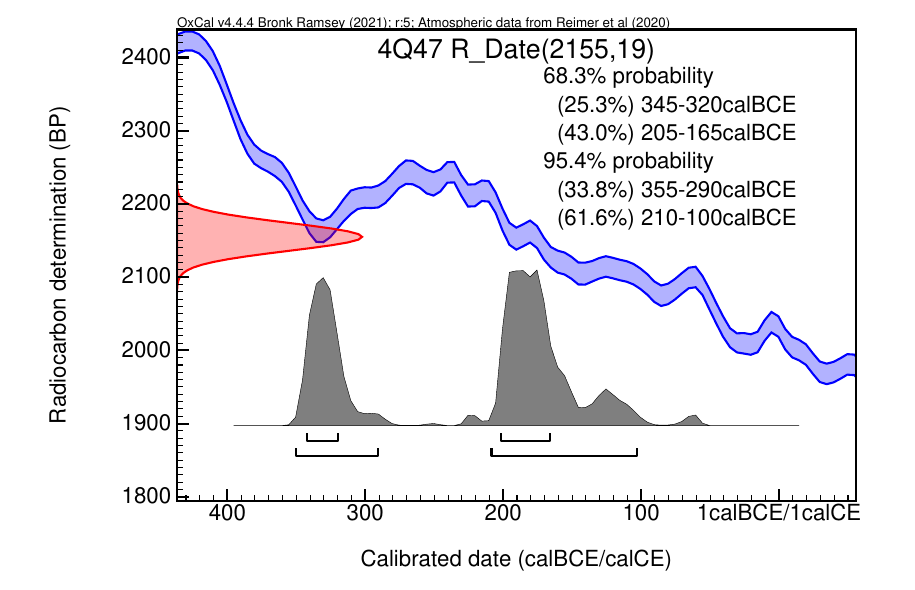} \\
    \includegraphics[width=0.49\textwidth]{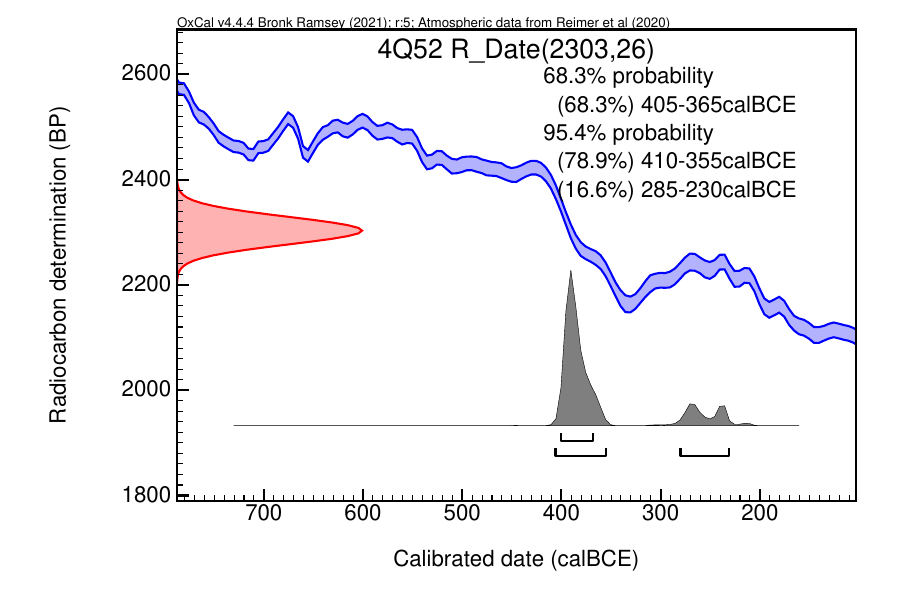} &
    \includegraphics[width=0.49\textwidth]{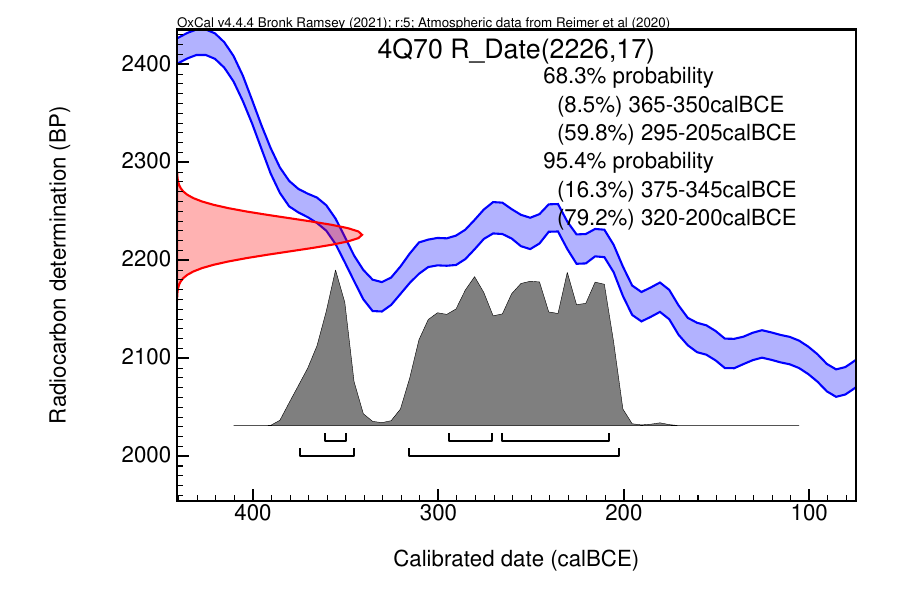} \\
    \includegraphics[width=0.49\textwidth]{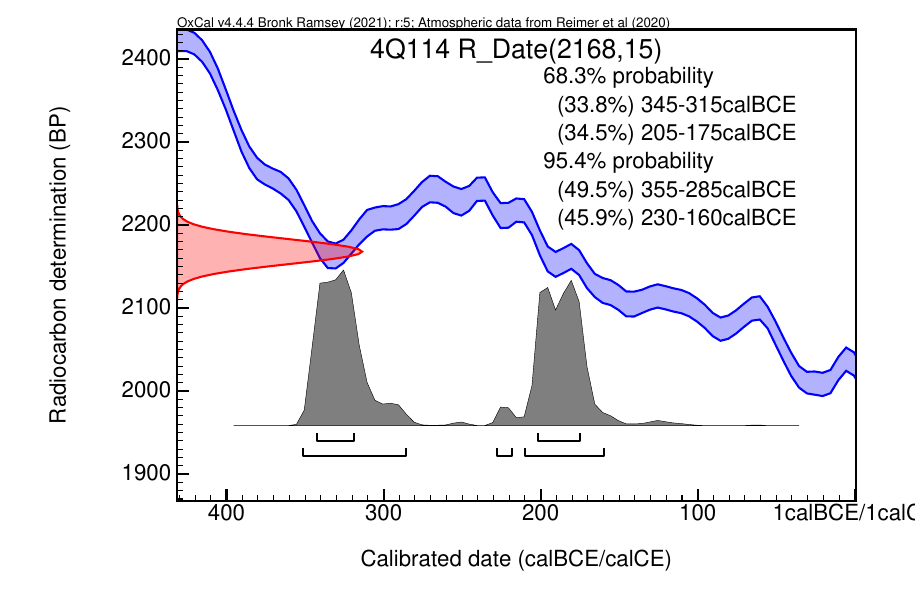} &
    \includegraphics[width=0.49\textwidth]{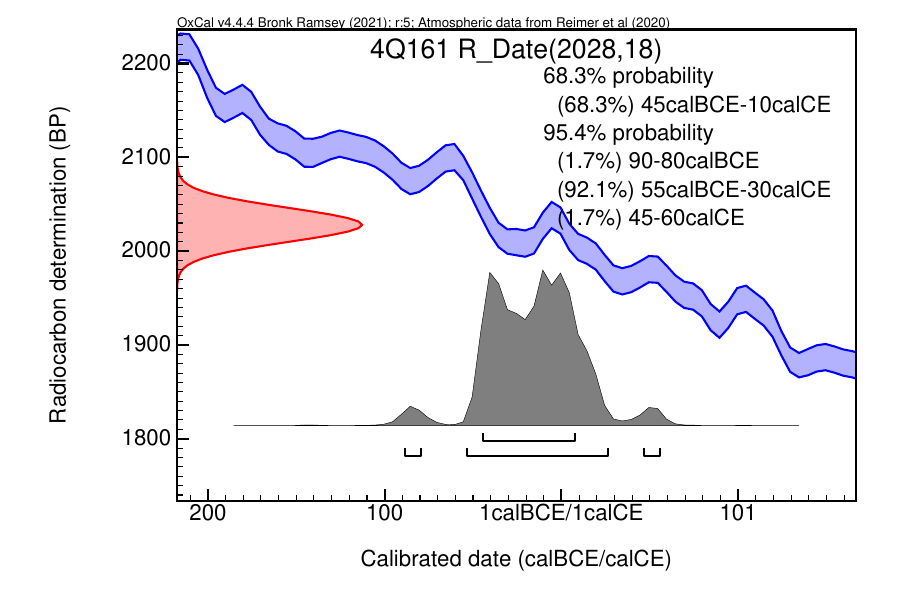} \\
    \includegraphics[width=0.49\textwidth]{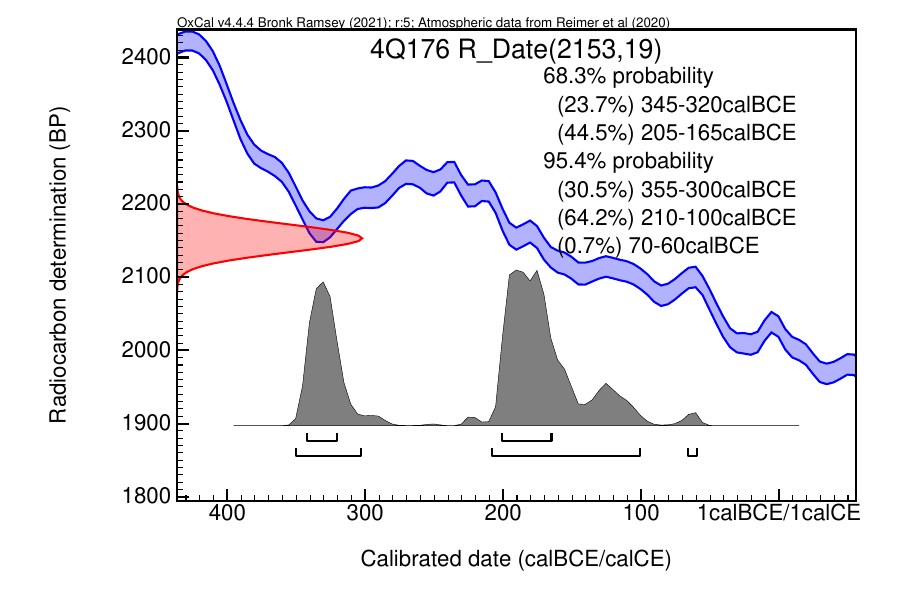} &
    \includegraphics[width=0.49\textwidth]{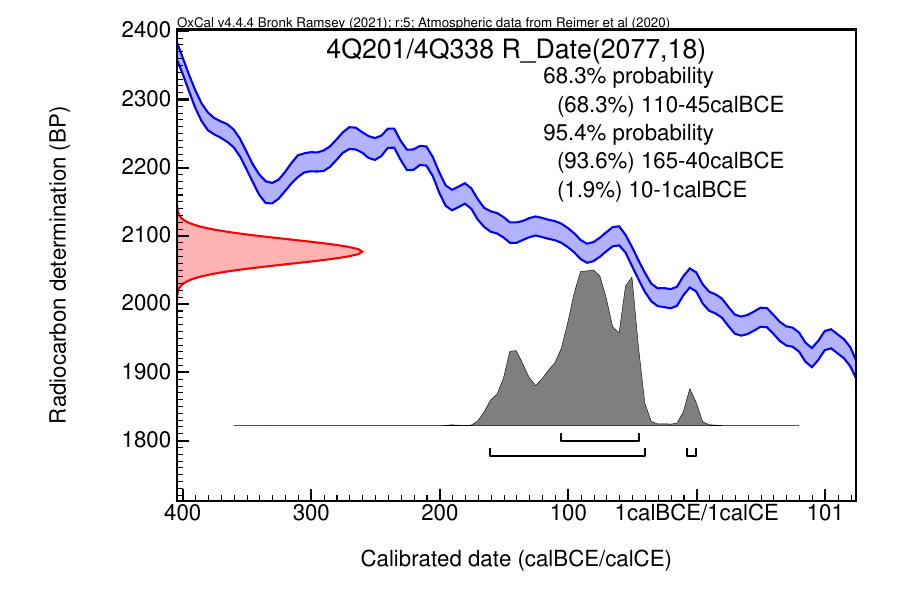} \\
    \includegraphics[width=0.49\textwidth]{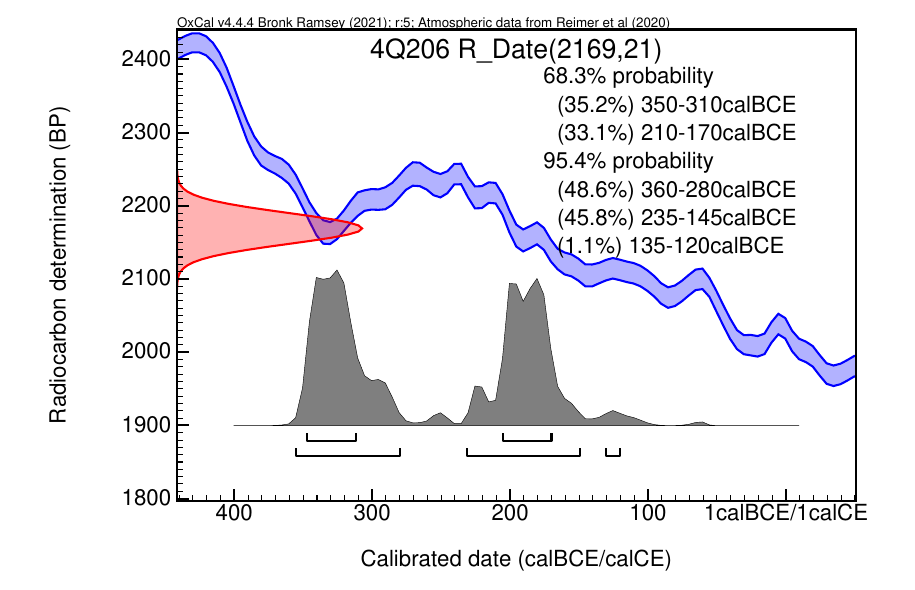} &
    \includegraphics[width=0.49\textwidth]{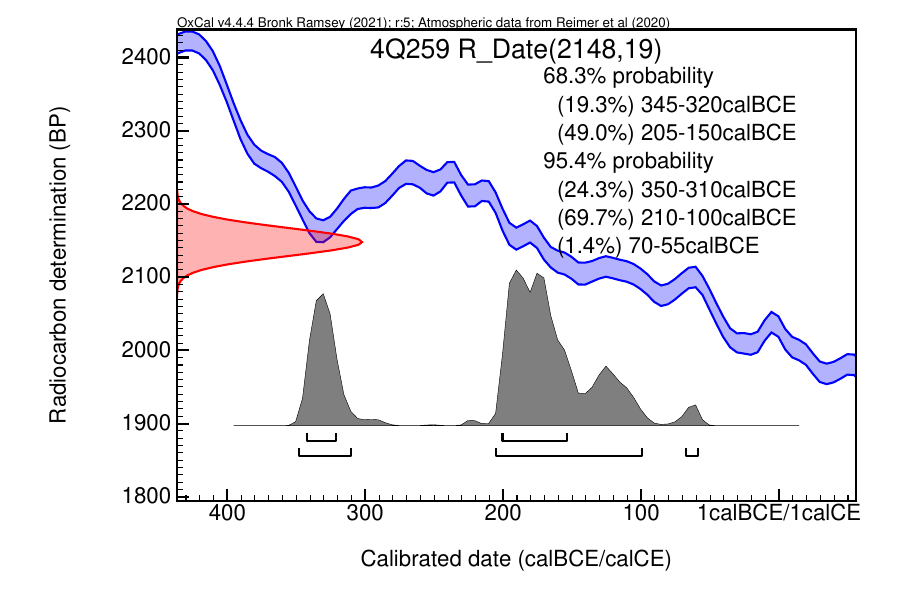} \\
    \includegraphics[width=0.49\textwidth]{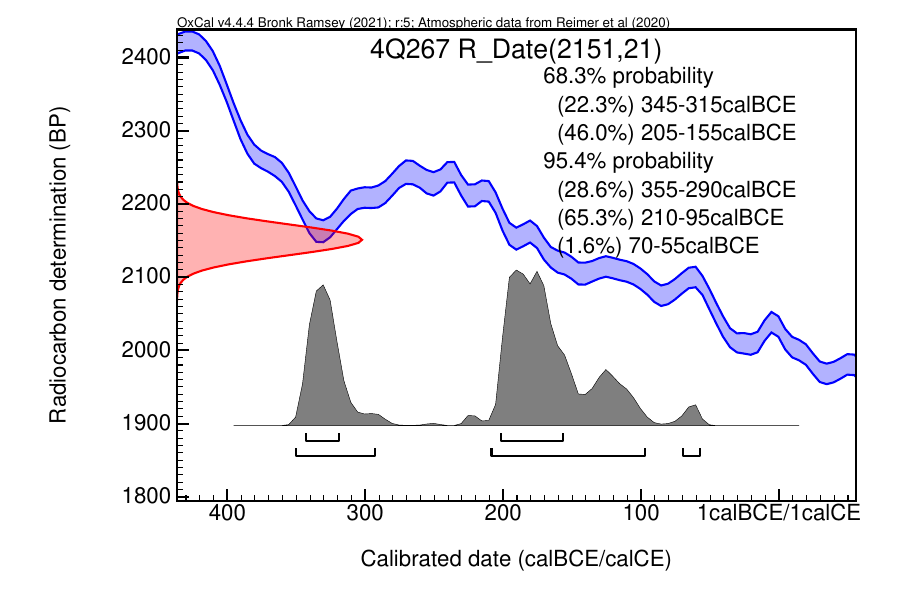} &
    \includegraphics[width=0.49\textwidth]{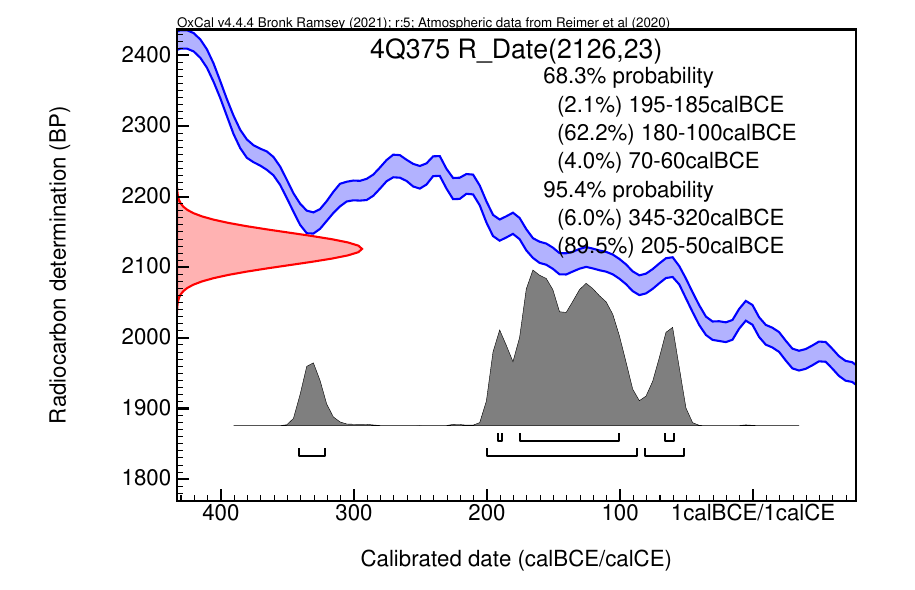} \\
    \includegraphics[width=0.49\textwidth]{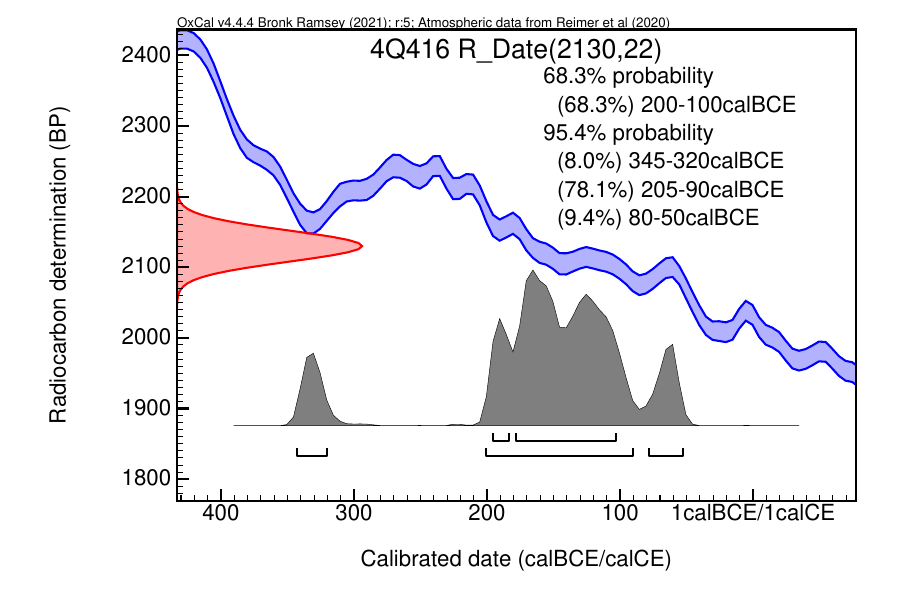} &
    \includegraphics[width=0.49\textwidth]{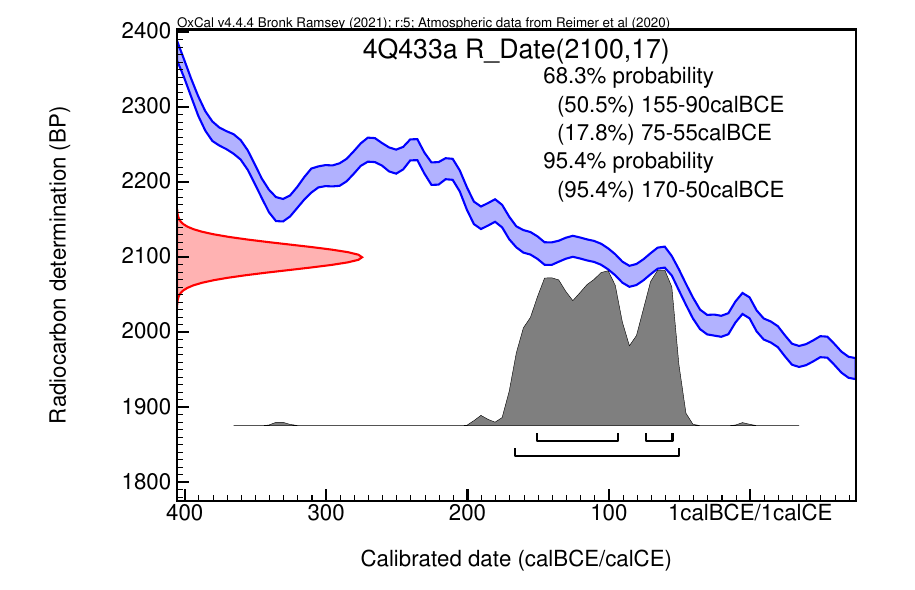} \\
    \includegraphics[width=0.49\textwidth]{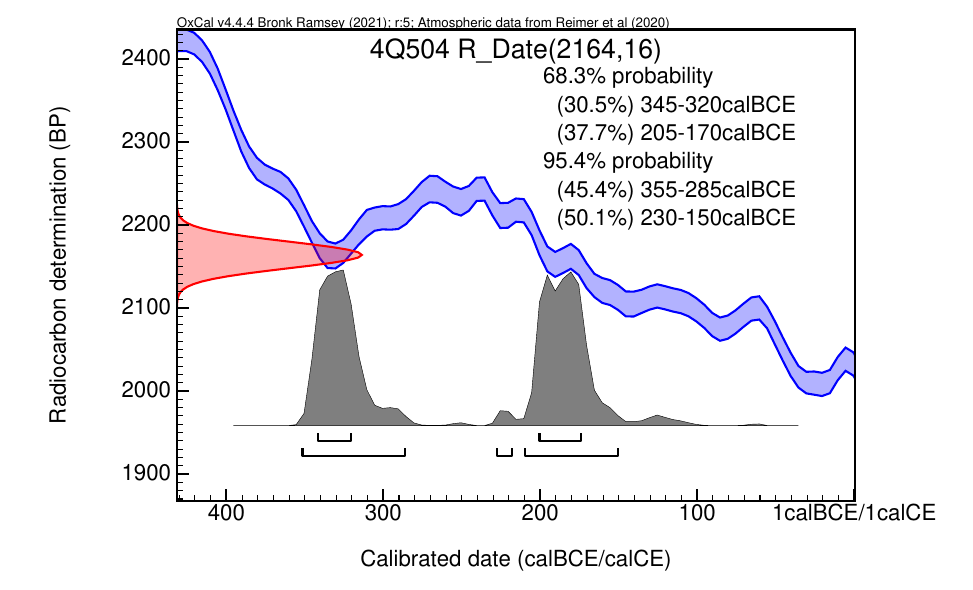} &
    \includegraphics[width=0.49\textwidth]{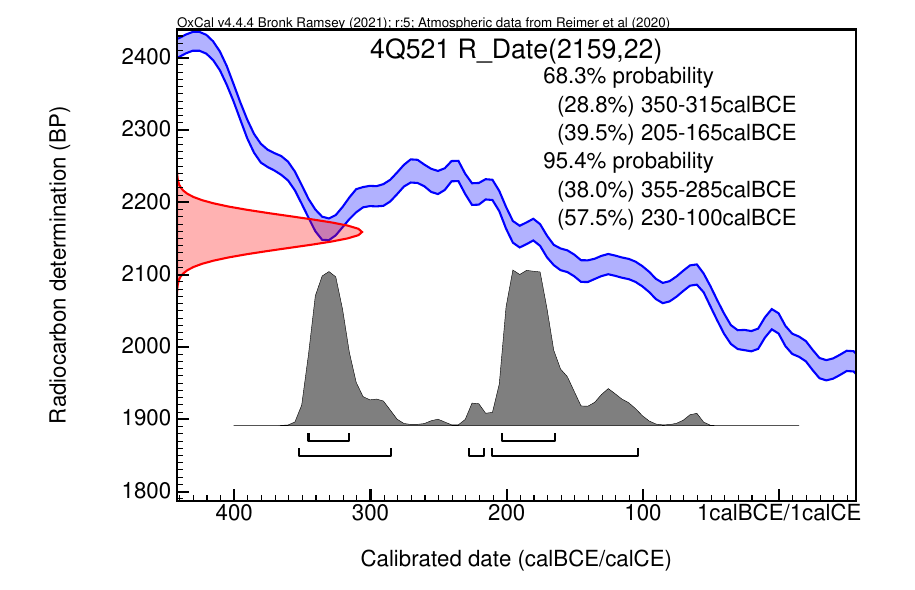} \\
    \includegraphics[width=0.49\textwidth]{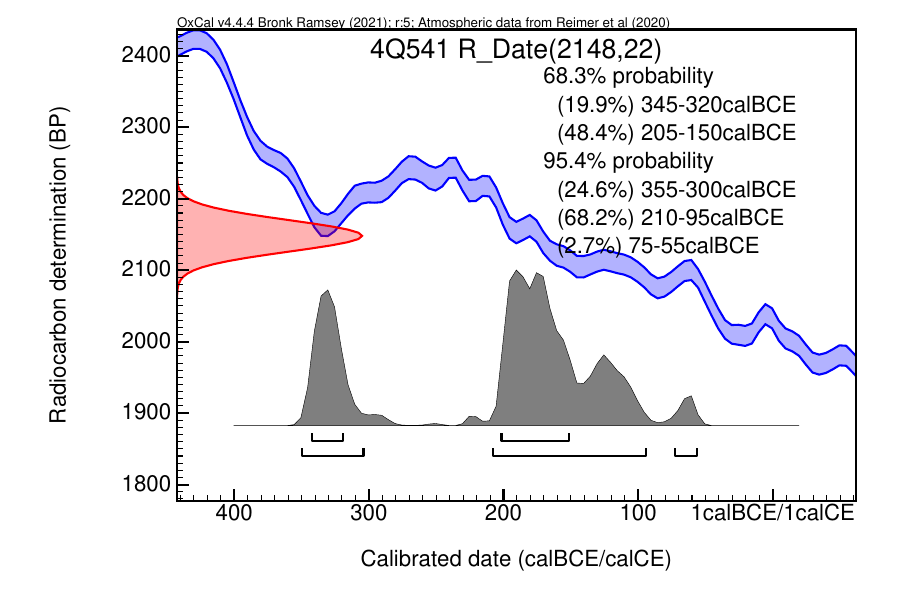} &
    \includegraphics[width=0.49\textwidth]{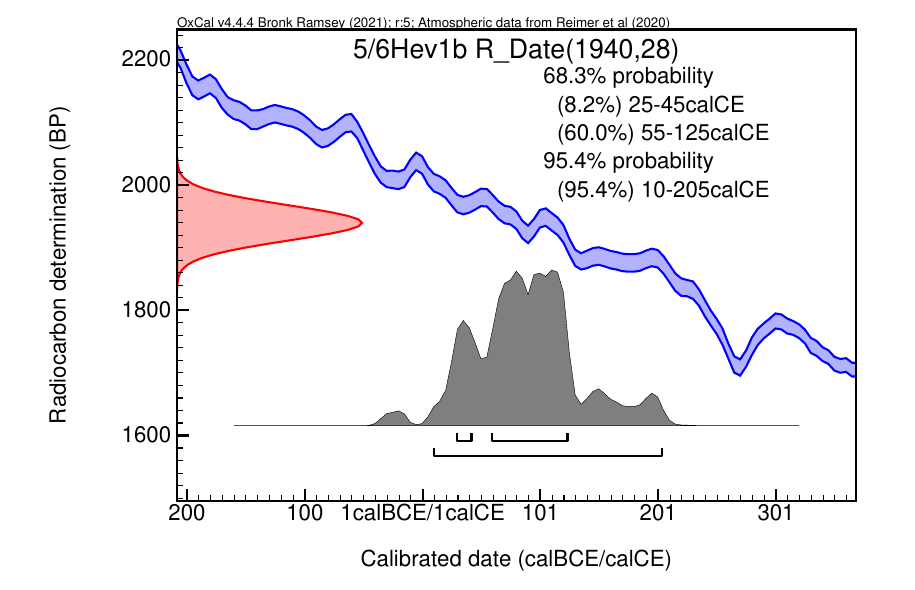} \\
    \includegraphics[width=0.49\textwidth]{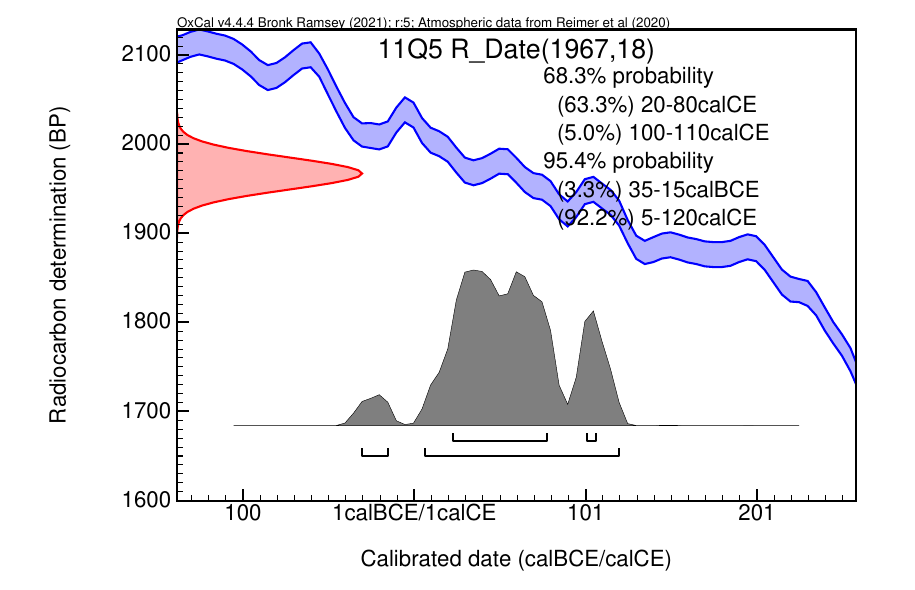} &
    \includegraphics[width=0.49\textwidth]{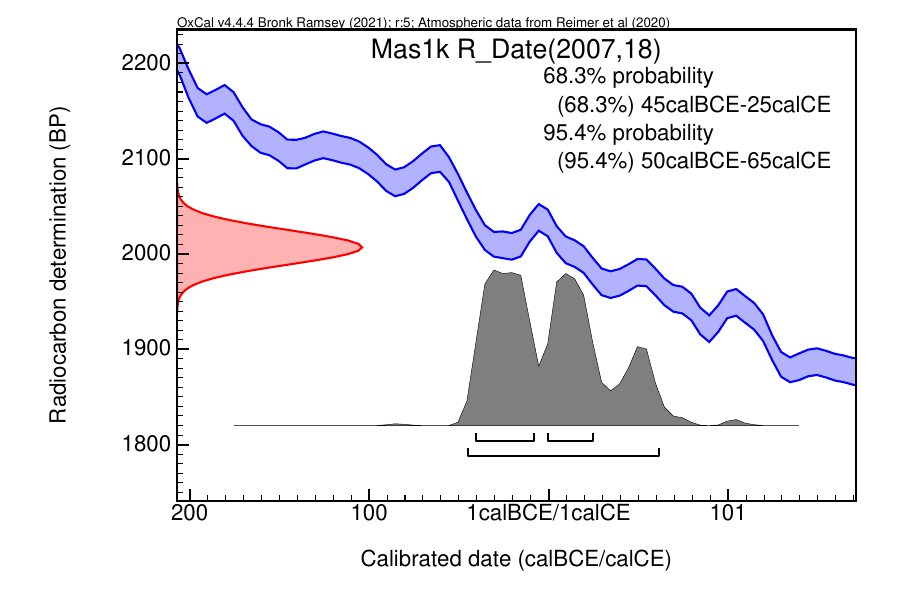} \\
    \includegraphics[width=0.49\textwidth]{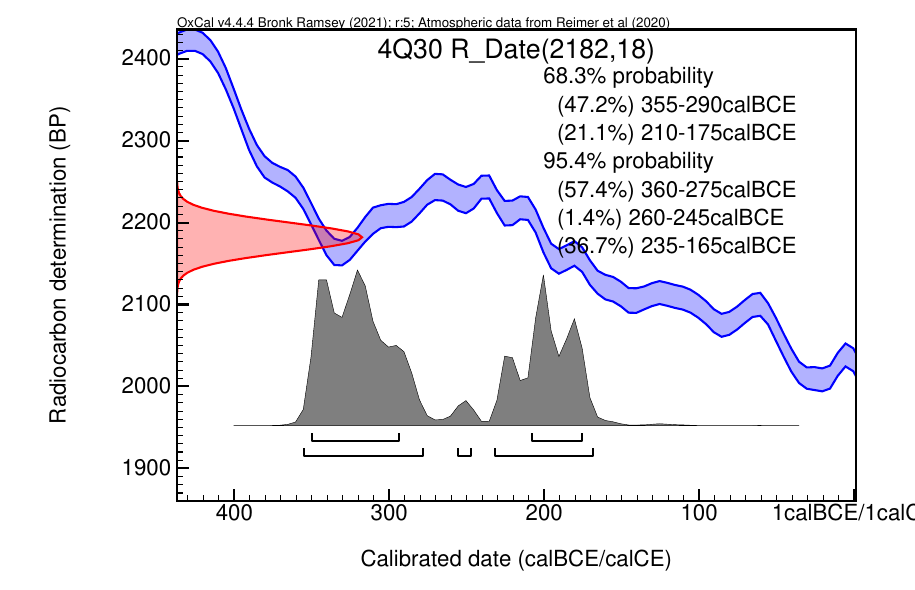} & 
    \includegraphics[width=0.49\textwidth]{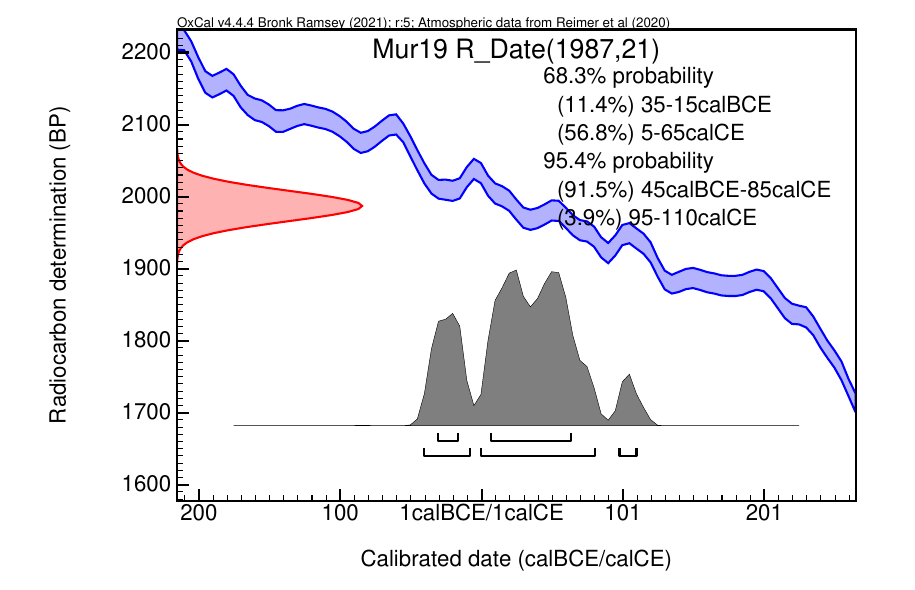} \\
\end{longtable}

%% file: tex-sections/D-palaeography-C14.tex
\section{Palaeography and radiocarbon dating of the Dead Sea Scrolls}\label{appen:C-new}

\subsection{Comparing radiocarbon results and palaeographic estimates} \label{appen:B8:comparison}

We make the comparison between the radiocarbon dates (Table~\ref{tab:summarized-c14} in Appendix~\ref{appen:B}) and previous palaeographic estimates on the basis of the estimates given in the official publication series, Discoveries in the Judaean Desert (DJD), as these are considered the standard in the field, but sometimes we include references to estimates of other scholars when relevant. 

However, we also critically assess previous palaeographic estimates. We do that on two levels. First, we reason according to the relative typology of the so-called Cross model and assess its application to individual manuscripts. This leads occasionally to palaeographic assessments that correct previous ones. Second, we desist from translating a relative typology to an absolute chronology. Because of the lack of date-bearing documents for the time-period one cannot impose the traditional framework’s unsubstantiated chronological limitations on when the so-called Hasmonaean and Herodian script features would have started to develop (see Section~\ref{appen:A2.2:unsub}). This also applies to chronological distinctions within the general indications of Hasmonaean-type and Herodian-type scripts. Cross suggested chronological ranges of 50 years, and sometimes even shorter ranges of 25–50 years, as he assumed a rapid development of the script from the Hasmonaean period onward, contrary to the presumed slow development in the third century BCE. However, this assumption of a rapid evolution remains unsubstantiated, too (see below). As more reliable time markers, our study’s \textsuperscript{14}C calibrated ranges demonstrate older date ranges than previously thought for individual manuscripts as well as for the beginnings of the Hasmonaean/Herodian scripts. 

We can compare the radiocarbon dates with previous palaeographic estimates only in a general sense, not as a rigid application of these estimates. The early 1990s guideline that editors of manuscripts in the DJD series would date according to the typological specimens of Cross’s 1961 article has proved unfortunate. One problem is that many of the palaeographic estimates offered in the DJD series since the 1990s suffer from an insufficient understanding of Cross’s model, producing unreliable estimates~\cite{Tigchelaar2020}. This unreliability is further exacerbated by the problems within Cross’s palaeographic model that conflates supposed historical and political developments with palaeographic style developments. 

Cross~\cite{Cross2003} presented few specimens for other scholars to work with. This makes it difficult to substantiate style developments within, for example, Hasmonaean formal script, and to account for the complexity of script in individual manuscripts. Moreover, Cross also suggested mutual influences between formal, semiformal, and semicursive in such a way that sometimes a typological development of an individual letter is thought to have occurred earlier in, e.g., semicursive than in formal script (e.g., for \textit{samek}). 

For the Hasmonaean formal script Cross~\cite{Cross2003, Cross1998} singled out only three manuscripts, assuming absolute dates between ca. 175–30 BCE: 4Q28, 4Q30, and 4Q51 (for a number of individual letters, Cross also referred to 1QIsa\textsuperscript{a} and 4Q1, as respectively middle and early Hasmonaean formal, as well as to 4Q109 and 4Q504 as early Hasmonaean semiformal). Thus, 4Q30 is said to be “a typical Hasmonaean” script, without explanation why that is so, from the middle of the period, 125–100 BCE (Cross might have used 1QIsa\textsuperscript{a} instead but because he deemed it to have more idiosyncratic forms he gave preference to 4Q30 which he understood to have been copied by a more conventional scribe; no further substantiation is provided for these claims). The other two manuscripts are at the outer ends of the Hasmonaean formal script spectrum, apparently for having script style elements in common with earlier and later periods. So 4Q28 is presented as transitional between Archaic and the beginning of the Hasmonaean development (175–150 BCE) and 4Q51 as a late transitional script from the end of the Hasmonaean period or the beginning of the Herodian period (50–25 BCE). 

For the Herodian formal script Cross singled out seven manuscripts, assuming absolute dates between 30 BCE and 70 CE: 1QM, 4Q27, 4Q37, 4Q85, 4Q113, 5/6Hev1b, and Mur24. In fact, only four manuscripts are singled out for the Herodian formal script. 1QM is presented as “a typical early Herodian formal script”  (ca. 30–1 BCE), while 4Q113 would represent “a developed Herodian formal script” (20–50 CE) and 4Q37 and 4Q85 late Herodian formal scripts from respectively ca. 50 CE and ca. 50–68 CE. 4Q27 is said to be “a typical exemplar of the extremely popular Round semiformal style” (also called rustic, and considered distinct from the Vulgar semiformal) from the early Herodian period, ca. 30 BCE–20 CE. The final two manuscripts are actually considered so-called post-Herodian, assuming absolute dates between 70–135 CE: 5/6Hev1b was estimated by Cross from 75–100 CE (Flint suggested 50–68 CE~\cite{DJD38}) and Mur24 is a date-bearing document from 133 CE. Cross saw in some Herodian scripts the types of individual letters mixed so that semiformal can “invade” formal or Vulgar semiformal “makes its way into the formal character”, e.g., \textit{mem}, 

Apart from evaluating individual letters, there is no method in the field for dating an entire manuscript on the basis of mixed evidence of ‘older’ and ‘later’ forms of individual letters.  Perhaps some scholars apply a form of quantification, weighing the instances of ‘older’ and ‘later’ forms, but this is never explicated. Rather, the assumption generally seems to be that ‘later’ forms cannot have developed earlier but ‘older’ forms can still have been in use at a later time, whether or not as a case of ‘archaizing’. While it may certainly be true that ‘older’ forms can have been in use for a long time, the claim that ‘later’ forms cannot have developed earlier remains unproven for lack of dated evidence. This means that what are perceived as, e.g., late Hasmonaean or early Herodian letter forms may have developed earlier than currently thought.  

Even if one adopts Cross’s typological development, the issue of the absolute dating or calibration of the types remains~\cite{Tigchelaar2020}. A mixture of ‘older’ (more ancient) and ‘later’ (more developed) forms can appear in one and the same manuscript. A focus on individual letters alone cannot be indicative for earlier or later chronology, whether relative or absolute. The study of individual manuscripts demonstrates a more complex development (see, e.g., for 4Q1~\cite{Tigchelaar2023}). There are examples of experienced palaeographers coming up with widely diverging dates for the same scrolls. Thus, a range of individual manuscripts cannot be fitted precisely in a sequence on the basis of traditional palaeography.  

The radiocarbon dates and the palaeographic estimates are two independent information sources about history, based on two different methodologies: one is a physically measured “yardstick of time”, the other is a cultural and qualitative assessment. At present, in the absence of an abundance of date-bearing manuscripts between the third century BCE and the first century CE, radiocarbon dates (\textsuperscript{14}C) derived from manuscript samples are more reliable time markers. The palaeographic estimates do not provide absolute or fixed dates.

With these caveats in mind, Figure~\ref{fig:AIvsC14PAL} in the main article shows the comparison between the (accepted) 2$\sigma$ calibrated ranges and previous palaeographic estimates (see the worksheet in Appendix~\ref{appen:I} for the specific data and information). Additional plots can be found in Appendix~\ref{appen:E-plots} where Figure~\ref{fig:AIvsC14PAL-wo-minorpeaks} and \ref{fig:AIvsC14PAL-continuous} presents the effect of including or excluding minor peaks to the 2$\sigma$ calibrated ranges and Figure~\ref{fig:AIvsC14PAL-1sigma} presents the outcome of selecting 1$\sigma$ calibrated range.


\subsubsection{Whole or partial overlap}\label{appen:B8.1:overlap}
Comparing previously given palaeographic estimates and our \textsuperscript{14}C 2$\sigma$ results, shows that 17 of the 26 sampled manuscripts in our project have whole or partial overlap. This applies to: 4Q23, 4Q47, 4Q52, 4Q70, 4Q161, 4Q176, 4Q201/4Q338, 4Q255/4Q433a, 4Q259, 4Q504, 4Q521, 4Q541, 11Q5, Mas1k, Mur19, 5/6Hev1b, XHev/Se2.

4Q47 is a good example of how palaeographic estimates cannot be precise or clearly substantiated. Ulrich~\cite{DJD14} reports that Cross had identified its script as Hasmonaean---thus dating it probably in the second half of the second century or the first half of the first century BCE---but refrained from offering a more precise estimate within the Hasmonaean period. Langlois~\cite{Langlois2011} and Puech~\cite{Puech2015} favoured the first half of the first century BCE. Langlois referred to some letters showing a typologically older form (\textit{bet}, \textit{dalet}, \textit{vav}, \textit{khet}, \textit{nun}), while others would have a form more in line with those seen in late Hasmonaean or early Herodian periods (\textit{aleph}, \textit{he}, \textit{tet}, \textit{samek}, \textit{pe}). However, considering, e.g., \textit{aleph} one can see two forms, one of them being a typologically older form where the left leg often connects to the middle of the diagonal instead of more toward the top of it; the same for \textit{samek} that appears in both closed (younger) and open (older) form. Also, \textit{ayin} is often small, considered an older form, while \textit{yod} shows the triangular head, seen as typical for the late Hasmonaean period. Instead of trying to fit this manuscript overall into a linear date estimate, the mixed typological evidence can be better explained as demonstrating overlapping or partly adjacent style developments. 

In Section~\ref{appen:B4:amsdating} in Appendix~\ref{appen:B} we noted that our calibrated results are often bimodal, especially for 2$\sigma$ distributions. 4Q47 is an example of such bimodal calibrated results, also for the 1$\sigma$ distribution. The \textsuperscript{14}C 2$\sigma$ calibrated range of 210–100 BCE (61.6\% probability) overlaps with the broad palaeographic estimate ‘Hasmonaean’—but less with the more specific ones of Puech and Langlois—and also allows for dating the script style of 4Q47 to the first half of the second century BCE. 

The older 2$\sigma$ calibrated range of 355–290 BCE (33.8\%) is far removed on the timeline from previous palaeographic estimates. Although the older 2$\sigma$ peak represents a mathematically valid solution of the dating process, the younger calibrated peak must be preferred for 4Q47 over the older calibrated peak. Following the palaeographic principle to compare the script of an undated manuscript to that of dated writings with a similar script (see Section~\ref{appen:A2.2:unsub} in Appendix~\ref{appen:A}), it should be noted that 4Q47 does not compare to the extant typological evidence from date-bearing Aramaic manuscripts from the fourth century BCE. Typologically, the script of 4Q47 does not correspond to that of the script in date-bearing documents from the Persian period such as those from Bactria or from Wadi Daliyeh from the same region as the Dead Sea Scrolls. So, from a palaeographic perspective, 4Q47 is clearly younger than where the older calibrated peaks appear on the timeline. 

Prior to the discovery of the Wadi Daliyeh documents, 4Q52 was argued by Cross to be the oldest manuscript among the Dead Sea Scrolls, and it certainly has the best cards for being the oldest biblical manuscript. In the official publication, Cross et al. estimated 4Q52 to ca. 250 BCE~\cite{DJD17}. 

The \textsuperscript{14}C evidence is bimodal for the 2$\sigma$ distribution. The younger peak of 285–230 BCE (16.6\% probability) agrees well with the palaeographic estimate. The older calibrated range is 410–355 (78.9\%). Although it cannot be ruled out completely from a palaeographic perspective, this older date seems typologically slightly too early for the script in 4Q52 in comparison to date-bearing documents from Elephantine from the late fifth century BCE and date-bearing documents from Bactria from 353 to 324 BCE, although it is difficult to factor in consequences of geographical variance for script variations. A date range in the second half of the fourth century BCE would seem more suitable for 4Q52. Following palaeographic principle, 4Q52 would have to be dated chronologically nearer to the Wadi Daliyeh manuscripts, especially WDSP 1 from 335 BCE (see Section~\ref{appen:A2.2:unsub}). But for that date range there is no \textsuperscript{14}C result. 

Hence, from a palaeographic perspective a clear preference for one of the two peaks in the probability distribution cannot be substantiated. The 2$\sigma$ range of 410–355 is perhaps only a few decades too old and not one to two centuries as for most other bimodal results of our \textsuperscript{14}C measurements. So, in the case of 4Q52, the older peak cannot be rejected as a possible solution with as much confidence as for most other \textsuperscript{14}C samples with bimodal evidence.

4Q176 has two script styles (plausibly from two scribes): the script of fragment 1–2 i looks entirely different from 1-2 ii. The \textsuperscript{14}C sample in this study was taken from fragment 1-2 ii. Strugnell~\cite{Strugnell1970} and Tigchelaar~\cite{Tigchelaar2019} characterized its script as ‘middle Hasmonaean’, i.e., ca. 125–75 BCE. Strugnell’s palaeographic analysis can be easily misunderstood. He explains that many of the letter forms of the second script style seem to be Herodian, such as \textit{bet}, \textit{tet}, \textit{mem}, and \textit{qoph}. Yet, because the script is not formal but semiformal these forms must be dated to the middle Hasmonaean period. Fragment 1-2 ii shows less uniformity in size than fragment 1-2 i, e.g., \textit{kaph} or medial \textit{mem}. This can be understood as a typologically older feature where \textit{kaph} and medial \textit{mem} are still larger other than letters. The ideal of a base line seems not yet well developed. The three-stroke \textit{he} and the small-sized \textit{ayin} seem archaic. On the other hand, the \textit{bet} has a broad base stroke and protrudes to the right, in formal script generally typologically connected to late Hasmonaean or early Herodian. But if the distinction between formal and semiformal cannot be clearly made, 4Q176 is another example of mixed evidence. 

4Q176 is another example of bimodal calibrated results, having in addition also minor peaks of low probability. The 2$\sigma$ calibrated range of 210–100 BCE (64.2\% probability) and the minor peak of 70–60 BCE (0.7\%) are consistent with previous palaeographic assessments. The 2$\sigma$ calibrated range of 210–100 BCE also makes an older dating of the script style possible.

These assessments for 4Q47, 4Q52, and 4Q176 also apply to 4Q23, 4Q70, 4Q161, 4Q255/4Q433a, 4Q259, 4Q504, 4Q521, 4Q541, Mas1k, and XHev/Se2. Only in the case of 4Q201 and 11Q5 do the \textsuperscript{14}C results indicate a date range that goes in the direction of a younger possible date, whereas in almost all cases the direction is toward an older possible date range. 

Regarding 4Q201, Milik’s edition~\cite{Milik1976} suggested the first half of the second century BCE, and most scholars have accepted this estimate. He considered its script to be quite archaic and connected to the third and second-century BCE semicursive or semiformal scripts, perhaps more dependent on the Aramaic writing of northern Syria or Mesopotamia than on those of Judaea or Egypt. Similar comparisons with northern Syria have been made for 4Q17 and 4Q109, but concrete connections cannot be substantiated. Puech~\cite{Puech2017} also saw the script as semiformal/semicursive, dating from ca. 200 BCE, while Langlois~\cite{Langlois2011} gave an estimate of ca. 150 BCE. 

4Q201 has a 2$\sigma$ calibrated range of 165–40 BCE (93.6\%) and a minor peak of 10–1 BCE (1.9\%). This overlaps with the palaeographic estimates, but instead of an older date, a younger date than previously considered is also possible. 

The script of 4Q201 is hard to assess, in part because the scribe used a pen with a thick, worn nib to write small letters, which may account for the atypical \textit{aleph}. Yet, apart from archaic forms of \textit{samek} and \textit{shin} nothing is typologically incongruent with the early Hasmonaean script. 

As for 11Q5, Sanders~\cite{DJD4} understood its script as transitional from early to late Herodian, comparing it to 4Q113 and also 1QM, 4Q27, 4Q37, and 4Q51. He estimated its script to the first half of the first century CE, possibly slightly earlier than 4Q113, Cross’s specimen for “a developed Herodian formal script”. However, clear typological distinctions on the level of individual letters between ‘early’, ‘developed’, and ‘late’ Herodian according to Cross’s specimens are not that easily made. For example, one may consider \textit{aleph} which from early to late Herodian would advance to an inverted “v” form of the left leg and oblique axis, or \textit{dalet}, where the horizontal stroke breaks through the right leg, and see that there is no difference here between the ‘developed’ and ‘late’ Herodian specimens of 4Q113, 4Q37, and 4Q85. On the other hand, the sharp bent in the right leg of \textit{ayin} and \textit{sin}/\textit{shin} may be seen in early as well as developed and late Herodian exemplars, whereas in some manuscripts that are considered to be late Herodian the sharp bent is not clearly shown, e.g., Mur88 and 5/6Hev1b. As to more general features of the Herodian formal script, one may consider a generally uniform letter size, a base line, ligatures, and the development of \textit{keraiai} or serifs. But beyond a general impression, these features are difficult to use for a clear typological differentiation of manuscripts within the Herodian formal script. 

11Q5 has a 2$\sigma$ calibrated range of 5–120 CE (92.2\%) and a minor peak of 35–15 BCE (3.3\%), showing clear overlap with the different presumed Herodian palaeographic periods, even post-Herodian. The measurement has a standard deviation of only 18 in \textsuperscript{14}C years (BP). The length of the 2$\sigma$ calibrated range, 35 BCE–120 CE, is caused by the shape of the calibration curve in this period when converting the BP dates to calendar dates. Scholars of the Dead Sea Scrolls may consider a date later than 70 CE for 11Q5 unlikely because the scrolls found in the Qumran caves are assumed to have been hidden in the summer of 68 CE~\cite{Popovic2012}. 

4Q259 is notorious for its widely varying palaeographic estimates in the second-first centuries BCE. Cross~\cite{Charlesworth1994} described 4Q259 as written in an unusual semicursive with mixed semicursive and semiformal script features. He gave 50–25 BCE as a date estimate. Earlier, Milik~\cite{Milik1976} had suggested the second half of the second century BCE (Milik used the older reference number 4Q260), while later Puech~\cite{Puech1998} suggested the first half of the first century BCE, preferably shortly after 100 BCE. Puech argued for this date on a combined basis of a palaeographic analysis of Cryptic A script (compared to 4Q298 and especially to 4Q249 and 4Q317) and the \textsuperscript{14}C dating of 4Q317~\cite{Jull1995}. 

4Q259 has a 2$\sigma$ calibrated range of 210–100 BCE (69.7\%) and a minor peak of 70–55 BCE (1.4\%). The 2$\sigma$ calibrated range of 210–100 BCE agrees with the two older palaeographic estimates of Milik and Puech, whereas the minor peak of 70–55 BCE is nearer to Cross’s estimate. The bimodal evidence for 4Q259 shows an older 2$\sigma$ calibrated range of 350–310 BCE (24.3\%), but, as for 4Q47, this older peak can be rejected as possible solutions based on typological comparison with date-bearing Aramaic manuscripts from the fourth century BCE. 

Following Cross’s typology, Puech~\cite{Puech1998B} analysed 4Q521 as a Hasmonaean formal script and estimated it between 100–80 BCE. This manuscript was also radiocarbon dated in the 1990s~\cite{Jull1995}. That BP date (1984 ± 33) now has to be recalibrated according to the IntCal20 calibration curve (\cite{reimer2020intcal20}), which results in a 2$\sigma$ date range of 45 BCE–120 CE. According to the bimodal evidence of our study, the younger 2$\sigma$ calibrated range is 230–100 BCE (57.5\%), while the older peak in the 2$\sigma$ range of 355–285 BCE (38.0\%) can be rejected as a possible solution due to comparative typological evidence from date-bearing Aramaic manuscripts from that period. The difference in age between the two radiocarbon tests may be due to the Soxhlet procedure cleaning castor oil from the sample, but it is not possible to quantify or ascertain that. The palaeographic estimate of 100–80 BCE and our 2$\sigma$ calibrated range of 230–100 BCE connect in the year 100 BCE. So, considering measurement uncertainties, 4Q521 can be taken as a partial overlap. 

The script of 5/6Hev1b was considered by Cross~\cite{Cross2003} to be a post-Herodian formal, estimated from 75–100 CE (Flint~\cite{DJD38} suggested 50–68 CE). This sample was the least precise \textsuperscript{14}C result in our study, with a standard deviation of 28 years in BP, and calibrated in 2$\sigma$ to 10–205 CE. The large calibrated date range, caused by the shape of the calibration curve in this period, clearly encompasses previous palaeographic estimates, but also moves in both a much older and a much younger direction of possible dates. 

\subsubsection{No overlap}\label{appen:B8.2:nooverlap}
Nine out of 26 samples yield (accepted) 2$\sigma$ calibrated ages that do not overlap with previous palaeographic estimates. In all 9 cases, the \textsuperscript{14}C results give calibrated age ranges that are older than previous palaeographic estimates. Yet, in light of our critical assessment, the older \textsuperscript{14}C age ranges are in most cases also palaeographically possible and realistic. This applies to: 4Q2, 4Q3, 4Q27, 4Q30, 4Q114, 4Q206, 4Q267, 4Q375, 4Q416. 

4Q30 was Cross’s “typical Hasmonaean” script specimen from the middle of the period, 125–100 BCE, like 1QIsa\textsuperscript{a}~\cite{Cross2003}. The calibrated result for this sample in our study is bimodal. According to the \textsuperscript{14}C measurement, 4Q30 has a 2$\sigma$ calibrated range of 235–165 BCE (36.7\%) and a minor peak of 260–245 BCE (1.4\%). The older 2$\sigma$ peak of 360–275 BCE (57.4\%) can be rejected as a possible solution based on palaeographic comparison with date-bearing manuscripts in Aramaic script from the period. Though Cross gave the more narrow estimate from 125–100 BCE, White Crawford estimated more broadly from 150–100 BCE~\cite{DJD14}. An earlier date range, say in the first half of the second century BCE, as indicated by \textsuperscript{14}C, is realistic and possible. In general, there is no reason to chronologically limit the script identified as Hasmonaean to the upper range of the political-historical period of the same name in the mid-second century BCE (see Section~\ref{appen:A2.2:unsub}). The sequence of relative typology can chronologically easily be moved to an older age range. Though in general 4Q30 shows a more uniform letter size, at the level of individual letters, the often not yet ‘standard’ letter size of \textit{aleph} and the often small \textit{ayin} point to earlier typology in the Hasmonaean script. As we also argued for, for example, 4Q47 and 4Q176 (Section~\ref{appen:B8.1:overlap}), 4Q30 shows mixed typological evidence. 

4Q27 was Cross’s “typical exemplar of the extremely popular Round semiformal style”, initially estimated by him to be early Herodian (ca. 30 BCE–20 CE) but later slightly revised by Jastram and Cross to the latter half of the first century BCE~\cite{DJD12}. 4Q27 has a 2$\sigma$ calibrated range of 200–50 BCE (94.2\%) and a minor peak of 340–330 BCE (1.3\%) that can be rejected as a possible solution for palaeographic reasons. The 2$\sigma$ calibrated range of 200–50 BCE comes near the revised palaeographic estimate. The calibrated date has a large range. This is caused by the measurement’s standard deviation of 26 \textsuperscript{14}C years (BP) in combination with the shape of the calibration curve in this period. 

Interestingly, the 2$\sigma$ calibrated range for another specimen of the Herodian round semiformal, 4Q161, is 55 BCE–30 CE (92.1\%), with two minor peaks of 90–80 BCE (1.7\%) and 45–60 CE (1.7\%). This may suggest a longer and somewhat older age range for this Herodian-type script than only the latter half of the first century BCE. Palaeographically, there are also many differences between 4Q27 and 4Q161. In 4Q161 the long extending base strokes of \textit{kaph}, broad \textit{dalet}, ligatures, and strikingly penned \textit{tet} and \textit{shin} stand out, whereas 4Q27 shows less tendency to broadening of letters. Although possible from a \textsuperscript{14}C perspective, a date in the first half of the second century BCE for 4Q27 seems unlikely from a typological perspective in comparison with other manuscripts. Nonetheless, there are four more Herodian-type manuscripts dated to that range by \textsuperscript{14}C in our study: 4Q3, 4Q267, 4Q375, and 4Q416. 

4Q267 is another example of Cross’s early Herodian round semiformal. Yardeni related 4Q267 to 4Q397 as possibly written by the same scribe and estimated it from 30 BCE–20 CE~\cite{DJD18} (Yardeni did not take over Cross’s round semiformal categorization and understood its script as formal). 4Q267 was also radiocarbon dated in the 1990s~\cite{Jull1995}. That BP date (2094 ± 29) now has to be recalibrated according to the IntCal20 calibration curve (\cite{reimer2020intcal20}), which results in a 2$\sigma$ calibrated range of 200–40 BCE (94.0\%) and a minor peak of 10 BCE–5 CE (1.5\%). According to our study, 4Q267 has a 2$\sigma$ calibrated range of 210–95 BCE (65.3\%) and a minor peak of 70–55 BCE (1.6\%), whereas the older 2$\sigma$ range of 355–290 BCE (28.6\%) can be rejected as a possible solution due to comparative typological evidence from date-bearing Aramaic manuscripts from that period. The difference in age between the two radiocarbon tests may by due to the Soxhlet procedure cleaning castor oil from the sample, but it is not possible to quantify or ascertain that. 

From a typological perspective it is difficult to understand the script of 4Q267 being chronologically so near to quite different typological specimens in the second century BCE. We may have to reckon with overlapping or partly adjacent style developments but in this case it would severely impact the relative typology dominant in the field, not just moving it chronologically and keeping the relative typology intact. 

4Q267 might be an outlier, yet this \textsuperscript{14}C result raises the fundamental issue of how the absolute, chronological dating of typological differences in a linear sequence has been substantiated. Cross assumed a slow development of the Aramaic/Hebrew script in the third century BCE and he assumed a rapid evolution of the script in the Hasmonaean and Herodian eras, but he could not substantiate either assumption, due to the lack of date-bearing documents. He assumed but did not demonstrate that the finer typological distinctions had to be chronologically sequenced one after the other instead of existing partially next to each other (see, similarly,~\cite{Sirat1986}).

Cross wavered with his palaeographic estimate of the ‘semicursive’ 4Q114 from the late second century BCE (125–100) to ca. 100–50 BCE, and, under influence of the finds of Wadi Daliyeh, back to the late second century BCE~\cite{DJD16}, “no more than about a half century younger than the autograph”, Cross said~\cite{Cross1961B}. Interestingly, Cross dated 4Q114 contemporary to the formal hand of 4Q30. 4Q114 preserves Daniel 8–11, a part of the book which scholars argue on literary-historical grounds to have been composed in the 160s BCE. 4Q114 has a 2$\sigma$ calibrated range of 230–160 (45.9\%) and an older 2$\sigma$ range of 355–285 (49.5\%) that can be rejected as a possible solution based on comparative typological evidence from date-bearing Aramaic manuscripts from that period. 

Because of its scribal errors, it is unlikely that the scribe of 4Q114 was the author. But the early date and low scribal quality of 4Q114 shed new light on the production and circulation of literature in ancient Judaea: its date is indicative for the speed of the text’s spread, and the low quality of the manuscript may indicate it originated in a social context close to the original author \cite{Popovic2023}; future research may further validate this. 4Q114 would then have been copied very soon after the assumed composition of Daniel 8–11. The \textsuperscript{14}C 2$\sigma$ date of 230–160 BCE for 4Q114 is matched by a very much comparable older \textsuperscript{14}C date of 4Q30. 

For 4Q206, Milik~\cite{Milik1976} gave an estimate from the first half of the first century BCE, and simply referred to four of the exemplary Hasmonaean manuscripts given by Cross (4Q30, 4Q51, 4Q114, 4Q398), apparently with no concern for their differences in style and for Cross dating these quite differently. In his recent edition in consultation with Puech, Drawnel~\cite{Drawnel2019} estimated 4Q206 to be from the middle of the first century BCE. It is interesting that two of Milik’s typological comparanda, 4Q30 and 4Q114, have \textsuperscript{14}C results in our study similar to 4Q206: the 2$\sigma$ calibrated range for 4Q206 is 235–145 BCE (45.8\%) with a minor peak of 135–120 BCE (1.1\%); the older 2$\sigma$ range of 360–280 BCE (48.6\%) can be rejected as a possible solution for palaeographic reasons. In each of these cases the \textsuperscript{14}C results indicate an earlier chronological date than the palaeographic estimates. But typologically some letters are slightly different and commonly seen as a later development of the letter form, e.g., \textit{bet}, \textit{mem}, and \textit{ayin}. Yet, other letters show varied forms within 4Q206 and some compare well with instances from 4Q30, e.g., \textit{aleph}, \textit{he}. So 4Q206 may be another example of mixed typological evidence.

Then there are four Herodian-type manuscripts whose \textsuperscript{14}C dates extend into the second century BCE: 4Q2, 4Q3 and 4Q375 and 4Q416.

The script of 4Q2 has been described as late Herodian or even post-Herodian (ca. 50–68+ CE), in part because of the increasing use of \textit{keraiai}~\cite{DJD12}. While the script is typologically certainly Herodian, the assumption that calligraphic features are typical for its latest period cannot be substantiated. 4Q2 has a 2$\sigma$ calibrated range of 125 BCE–10 CE (90.3\%) with a minor peak of 155–130 BCE (5.2\%), providing a date range up to 10 CE, which seems realistic to us. 

The case of 4Q3 is more difficult. Its script was tersely described as “an Herodian formal hand dating from the middle to end of that period (c. 20-68 CE)”~\cite{DJD12}. Indeed, the script of 4Q3 features several letters and elements generally regarded to be developed Herodian, like the small tick above the crossbar of the final \textit{mem}. Yet, some letters have older shapes which are uncommon in those developed Herodian formal hands, such as the ‘horned’ \textit{dalet}. 4Q3 has a 2$\sigma$ calibrated range of 200–50 BCE (92.0\%) and a minor peak of 340–325 BCE (3.5\%) that can be rejected as a possible solution for palaeographic reasons. The palaeographic rule of thumb that the latest forms are indicative for its age would militate against the 2$\sigma$ range of 200–50 BCE. Yet, the exact moment when those latest forms have arisen has not been substantiated in the field. Future evidence may further validate this.

Strugnell provided a judicious analysis of the palaeography of 4Q375, comparing its style to that of the round or rustic semiformal series which is generally associated with early Herodian, but also arguing that, typologically, it must be an early exemplar since some letters do not yet have the typically Herodian forms~\cite{DJD19}. True to his custom, he did not translate this typological assessment into a calendar date, but in Cross’s correspondence between hand and style this would amount to ca. 50–25 BCE, which would nearly agree with the 2$\sigma$ calibrated range of 205–50 BCE (89.5\%). Considering the uncertainties in the palaeographic estimate, this is acceptable. The 2$\sigma$ peak of 340–320 BCE (6.0\%) can be rejected as a possible solution for palaeographic reasons. 

Also for 4Q416, Strugnell carefully analysed its individual letters, arguing that in most cases these should be placed between 4Q51 and 1QM, hence “in a date transitional between the late Hasmonaean and the earliest Herodian hands”~\cite{DJD34}. He judged the script of 4Q416 to be earlier than those of 4Q415, 4Q417, and 4Q418 by some twenty-five years so that a palaeographic estimate of 50–25 BCE presents itself. 4Q416 has a 2$\sigma$ calibrated range of 205–90 BCE (78.1\%) and a smaller peak of 80–50 BCE (9.4\%). Considering the uncertainties in the palaeographic estimate, this is acceptable. The 2$\sigma$ peak of 345–320 BCE (8.0\%) can be rejected as a possible solution for palaeographic reasons.

\subsubsection{Concluding the comparison between radiocarbon results and palaeographic estimates}\label{appen:B8.3:conclcomp}

Based on this comparison between (accepted) 2$\sigma$ calibrated dates and previous palaeographic estimates we make the following concluding observations.

Overall, the \textsuperscript{14}C results indicate older date ranges for individual manuscripts. Only two manuscripts, 4Q201 and 11Q5, have date ranges that go in the direction of a younger possible range (5/6Hev1b has a range both a bit older and much younger). Thus, Hasmonaean-type manuscripts have \textsuperscript{14}C date ranges that allow for older dates in the first half of the second century BCE, and sometimes also up to the latter part of the third century BCE, instead of the late second century or early first century BCE. There are no compelling palaeographic or historical reasons that preclude these older dates as reliable time markers for the Hasmonaean script (this also applies to the solid third-century BCE range for 4Q70 and its Archaic-type script). 

The \textsuperscript{14}C results for most manuscripts confirm the basic distinction between Hasmonaean-type manuscripts that are older, and Herodian-style manuscripts that are younger, and, for that matter, also between Archaic-type (4Q52 and 4Q70) and Hasmonaean-type manuscripts. However, the \textsuperscript{14}C date ranges for manuscripts that are traditionally considered Hasmonaean and Herodian are quite differently distributed across the timeline. 

As can be seen in Figure~\ref{fig:AIvsC14PAL} in the main article, the twelve Hasmonaean-type manuscripts in our sample set have (accepted) 2$\sigma$ calibrated date ranges from the second and first century BCE, as expected, and most extend also into the late third century BCE. Three Herodian-type manuscripts (4Q161, Mas1k, XHev/Se2) have (accepted) 2$\sigma$ calibrated date ranges from the latter half of the first century BCE and the first century CE, as expected. Two Herodian-type manuscripts have date ranges in the first century CE, as expected, but also extend into the second century CE (11Q5 and 5/6Hev1b, the latter even into the early third century CE). 4Q2 has a date range extending from the early first century CE back to the early second century BCE. And five Herodian-type manuscripts have (accepted) 2$\sigma$ calibrated date ranges in the second century BCE (4Q3, 4Q27, 4Q267, 4Q375, and 4Q416), though 4Q27 extends into the first century BCE. 

This adds a third component to our critical assessment. In addition to critiquing the application of traditional typology to individual manuscripts and dismantling unsubstantiated historical suppositions and chronological limitations, the results of this study also question the validity of the relative typology as such. The traditional relative typology can be maintained but not in all cases. The spread of the Hasmonaean-type manuscripts over the timeline does not affect Cross’s relative typology in a major way but the older, second-century BCE date ranges of the Herodian-type manuscripts do affect the relative typology potentially in a major way. 

Individual manuscripts frequently show mixed typological evidence: a manuscript can have different forms of individual letters that are considered ‘older’ or ‘younger’ according to the traditional palaeographic framework. There is, however, not a good method to assess what this means in terms of relative placement of the individual manuscript, nor, for that matter, what it means for relative typology in general. The rule of thumb that the typologically latest forms should determine one’s palaeographical estimate of a manuscript, presupposes existing palaeographical date markers and a decision which features are typologically important or indicative. The dated Wadi Daliyeh discoveries showed that features that were supposed to be significantly later, already appeared many decades earlier than previously expected. Moreover, it is assumed but not substantiated that typological differences must be translated to chronological linear sequences. Instead of a linear development, as Cross and others have assumed, the possibility of overlapping or partly adjacent style developments must be considered.

So, the so-called Hasmonaean script can indeed be regarded older than the so-called Herodian script but the \textsuperscript{14}C results of this study indicate that the Herodian script was present earlier than previously thought. This suggests that these scripts were not transitioning from the mid-first century BCE onward (the so-called late Hasmonaean/early Herodian category of manuscripts) but that much earlier they already existed partially next to each other. 

This study shows that there are no cogent reasons for limiting the palaeographic dating of style developments to political-historical periods such as Hasmonaean or Herodian. The terms ‘Hasmonaean’ and ‘Herodian’ might still be employed for types of script, but these cannot be converted to specific date ranges. For date estimations of individual manuscripts, one should rather use concrete age ranges.

\subsection{Combining palaeography and radiocarbon data to train the artificial intelligence-based date-prediction model}\label{appen:C2:combipalaeoradiocarbon}

In this study, we combine palaeography and radiocarbon dating to train our date-prediction model. It should be stressed that this means a combination of qualitative and quantitative approaches and methods. Palaeography is a qualitative approach, based on expert knowledge, which is similar to the role of, for example, epigraphy as expert knowledge in~\cite{assael2022restoring}. This is different from, for example, an archaeological or geological stratigraphy that often can be quantified on the timeline. From a palaeographic approach alone it is not possible to pinpoint an exact date, date range, or date limit on the timeline of the period of the Dead Sea Scrolls, because palaeographic dates are estimates and do not provide absolute or fixed dates. For example, in our research, palaeography tells us that most scrolls cannot date to the fourth century BCE but we cannot assign a limiting number like 300 BCE. Moreover, typological script development is a gradual process, not necessarily following a linear time trajectory. 

Most Dead Sea Scrolls are typologically younger when compared to the script in Aramaic date-bearing documents from the fourth century BCE. Therefore, when any knowledgeable palaeographer is presented with the bimodal evidence in the 2$\sigma$ range as is the case with our study, then they certainly will reject the older peak as a possible solution, as we have already explained (Sections~\ref{appen:B8.1:overlap} and~\ref{appen:B8.2:nooverlap}), a principle also tacitly applied by~\cite{Bonani1992,Jull1995}. Hence, in these cases of bimodal evidence typologically younger is also chronologically younger.

However, it is an open research question whether some of the oldest Dead Sea Scrolls might be older than their palaeographic estimated date in the mid-third century BCE. Following the palaeographic principle to compare the script of an undated manuscript to that of dated writings with a similar script (see Section~\ref{appen:A2.2:unsub} in Appendix~\ref{appen:A}), we have already argued that 4Q52 would have to be dated chronologically nearer to the Wadi Daliyeh manuscripts, especially WDSP 1 from 335 BCE (see Section~\ref{appen:B8.1:overlap}). This may also apply to 4Q70, which has 2$\sigma$ calibrated ranges of 320–200 BCE (79.2\%) and 375–345 BCE (16.3\%). The older 2$\sigma$ peak can be rejected as a possible solution based on typological comparison but the younger 2$\sigma$ peak’s extending into the fourth century BCE cannot be completely ruled out, although 4Q70 is typologically further removed from the date-bearing Aramaic manuscripts from the fourth century BCE than 4Q52 and a date in the third century BCE for 4Q70 is more likely from a palaeographic perspective. However, such a qualitative assessment cannot be characterised as a specific quantitative prior (an expected date with mean and standard deviation) in the timeline. Therefore, palaeographers cannot give an exact date such as 280, 300, or 320 BCE as the year before which none of the Dead Sea Scrolls can be dated.

In order to train our artificial intelligence-based date-prediction model, we use the accepted 2$\sigma$ calibrated data from 24 of the 26 accepted \textsuperscript{14}C results (Table~\ref{tab:summarized-c14}). For the training of Enoch, the data from two manuscript samples are not used: Mur19 and 4Q52. Because of its cursive script, the papyrus fragment Mur19 is, at the moment, not relevant for Enoch. We also leave 4Q52 out of consideration because, in this case, we cannot decide between the two peaks in the probability distribution (see Section~\ref{appen:B8.1:overlap}). This is why we work with the tentative addition or deletion of 4Q52 in the training of our algorithm. This is how we get from 26 accepted \textsuperscript{14}C results to 24 manuscripts used as the primary training set for Enoch.

%% file: tex-sections/E-ai.tex
\section{Artificial intelligence (AI) in dating the scrolls}\label{appen:C}

In this project, we do use deep learning for image processing (binarisation) but have refrained from properly using it for the date prediction. See Appendix~\ref{appen:J:transferdeeplearningillustration}, explaining the objections to the use of deep learning for the date prediction, including an analysis of an experiment we executed, using transfer learning starting with a state-of-the-art foundational deep-learning model.

\subsection{Data preparation}
\begin{figure}[!ht]
    \centering
    \includegraphics[width=.8\textwidth]{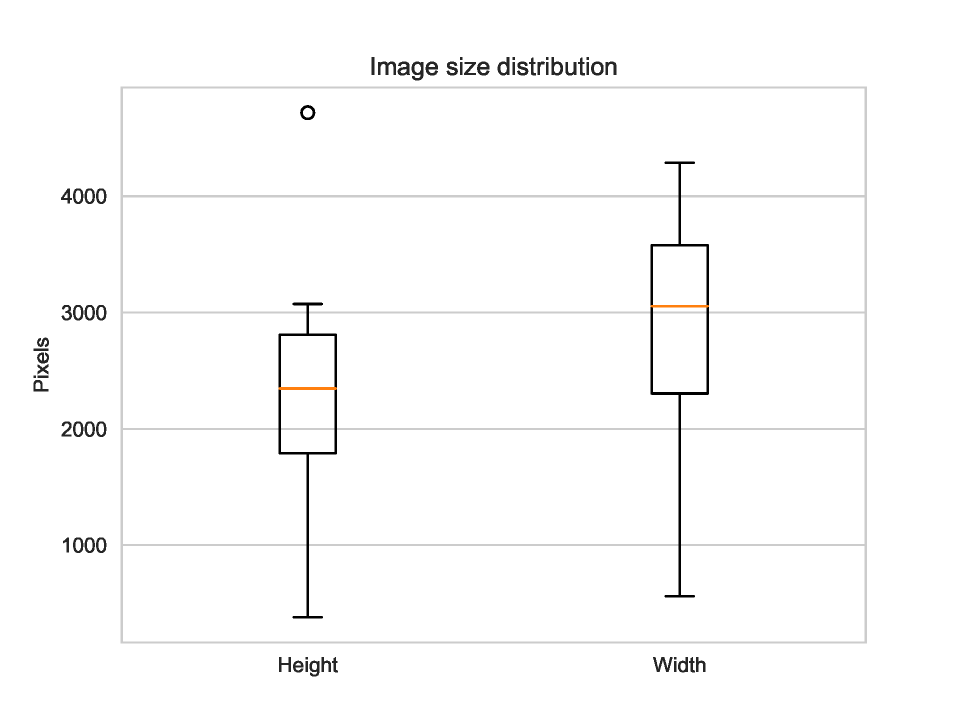}
	\caption{Box-plot showing the height and width spreads of the 62 training images.}
	\label{fig:happen:C:trainddist}
\end{figure}

Our first step is to collect and prepare the data for the date prediction model. We collect the images of the manuscripts for each of the \textsuperscript{14}C samples with accepted dates. We have used 24 manuscripts as the primary training set for our date-prediction model (a complete list can be found in Appendix~\ref{appen:D} in the supplementary materials). The physical 24 radiocarbon-dated manuscripts are visually spread out on many individual fragment images of the IAA’s Leon Levy Dead Sea Scrolls Digital Library collection~\cite{dssllweb}. In addition to this primary training set, we have created different combinations of training data to perform comparative analyses and further check the robustness of the model (see Subsection~\ref{appen:C:training} for details). We obtained a data set of 75 images from the 24 radiocarbon-dated manuscripts. We use 62 of these images to train our model (Figure~\ref{fig:happen:C:trainddist} shows the size distribution of the training images after the preprocessing steps). The remaining 13 images, chosen deliberately and randomly, are passed as unseen test data to validate the robustness and reliability of the model’s performance. We also select a large number of images to perform tests on the date prediction model. Once the images are selected, we start with the preprocessing task, where we use BiNet, the neural network architecture, to extract the characters. The binarization, along with alignment correction and fragment arrangement, provides better-quality images (see Subsection~\ref{appen:C:binarize}). It is extremely important to obtain the highest quality of binarized images. This is because the image quality determines the success of the feature computation and the ultimate date regression model.

\subsubsection{Preprocessing: binarization, alignment and arrangement correction} \label{appen:C:binarize}
\begin{figure}[!ht]
	\centering
	\includegraphics[width=\textwidth]{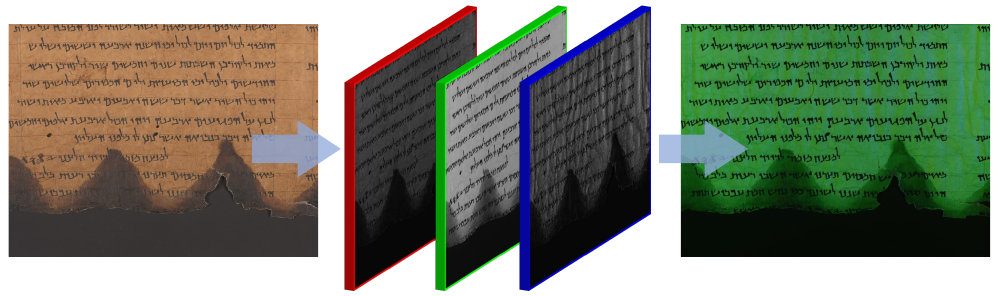}
	\caption{An illustration of creating a pseudo-colour image. The first image from the left is the full spectrum colour image of 11Q5 from plate $974$. The next three are the band images (for the formation of three channels using multispectral images) with wavelengths of $595nm$, $924nm$, and $638nm$, respectively. On the right side is the resultant pseudo-colour image (fused image).}
	\label{fig:expFused}
\end{figure}

\begin{figure*}[!ht]
	\centering
	\resizebox{\textwidth}{!}{%
		\includegraphics[width=\textwidth]{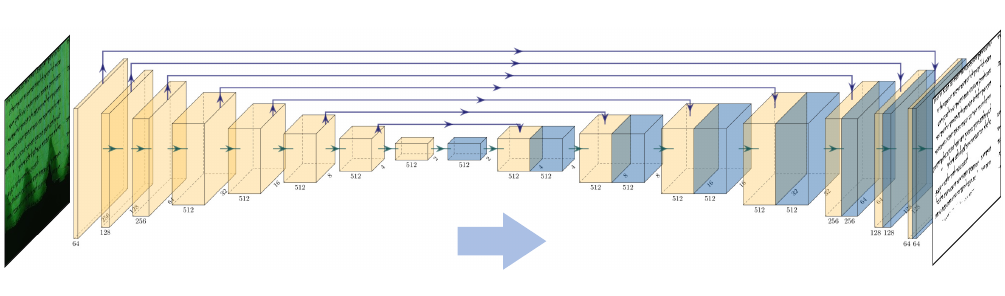}}
	\caption{The BiNet architecture shows the encoder (contracting path) at the left half and the decoder (expanding path) at the right half of the image. Each step in the decoder part receives a concatenation with the corresponding feature map from the encoder part through the skip connections. This concatenation circumvents the bottleneck issue at the deepest layers of the encoder and ensures the precise localization of the foreground-background pixels. The example shows a pseudo-colour image of 11Q5 (Plate 974) as input on the left and the output binarized image on the right.}
	\label{fig:binet1}
\end{figure*}

We start with the multispectral band images for each fragment to use an image fusion technique designed in-house~\cite{dhali2019binet} to create pseudo-colour images from a weighted combination of the band images (see Figure~\ref{fig:expFused}). The resultant pseudo-colour images offer high contrast and facilitate better separation of ink from backgrounds, a task commonly known as binarization. Both training and test images go through the same preprocessing techniques. It is important to note that although many modern deep-learning methods can be trained directly using the colour/grayscale without binarization, this approach is not suitable for dating the scrolls. Direct end-to-end solutions, i.e., classification or clustering of the training images with testing images, may seem feasible, but there is a risk of obtaining completely inaccurate results. Artificial neural networks, for instance, may make decisions based on superficial correlations with the texture of the parchment, leading to erroneous outcomes. Therefore, isolating only the ink traces (foreground) and excluding any other material features in the images (background) is crucial. BiNet is a deep-learning-based method specially designed to binarize scroll images. Instead of using a simple filtering technique, BiNet uses a neural network architecture for the binarization task and therefore yields better output \cite{dhali2019binet} (see Figure~\ref{fig:binet1}). 

\begin{figure}[!ht]
    \centering
    \includegraphics[width=\textwidth]{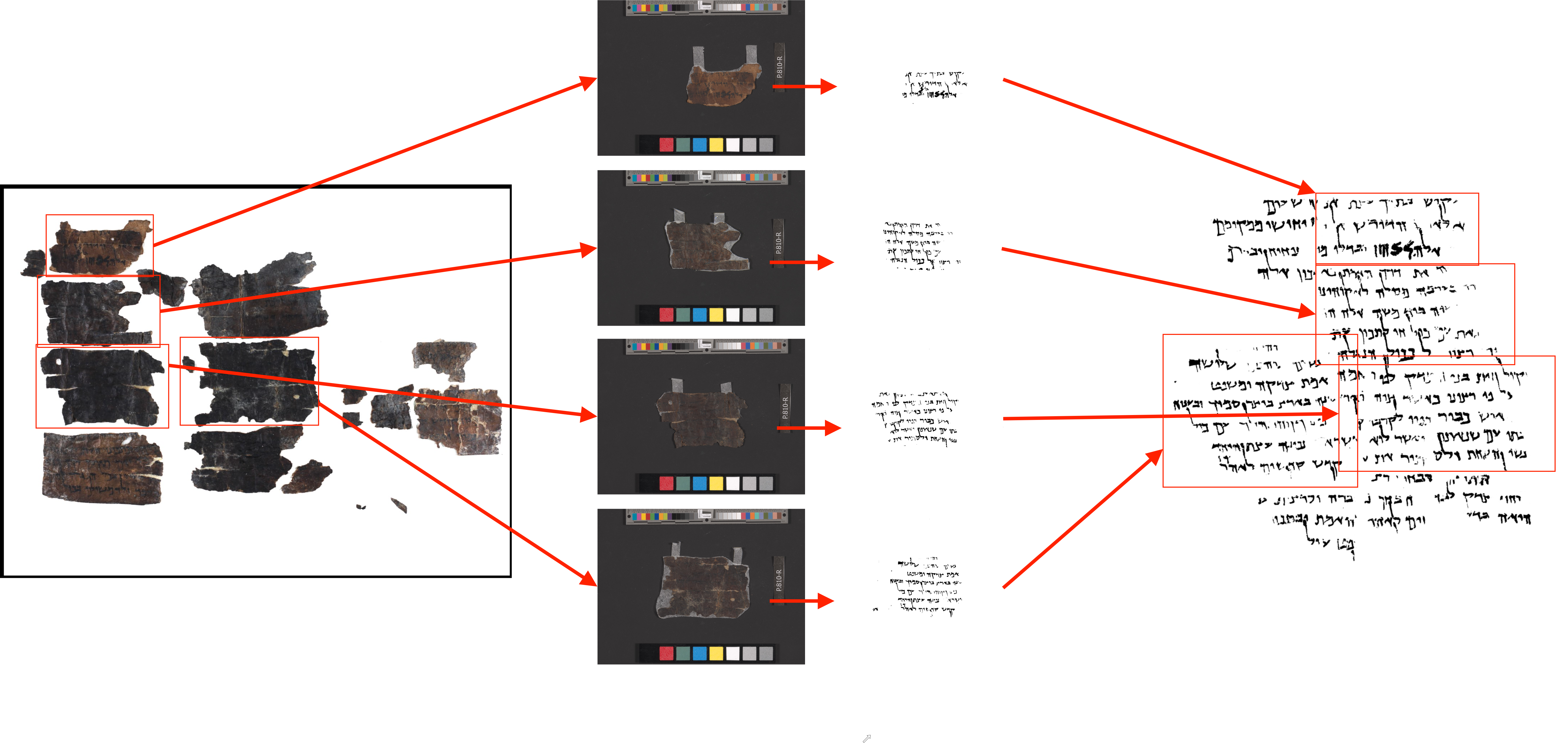}
    \caption{An example of image preparation for 4Q319: a full plate image from IAA is shown on the left side (Plate 810). Then in the middle column, four different fragment images (full spectrum colour images) and their binarized outputs are presented. Finally, further cleaning, alignment correction, and arrangement are performed to produce the final image of 4Q319 on the right.}
    \label{fig:4Q319proccess}
\end{figure}

\begin{figure}[!ht]
    \centering
    \includegraphics[width=0.8\textwidth]{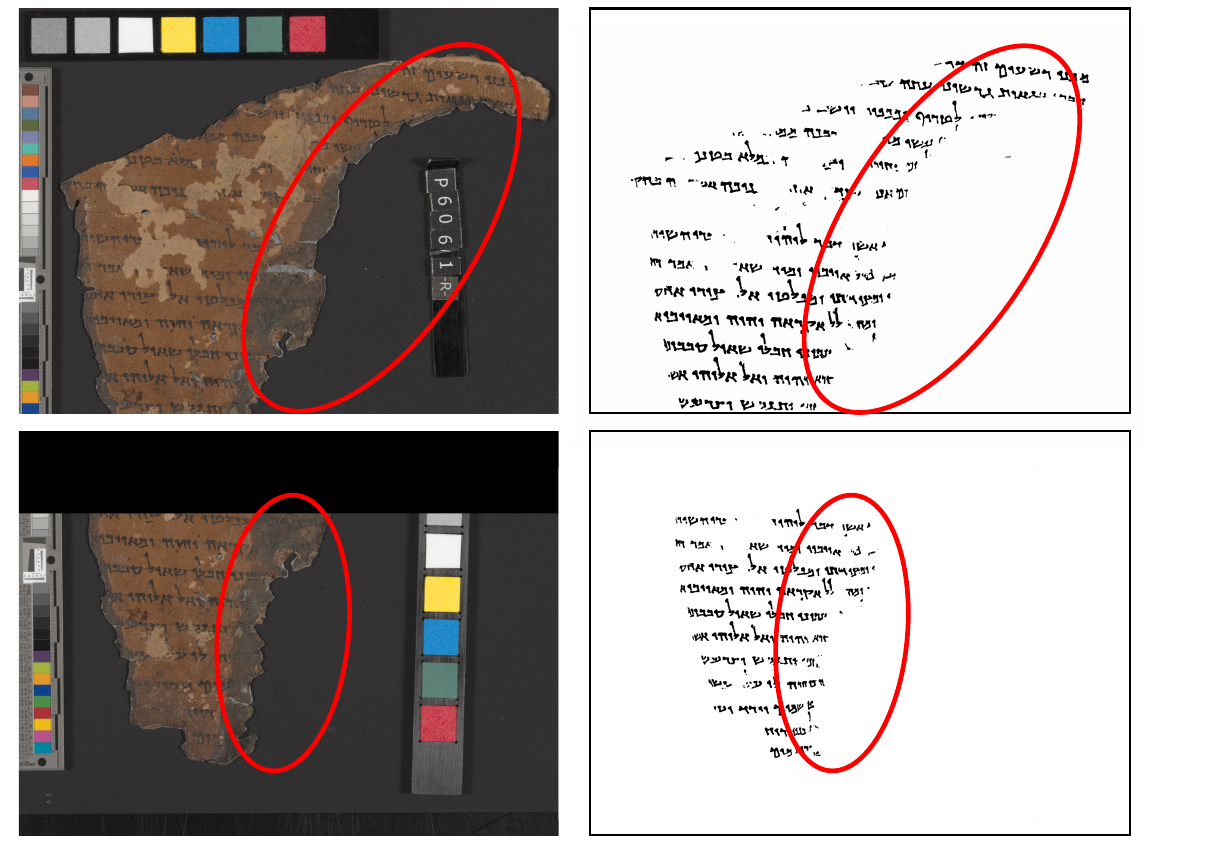}
    \caption{Characters are deformed (marked in red) near the edges of the physical fragments from one of the test manuscripts, 11Q7 (IAA plate 606-1)—these deformities in the binarized images (with slanted or skewed characters) affect the textural and allographic feature calculations.}
    \label{fig:11q7issues}
\end{figure}

Once the binarization process is complete, additional cleaning of the images is performed. This cleaning step aims to remove any extra noise or speckles that were not completely removed by the binarization technique. This is a crucial procedure to ensure that features are extracted only from the characters of each image. Subsequently, rotation and alignment correction are also performed. If the images are rotated at some angle to the horizontal axis, it can affect feature calculations that rely on rotation invariance. Therefore, rotation correction is applied to align the text lines horizontally. In some cases, a minor affine transformation and stretching correction are executed in a selective manner. These corrections are specifically intended to align the twisted text lines caused by the degradation of the parchment. In many cases, one manuscript contains multiple fragments. In these cases, we put the fragments together and arrange them into a single image (see Figure~\ref{fig:4Q319proccess}). The GIMP tool, a free and open-source graphics editor, is used for rotation and arrangement correction \cite{gimp}. It is important to note that the alignment and arrangement corrections are mostly done with the training images to obtain accurate feature extractions for the style periods represented by those images. However, these corrections are only done for some of the test images due to the limitation of time and resources. Most test images are used directly after binarization, sometimes leading to an unrealistic prediction due to damaged and deformed characters (see Figure~\ref{fig:11q7issues}). If any test images need special attention in the future, extra steps can be performed to obtain a better image for a better prediction from Enoch. 

\subsection{Data augmentation}
We have a very limited number of radiocarbon-dated manuscripts from which we derive the training images. During the writing process of any document, writers naturally introduce variability even within the same time period. In order to address both issues of data scarcity and writing variations within a period, we perform data augmentation by introducing acceptable variation to the data. The small random shape perturbations will, on the one hand, ensure the system's robustness and, on the other hand, consider variations of writing styles within a particular period.
In machine learning, augmentation is an often-used method to counteract the effects of lack of data and imbalance in sampling~\cite{MUMUNI2022augmentation}. We augment training and testing data by generating synthetic images using random geometric distortions \cite{bulacu2009morph}. 

\begin{figure}[!ht]
    \centering
    \includegraphics[width=\textwidth]{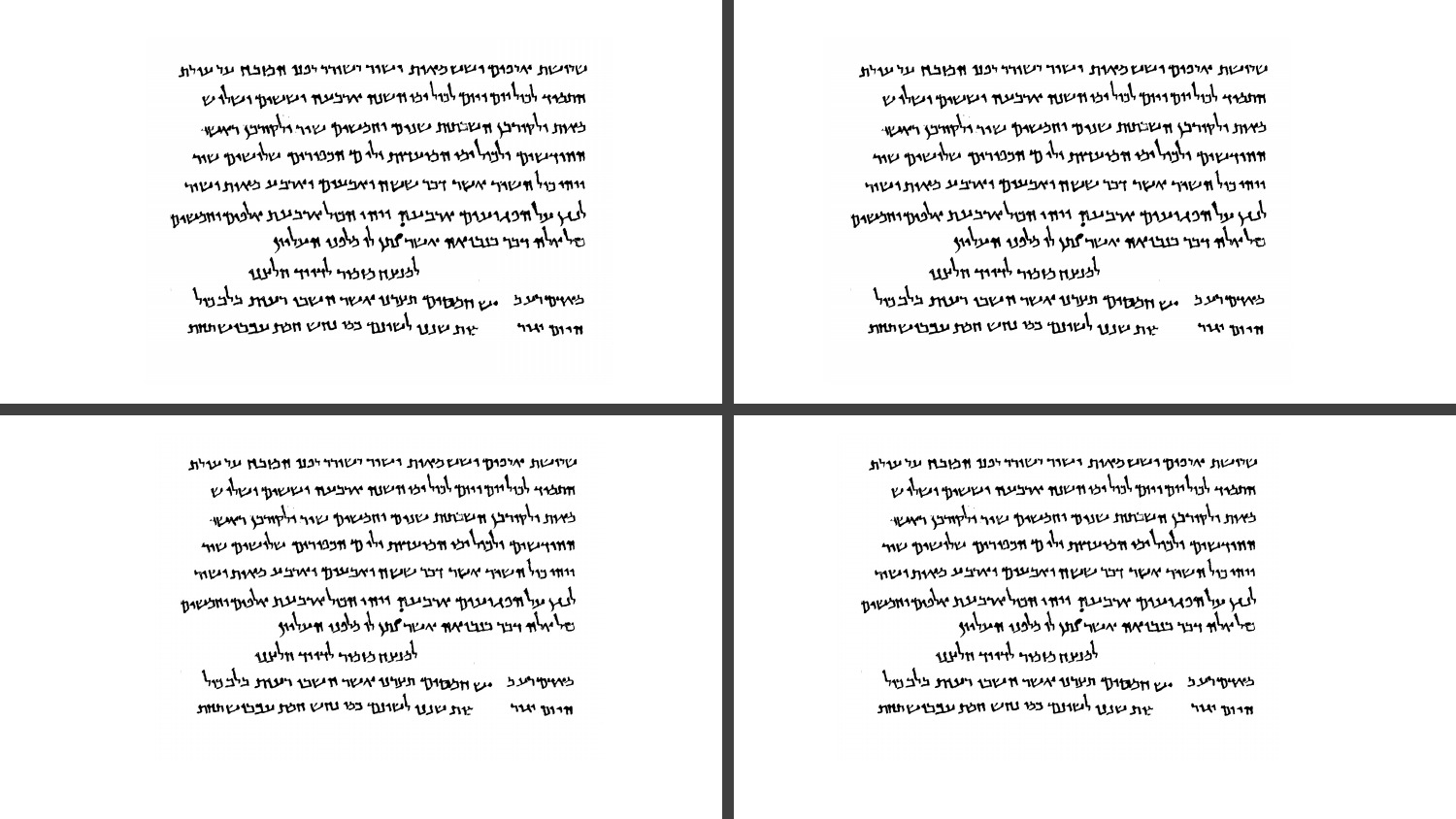}
  \caption{The original binarized image of 11Q5 (Plate 974) on top left and three randomly augmented morphed images. A close inspection of the images shows small geometric distortion introduced to the characters using the elastic-morphing technique.}
	\label{fig:supp:morphed}
\end{figure}

We perform data augmentation using applying random elastic `rubber-sheet' transforms. For each pixel $(i, j)$ of the column images, a random displacement vector $(\Delta x, \Delta y)$ is generated. The complete image's displacement field is smoothed using a Gaussian convolution kernel with a standard deviation $\sigma$. We then rescale the field to an average amplitude $A$. The new morphed image $(i' , j')$ is generated using the displacement field and bilinear interpolation:
\begin{equation}
i' = i + \Delta x, j' = j + \Delta y.    
\end{equation}
Two parameters control this morphing process: the smoothing radius $\sigma$ and the average pixel displacement $A$. Both parameters are measured in units of pixels. In our experiment, we empirically chose a displacement value of $1.0$ and a smoothing radius of $8.0$ (see Figure~\ref{fig:supp:morphed}).

\subsection{Allographic codebook with neural networks}
\begin{figure}[!ht]
    \centering
    \includegraphics[width=.8\textwidth]{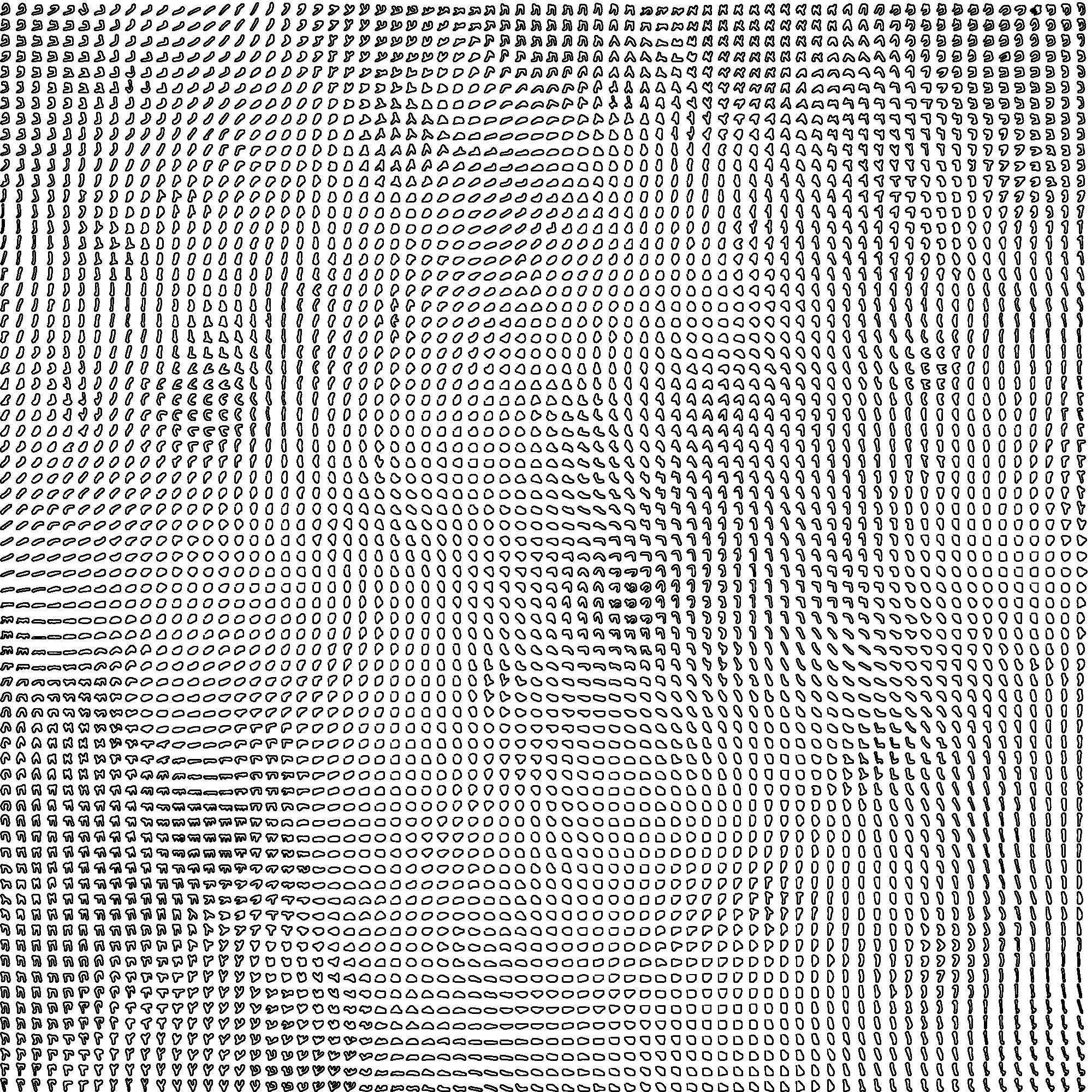}
	\caption{A visualization of 70x70 Kohonen map of fragmented connected components (200 x,y points per contour centroid) from the Dead Sea Scrolls collection. Image adapted from \cite{PlosOne}.}
	\label{fig:app:allo}
\end{figure}

After binarization with BiNet~\cite{dhali2019binet}, connected components of ink were fragmented on Y-minima, to prevent large blobs of multi-character components, yielding ‘fraglets’. For each fraglet, the contour curve was determined, running over the edge of a connected component in a counter-clockwise manner. Each contour-pixel sequence is ‘time’ normalized to 200 samples, (cosine, sine) pairs, yielding a feature vector of 400 values. Using the Kohonen~\cite{Kohonen1982} self-organizing map neural network, codebooks of $70\times70$ and $80\times80$ prototypical contours were computed~\cite{PlosOne} (see Figure~\ref{fig:app:allo}). As a proof of concept, $590$ manuscripts from the Dead Sea Scrolls collection were manually labeled as ‘Hasmonaean’ ($Nhas=307$) or ‘Herodian’ ($Nher=283$) by a palaeographer. During {\em training} on half of the data, each codebook element obtained the counts for its occurrence in ‘Hasmonaean’ or ‘Herodian’ manuscripts, respectively. During {\em testing}, each manuscript is characterized by its relative occurrence of Herodian-like vs. Hasmonaean-like fraglets. Using the 80x80 map and applying a linear SVM~\cite{SVMlite} on this 2D feature representation, a classification accuracy of 93 ($\pm$ 2.3\%) was obtained, computed over 20 random odd/even splits of the 590 manuscripts. Individual accuracy test results: 90.9, 89.2, 93.9, 91.2, 94.2, 90.2, 93.2, 94.2, 91.2, 95.3,	92.9, 96.3, 94.6, 92.2,	95.3, 92.5, 96.6, 88.8, 94.9, 93.2 (\%). On the basis of this pilot experiment and earlier work~\cite{PlosOne}, the allographic feature was deemed a usable candidate for the more fine-grained manuscript-dating algorithm using the carbon-dated training samples.

\subsection{Textural-level features}
\begin{figure}[!ht]
    \centering
    \includegraphics[width=.6\textwidth]{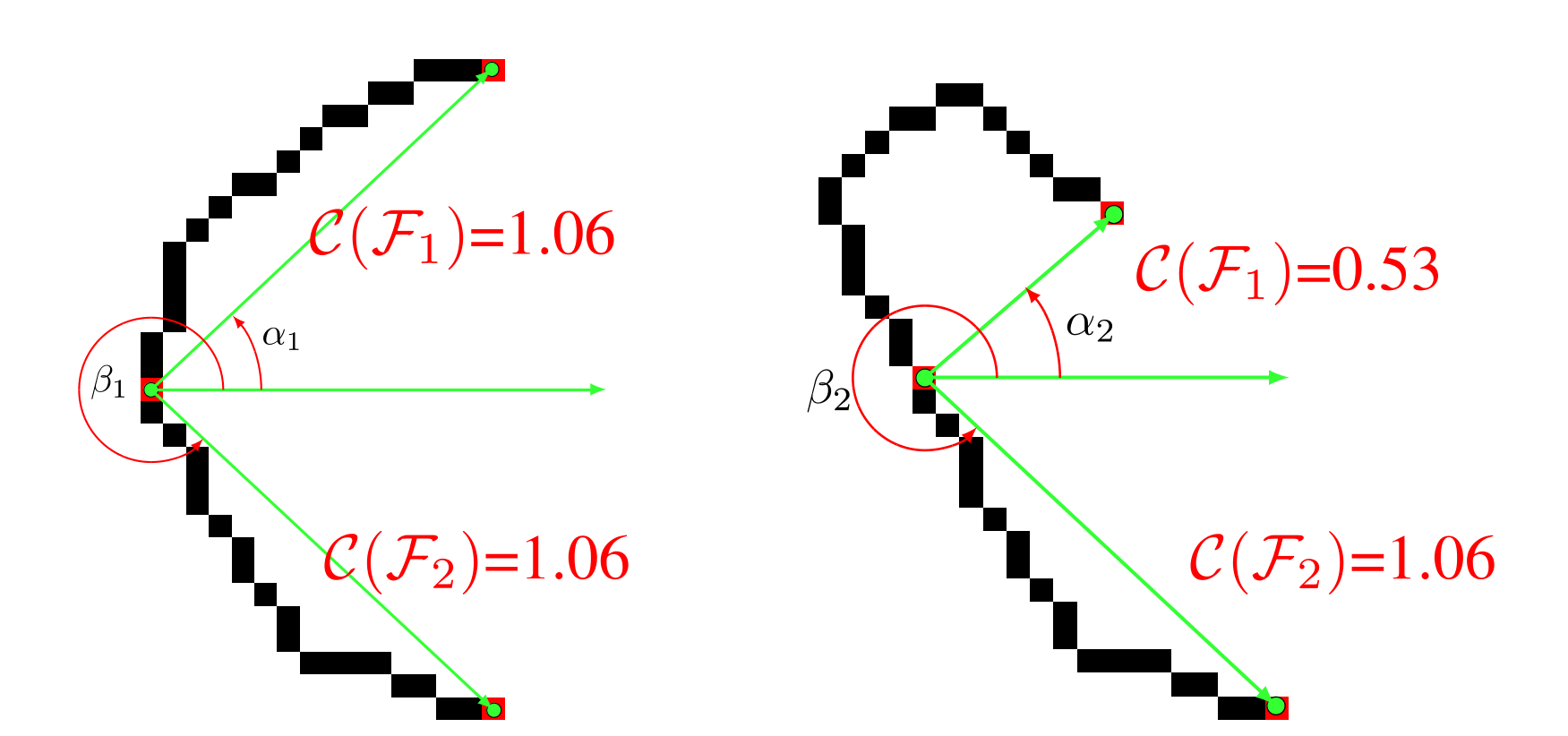}
	\caption{Hinge kernel; the angles and leg-lengths for two different character shapes. Image adapted from \cite{PlosOne}.}
	\label{fig:app:hinge}
\end{figure}

Similarly, the ‘Hinge’ feature~\cite{Bulacu2007, PlosOne} was chosen because of its ability to capture curvature-related differences between different samples of handwriting (see Figure~\ref{fig:app:hinge}). It addresses the occurrence of different degrees of roundness or sharpness of the path described by the edge between ink traces and paper. Its ability to classify between ‘Hasmonean’ and ‘Herodian’ styles is less powerful than in the case of allographic fraglets. Using the nearest mean and the Chi-square distance on a 195-dim hinge feature delivers 63.5\% accuracy ($\pm$2.9\%). This dimensionality is still high, and collinearity problems due to feature correlation need to be avoided. Subsequently using PCA, selecting the 15 largest eigenvectors and applying a linear SVM for this binary classification task yields 73.1\% accuracy ($\pm$ 0.24\%). Still, on the basis of the complementary nature of the allographic and textural feature methods, it was decided to include the Hinge feature for the manuscript dating problem.

\subsection{Adjoined feature}
As shown in~\cite{bulacu2006}, the combination of a fraglet codebook and the hinge feature proved to be very effective in writer identification. The assumption in the current study is that different historical style periods are revealed by the statistical characteristics both of allographic shape fragments and of angular distributions. Consider, for instance, a manuscript with predominantly vertical and horizontal strokes (‘formal’) and a manuscript written in a more informal (‘cursive’) style, both containing their characteristic shape elements in individual characters. The use of the two feature methods together will capture the underlying shape differences. The feature combination is realized by an adjoining of the two feature vectors: the arrays of feature values are combined in a single array containing the combined descriptor. Adjoined features are the weighted combination of both Hinge and Fraglet. The adjoining results in a feature vector of $5365$ dimensions, preserving the handwriting style description from both feature levels.

\input{tex-sections/E-ai2} 

\subsection{Data balancing} \label{subsec:data-balance}
\begin{figure}[!ht]
    \centering
    \includegraphics[width=\textwidth]{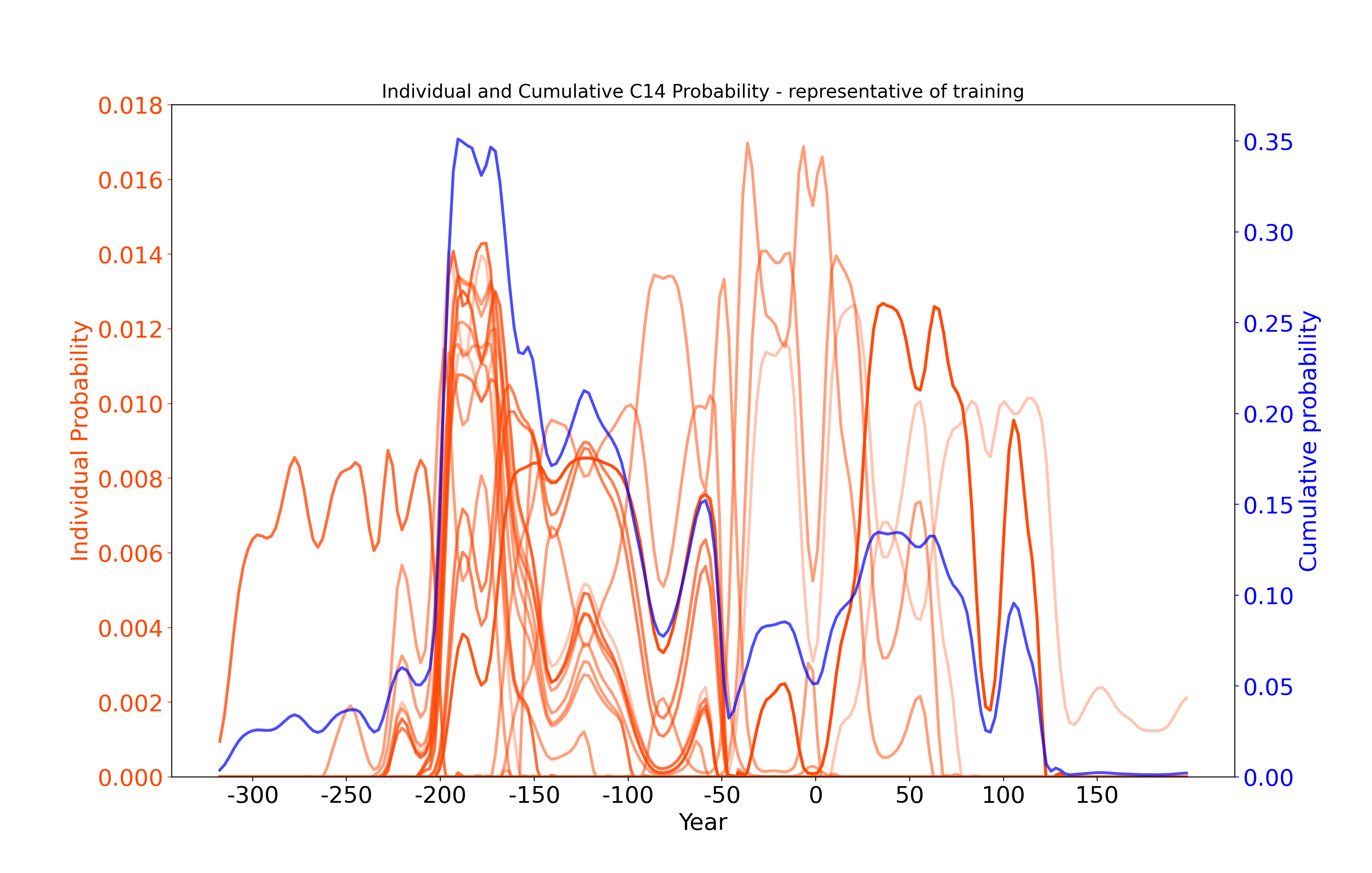}
	\caption{Distributions in the training data (orange) and total accumulated distribution of the C14 training data (blue), \textbf{including} all minor peaks of the (accepted) 2-sigma ranges.}
	\label{fig:cum_c14-wMinor}
\end{figure}

\begin{figure}[!ht]
    \centering
    \includegraphics[width=\textwidth]{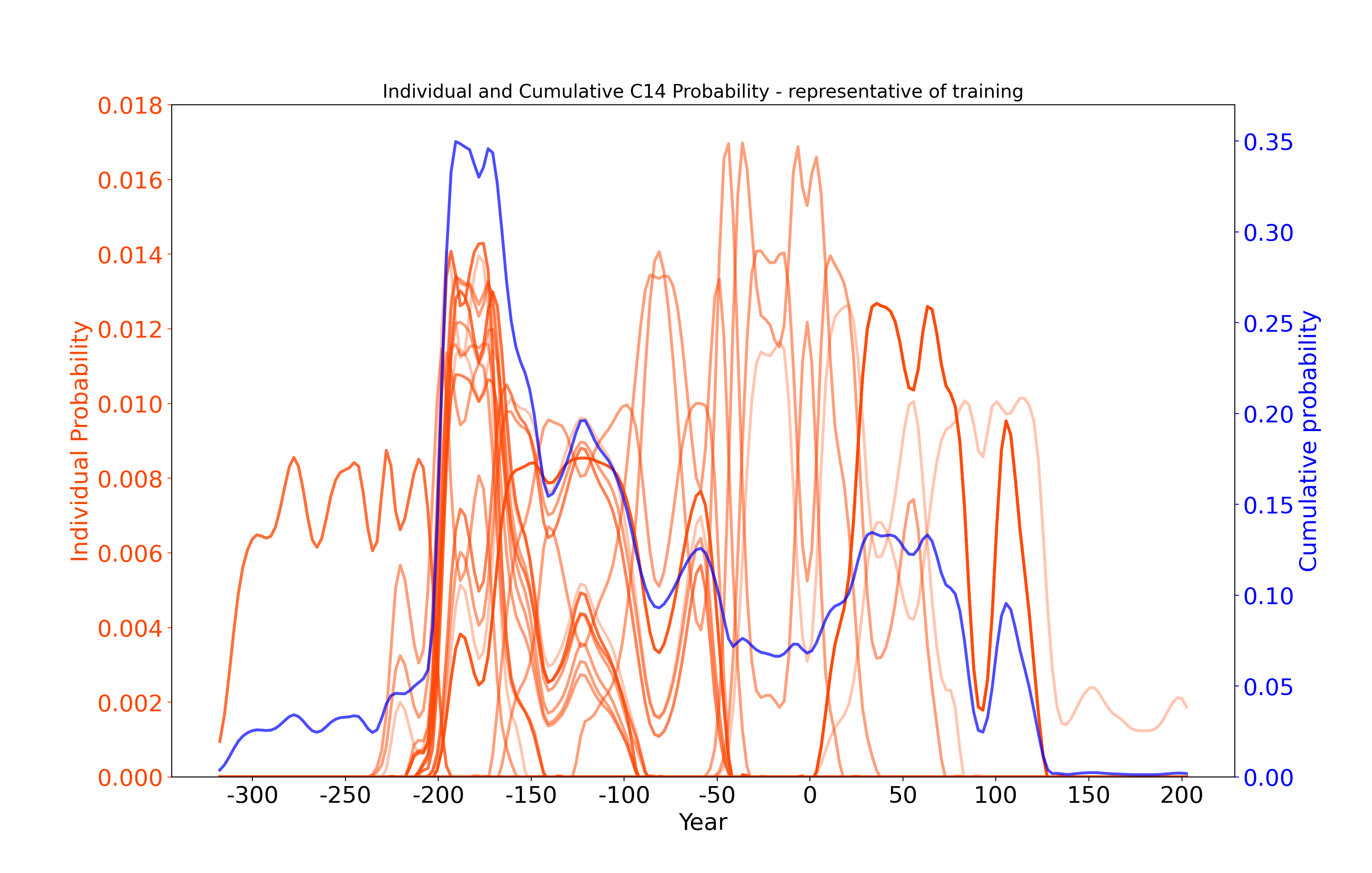}
	\caption{Distributions in the training data (orange) and total accumulated distribution of the C14 training data (blue), \textbf{excluding} minor peaks.}
	\label{fig:cum_c14}
\end{figure}

In addition to our original training data and the date prediction model described in the previous sections, we also employ two types of data balancing techniques to help reduce the time-axis bias in the training data. As can be seen in Figures~\ref{fig:cum_c14-wMinor} and \ref{fig:cum_c14}, the training data is biased in the sense that there are many more high probabilities in the -200 to the -150 region. This creates much higher priors in that region. However, this bias is caused only by the samples that were chosen to be radiocarbon-dated and are not representative of the actual prior probability for the whole Dead Sea Scrolls collection. In order to make the predictions less dependent on the priors within the training data, two data-balancing strategies were implemented.

The first method concerns {\em balancing using weights}, where the output probabilities from the model are dampened or boosted based on the weights provided by the overall accumulated distribution seen in Figures~\ref{fig:cum_c14-wMinor} and \ref{fig:cum_c14}. 

The other data-balancing implementation was through {\em augmentation}, where underrepresented training data was compensatorily oversampled based on the overall accumulated distribution. The technical details for both implementations will be described in the following subsections.

\begin{figure}[!ht]
    \centering
    \includegraphics[width=1.1\textwidth]{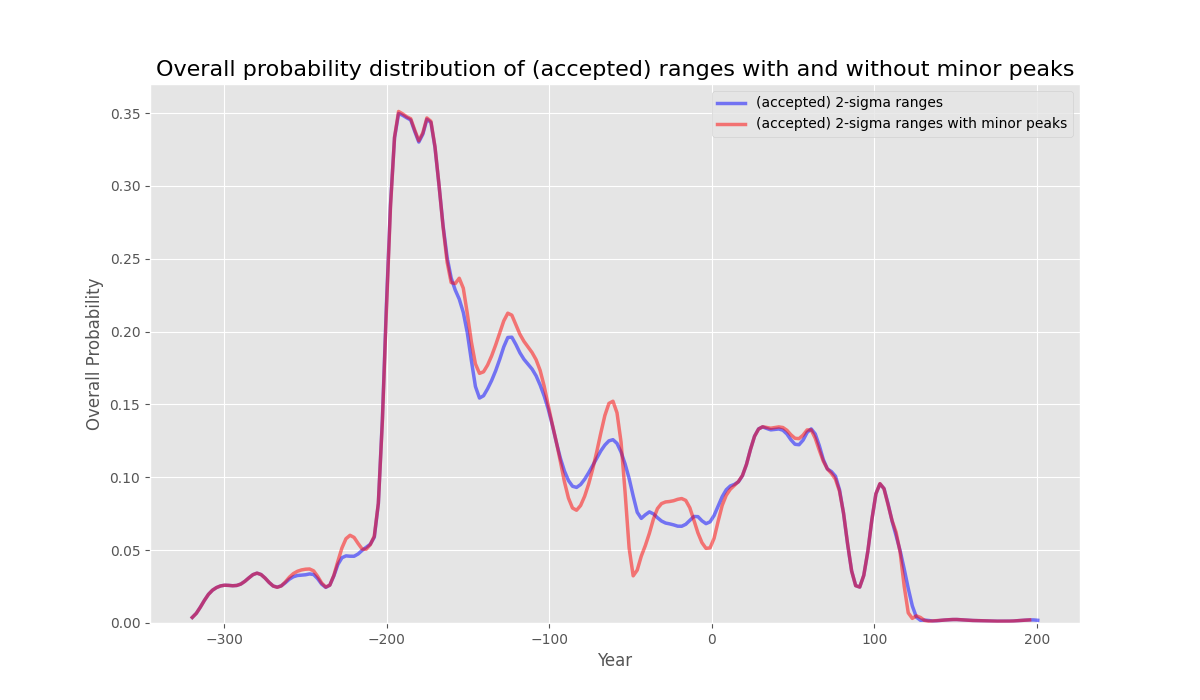}
	\caption{Comparison between overall probability distributions of (accepted) ranges with the inclusion and exclusion of minor peaks, as mentioned in the procedure in Section~\ref{Ai-2sigma}.}
	\label{fig:cum_compare}
\end{figure}

Please note that in addition to cross-validation and leave-one-out statistical tests (see Section~\ref{sec:loo}), we also check the sensitivity of the model with the inclusion and exclusion of minor peaks on the (accepted) 2$\sigma$ ranges. Figures~\ref{fig:cum_c14-wMinor} and \ref{fig:cum_c14} already show minimum changes over the overall probability distribution. This can be better visualized from Figure~\ref{fig:cum_compare}. The Euclidean distance calculated over the whole range sampled by five years is $0.104$ between the two (accepted) ranges (with and without minor peaks). The chi-square and Bhattacharyya distances are $0.124$ and $0.044$, respectively, showing no significant changes in the overall probability distribution. The predicted test results also remain unchanged. It is important to note, that incorporating the minor peaks did not lead to horizontal time shifts in existing high-probability peaks, at all (see Figure~\ref{fig:cum_compare}).

\subsubsection{Balance using weights}

Given probability $p_{i}$ where $i$ is a given year, threshold $T$, (binned) accumulated C14 probability $cum\_c14$, maximum accumulated C14 probability $M = \max(cum\_c14)$, and the number of summations that generated the accumulated probability in a bin $n_{cum\_c14_i}$. The weighted probability of each bin $w_{p_i}$ is calculated as:

\begin{equation}
    w_{p_i} = 
    \begin{cases}
        \frac{p_{i}}{cum\_c14_i} & \text{if } (p_{i} > T \cdot M) \text{ and } (n_{cum\_c14_i} > 2) \\
        p_{i} & \text{otherwise}
    \end{cases}
\end{equation}

The weighted probabilities are then normalized to ensure that the scale of the weighted predictions is consistent with the original predictions. The process is described below.

Given the global maximum probability values in the original predictions, $max\_p$, and in the weighted predictions, $max\_weighted\_p$, the normalization process for each probability value $w_{p_i}$ in the weighted predictions is described as follows:

Calculate the normalized probability $w_{p\_norm_i}$:

\begin{equation}
    w_{p\_norm_i} = \frac{w_{p_i}}{max\_weighted\_p} \cdot max\_p
\end{equation}

\subsection{Balance using augmentation}
\begin{figure}[!ht]
    \centering
    \includegraphics[width=\textwidth]{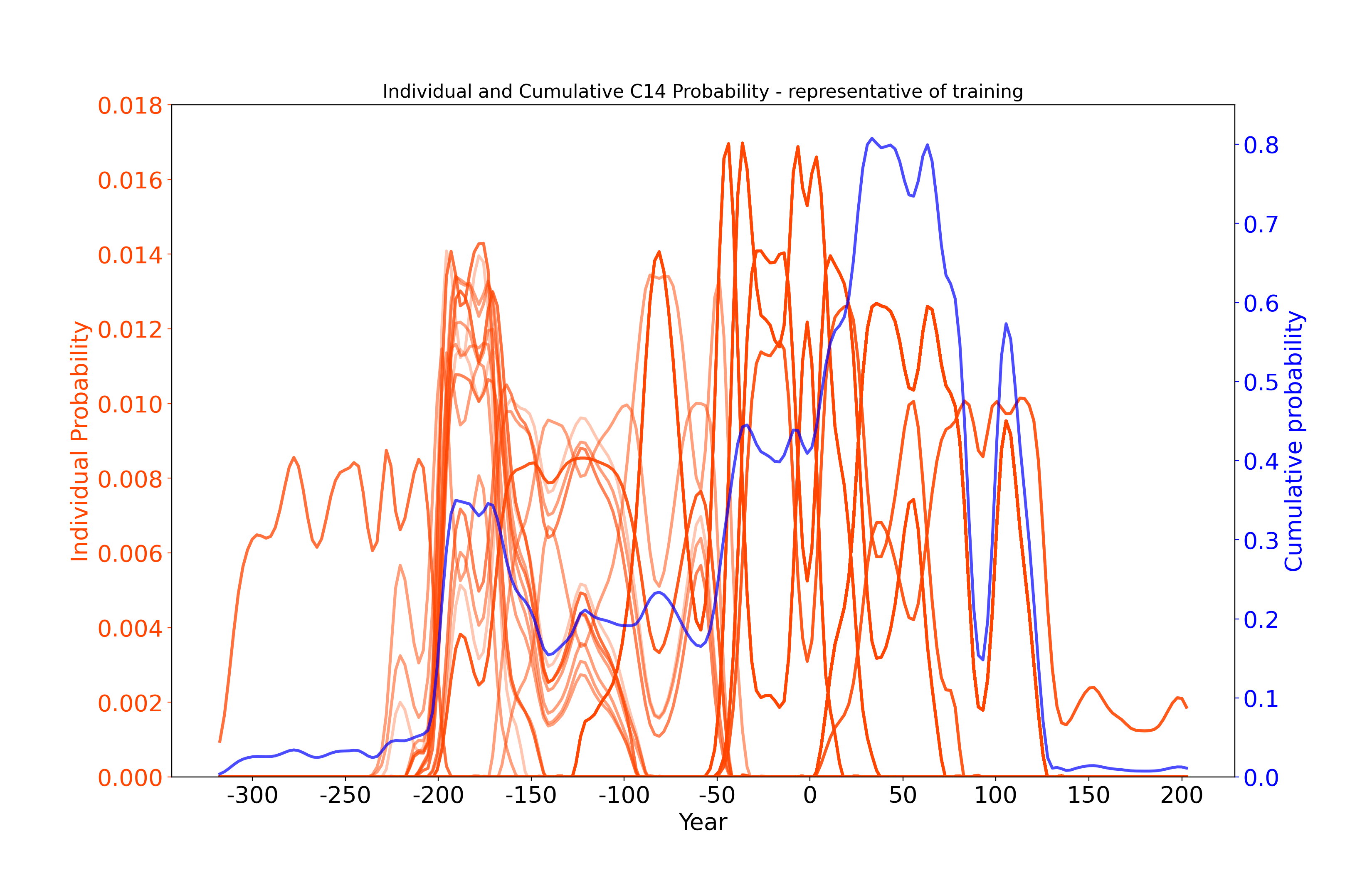}
    \caption{Distributions in the training data (orange) and overall accumulated distribution (blue) of the C14 training data \textbf{after augmentation}.}
    \label{fig:aug}
    
\end{figure}

For augmentation, five scrolls were chosen to be duplicated in the training data in order to boost the underrepresented prior probabilities within the training data. Table \ref{tab:aug} details the scrolls and number of fragments after the augmentation was applied, and Figure~\ref{fig:aug} shows the effect this had on the overall accumulated probabilities.

\begin{table}[ht]
\centering
\caption{Table detailing number of duplications after augmentation procedure}
\label{tab:aug}
\begin{tabular}{l|l|l}
\hline \hline
Scroll &
  \begin{tabular}[c]{@{}l@{}}Number of fragments \\ originally in the training data\end{tabular} &
  \begin{tabular}[c]{@{}l@{}}Number of fragments \\ after augmentation\end{tabular} \\ \hline
4Q2       & 2 & 12 \\
4Q161     & 2 & 12 \\
5\_6Hev1b & 1 & 6  \\
11Q5      & 9 & 54 \\
Mas1k     & 2 & 12 \\
XHev\_Se2 & 1 & 6  \\ \hline \hline
\end{tabular}
\end{table}

\subsection{Training options} \label{appen:C:training}
For training, we create three main pools of data:

\begin{itemize}
    \item \textsuperscript{14}C dated manuscripts
    \item \textsuperscript{14}C dated manuscripts with the addition of the old \textsuperscript{14}C from the 1990s
    \item \textsuperscript{14}C dated manuscripts with augmentation
\end{itemize}

Within each of these training datasets, there are different possible subsets of the training data that are usable (all the following subsets include the \textsuperscript{14}C dated manuscripts):

\begin{itemize}
    \item 4Q52 can be included or excluded
    \item Internally dated scrolls can be included or excluded
    \item Maresha Ostracon can be included or excluded
\end{itemize}

\subsubsection{Leave-one-out statistical test} \label{sec:loo}
In order to test whether the model predictions are robust, we performed a leave-one-out statistical test. The ‘leave-one-out’ (LOO) statistical test is a resampling technique used to evaluate the performance or robustness of a statistical model. In LOO, each sample in our training data is sequentially removed (left out), and the model is then trained on the remaining data points. The left-out observation is then used to evaluate the model's performance or make predictions. This process is repeated for each observation in the training data. The goal of this statistical test is to get an indication of the variability of the model predictions by checking whether the model is overfitting the training data and the impact of outliers on the model’s performance.

LOO is commonly employed when the dataset is small (which is the case for our dataset) or when it is important to assess the model’s performance on each individual observation. The LOO test is a valuable tool for assessing the reliability and generalizability of a statistical model by iteratively evaluating its performance on subsets of the training data while systematically excluding each sample. Given a training set of size $N$, we train $N$ models by leaving out one different data point for each model and training with the remaining $N-1$ data points. Then, on the test set, we make predictions using all $N$ models and overlap the resulting predictions obtained from each model. This gives a visual representation of the amount of variation between the predictions of the different models. 

\subsubsection{Gaussian of Gaussian}
We obtain the probability and error margin for each 10-year bin from the date-prediction model. This gives us vertical Gaussian for each 10-year over the entire timeline. To convert them into a single Gaussian on the horizontal time axis, we perform a tool, dubbed ‘Gaussian of Gaussian’ on the predicted dates.

This program generates $1000$ iterative attempts of randomly drawing a wave shape instance from the $n-bins$ distribution (over the entire timeline), assuming Gaussians $(\mu,\sigma)$ per bin. The max $y-peak$ position of the wave shape can be detected along the $x-axis$: For our manuscript dating problem, $x$ represents the year value. In this manner, it becomes possible to estimate the uncertainty of peak detection in the style-based OxCal approximation, and its effect on the date estimation. This addition to the Enoch method allows to obtain an estimate of date variability, similar to the output of OxCal itself. We explored whether smoothed distribution shapes were needed for this, but a detailed analysis fortunately revealed that the method could be kept simple: Smoothing of the shape often led to an $x-axis$ shift and an increase of the $x$ variability. The shape asymmetry of the (assumed) peak shape causes this time bias. Hence, we avoided any smoothing and {\em used the raw, unfiltered generated histograms}. The implicit assumption is that the `maximum' co-occurs with a peak. Comparative plots for different information sources are obtained using the ‘Gaussian of Gaussian’ (see Appendix~\ref{appen:E-plots}).

%% file: tex-sections/E-ai2.tex
\subsection{Date-prediction model} \label{appen:E6:datepredictionmodel}
We employ our date-prediction model once the features are calculated from the images. Given, for each manuscript, a style feature vector, we now address the transformation of this representation into an OxCal-type curve, i.e., a vector containing the estimated date probabilities for the sample. Because of the small size of the data set, high-parametric models such as period-specific temporal codebooks~\cite{He2016} cannot be used here. We use conditional modeling using Bayesian Ridge regression~\cite{Hoerl2000} that applies Bayesian inference to estimate the model parameters for date prediction. First, a prior distribution is placed on the model parameters, which expresses known constraints on the values of the parameters. The prior distribution is then updated with the observed data using Bayes’ rule to obtain the parameters' posterior distribution and predicted dates. 

We propose the Bayesian approach due to the nature of our target output data: the \textsuperscript{14}C data are not a single point on the timeline but are given as a distribution of probable dates within sigma ($\sigma$) ranges. Hence, the probabilistic approach allows the use of all available information while remaining explainable. Furthermore, we can observe a full posterior distribution, which is used to assess the uncertainty of the estimated dates. Finally, the Bayesian approach also allows the model to indicate error margins for predictions on unseen data.

\subsubsection{Unmodelled values from OxCal} \label{OxCal-rawdata-acquire}
The input of the date prediction model is the feature vectors of the training images, along with their probability distribution from radiocarbon dating as labels. We obtain the probability distribution as unmodelled raw values of 5-year resolution after the radiocarbon calibration was performed using OxCal v4.4.2~\cite{Oxcal, Oxcal2}. We created a new project code for the 26 sample manuscripts using the \textsuperscript{14}C age (BP) and sigma (BP) values (see Table~\ref{tab:summarized-c14} for the BP values. For the code, please check the Zenodo repository (\href{https://doi.org/10.5281/zenodo.11371749}{https://doi.org/10.5281/zenodo.11371749})). Using the measured BP values, the code (entirely reproducible) uses the simple format:
\begin{verbatim}
Plot()
  {
  R_Date("Q-number", age(BP), sigma(BP));
   };
\end{verbatim}

\noindent OxCal generates the clean unmodelled (BCE/CE) probability values in a \verb+.csv+ file once the code is run. We obtain these values for each individual sample from OxCal options: \textit{View} $\gg$ \textit{Raw output}. Please note that we do not specify any resolution in our code. Hence, our raw data are in the default resolution of 5 years, which is the same as the resolution of the IntCal curves, so no interpolation or binning is needed. It is possible to set the resolution to less than 5. Then, the curve will be interpolated by a cubic (or linear if that option is set) function by OxCal (as explained in Section~\ref{appen:B4:amsdating} in Appendix~\ref{appen:B}).

\subsubsection{Calibrated dates from 2-sigma ranges} \label{Ai-2sigma}
Having performed the radiocarbon dating in its entirety (Appendix~\ref{appen:B}) and then the palaeographic evaluation of the calibrated dates (Appendix~\ref{appen:C-new}), as was also done in~\cite{Bonani1992,Jull1995}, we use the calibrated dates from the 2$\sigma$ range for firmer grounding of our date-prediction model. In the case of bimodal evidence, palaeography determines that in most cases the younger 2$\sigma$ peak should be used for analysis. But, palaeography cannot be characterised as a specific quantitative prior (an expected date with mean and standard deviation) in the timeline. The issue is that we cannot assume a single point-spread density (Gaussian) along the time axis. Palaeography, in this sense, does not deliver a point-wise prior. What palaeography can deliver, is the identification of a point on the timeline that represents the historical impossibility of a range of dates on the left or right. This knowledge is based on intersubjective expert knowledge (Section~\ref{appen:C2:combipalaeoradiocarbon}). Therefore, the use of expert knowledge in our case is not based on the usual Bayesian-plus-Gaussian method but on a more direct use of existing, qualitative, and intersubjective palaeographic knowledge, which allows splitting the 2$\sigma$ calibrated date range into a ‘left-half’ vs. ‘right-half’ time region of interest. 

Specifically, palaeographic knowledge allows to make a binary split in the OxCal distribution of bimodal 2$\sigma$ ranges using the Heaviside function with the position of the step being placed at an innocuous low-probability point on the curve, where the probability has a plateau around zero. Applying a Heaviside multiplicative bias on the empirical density function is a valid Bayesian approach to perform peak selection. Another, e.g., smooth logistic variant of the step function or a cumulative Gaussian could have been used. In either case this would require the specification of a steepness parameter, the value of which is unknown. Similarly, from the palaeographic constraints, the standard deviation which would be needed under a Gaussian, i.e., point-localized density assumption in the Bayesian reasoning process, is not known. Note that apart from the Gaussian, many other distribution functions exist, e.g., Poisson, Weibull, Gamma, etc., for other applications in Bayesian reasoning. Such distribution functions could be used, \textit{if} there are reasons to assume them. We do not have any reason for a detailed distribution choice and can only choose to disregard ‘impossible time regions’. In our case, for most bimodal 2$\sigma$ calibrated ranges we assume that the palaeographic evaluation that a left-most or right-most) region is impossible is correct (see Section~\ref{appen:C2:combipalaeoradiocarbon} in Appendix~\ref{appen:C-new}), leading to a collapse of the probabilities in that range. The shape of the remaining distribution reflects the likelihood of dates. 

Thus, the procedure we use is as follows: 

\begin{enumerate}

\item We perform the radiocarbon dating in its entirety, with the calibrated dates having been generated using OxCal data with only \textsuperscript{14}C (BP) and $\sigma$ (BP), see Appendix~\ref{appen:B}. 

\item We use the calibrated dates from the 2$\sigma$ range for firmer grounding of our date-prediction model. Only in the case of bimodal evidence in these 2$\sigma$ ranges do we apply the Heaviside-function at a near-zero probability point on the curve to reject older peaks and accept younger/‘right-hand’ peaks as a possible solution, on the basis of expert palaeographic knowledge (Appendix~\ref{appen:C-new}).

\item From OxCal, we obtain the raw data of the probability densities of the 2$\sigma$ ranges, which are used as input in our date-prediction model. 

\item We work with the inclusion and exclusion of so-called minor or smaller probability peaks, which in 12 out of 14 instances have a probability of less than $4\%$; in the remaining two cases, it is 5.2\% and 9.4\%. The inclusion or exclusion of these peaks has minimal and insignificant consequences for the interpretation of the results (see Section~\ref{subsec:data-balance}).

\item Because applying the Heaviside function for bimodal evidence leaves less than $95.4\%$ of the entire 2-$\sigma$ probability for each sample, we normalise the accepted 2$\sigma$ calibrated probabilities. The output probability predictions of the dating model are also balanced and normalised using both weights and data augmentation (see Section~\ref{subsec:data-balance}).

\item Within the accepted part of the 95.4\% confidence range, the points in the probability distribution curve as calculated by OxCal are used as target values in the training of Enoch. The output distribution delivered by the Enoch model is a mixture of Gaussians approximating the shape of the OxCal curve. From those outputs, we select the 1$\sigma$ range for a clear, narrow visualization of the predicted date ranges (see Figure~\ref{fig:AIvsC14PAL} in the main article). This choice is independent of the original selection of OxCal calibration ranges because the two methods are fundamentally different.

\end{enumerate}

We emphasise that we do not perform modelling within the OxCal programme as is commonly used in radiocarbon dating practice. This does not compromise the transparency and reproducibility of our procedure. Using the \textsuperscript{14}C dates in BP and their measurement uncertainties ($\sigma$), all plots can be reproduced using OxCal. Our reasoning for rejecting part of the bimodal data (see Appendix~\ref{appen:C-new}) and the exact 2$\sigma$ probabilities we use (see Table~\ref{Ai-2sigma}) are provided. Yet, other researchers can also use all our \textsuperscript{14}C data instead of following our reasoning for accepting part of the bimodal data, and justify their reasoning. 

\input{tables/Table-accept}

In the following subsections, we present the mathematical derivation of the Bayesian regression from simple linear regression as used in Enoch, our date prediction model. 

\subsubsection{Linear regression}
Given a set of training data $\{(\mathbf{x_n}, t_n)\}^{N}_{n=1}$ comprising $N$ observations of dimensionality $M$, where $\mathbf{x}_n \in \RR^M$,  $t_n \in \RR$, the goal in a regression model is to find a linear mapping $f: \RR^M \rightarrow \RR$ which approximates $t_n$ given $\mathbf{x_n}$ as close as possible. Furthermore, the mapping should generalize to values outside the training data. From a probabilistic perspective, the aim is to model the \textit{predictive distribution} $p(t_n \mid \mathbf{x}_n)$. In a linear regression model, the assumption is that the target variable $t$ is given by a deterministic function $f(\mathbf{x}_n,\mathbf{w})$ with added Gaussian noise, such that:

\begin{equation} \label{eq:target-def}
	t_n = f(\mathbf{x}_n,\mathbf{w}) + \epsilon
\end{equation}
where $\epsilon$ is a Gaussian random variable with mean 0 and inverse variance parameter $\beta$, also called the \textit{precision}. In a linear model where $f(\mathbf{x}_n,\mathbf{w}) = \mathbf{w}^T\mathbf{x}$, the predictive distribution takes the form

\begin{equation} \label{eq:pred-dist}
	p(t_n\mid \mathbf{x}_n) = \mathcal{N}(\mathbf{w}^T\mathbf{x}_n, \beta^{-1}).
\end{equation}
Even though the predictive distribution is indirectly used to optimize the parameters of the linear regression model, we do not explicitly model this distribution. We will see in the next section that in the Bayesian interpretation of linear regression, we instead stay within a probabilistic framework and model the full predictive distribution, leading to several advantages over the standard linear regression model. 

We first turn to parameter estimation for a linear model. This means estimating a value for the weight vector $\mathbf{w}$ that fits the data well. Most commonly, the \textit{least-squares} criterion is used to estimate the weight vector $\mathbf{w}$: 

\begin{equation} \label{eq:least-squares}
	\mathbf{w}^* = 	\argmin_{\mathbf{w}^*} \sum_{n=1}^{N} (t_n - \mathbf{w}^T\mathbf{x}_n)^2
\end{equation}

This can be justified using maximum likelihood estimation if we assume that the training data is independent and identically distributed (i.i.d.). This works as follows. Let $\mathbf{X} = (\mathbf{x}_1,\dots,\mathbf{x}_N)^T$ and $\mathbf{t} = (t_1,\dots,t_N)^T$. The log-likelihood of the training data can then be written as 

\begin{align}
	\ln p(\mathbf{t}\mid \mathbf{X},\mathbf{w},\beta)
	\nonumber &= 
	\ln \prod^N_{n=1} p(t_n \mid  \mathbf{x}_n, \mathbf{w}, \beta) \\
	\nonumber &=
	\ln \prod^N_{n=1} \mathcal{N}(t_n\mid \mathbf{w}^T\mathbf{x}_n, \beta^{-1}) \\
	\nonumber &=  
	\sum^N_{n=1}\ln \mathcal{N}(t_n\mid \mathbf{w}^T\mathbf{x}_n, \beta^{-1}) \\
	\nonumber &=
	\sum_{n=1}^{N} \ln \{(2\pi)^{-1/2}\beta^{1/2}
	\text{exp}(-\frac{\beta}{2} (t_n - \mathbf{w}^T\mathbf{x}_n)^2)\} \\
	&=
	\frac{N}{2}\ln \beta - \frac{N}{2}\ln 2\pi - \beta E_D(\mathbf{w}),
	\label{eq:log-likelihood}
\end{align}

where we make use of (\ref{eq:pred-dist}). The $E_D(\mathbf{w})$ term represents a sum-of-squares function, defined as

\begin{equation}
	E_D(\mathbf{w}) = \frac{1}{2} \sum_{n=1}^{N} (t_n - \mathbf{w}^T\mathbf{x}_n)^2.
\end{equation}

Considering that maximizing the likelihood function with respect to $\mathbf{w}$ only depends on  $E_D(\mathbf{w})$, expression (\ref{eq:log-likelihood}) can be maximized by maximizing $-E_D(\mathbf{w})$, or equivalently, minimizing $E_D(\mathbf{w})$. This corresponds to the least-squares objective shown in (\ref{eq:least-squares}).

\subsubsection{Ridge regression}
We now turn to the ridge regression model, an extension of the linear regression model with more desirable properties, such as mitigating over-fitting. Concretely, we add a prior distribution over the weights $\mathbf{w}$, leading to the following log-likelihood function:

\begin{equation} \label{eq:log-likelihood-ridge}
	\ln p(\mathbf{t}\mid \mathbf{X},\mathbf{w},\beta) + \ln p(\mathbf{w} \mid  \alpha).
\end{equation}

The prior distribution over the weights can be interpreted with the Bayes rule, showing the relationship to a posterior distribution over $\mathbf{w}$: 

\begin{equation} \label{eq:propto}
	p(\mathbf{w}\mid \mathbf{t}) \propto p(\mathbf{t}\mid \mathbf{w}) p(\mathbf{w}),
\end{equation}

where we omit the $\mathbf{X}$, $\alpha$, and $\beta$ terms to keep the notation uncluttered. In other words, maximizing (\ref{eq:log-likelihood-ridge}) corresponds to maximizing a posterior distribution over $\mathbf{w}$. The question now arises what is a suitable form of the prior distribution $p(\mathbf{w})$? To ensure that $p(\mathbf{w}\mid \mathbf{t})$ has the same functional form as $p(\mathbf{t}\mid \mathbf{w})$, we choose $p(\mathbf{w})$ to be a conjugate prior of $p(\mathbf{t}\mid \mathbf{w})$, namely a multivariate isotropic Gaussian distribution, taken to be zero-centered with precision parameter $\alpha$. The log-likelihood then becomes

\begin{align} 
	\ln p(\mathbf{t}\mid \mathbf{X},\mathbf{w},\beta) + \ln p(\mathbf{w} \mid  \alpha)
	\nonumber &= \\
	\sum^N_{n=1} \{\ln \mathcal{N}(t_n\mid \mathbf{w}^T\mathbf{x}_n, \beta^{-1})\} + \ln
	\mathcal{N}(\mathbf{w}\mid 0, \alpha^{-1} \mathbf{I}) \\
	&= \\
	\frac{N}{2}\ln \beta - \frac{N}{2}\ln 2\pi - \beta E_D(\mathbf{w}) +
	\frac{M}{2} \ln \alpha - \frac{M}{2} \ln 2\pi - \alpha E_W(\mathbf{w}),
 \label{eq:log-likelihood-ridge2}
\end{align}

where $M$ denotes the number of dimensions of the weight parameter $\mathbf{w}$ and $I$ denotes the identity matrix. Note that the first three summands of (\ref{eq:log-likelihood-ridge2}) correspond to (\ref{eq:log-likelihood}). The $E_W(\mathbf{w})$ term represent a regularization term, defined by 

\begin{equation}
	E_W(\mathbf{w}) = \frac{1}{2} \mathbf{w}^T\mathbf{w}.
\end{equation}

By removing terms from (\ref{eq:log-likelihood-ridge2}) that do not depend on $\mathbf{w}$, we end up minimizing the sum of two terms, $E_D(\mathbf{w})$ and $E_W(\mathbf{w})$, denoting the data-dependent error and the regularization error, respectively. The relative importance of both terms is controlled by the $\alpha$ and $\beta$ hyperparameters. Equivalently, we minimize: 

\begin{equation}
	\beta E_D(\mathbf{w}) + \alpha E_W(\mathbf{w}).
\end{equation}

If we combine $\alpha$ and $\beta$ into one hyperparameter $\lambda = \alpha / \beta$, we can equivalently write

\begin{equation}
	E_D(\mathbf{w}) + \lambda E_W(\mathbf{w}),
\end{equation}

Which corresponds to 

\begin{equation} \label{eq:ridge-objective}
	\frac{1}{2}\sum_{n=1}^{N} (\mathbf{w}^T\mathbf{x}_n - t_n)^2 + \frac{\lambda}{2} \mathbf{w}^T\mathbf{w},
\end{equation}

Which forms the ridge-regression objective function. The $\lambda$ hyperparameter can control the degree of \textit{parameter shrinkage} \cite{hastie2009elements}, whereby the weight parameters are shrunk by imposing a penalty on their size. This brings the additional task of setting $\lambda$, generally done using cross-validation. By setting the gradient of  (\ref{eq:ridge-objective}) with respect to $\mathbf{w}$ to 0 and solving for $\mathbf{w}$, the approximate solution can be expressed in closed form using the standard equations:

\begin{equation} \label{eq:w-point}
	\mathbf{w} = (\mathbf{X}^T\mathbf{X} + \lambda \mathbf{I})^{-1} \mathbf{X}^T
	\mathbf{t}.
\end{equation}

\subsubsection{Bayesian regression}
We now turn to a Bayesian treatment of the ridge regression model discussed in the previous subsection. First, consider the relationship we established between the posterior over $\mathbf{w}$ and the product of the likelihood and prior, as shown in (\ref{eq:propto}). This is the point at which the ridge and Bayesian regression models diverge in their approach. For the ridge regression model, a point estimate for the weight vector $\mathbf{w}$ is obtained by using maximum a posteriori estimation (MAP), which involves maximizing the right-hand side of (\ref{eq:propto}). We now discuss the alternative, fully Bayesian treatment, which explicitly models the posterior distribution on the left-hand side of (\ref{eq:propto}).

Recall that we defined the prior distribution $p(\mathbf{w})$ as a conjugate prior to the likelihood function, leading to a multivariate Gaussian distribution. The result is that the posterior $p(\mathbf{w}\mid \mathbf{t}, \mathbf{X})$ also will have a Gaussian distribution. We can thus rewrite (\ref{eq:propto}) to: 

\begin{equation}
	\mathcal{N}(\mathbf{m}_N, \mathbf{S}_N) \, \propto \,
	\mathcal{N}(\mathbf{X}\mathbf{w}, \beta^{-1}\mathbf{I}) 
	\mathcal{N}(0, \alpha^{-1}\mathbf{I}),
\end{equation}

where the posterior is a Gaussian with mean $\mathbf{m}_N$ and covariance $\mathbf{S}_N$.  We can use the Bayes theorem for Gaussian random variables to find $\mathbf{m}_N$ and $\mathbf{S}_N$. From this, it follows:

\begin{align}
	\mathbf{m}_N &= \beta \mathbf{S}_N \mathbf{X}^T \mathbf{t}\\
	\mathbf{S}_N^{-1} &= \beta \mathbf{X}^T \mathbf{X} + \alpha \mathbf{I}.
\end{align}

It is worth noting the correspondence between the point estimate of $\mathbf{w}$ obtained in the ridge regression solution (\ref{eq:w-point}) and the mean of the posterior $\mathbf{m}_N$. If we fully write out $\mathbf{m}_N$, we see that

\begin{align*}
	\mathbf{m}_N 
	&= 
	\beta (\beta \mathbf{X}^T \mathbf{X} + \alpha \mathbf{I})^{-1} \mathbf{X}^T \mathbf{t} \\
	&= 
	\beta (\beta (\mathbf{X}^T \mathbf{X} + \lambda \mathbf{I}))^{-1} \mathbf{X}^T \mathbf{t} \\
	&=
	\beta (\beta^{-1} (\mathbf{X}^T \mathbf{X} + \lambda \mathbf{I})^{-1}) \mathbf{X}^T \mathbf{t} \\
	&=
	(\mathbf{X}^T \mathbf{X} + \lambda \mathbf{I})^{-1} \mathbf{X}^T \mathbf{t},
\end{align*}

which corresponds to (\ref{eq:w-point}). This means that the mode of the posterior distribution corresponds to the ridge regression solution. However, we use the full posterior distribution over $\mathbf{w}$ in the Bayesian regression approach rather than taking the mean $\mathbf{m}_N$ as a point estimate. This works as follows. We first note that once we obtained the posterior distribution over $\mathbf{w}$, the predictive distribution informed by the training data can now be written as

\begin{equation} \label{eq:pred-dist2}
	p(t\mid \mathbf{x}, \mathbf{t}, \mathbf{X}) 
	= 
	\int p(t\mid \mathbf{x}, \mathbf{w}) p(\mathbf{w}\mid \mathbf{t}, \mathbf{X}) \, d\mathbf{w}
\end{equation}

for an input $\mathbf{x}$, where we once again omit the $\alpha$ and $\beta$ terms for readability. Noting that (\ref{eq:pred-dist2}) is a marginal distribution and a convolution of two Gaussians, we can once again make use of Bayes theorem for Gaussian variables, resulting in the predictive distribution

\begin{equation}
	p(t\mid \mathbf{x}, \mathbf{t}, \mathbf{X}, \alpha, \beta)  
	=
	\mathcal{N}(t\mid \mathbf{m}_N^T\mathbf{x}, \beta^{-1} + \mathbf{x}^T\mathbf{S}_N\mathbf{x}).
\end{equation}

The mean of this distribution is simply the mean of the posterior distribution multiplied by the input vector. As can be seen from the variance parameter of this equation, the predictive variance associated with an input $\mathbf{x}$ consists of a sum of two terms, which can be understood as follows. The first term expresses variance due to the noise in the training data. The second term describes the uncertainty associated with $\mathbf{w}$, which varies according to the input $\mathbf{x}$.

Given this predictive distribution, we can make predictions for new input values by calculating the conditional expectation,

\begin{equation}
	\EE[t\mid \mathbf{x}, \mathbf{t}, \mathbf{X}] 
	=
	\int t \, p(t\mid \mathbf{x}, \mathbf{t}, \mathbf{X}) \, dt.
\end{equation}

An alternative is to directly take the mean $\mathbf{m}_N$ of the posterior as an estimate for $\mathbf{w}$, which is used in some implementations of the Bayesian regression model \cite{scikit-learn-bayesian-ridge}.

\subsubsection{Hyperparameter selection}
In a Bayesian framework, defining a prior distribution over one or both hyperparameters, also known as a \textit{hyperprior}, can be used in finding hyperparameters using cross-validation. We can then marginalize all the parameters, which leads to a predictive distribution of the form.

\begin{equation}
	p(t\mid \mathbf{x}, \mathbf{t}, \mathbf{X}) 
	=
	\iiint (t\mid \mathbf{x}, \mathbf{w}) p(\mathbf{w}\mid \mathbf{t}, \mathbf{X},
	\alpha, \beta) p(\alpha, \beta\mid \mathbf{t}, \mathbf{X}) \, d\mathbf{w} \,
	d\alpha \, d\beta.
\end{equation}

Unfortunately, this expression is analytically intractable. Nevertheless, a framework for calculating an approximation, named \textit{evidence approximation} \cite{bishop2006pattern}, can compute estimates for $\alpha$ and $\beta$. This framework is also referred to as \textit{type II maximum likelihood}, which involves maximizing the \textit{marginal likelihood function} $p(\mathbf{t}\mid \alpha, \beta)$, where $\mathbf{w}$ has been integrated out. We will not go into depth into the evidence approximation framework. For more extensive treatment, the reader is referred to the statistical book\cite{bishop2006pattern}. It should be noted that this approach, where we include the hyperparameters as part of the training process by regarding them as random variables, does not necessarily lead to better estimates than those obtained with cross-validation. Nevertheless, automatically finding hyperparameters as part of the training process can be helpful in certain situations, for example, if cross-validation is not feasible. 

The output of the date prediction model is a probability estimate for each 10-year bin in our timeline, along with error margins to estimate uncertainty. Within the Bayesian regression, we apply parameter constraints to restrict the uncertainties to non-negative values as they do not impact the probability estimation of our model, and the final results and interpretation. However, future research can explore the feasibility of an asymmetric error estimation above the x-axis. The choice of the 10-year bin is made empirically, and we keep the option of changing the bins to 5 or 15 for either thinner or thicker plots. 

%% file: tables/Table-accept.tex
\footnotesize
\LTcapwidth=\textwidth
{\tabcolsep=3.3pt

\begin{longtable}{llrrlrrllllll}
\caption{Unmodelled radiocarbon calibrated dates for 2$\sigma$ ranges. Please note that the 2$\sigma$ values are the same as Table~\ref{tab:summarized-c14} in Appendix~\ref{appen:B}. However, in this table, the highlighted date ranges indicate each sample's accepted 2$\sigma$ intervals for the Enoch model. }
\label{tab:2-sigma-accepted}\\
\hline \hline
\textbf{Q-number}      &  & \multicolumn{2}{l}{\textbf{2$\sigma$ range}}                &                          & \multicolumn{2}{l}{\textbf{2$\sigma$ range}}                &                          & \multicolumn{2}{l}{\textbf{2$\sigma$ range}}                                                        &                          & \multicolumn{2}{l}{\textbf{2$\sigma$ range}}                                                      \\ \cline{1-1} \cline{3-4} \cline{6-7} \cline{9-10} \cline{12-13} 
\endfirsthead
\endhead
\textbf{4Q504}         &  & -355                         & -285                         &                          & \cellcolor[HTML]{FFFC9E}-230 & \cellcolor[HTML]{FFFC9E}-150 &                          &                                                  &                                                  &                          &                                                 &                                                 \\ \cline{1-1} \cline{3-4} \cline{6-7} \cline{9-10} \cline{12-13} 
\textbf{4Q52}          &  & -410                         & -355                         &                          & \cellcolor[HTML]{FFFC9E}-285 & \cellcolor[HTML]{FFFC9E}-230 &                          &                                                  &                                                  &                          &                                                 &                                                 \\ \cline{1-1} \cline{3-4} \cline{6-7} \cline{9-10} \cline{12-13} 
\textbf{4Q176}         &  & -355                         & -300                         &                          & \cellcolor[HTML]{FFFC9E}-210 & \cellcolor[HTML]{FFFC9E}-100 & \cellcolor[HTML]{FFFC9E} & \multicolumn{1}{r}{\cellcolor[HTML]{FFFC9E}-70}  & \multicolumn{1}{r}{\cellcolor[HTML]{FFFC9E}-60}  &                          &                                                 &                                                 \\ \cline{1-1} \cline{3-4} \cline{6-7} \cline{9-10} \cline{12-13} 
\textbf{4Q114}         &  & -355                         & -285                         &                          & \cellcolor[HTML]{FFFC9E}-230 & \cellcolor[HTML]{FFFC9E}-160 &                          &                                                  &                                                  &                          &                                                 &                                                 \\ \cline{1-1} \cline{3-4} \cline{6-7} \cline{9-10} \cline{12-13} 
\textbf{5\_6Hev1b}     &  & \cellcolor[HTML]{FFFC9E}10   & \cellcolor[HTML]{FFFC9E}205  &                          & \multicolumn{1}{l}{}         & \multicolumn{1}{l}{}         &                          &                                                  &                                                  &                          &                                                 &                                                 \\ \cline{1-1} \cline{3-4} \cline{6-7} \cline{9-10} \cline{12-13} 
\textbf{4Q161}         &  & \cellcolor[HTML]{FFFC9E}-90  & \cellcolor[HTML]{FFFC9E}-80  & \cellcolor[HTML]{FFFC9E} & \cellcolor[HTML]{FFFC9E}-55  & \cellcolor[HTML]{FFFC9E}30   & \cellcolor[HTML]{FFFC9E} & \multicolumn{1}{r}{\cellcolor[HTML]{FFFC9E}45}   & \multicolumn{1}{r}{\cellcolor[HTML]{FFFC9E}60}   &                          &                                                 &                                                 \\ \cline{1-1} \cline{3-4} \cline{6-7} \cline{9-10} \cline{12-13} 
\textbf{4Q70}          &  & -375                         & -345                         &                          & \cellcolor[HTML]{FFFC9E}-320 & \cellcolor[HTML]{FFFC9E}-200 &                          &                                                  &                                                  &                          &                                                 &                                                 \\ \cline{1-1} \cline{3-4} \cline{6-7} \cline{9-10} \cline{12-13} 
\textbf{4Q47}          &  & -355                         & -290                         &                          & \cellcolor[HTML]{FFFC9E}-210 & \cellcolor[HTML]{FFFC9E}-100 &                          &                                                  &                                                  &                          &                                                 &                                                 \\ \cline{1-1} \cline{3-4} \cline{6-7} \cline{9-10} \cline{12-13} 
\textbf{4Q23}          &  & -355                         & -285                         &                          & \cellcolor[HTML]{FFFC9E}-230 & \cellcolor[HTML]{FFFC9E}-220 & \cellcolor[HTML]{FFFC9E} & \multicolumn{1}{r}{\cellcolor[HTML]{FFFC9E}-210} & \multicolumn{1}{r}{\cellcolor[HTML]{FFFC9E}-95}  & \cellcolor[HTML]{FFFC9E} & \multicolumn{1}{r}{\cellcolor[HTML]{FFFC9E}-75} & \multicolumn{1}{r}{\cellcolor[HTML]{FFFC9E}-55} \\ \cline{1-1} \cline{3-4} \cline{6-7} \cline{9-10} \cline{12-13} 
\textbf{4Q255\_4Q433a} &  & \cellcolor[HTML]{FFFC9E}-170 & \cellcolor[HTML]{FFFC9E}-50  &                          & \multicolumn{1}{l}{}         & \multicolumn{1}{l}{}         &                          &                                                  &                                                  &                          &                                                 &                                                 \\ \cline{1-1} \cline{3-4} \cline{6-7} \cline{9-10} \cline{12-13} 
\textbf{11Q5}          &  & \cellcolor[HTML]{FFFC9E}-35  & \cellcolor[HTML]{FFFC9E}-15  & \cellcolor[HTML]{FFFC9E} & \cellcolor[HTML]{FFFC9E}5    & \cellcolor[HTML]{FFFC9E}120  &                          &                                                  &                                                  &                          &                                                 &                                                 \\ \cline{1-1} \cline{3-4} \cline{6-7} \cline{9-10} \cline{12-13} 
\textbf{4Q3}           &  & -340                         & -325                         &                          & \cellcolor[HTML]{FFFC9E}-200 & \cellcolor[HTML]{FFFC9E}-50  &                          &                                                  &                                                  &                          &                                                 &                                                 \\ \cline{1-1} \cline{3-4} \cline{6-7} \cline{9-10} \cline{12-13} 
\textbf{4Q27}          &  & -340                         & -330                         &                          & \cellcolor[HTML]{FFFC9E}-200 & \cellcolor[HTML]{FFFC9E}-50  &                          &                                                  &                                                  &                          &                                                 &                                                 \\ \cline{1-1} \cline{3-4} \cline{6-7} \cline{9-10} \cline{12-13} 
\textbf{Mas1k}         &  & \cellcolor[HTML]{FFFC9E}-50  & \cellcolor[HTML]{FFFC9E}65   &                          & \multicolumn{1}{l}{}         & \multicolumn{1}{l}{}         &                          &                                                  &                                                  &                          &                                                 &                                                 \\ \cline{1-1} \cline{3-4} \cline{6-7} \cline{9-10} \cline{12-13} 
\textbf{4Q206}         &  & -360                         & -280                         &                          & \cellcolor[HTML]{FFFC9E}-235 & \cellcolor[HTML]{FFFC9E}-145 & \cellcolor[HTML]{FFFC9E} & \multicolumn{1}{r}{\cellcolor[HTML]{FFFC9E}-135} & \multicolumn{1}{r}{\cellcolor[HTML]{FFFC9E}-120} &                          &                                                 &                                                 \\ \cline{1-1} \cline{3-4} \cline{6-7} \cline{9-10} \cline{12-13} 
\textbf{4Q30}          &  & -360                         & -275                         &                          & \cellcolor[HTML]{FFFC9E}-260 & \cellcolor[HTML]{FFFC9E}-245 & \cellcolor[HTML]{FFFC9E} & \multicolumn{1}{r}{\cellcolor[HTML]{FFFC9E}-235} & \multicolumn{1}{r}{\cellcolor[HTML]{FFFC9E}-165} &                          &                                                 &                                                 \\ \cline{1-1} \cline{3-4} \cline{6-7} \cline{9-10} \cline{12-13} 
\textbf{4Q201\_4Q338}  &  & \cellcolor[HTML]{FFFC9E}-165 & \cellcolor[HTML]{FFFC9E}-40  & \cellcolor[HTML]{FFFC9E} & \cellcolor[HTML]{FFFC9E}-10  & \cellcolor[HTML]{FFFC9E}-1   &                          &                                                  &                                                  &                          &                                                 &                                                 \\ \cline{1-1} \cline{3-4} \cline{6-7} \cline{9-10} \cline{12-13} 
\textbf{4Q259}         &  & -350                         & -310                         &                          & \cellcolor[HTML]{FFFC9E}-210 & \cellcolor[HTML]{FFFC9E}-100 & \cellcolor[HTML]{FFFC9E} & \multicolumn{1}{r}{\cellcolor[HTML]{FFFC9E}-70}  & \multicolumn{1}{r}{\cellcolor[HTML]{FFFC9E}-55}  &                          &                                                 &                                                 \\ \cline{1-1} \cline{3-4} \cline{6-7} \cline{9-10} \cline{12-13} 
\textbf{4Q416}         &  & -345                         & -320                         &                          & \cellcolor[HTML]{FFFC9E}-205 & \cellcolor[HTML]{FFFC9E}-90  & \cellcolor[HTML]{FFFC9E} & \multicolumn{1}{r}{\cellcolor[HTML]{FFFC9E}-80}  & \multicolumn{1}{r}{\cellcolor[HTML]{FFFC9E}-50}  &                          &                                                 &                                                 \\ \cline{1-1} \cline{3-4} \cline{6-7} \cline{9-10} \cline{12-13} 
\textbf{4Q2}           &  & \cellcolor[HTML]{FFFC9E}-155 & \cellcolor[HTML]{FFFC9E}-130 & \cellcolor[HTML]{FFFC9E} & \cellcolor[HTML]{FFFC9E}-125 & \cellcolor[HTML]{FFFC9E}10   &                          &                                                  &                                                  &                          &                                                 &                                                 \\ \cline{1-1} \cline{3-4} \cline{6-7} \cline{9-10} \cline{12-13} 
\textbf{4Q375}         &  & -345                         & -320                         &                          & \cellcolor[HTML]{FFFC9E}-205 & \cellcolor[HTML]{FFFC9E}-50  &                          &                                                  &                                                  &                          &                                                 &                                                 \\ \cline{1-1} \cline{3-4} \cline{6-7} \cline{9-10} \cline{12-13} 
\textbf{Xhev\_Se2}     &  & \cellcolor[HTML]{FFFC9E}-45  & \cellcolor[HTML]{FFFC9E}75   &                          & \multicolumn{1}{l}{}         & \multicolumn{1}{l}{}         &                          &                                                  &                                                  &                          &                                                 &                                                 \\ \cline{1-1} \cline{3-4} \cline{6-7} \cline{9-10} \cline{12-13} 
\textbf{4Q541}         &  & -355                         & -300                         &                          & \cellcolor[HTML]{FFFC9E}-210 & \cellcolor[HTML]{FFFC9E}-95  & \cellcolor[HTML]{FFFC9E} & \multicolumn{1}{r}{\cellcolor[HTML]{FFFC9E}-75}  & \multicolumn{1}{r}{\cellcolor[HTML]{FFFC9E}-55}  &                          &                                                 &                                                 \\ \cline{1-1} \cline{3-4} \cline{6-7} \cline{9-10} \cline{12-13} 
\textbf{4Q521}         &  & -355                         & -285                         &                          & \cellcolor[HTML]{FFFC9E}-230 & \cellcolor[HTML]{FFFC9E}-100 &                          &                                                  &                                                  &                          &                                                 &                                                 \\ \cline{1-1} \cline{3-4} \cline{6-7} \cline{9-10} \cline{12-13} 
\textbf{4Q267}         &  & -355                         & -290                         &                          & \cellcolor[HTML]{FFFC9E}-210 & \cellcolor[HTML]{FFFC9E}-95  & \cellcolor[HTML]{FFFC9E} & \multicolumn{1}{r}{\cellcolor[HTML]{FFFC9E}-70}  & \multicolumn{1}{r}{\cellcolor[HTML]{FFFC9E}-55}  &                          &                                                 &                                                 \\ \cline{1-1} \cline{3-4} \cline{6-7} \cline{9-10} \cline{12-13} 
\textbf{Mur19}         &  & \cellcolor[HTML]{FFFC9E}-45  & \cellcolor[HTML]{FFFC9E}85   & \cellcolor[HTML]{FFFC9E} & \cellcolor[HTML]{FFFC9E}95   & \cellcolor[HTML]{FFFC9E}110  &                          &                                                  &                                                  &                          &                                                 &                                                 \\ 
\hline \hline
\end{longtable}
}

\normalsize

%% file: tex-sections/F-deep-learning.tex
\section{On the use of pre-trained deep learning methods for image-based dating}
\label{appen:J:transferdeeplearningillustration}

\subsection{Considerations on the use of training deep learning neural networks on a problem with only 24 examples}
Since the mathematical proof by Hornik~\cite{Hornik1989}, it took some time but today, deep multilayer neural networks have excelled in many applications, especially since the advent of large data sets and the increase in computational power. However, as observed in the introduction of this article, the likelihood of success is low when training a deep network with too many parameters on a tiny data set. There is a serious risk of an `overfit', i.e., a computed solution that appears to be performant on a training data set but fails to generalize (interpolate) properly when presented with unseen data~\cite{Vapnik2000}. We have looked at a list of 44 modern deep-learning vision models that were published since 2010 and were cited minimally 100 times. Such models have, on average, 454 million weights (coefficients) which are computed from 715 million data points in training, on average, i.e., per single model. A ‘data point’ is a tuple of an image and its corresponding desired model output vector for classification, regression or generative task. The meta-analysis table is kindly provided by~\cite{villalobos2022run,epochMachineLearningData2022}. The most recent, transformer-based, models will even have billions of parameters and data points. It is evident that such large modern models can never be trained from scratch on a data set with just a few dozen, i.e., 24, radiocarbon-dated images as data points, for our problem. 

An alternative approach would be the use of deep {\em transfer} learning~\cite{TransferZhuang2021,ribani2019survey} where an existing deep neural network, trained on a sufficiently large image data set~\cite{Krizhevsky2017}, is fine-tuned on a smaller set of \textsuperscript{14}C-based dated images. In such a case, a hidden layer from a frozen pre-trained network is chosen as the shape-feature vector, and a new post-processing multilayer perceptron or dense network layer is trained to transform that feature vector to produce the output vector required by the actual task, for a given input image sample. We will mention five objectionable points to the use of deep transfer learning for the date prediction task.

\medskip\noindent
{\bf Point 1}. It is questionable whether currently common networks that are trained and designed for natural full-colour RGB photographic image classification will deliver a shape feature vector in their penultimate layer that is optimal for writer identification in bi-tonal manuscript images. Bitonal manuscript images have a flat-white background, and the interesting patterns are in the ink traces only. Such material is rarely present in generic photographic image collections. At the very least, there will be serious worries concerning efficiency, because about two-thirds of the connection weights are likely to be superfluous.

\medskip\noindent
{\bf Point 2}. Even if the colour-channel argument is dismissed, end users may argue that an opaque neural~\cite{Krizhevsky2017} network or vision transformer~\cite{ViTpaperDosivitskiy2020} method that is pretrained on non-representative image material ('photos of cats, dogs and urban scenes, etc.') would not be acceptable for answering scholarly questions. To put this in comparison, current deep foundation models are not considered a good basis for the serious application in medical diagnostics in radiology yet and massive data would need to be collected in order to achieve such a status~\cite{WilleminkFoundationRadiology2022}.

\medskip\noindent
{\bf Point 3}. Current deep-learning methods rely on images that are often very small, i.e., 224x224 or 512x512 pixel images. Only recently, with increased memory capacity in GPUs, images of 768x768 pixels can be used. This leads to many problems in the real-world application of deep learning, e.g., in a medical context~\cite{Thambawita2021}. Our manuscript-image sizes are large and of variable aspect ratio, with widths of $\mu_{w}=3871$ ($\pm \sigma_{w}=1069$) pixels and heights of $\mu_{h}=3857$ ($\pm \sigma_{h}=740$) pixels. On top of the other restrictions mentioned here, using current deep learning would require a downscaling of the high-quality manuscript images with a factor of 5 to 7, with considerable loss of information. Whereas recent vision transformers~\cite{ViTpaperDosivitskiy2020} are better suited to deal with large images, they are based on extraneous very large image and photographic collections (cf. Point 2).

\medskip\noindent
{\bf Point 4}. Alternatively, tiling~\cite{Haja2021} would unnecessarily complicate the analyses because of the likely imbalance of character content between the tiles and damage to the original character appearance at the tile margins. In spite of the success and allure of deep learning, dealing with large, variable-sized images has not been fundamentally solved. This puts a limit on their applicability in several scientific domains. In microscopy~\cite{Campanella2019} and astronomy~\cite{ivezic2019lsst}, multi-gigapixel, terabyte images are already common. As in our case, current convolutional neural networks, per se, cannot process an original whole image in its unscaled entirety without information loss, e.g., for a prediction task. 

\medskip\noindent
{\bf Point 5}. Regardless of the methodological problems in the face of sparse data, an end-to-end deep learning
approach, i.e., transforming image pixels into a date prediction directly, has the disadvantage of limited explainability. If dedicated features can be used that are explainable and a regression model can be trained that requires limited data, such a modular approach has a distinct advantage. Still, it is worthwhile to explore a deep-learned variant for date prediction, as more (radiocarbon-)dated samples will become available.


\subsection{An attempt in using state-of-the-art deep-learning methods, PNASNet}
\begin{figure}[!ht]
    \centering
        \begin{subfigure}{0.49\textwidth}
        \includegraphics[width=\textwidth]{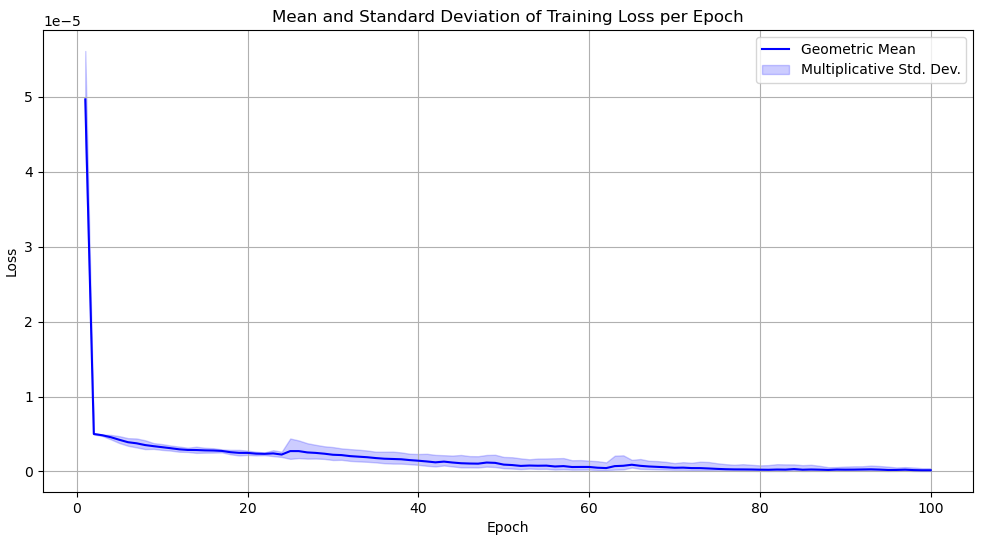}
        \phantomsubcaption
        \end{subfigure}
    \hfill
        \begin{subfigure}{0.49\textwidth}
        \includegraphics[width=\textwidth]{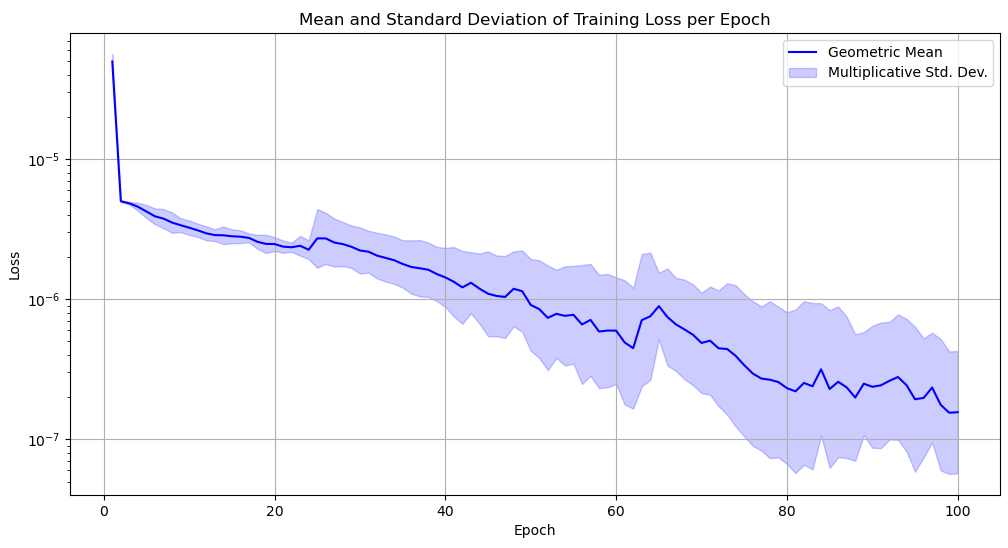}
        \phantomsubcaption
        \end{subfigure}
    \caption{PNASNet training loss per epoch for 4-fold cross-validation (log-scale on the right side).}
    \label{fig:pnasnet-training-epoch}
\end{figure}

\begin{figure}[!ht]
    \centering
        \begin{subfigure}{0.49\textwidth}
    \includegraphics[width=\textwidth]{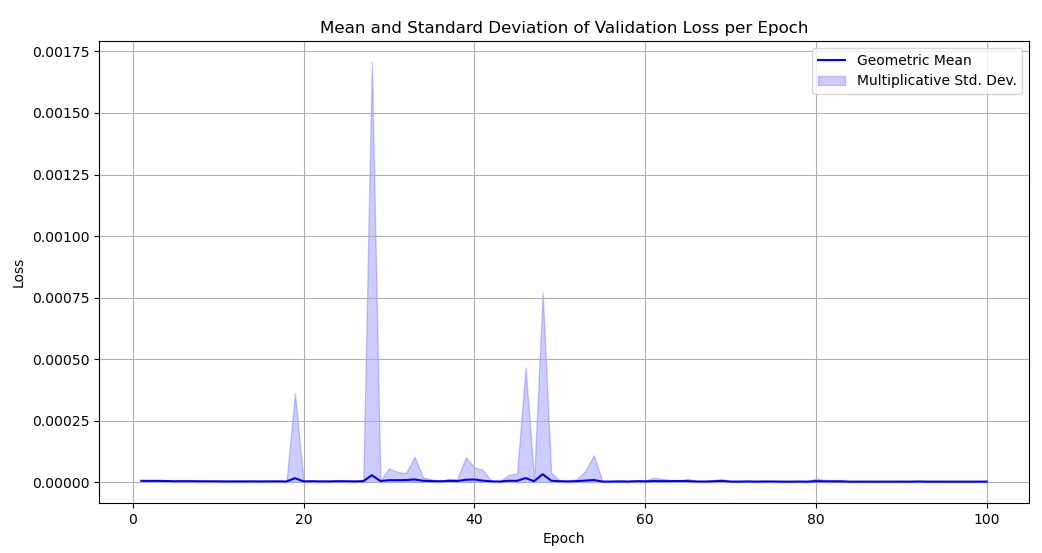}
        \phantomsubcaption
        \end{subfigure}
    \hfill
        \begin{subfigure}{0.49\textwidth}
    \includegraphics[width=\textwidth]{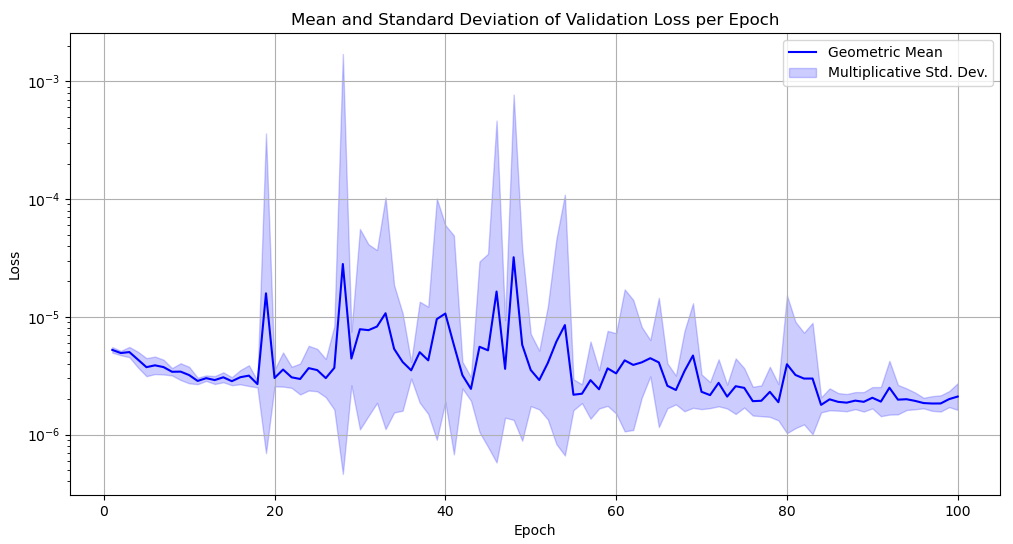}
        \phantomsubcaption
        \end{subfigure}
    \caption{PNASNet validation loss per epoch for 4-fold cross-validation (log-scale on the right side).}
  \label{fig:pnasnet-validation-epoch}
\end{figure}

\begin{figure}[!ht]
    \centering
        \begin{subfigure}{0.49\textwidth}
    \includegraphics[width=\textwidth]{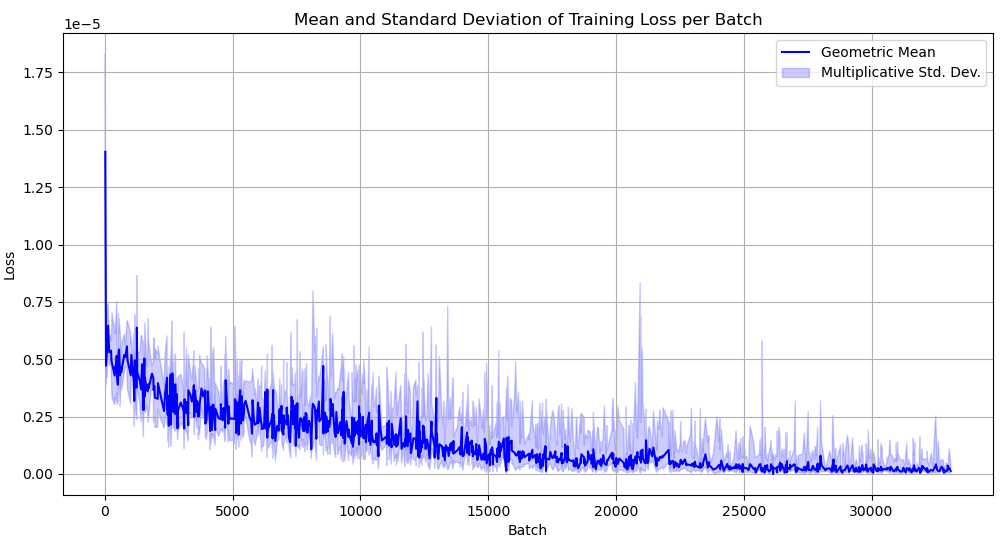}
        \phantomsubcaption
        \end{subfigure}
    \hfill
        \begin{subfigure}{0.49\textwidth}
    \includegraphics[width=\textwidth]{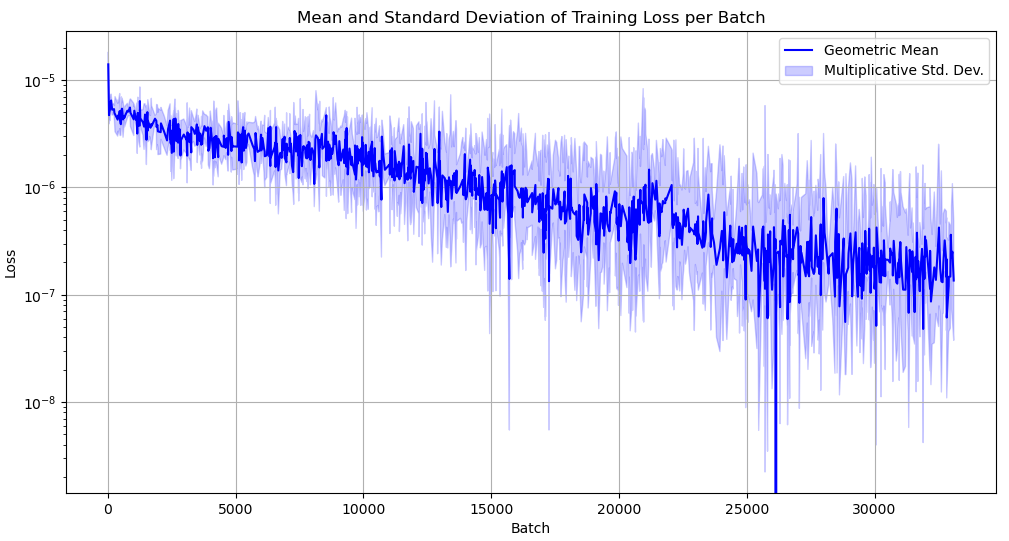}
        \phantomsubcaption
        \end{subfigure}
    \caption{PNASNet training loss per batch for 4-fold cross-validation (log-scale on the right side).}
  \label{fig:pnasnet-training-batch}
\end{figure}

\begin{figure}[!ht]
    \centering
        \begin{subfigure}{0.49\textwidth}
    \includegraphics[width=\textwidth]{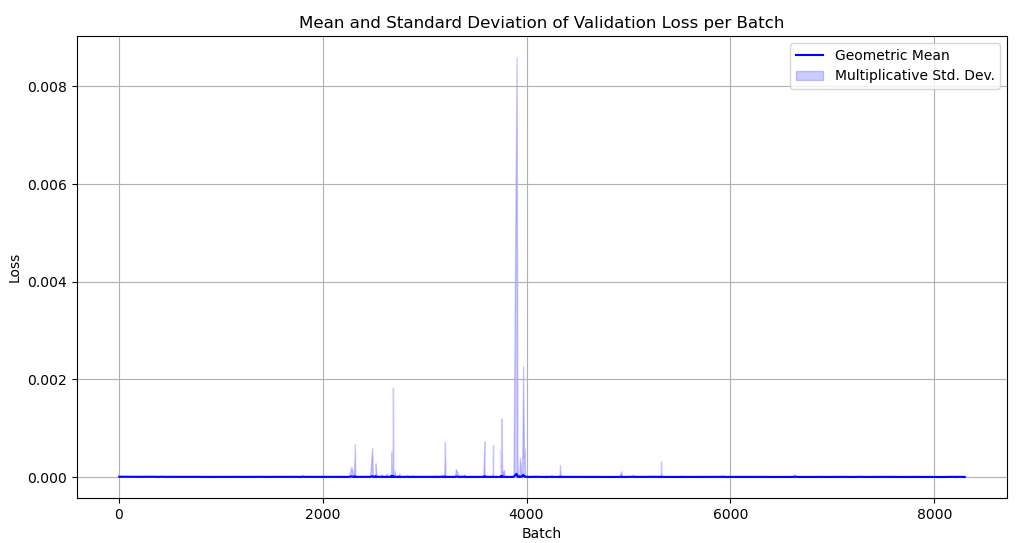}
        \phantomsubcaption
        \end{subfigure}
    \hfill
        \begin{subfigure}{0.49\textwidth}
    \includegraphics[width=\textwidth]{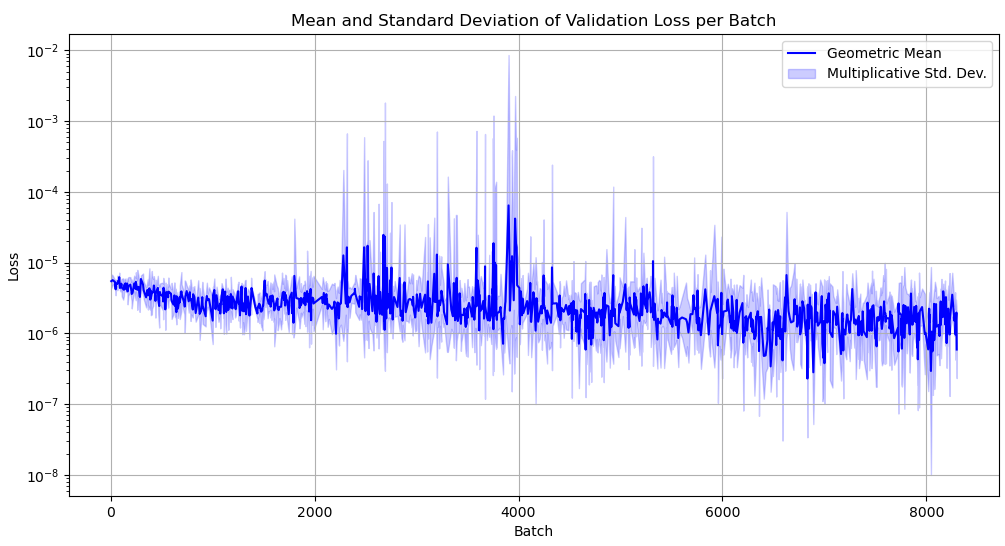}
        \phantomsubcaption
        \end{subfigure}
    \caption{PNASNet validation loss per batch for 4-fold cross-validation. Validation is performed after every fourth epoch, so the curve is aligned with Figure ~\ref{fig:pnasnet-training-batch} (log-scale on the right side).}
  \label{fig:pnasnet-validation-batch}
\end{figure}

However, in order to empirically illustrate the problems with current deep learning, even when used in a transfer-learning setup, we have used a common foundational model (PNASNet) and used its output to estimate a date probability distribution. Using a pretrained PNASNet~\cite{PNASnet5-Liu2018}, we rescaled each high-resolution manuscript image to the 'passport-photo' size, which is customary for these models, i.e., $331\times331$ pixels. We then used the penultimate layer ($N_{hidden}=4320$) of this existing network as a pretrained feature vector for a date-estimation output layer with an 'OxCal' format probability target function. Figures~\ref{fig:pnasnet-training-epoch} and \ref{fig:pnasnet-training-batch} show the evolution of the loss curve in a typical training session. Although there is some variance present, the model seems to converge more or less after the presentation of 32,000 batches. However, when looking at the validation curve (Figure~\ref{fig:pnasnet-validation-epoch} and \ref{fig:pnasnet-validation-batch}), we can see that the loss remains highly irregular. The most likely reason for this behavior is that, in spite of the frozen pre-trained mass of PNASNet, the number of fluid weights that need to be estimated for the transfer task is still in the order of 457k ($N_{hidden}=4320$ x $N_{OxCal}=106$). This number is irresponsibly high, in relation to the small number of images in the data set.
The subpar loss level in comparison to training, the horizontality and irregularity of the validation loss curves gives us very strong support for the decision to avoid this pathway. At the very least, it can be concluded that considerable additional research would be needed to improve the DL-transfer performance, from here. Given the particular conditions of this study, we have avoided the use of deep learning for the regression task, waiting for the data sparsity to be solved.

~
\vfill

%% file: tex-sections/G-test-135.tex
\section{Enoch’s date predictions for 135 previously undated manuscripts}\label{appen:H}
Before we discuss the results of the palaeographic post-hoc evaluation of the 135 unseen samples (Section~\ref{appen:H1:firsteval}), we elaborate on the physical and image quality of the data, as well as explain how to read Enoch’s prediction plots and elaborate on how Enoch differs from traditional palaeography.

\subsection{On the physical and image quality of the data}\label{appen:H2:dataquality}
In order to appreciate how the Enoch model works, it should be noted what challenges the data pose, physically and image-wise. 

As we have mentioned before, the Dead Sea Scrolls are extremely delicate material (see Section~\ref{sec:c14} in the main article and Section~\ref{appen:B2:pretreat} in Appendix~\ref{appen:B}). In a few cases, the physical evidence consists of largely intact bookrolls of several meters in length, such as the Great Isaiah Scroll (see Section~\ref{harvest-enoch} in the main article). But in most cases, what were once large and small bookrolls are now only extant as fragmentary, deteriorated remains of various sizes and shapes. This means that the Enoch model has to deal with very diverse material remains that are available as digital images (see Section~\ref{deepneuralnetworksmainarticle} in the main article). 

The physical state of the data affects the image quality in various manners. For example, papyrus fragments often have damage patterns that affect the ink remains of the letters differently than fragments of animal skin remains do. Or, some manuscripts are represented by large, relatively well-preserved fragments, whereas others only have one small, badly damaged fragment left. Our binarized images for Enoch sometimes combine different fragments of a manuscript that are available on separate image plates of the IAA (e.g., 4Q86). Thus, the data for Enoch consists of diverse image types. To elaborate on what was briefly mentioned in Section~\ref{discussion-Enochapproach} in the main article, image preparation treatment is important to further improve Enoch’s prediction results. The model does not change its prediction with the same set of training and testing data, but predictions can change (read “improve”) because of better-cleaned images. 

This also means that two or more predictions for the same manuscript can have different results because the underlying data consists of diverse image types that warrant a diverse spread in the plots. Unlike MPS~\cite{He2016} (or other historical manuscripts), the Dead Sea Scrolls images are all different in shape, orientation, number of characters, ink thickness, etc. Considering 4Q57, for example, there are nine “curves” (plots), which the palaeographers used in the first evaluation, because there are nine images. It should be noted that these are nine different individual images, due to improved/updated preprocessing performed over time, of 4Q57, often of different fragments. As the model receives different images (features), it produces different curves (plots). Each image represents a different set of evidence (from character shapes/features) for each bin. If all nine plots of 4Q57 were exactly the same, then that would be problematic because each image fragment is different even though they are from the same manuscript. 

\begin{figure}[!ht]
    \centering
    \includegraphics[width=0.9\textwidth]{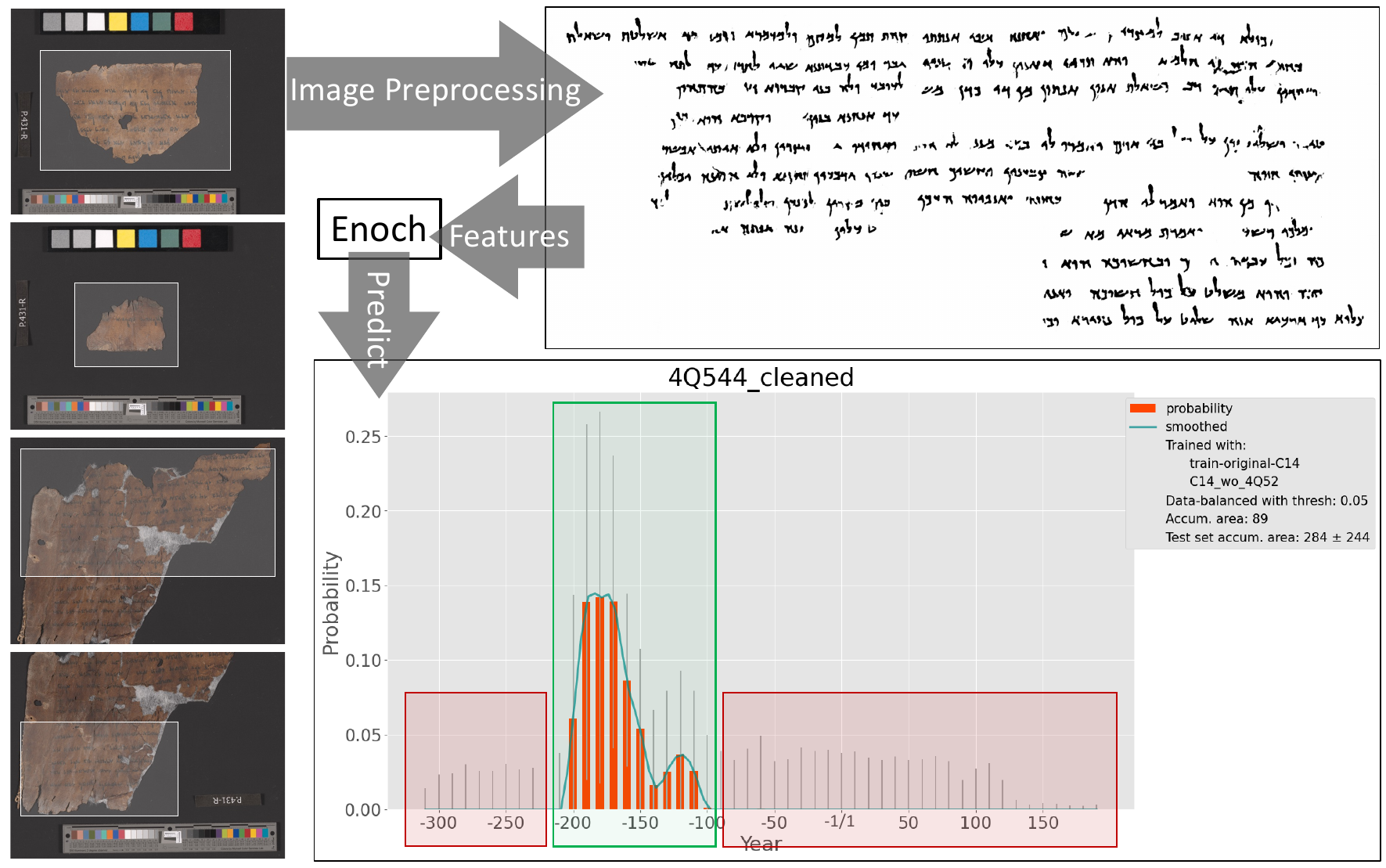}
	\caption{Enoch's prediction for 4Q544 as an example: Four fragment images from IAA plate 431 are on the left side. These fragment images are preprocessed using multispectral image fusion, neural network-based binarization (BiNet), noise reduction, and alignment correction to obtain the image on the top right. The features extracted from the preprocessed image are passed through the trained Enoch model to produce the prediction on the bottom right. This is a simple, unimodal prediction. The \textcolor{ForestGreen}{\textbf{green}} box indicates the probable date range with high mean values (in this case, 200–100 BCE), and the \textcolor{red}{\textbf{red}} box indicates no mean with high uncertainty areas. }
	\label{fig:howto-4q544}
\end{figure}

\subsection{How to read a prediction plot} \label{subsec:howtoread}
Each prediction plot produced by Enoch presents the output of the Bayesian regression model and pertains to an individual manuscript test sample. In the plot, the X-axis delineates the chronological timeline, partitioned into 10-year bins, while the Y-axis conveys probability values in the form of means with error bars. This representation encapsulates the model’s endeavour to infer the probable dating of the manuscripts, each within a 10-year interval, across a temporal expanse from 310 BCE to 200 CE years.

Within each 10-year bin, a pair of values is obtained: the mean and its corresponding error usually expressed as standard deviation. The mean is a point estimate, indicating the central tendency of the predicted manuscript dates for the given bin, thereby proposing an approximation of the most plausible date within the specific timeframe. The associated error, or standard deviation, serves as a critical metric showing the magnitude of variability inherent in the predicted dates and concurrently serves as a measure of uncertainty. An ‘ideal’ date prediction has a high probability and a low error.

The plot looks like a series of bars, like a histogram. By looking at these bars, we look for any patterns in the dates over time and gauge how confident or uncertain we are about these estimates. So, for individual plots, we look at the level of the mean value and the size of the error bars around it, to decide the most probable date or date range for that individual manuscript.

\begin{figure}[!ht]
    \centering
    \includegraphics[width=0.9\textwidth]{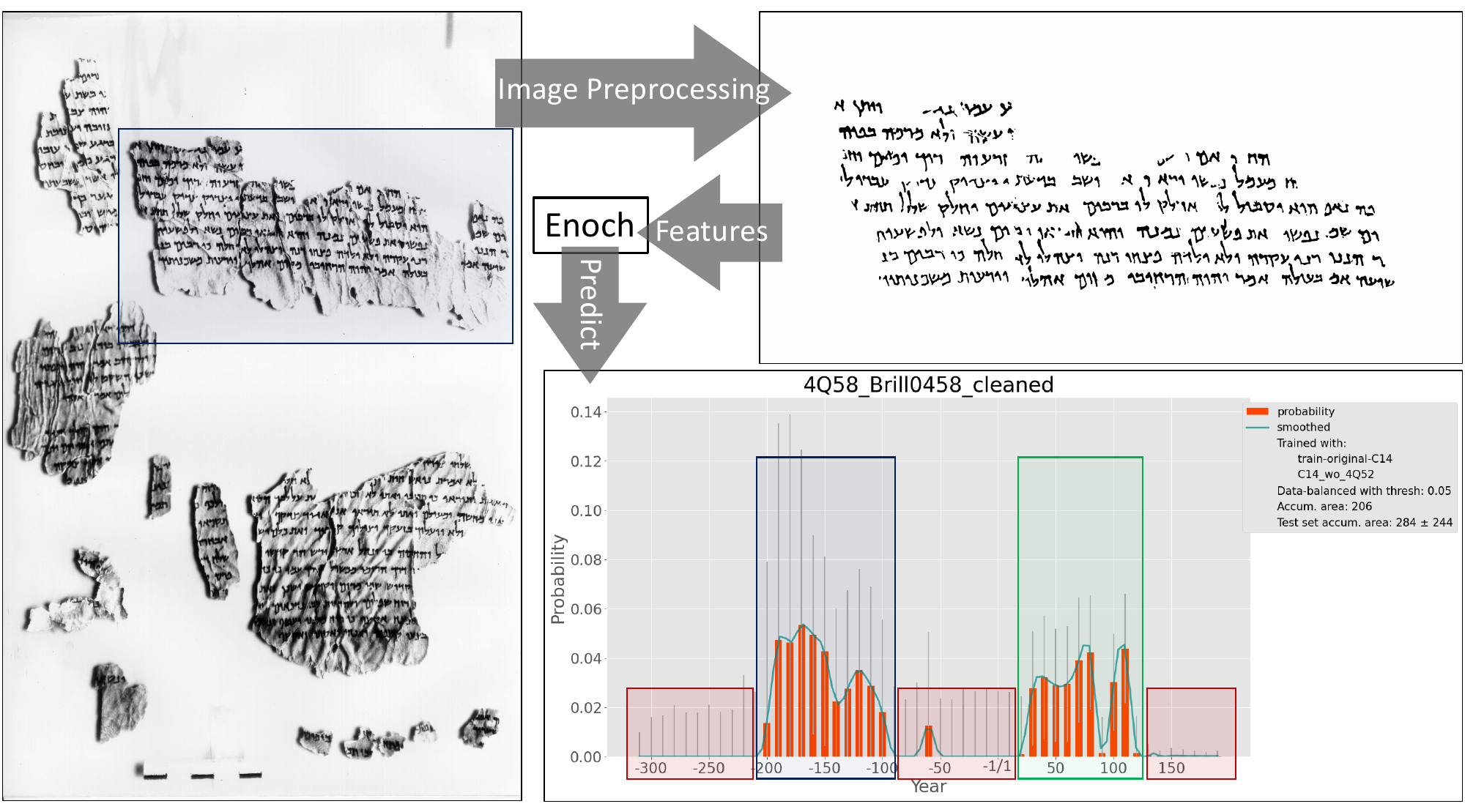}
	\caption{Enoch’s prediction for 4Q58 as an example: Brill scan 431 is on the left side. The top right fragment (marked in \textcolor{RoyalBlue}{\textbf{blue}}) is preprocessed using neural network-based binarization (BiNet), noise reduction, and alignment correction to obtain the image on the top right. The features extracted from the preprocessed image are passed through the trained Enoch model to produce the prediction on the bottom right. This is a bimodal prediction. In this prediction plot, the \textcolor{RoyalBlue}{\textbf{blue}} and \textcolor{ForestGreen}{\textbf{green}} boxes both indicate the probable date ranges with high mean values, and the \textcolor{red}{\textbf{red}} boxes indicate no significant mean with high uncertainty areas. Now, if the reader needs to choose one of the ranges from the \textcolor{RoyalBlue}{\textbf{blue}} or the \textcolor{ForestGreen}{\textbf{green}}, then the \textcolor{ForestGreen}{\textbf{green}} is the more probable range (in this case, 30– 120 CE) because of smaller error bars than the \textcolor{RoyalBlue}{\textbf{blue}} range.}
	\label{fig:howto-4q58}
\end{figure}

The discrete prediction bars can be mathematically smoothed into continuous curves, yielding Gaussian Mixture Models (GMMs) as a representation. This transformation allows for a more nuanced and probabilistic portrayal of the underlying distribution of predicted manuscript dates. If the smoothed prediction depicts a unimodal distribution, choosing the probable date range is easy (see Figure~\ref{fig:howto-4q544}). However, it requires more attention when the prediction is bimodal. The reader then needs to pay more attention to the error bars and the means for each 10-year bin (see Figure~\ref{fig:howto-4q58}). This cannot be easily solved with an algorithm: A high probability value is ‘good’, but not if it is accompanied by a large uncertainty. In that case, the choice of a stable estimate with a slightly lower mean probability may be advisable.


\begin{figure}[!ht]
    \centering
    \includegraphics[width=\textwidth]{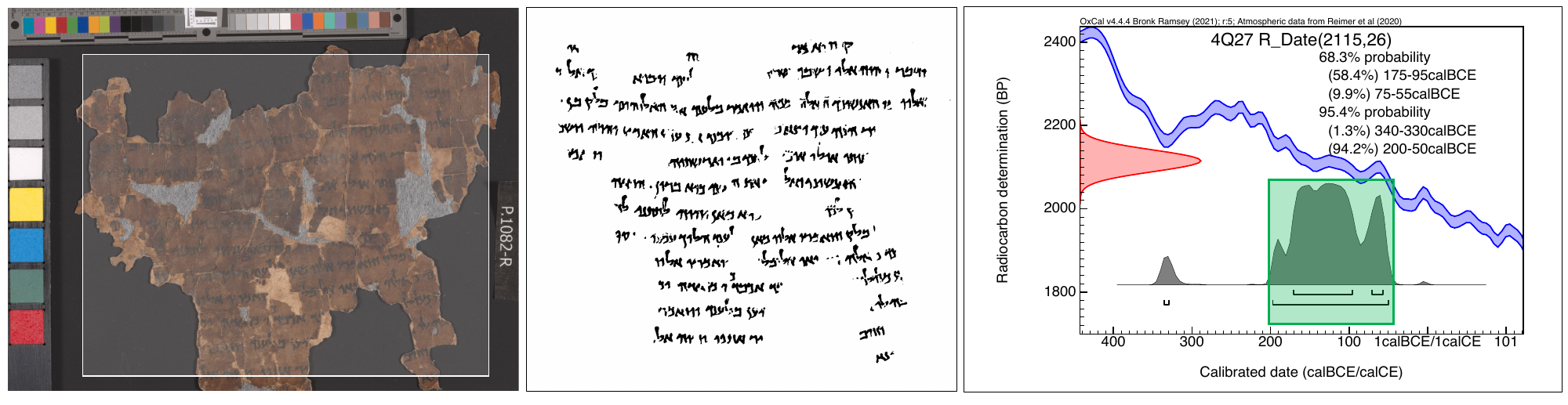}
	\caption{4Q27 as a training sample for Enoch’s date prediction (4Q27 has six images in training and two in testing. This is one of the six training images): Fragment 1 from IAA plate 1082 is on the left side. This fragment image is preprocessed using multispectral image fusion, neural network-based binarization (BiNet), noise reduction, and alignment correction to obtain the image in the middle. On the right is the OxCal data from radiocarbon dating for 4Q27, which is the target output for the training of Enoch. The \textcolor{ForestGreen}{\textbf{green}} area indicates the accepted part of the 2$\sigma$ calibrated bimodal data.}
	\label{fig:4Q27-train-visual}
\end{figure}

\subsection{On shared characteristics and finding matches elsewhere}\label{appen:H3:shared characteristics}
In order to appreciate how the Enoch model differs from traditional palaeographic approaches, we elaborate on what was briefly mentioned in Section~\ref{discussion-Enochapproach} in the main article, namely that Enoch emphasizes shared characteristics and similarity matching, whereas traditional palaeography focuses on dissimilarities that are assumed indicative for style development. 

\begin{figure}[!ht]
    \centering
    \includegraphics[width=\textwidth]{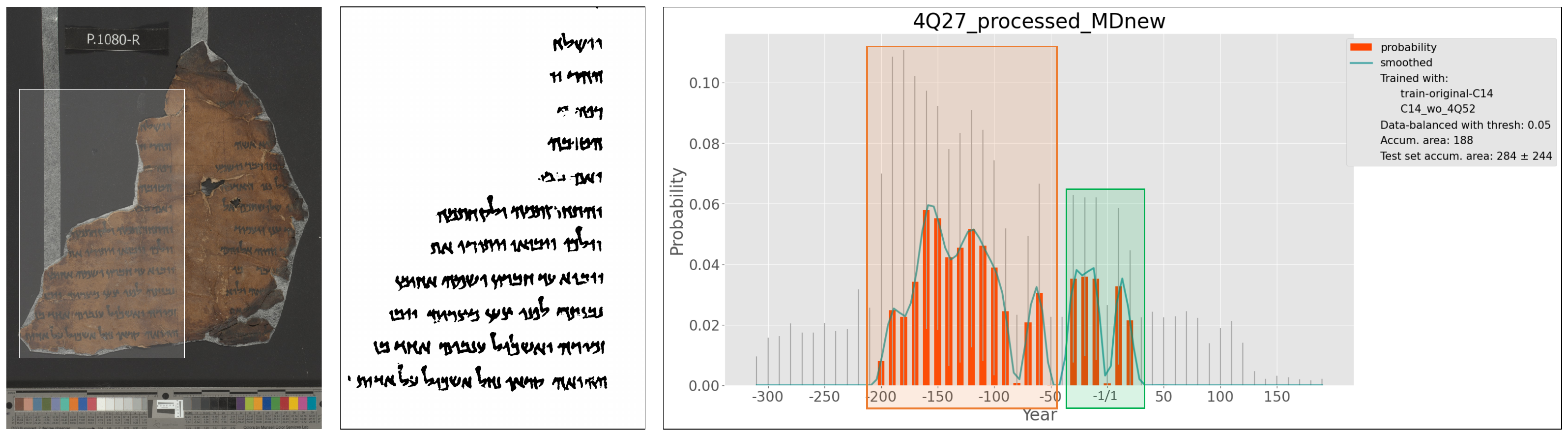}
	\caption{Enoch’s prediction for 4Q27 (4Q27 has six images in training and two in testing; this is one of the two test images): Fragment 6 from IAA plate 1080 is on the left side. This fragment image is preprocessed using multispectral image fusion, neural network-based binarization (BiNet), noise reduction, and alignment correction to obtain the test image in the middle. The features extracted from the preprocessed image are passed through the trained Enoch model to produce the prediction on the right. Here, the \textcolor{BurntOrange}{\textbf{orange}} area depicts a similar shape to the target OxCal for 4Q27 in training (see Figure~\ref{fig:4Q27-train-visual}), as expected from a regression model. In addition, Enoch also looks for similarity over all the training samples and finds an additional probability range (in \textcolor{ForestGreen}{\textbf{green}}). Due to the low error values, i.e., smaller error bars, the \textcolor{ForestGreen}{\textbf{green}} area is more probable (here, 30 BCE to 20 CE) than the \textcolor{BurntOrange}{\textbf{orange}} area in this prediction plot.}
	\label{fig:4Q27-test-visual}
\end{figure}

Enoch’s Bayesian regression model performs the quantitative analysis of textural and allographic feature vectors. These feature vectors encapsulate various handwriting characteristics, offering a systematic representation for predictive modeling. Unlike the traditional approach of human palaeographers, who often seek dissimilarities, this model employs a similarity-based strategy. It strives to uncover patterns and relationships within the feature space by quantifying the resemblance of test images to the training set with known \textsuperscript{14}C date distribution. Leveraging a Bayesian framework, the model offers a probabilistic and data-driven means of attributing dates to unseen manuscripts. It thus complements the qualitative expertise of human palaeographers with a quantitative approach that can reveal subtle patterns and associations within the data. 4Q27 provides an excellent example of this approach, with six images in training Enoch and two in test prediction. Figure~\ref{fig:4Q27-train-visual} shows one of the training images, and Figure~\ref{fig:4Q27-test-visual} shows the prediction plot for one of the two test images. 

Another example is that of 4Q319 (see Section \ref{harvest-enoch} and Figure \ref{fig:4Q319result} in the main article). Here, the AI experts’ preferred reading of the prediction plot (see Table~\ref{tab:expert-AI-135undated}) is for the younger range, against the so-called ‘biased range’ of 200–100 BCE. The occurrence of ‘young’ peaks can be seen as shape information suggesting an alternative to the bias (that is present for the 200-100 BCE range). The overall shape on the left-hand side is likely due to the OxCal-based training. However, in spite of the lesser occurrence of younger fragments in the training data, the right-hand part shows that from the style-based analysis a younger date is possible due to the stable results (with small error bars).

\subsection{Palaeographic post-hoc evaluation}\label{appen:H1:firsteval}
The 135 unseen samples were chosen for various palaeographic and historical reasons, such as a diachronic cross-section of a biblical book (Psalms), manuscripts that share the same writing style, or for no particular reason at all. 

Second, for the evaluation, we took the best image for each manuscript in terms of image quality: better-cleaned images give better results (see Section~\ref{appen:H2:dataquality}). The AI experts among the article’s authors, M.D. and L.S., made visual evaluations of the scans in order to ensure that data is of sufficient quality with a sufficient number of characters in the used sample. The list of which specific image for each manuscript sample was used in the evaluation can be seen in Table~\ref{tab:135-images}. Still, there remain very poor and difficult images for a number of scrolls to work with but we kept them in the test and did not want to tweak the data. So, in cases where the best available image is still a very poor image, we worked with that (see some examples as illustrations of this very poor quality in Table~\ref{tab:bad25-images}). However, we kept all the images so the reader can see all the implications and improvements that can be obtained from careful preprocessing of the images. In Zenodo data repository (\href{https://doi.org/10.5281/zenodo.10629480}{https://doi.org/10.5281/zenodo.10629480}), the images are organized in three different directories: the first one with all 359 images for the 135 manuscripts, the second one with the selected 135 images, and the final one with 25 images to illustrate the poor quality of images.

Third, we did different balancing tests (see Section~\ref{subsec:data-balance} in Appendix~\ref{appen:C}) and so produced different prediction plots. Yet, in the final evaluation we only use the balanced 0.05 plots, which we also indicate in Zenodo (\href{https://doi.org/10.5281/zenodo.10629480}{https://doi.org/10.5281/zenodo.10629480}), in the description of \textit{Organization of the data}. All other prediction plots are also available for readers to see the different balancing tests that we have done (see inside Enoch-predictions.tar.gz file in the Zenodo data repository).

Fourth, the AI experts performed a blind reading of the balanced 0.05 prediction plots. They had no knowledge of the manuscript dates and only read the prediction plots, giving estimated minimum and maximum ranges (see Table~\ref{tab:expert-AI-135undated}; for more details on how to read a plot, see Sections~\ref{subsec:howtoread} and~\ref{appen:H3:shared characteristics}). These estimated minimum and maximum date ranges were then passed on to the palaeographers to assess the outcomes as “realistic” or “unrealistic”. 

It is even possible to provide an algorithm to read the plots but the design philosophy of our date prediction model is based on the assumption that it is better to stay close to the known systematics in dealing with OxCal curves with date probabilities than to reside to an ‘oracle’ approach where an algorithm proposes a hard date range. The user can inspect the output of our Enoch model in a similar way as the OxCal curve analysis would ensue.

Fifth, in their qualitative post-hoc evaluation, the palaeography experts among the article’s authors, M.P. and E.T., regarded a date prediction as “realistic” if a prediction corresponds (partly or wholly) with their palaeographic estimates, the basis for which was already explained in detail in Appendix~\ref{appen:A} and Section~\ref{appen:B8:comparison}, and “unrealistic” when it does not. In other words, if there is an overlap between our palaeographic estimates and the machine-learning-based dating, even if the overlap is minimal, we regard the model’s date prediction as “realistic”, and “unrealistic” when there is no overlap, i.e., when it is older or younger (“too old” or “too young”). We provide our palaeographic date estimates for each of the 135 manuscripts (see Table~\ref{tab:expert-AI-135undated}), with the general principle in mind that we work with a 50-years range and allow for +/- 25 years on either side. Sometimes, e.g., in the case of quite idiosyncratic handwriting, we allow for an even broader range of 100 years. It should also be noted that if the data are of poor quality, especially if only little material is left and therefore few characters to inspect, then palaeographic estimations are more difficult to make. In other words, the palaeographic dates are not hard date ranges, but expert estimates. Still, for the evaluation we used the 50-years range in a strict sense for reasons of clarity, so that if there was only a 5- or 10-year gap we deemed the prediction as “too old” or “too young”. 

Summarized, our post-hoc palaeographic assessment is based on the following considerations:

\begin{enumerate}

\item In line with Cross and all other palaeographers, we make a distinction between Hasmonaean- and Herodian-style writing;

\item Our palaeographic date estimates of these styles vis-à-vis one another are informed by the traditional view of the Hasmonaean script as being older than the Herodian script. Our \textsuperscript{14}C results confirm for most manuscripts the basic distinction between Hasmonaean-type manuscripts that are older, and Herodian-style manuscripts that are younger. Yet, for Herodian-type script, our \textsuperscript{14}C results indicate that Herodian script was present earlier than previously thought. Our evaluation of the implications of the \textsuperscript{14}C data for Hasmonaean-type script provides evidence for dates in the second century BCE and also allows for the late third century BCE, and for Herodian-type script to be already in existence earlier side by side with Hasmonaean-type script in the second century BCE (see Section~\ref{appen:B8.3:conclcomp}). Thus, we took into account the general tendency in the \textsuperscript{14}C results that date both individual manuscripts and the emergence of the ‘Hasmonaean’ and ‘Herodian’ scripts about 50–75 years earlier than according to traditional palaeography;

\item Linear typological developments within both Hasmonaean- and Herodian-type script have been stated by scholars, rather than substantiated with external date-bearing evidence. This makes traditional assumptions about “within script” linear typological development problematic, in our view even more so of ‘Herodian’ than of ‘Hasmonaean’. Especially for script generally seen as Late Herodian, we would not exclude a date around the turn of the era or somewhat earlier. We reckon with the possibility of a longevity of script types longer than traditionally assumed. Cross assumed a rapid development of the script from the Hasmonaean period onward. He suggested chronological ranges of 50 years, and sometimes even shorter ranges of 25–50 years for typological developments, but these assumptions remain unsubstantiated.

\end{enumerate}
\medskip
\noindent

It should be noted that other researchers can follow our evaluation by taking the range estimates (see Table~\ref{tab:expert-AI-135undated}) and/or look at the prediction plots (from Zenodo repository (\href{https://doi.org/10.5281/zenodo.8168210}{https://doi.org/10.5281/zenodo.8168210})), then consider the specific images of the manuscripts in question in the IAA’s Leon Levy Dead Sea Scrolls Digital Library collection~\cite{dssllweb} and/or consider our binarized images (from the previously mentioned Zenodo repository), and take into account our considerations (see Appendix~\ref{appen:A} and Section~\ref{appen:B8:comparison}). Or, instead of following our reasoning for a “realistic” or “unrealistic” assessment, they can make their own palaeographic post-hoc assessment, and justify their reasoning. 

Also, please note that there are different probability values for each 10-year bin’s prediction within these minimum and maximum ranges. So, AI experts’ minimum and maximum values limit a probable range, but the range is not the final estimated date. One needs to read the probability plots to better estimate within the minimum-maximum range. This means that the range can sometimes be wide, but by reading the probability values along with the uncertainty estimates (or error bars), a reader can even narrow down to a more precise date range if they wish to do so.

The blind range estimation by the AI experts shows the distributions of year ranges in Table~\ref{tab:blind-ranges}.

\begin{table}[!ht]
\centering
\caption{Spread estimation (blind-test) by the AI expert}
\label{tab:blind-ranges}
\begin{tabular}{l|c|r}
\hline \hline
\textbf{Range}  & \textbf{Count} & \textbf{Percentage} \\ \hline
280 years                                & 2                                                            & 1.48\%                                        \\ \hline
240 years                                & 1                                                            & 0.74\%                                        \\ \hline
210 years                                & 4                                                            & 2.96\%                                        \\ \hline
190 years                                & 2                                                            & 1.48\%                                        \\ \hline
170 years                                & 6                                                            & 4.44\%                                        \\ \hline
160 years                                & 5                                                            & 3.70\%                                        \\ \hline
150 years                                & 5                                                            & 3.70\%                                        \\ \hline
140 years                                & 4                                                            & 2.96\%                                        \\ \hline
130 years                                & 8                                                            & 5.93\%                                        \\ \hline
120 years                                & 5                                                            & 3.70\%                                        \\ \hline
110 years                                & 8                                                            & 5.93\%                                        \\ \hline
100 years                                & 7                                                            & 5.19\%                                        \\ \hline
90 years                                 & 18                                                           & 13.33\%                                       \\ \hline
80 years                                 & 9                                                            & 6.67\%                                        \\ \hline
70 years                                 & 4                                                            & 2.96\%                                        \\ \hline
60 years                                 & 11                                                           & 8.15\%                                        \\ \hline
50 years                                 & 22                                                           & 16.30\%                                       \\ \hline
40 years                                 & 8                                                            & 5.93\%                                        \\ \hline
30 years                                 & 5                                                            & 3.70\%                                        \\ \hline
20 years                                 & 1                                                            & 0.74\%                                        \\ \hline
\textbf{Total:} & \textbf{135}   & \textbf{100.00\%}   \\ \hline \hline
\end{tabular}
\end{table}

Some year ranges are so wide that the date prediction loses its effect of offering a limited number of quantified probability options within the time period under consideration. Fortunately, the instance of wide prediction ranges is limited within the 135 test samples. The definition for “wide range” is informed by the accepted 2$\sigma$ calibrated ranges which are the training data for the Enoch model and are on average 135 years, including the so-called minor peaks, or 110 years excluding the so-called minor peaks (see Figures~\ref{fig:cum_c14-wMinor} and \ref{fig:cum_c14}, and Table~\ref{tab:2-sigma-accepted}). Twenty-nine of the 135 test samples (21\%) have a date range of more than 130 years, whereas 42 of the 135 test samples (31\%) have a date range of more than 110 years. 

In most cases, the date prediction range is well below 135 or 110 years, often only ca. 50 years (16\%), which has the highest count of all the ranges (see Table~\ref{tab:blind-ranges}).

The current average year value is 69.35 years, excluding wide ranges above 110 years, and 76.32 years, excluding wide ranges above 135 years. If one were to indiscriminately include all ranges, then the current average year value would be 98.76. The median value is 90 years. It should be noted that these average year values, as well as the median value, can change if, in the future, more manuscripts are tested. Also, if the image quality is further improved, these numbers can also be affected and improved (see below). 

Most date ranges are indeed below or up to 90 years (78 out of 135 test samples). It should be noted that the possibility was claimed for the traditional palaeographic model to be able to fix a characteristic bookhand or the copying of a manuscript within 50 years or even 25–50 years, but that this was not substantiated with external date-bearing evidence (see Section~\ref{appen:B8:comparison}). Now, our Enoch model can produce prediction ranges of 50 years that are empirically based on physical evidence derived from \textsuperscript{14}C and geometric evidence from shape-based analysis. Enoch outperforms the \textsuperscript{14}C results: Enoch’s predictions are even narrower than the \textsuperscript{14}C date ranges in the time period 300–50 BCE, provide a more fine-grained distribution (as mentioned in Section~\ref{main-valid-enoch} in the main article). 

As can be seen in Table~\ref{tab:undated} in the main article, 107 (79\%) of the 135 undated manuscripts were judged to have obtained a realistic date prediction. Of course, the wider the range of years of prediction plots are, the more manuscripts show an overlap between our palaeographic estimates and the machine-learning-based dating. If we disregard the 42 date predictions with a spread wider than 110 years, then the percentage of realistic predictions drops to 50\% (68 out of 135) or to 73\% (68 out of 93). Thus, even with a stricter selection rule, only allowing the narrow-range estimates, still a decent percentage of palaeographically realistic evaluations can be obtained from the harvest of undated material. Moreover, if we would also take into account the image quality of the samples and choose instead not to use data of very poor quality then the performance of the Enoch model becomes even more impressive. Twenty-five images are of poor quality (see Table~\ref{tab:bad25}), leaving 110 images and samples in the test, of which 91\% have a realistic prediction. Again, from these 110 images, if we ignore the 36 date predictions with a spread wider than 110 years, then the percentage of realistic predictions amounts to 61\% out of 110, or 89\% out of 74.

In the post-hoc evaluation, the palaeographers refrained from a decision in 4 cases (“see comment 1–4” in Table~\ref{tab:expert-AI-135undated}). The comments are as follows:
\begin{enumerate}

\item \label{see-comment1} 4Q73: we consider this test sample a borderline as we would expect an older dating, ca. 100 BCE or ca. 75 BCE, in view of our considerations, especially the \textsuperscript{14}C results for Hasmonaean manuscripts (Section~\ref{appen:H1:firsteval}). The traditional palaeographic date estimation, middle of the first century BCE~\cite{DJD15}, comes close to Enoch’s date prediction;

\item \label{see-comment2} 4Q379: we consider the semicursive script in this manuscript difficult to date. Some semicursive manuscripts are easier to date, but this one is difficult, also according to the traditional palaeographic model there is too little to go on. Therefore, we refrain from a decision; 4Q379 could be around 100 BCE and then the prediction is realistic, but it could also be later. Cf. also~\cite{DJD22}: the general indication “Hasmonaean semicursive” (263) indicates the difficulty in dating; 

\item \label{see-comment3} 4Q398: this is again a manuscript in semicursive script, and difficult to date. Other palaeography experts gave the following dates: Puech, second quarter of the first century BCE~\cite{Puech2015b}; Yardeni, 50–1 BCE~\cite{DJD10}. The prediction plot would be compatible with the latter date; 

\item \label{see-comment4} 4Q522: typologically, we would characterize the script as late Hasmonaean, but the date of the prediction model seems slightly too old to us. We would expect a slightly younger date, ca. 100–75 BCE, in view of our considerations (Section~\ref{appen:H1:firsteval}). The traditional palaeographic date estimate by Puech is late Hasmonaean, second third of the first century BCE~\cite{Puech1998B}.

\end{enumerate}

\noindent
Two observations on the basis of these comments:
\begin{enumerate}

\item Outside nice formal bookhands, ordering Dead Sea Scrolls manuscripts according to typology can be difficult for palaeographers, especially for the semicursive script. In addition to the physical and image quality of the data (see Section~\ref{appen:H2:dataquality}), script diversity can also pose a challenge for the Enoch model. More specifically, Enoch can handle formal and semiformal scripts well in predicting their age range, but manuscripts written in semicursive script are more difficult to date at the current stage. This can be explained by the fact that Enoch was not yet trained enough on this (only two \textsuperscript{14}C samples, 4Q114 and 4Q255/4Q433a, are in semicursive script);

\item The range 100–50 BCE is underrepresented in Enoch’s date predictions. This can be explained by the distribution of \textsuperscript{14}C samples across the time line, having little evidence securely fixed for this part of the time line: 4Q201, 4Q255/4Q433a, 4Q27, and 4Q2 cover (part of) the range 100–50 BCE but all of them extend beyond the range as well. Roughly speaking, Enoch predicts Hasmonaean-type manuscripts before 100 BCE and Herodian-type manuscripts after 50 BCE. Still, it should be noted that the range 100–50 BCE is not completely left devoid of Enoch’s prediction plots, as the plots for 4Q185, 4Q554, and 11Q13 show, albeit with a wide range of 150 years for 4Q554. 
\end{enumerate}

\medskip
\noindent
From the machine-learning perspective, these problems can be sorted out as more samples from critical time periods are added to the training data.

\subsection{6 July 2021 test}\label{appen:6July2021Test}
Earlier in the project, a test was conducted on 6 July 2021. The test consisted of giving manuscripts with unseen \textsuperscript{14}C results to the AI experts to see whether Enoch would give date prediction estimates that match the \textsuperscript{14}C results. However, at the start of the test, it was unknown to the AI experts that the samples were chosen because of \textsuperscript{14}C results being available for them afterward. 

The \textsuperscript{14}C results were taken from the 1990s \textsuperscript{14}C dating of the Dead Sea Scrolls~\cite{Bonani1992,Jull1995}. The assumption was that the manuscripts chosen were not contaminated with castor oil as these manuscripts were not handled by the original team of editors in the 1950s~\cite{Doudna1998,carmi200214c,rasmussen2009effects}. This applies to 1QIsa\textsuperscript{a}, 1QpHab, 1QapGen, 1QS, 1QH\textsuperscript{a}, 11Q19, Mas1l. 

Two more manuscripts were added for other reasons. 4Q53 was added because scholars assume that it was written by the same scribe as 1QS. 4Q319 was added because it is actually the same manuscript as 4Q259~\cite{Hempel2020}, which was subjected to \textsuperscript{14}C dating by our own project. 

The test was filmed. The film captures the whole process that was conducted in one go. The film can be accessed here:
\href{https://doi.org/10.5281/zenodo.8167946}{https://doi.org/10.5281/zenodo.8167946}

\begin{longtable}{l|rr|r|rr}
\caption{AI experts' (blind) range estimation (est\_min and est\_max) and palaeography experts’ evaluation (pal\_eval) with year range (pal\_min and pal\_max).}
\label{tab:expert-AI-135undated}\\
\hline  \hline
\textbf{Q-num} & \textbf{est\_min} & \textbf{est\_max} & \textbf{pal\_eval} & \multicolumn{1}{l}{\textbf{pal\_min}} & \multicolumn{1}{l}{\textbf{pal\_max}} \\ \hline
\endfirsthead
\endhead
1QapGen        & -50               & -1                & realistic          & -50                                   & -1                                    \\ \hline
1QH\textsuperscript{a}            & -140              & -1                & realistic          & -50                                   & -1                                    \\ \hline
1QIsa\textsuperscript{a}         & -200              & -100              & realistic          & -175                                  & -125                                  \\ \hline
1QpHab         & -40               & 10                & realistic          & -25                                   & 25                                    \\ \hline
1QS            & -190              & -100              & realistic          & -150                                  & -100                                  \\ \hline
2Q3            & -40               & 130               & realistic          & -50                                   & -1                                    \\ \hline
2Q14           & -310              & -100              & too\_old           & -75                                   & -25                                   \\ \hline
2Q24           & -40               & 10                & realistic          & -50                                   & -1                                    \\
\hline \hline
\end{longtable}

{\tabcolsep=3pt
\footnotesize
\begin{longtable}{ll|ll}
\caption{List of images for each manuscript sample used in the post-hoc evaluation}
\label{tab:135-images}\\
\hline \hline
\textbf{Q-number} & \textbf{Image-name}            & \textbf{Q-number} & \textbf{Image-name}              \\ \hline 
\endfirsthead
\endhead
1QapGen           & 1QapGen\_4\_crpcln             & 4Q301             & 4Q301\_2\_processed              \\ \hline
1QH\textsuperscript{a}               & 1QHa\_QIrug-1668\_cln...Lotte  & 4Q303             & 4Q303\_processed                 \\ \hline
1QIsa\textsuperscript{a}            & 1QIsaa\_col02\_cleaned         & 4Q319             & 4Q319\_1\_crpcln                 \\ \hline
1QpHab            & 1QpHab\_Brill2307\_cleaned\_MD & 4Q325             & 4Q325\_processed                 \\ \hline
1QS               & 1Qs\_QIrug-2153\_cln...Lotte   & 4Q373             & 4Q373\_processed                 \\ \hline
2Q3               & 2Q3\_processed                 & 4Q375             & 4Q375\_processed                 \\ \hline
2Q14              & 2Q14\_1\_processed             & 4Q379             & 4Q379\_1\_processed              \\ \hline
2Q24              & 2Q24\_processed                & 4Q390             & 4Q390\_processed                 \\
\hline \hline
\end{longtable}
}

\begin{longtable}{l|l|l}
\caption{List of twenty-five images of poor quality}
\label{tab:bad25}\\
\hline \hline
\textbf{Image-name} & \textbf{Image-name}       & \textbf{Image-name} \\ \hline 
\endfirsthead
\endhead
4Q540\_processed    & 4Q53\_405\_part2\_cleaned & 4Q577\_processed    \\ \hline
4Q196\_cleaned      & 4Q457\_processed          & 4Q508\_processed    \\ \hline
4Q86\_processed     & 4Q98\_processed           & 4Q98f\_processed    \\ \hline
4Q492\_processed    & 4Q448\_processed          & Mas1f\_processed    \\ \hline
4Q88\_3\_processed  & 4Q530\_processed          & 4Q98c\_processed    \\ \hline
2Q14\_1\_processed  & 4Q98g\_processed          & 4Q98b\_processed    \\ \hline
4Q284a\_processed   & 5Q5\_processed            & 4Q398\_1\_processed \\ \hline
4Q432\_processed    & 4Q373\_processed          & 4Q379\_1\_processed \\ \hline
6Q18\_processed     &                           &                     \\ \hline \hline
\end{longtable}

\newpage
\begin{longtable}{|l|l|}
\caption{A few examples of poor quality images (after binarization and cleaning).}
\label{tab:bad25-images}\\
\hline
\begin{adjustbox}{valign=c}\includegraphics[width=0.30\textwidth]{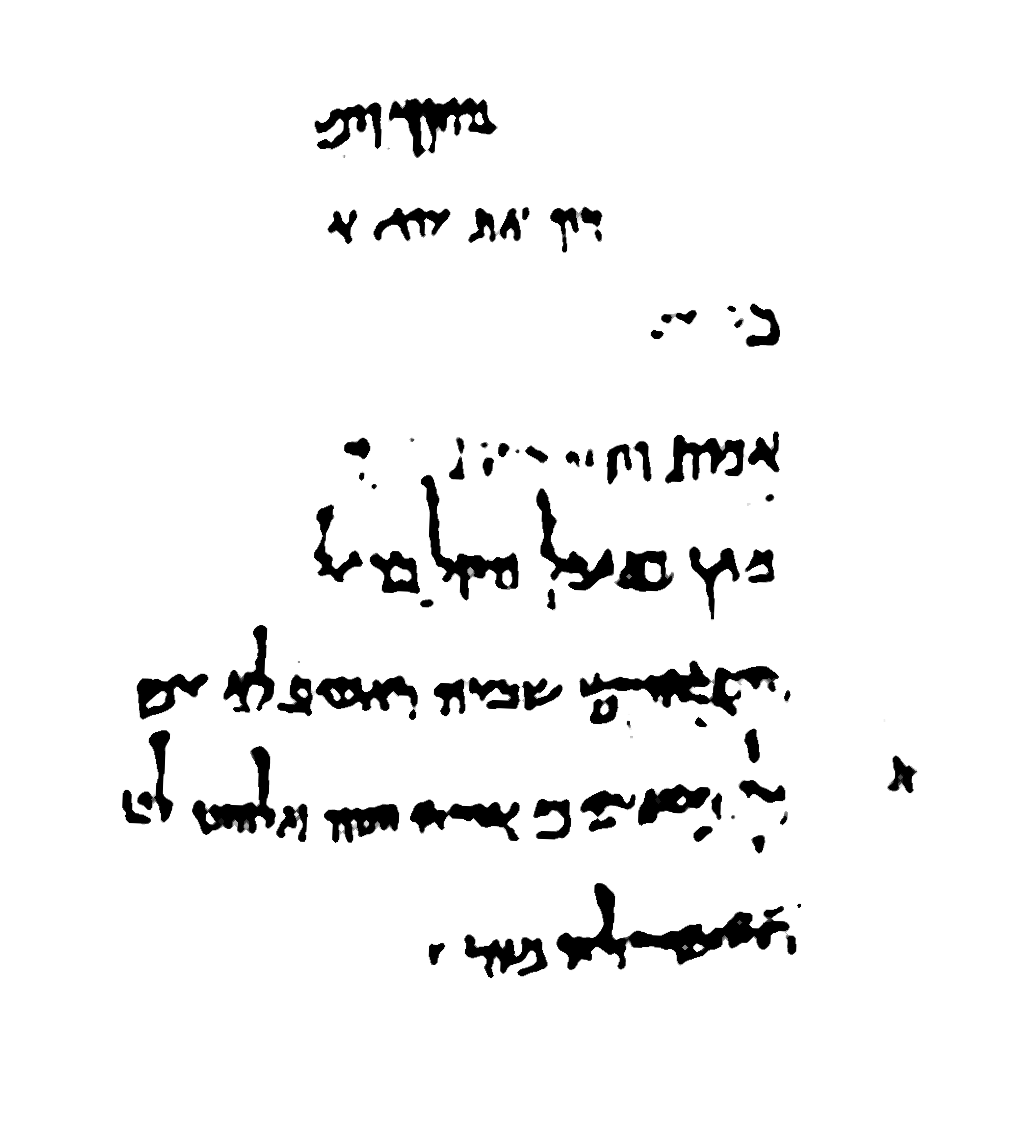}\end{adjustbox} & \begin{adjustbox}{valign=c}\includegraphics[width=0.30\textwidth]{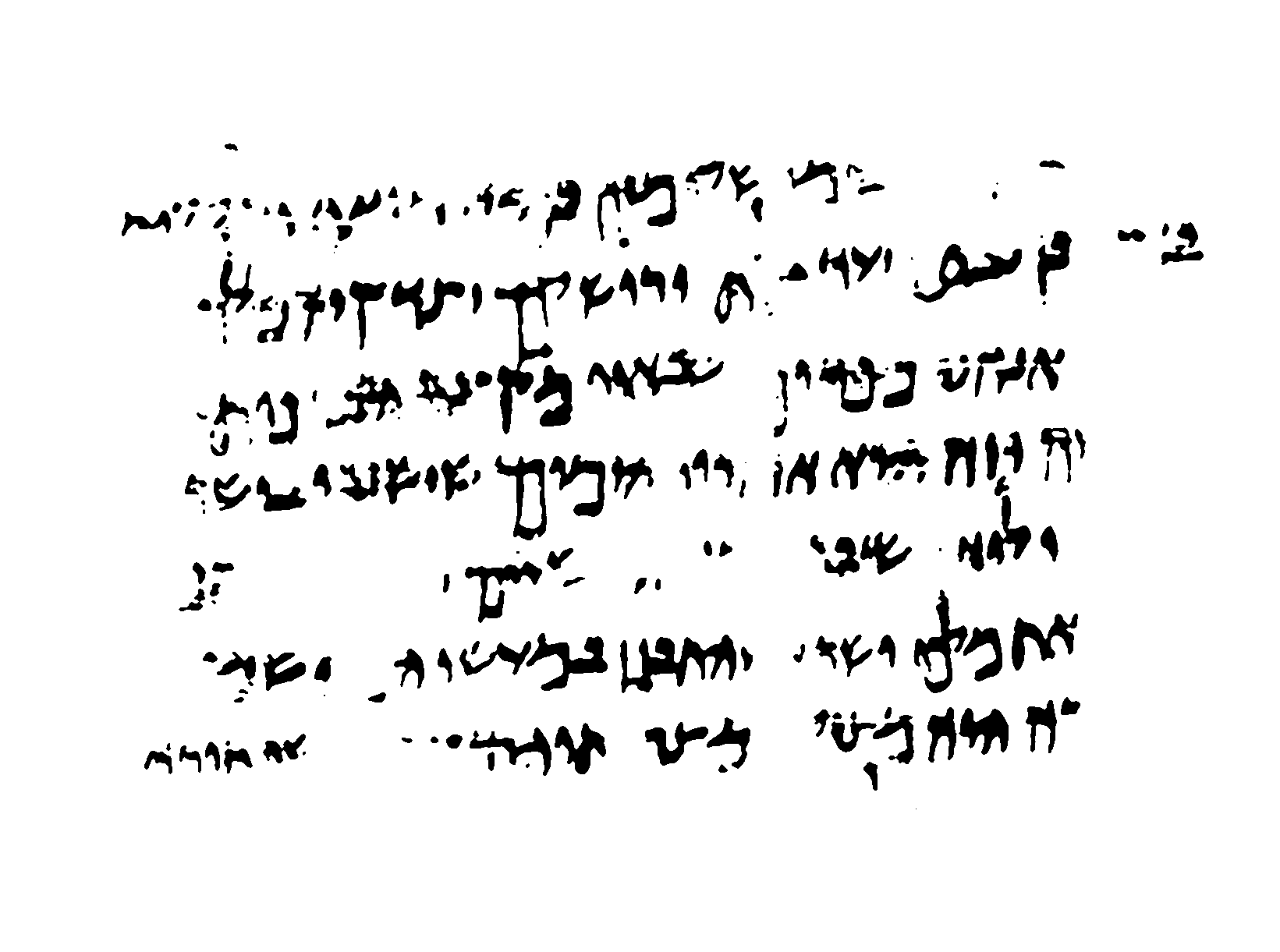}\end{adjustbox}  \\ \hline
\begin{adjustbox}{valign=c}\includegraphics[width=0.30\textwidth]{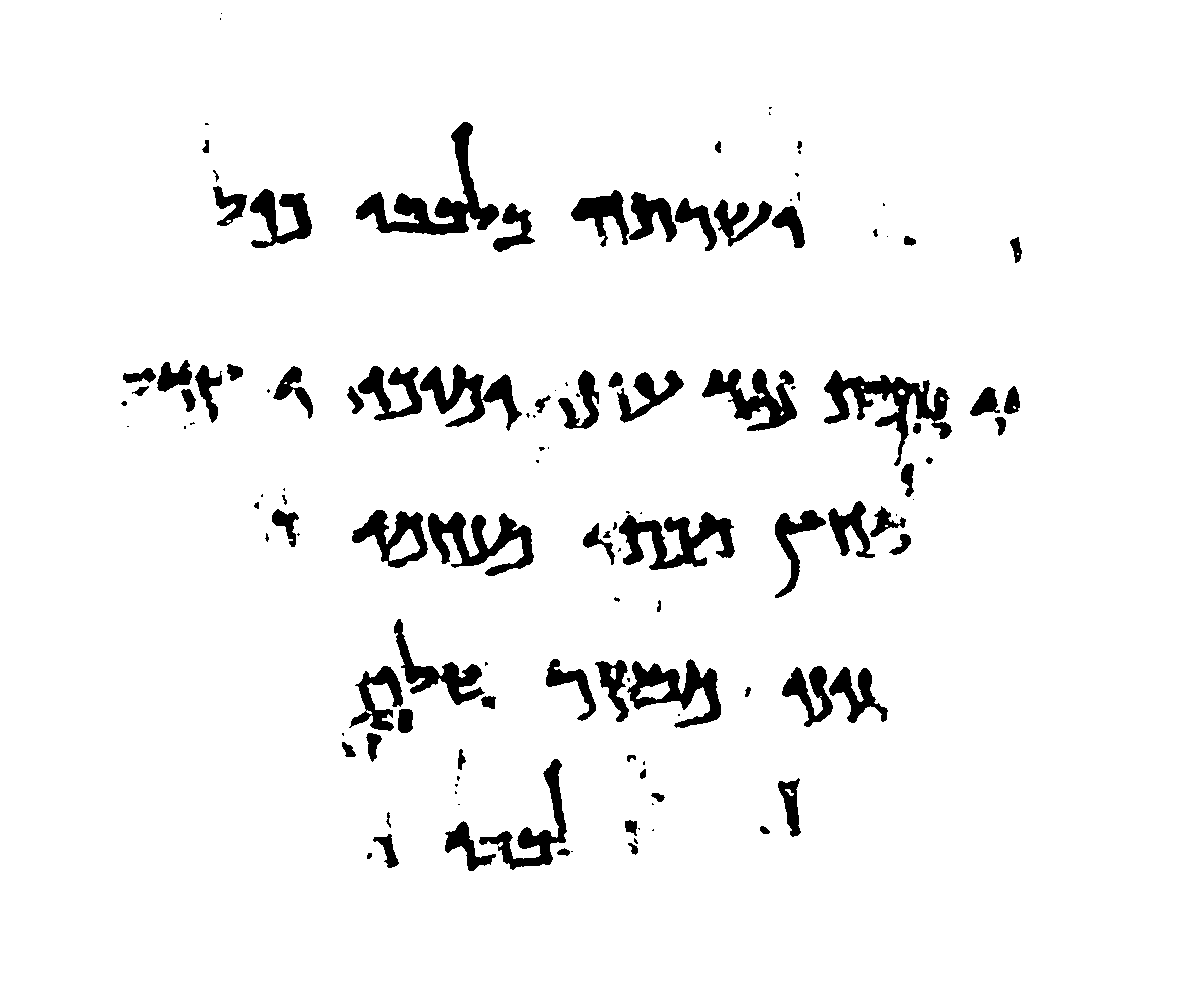}\end{adjustbox}  &  \begin{adjustbox}{valign=c}\includegraphics[width=0.30\textwidth]{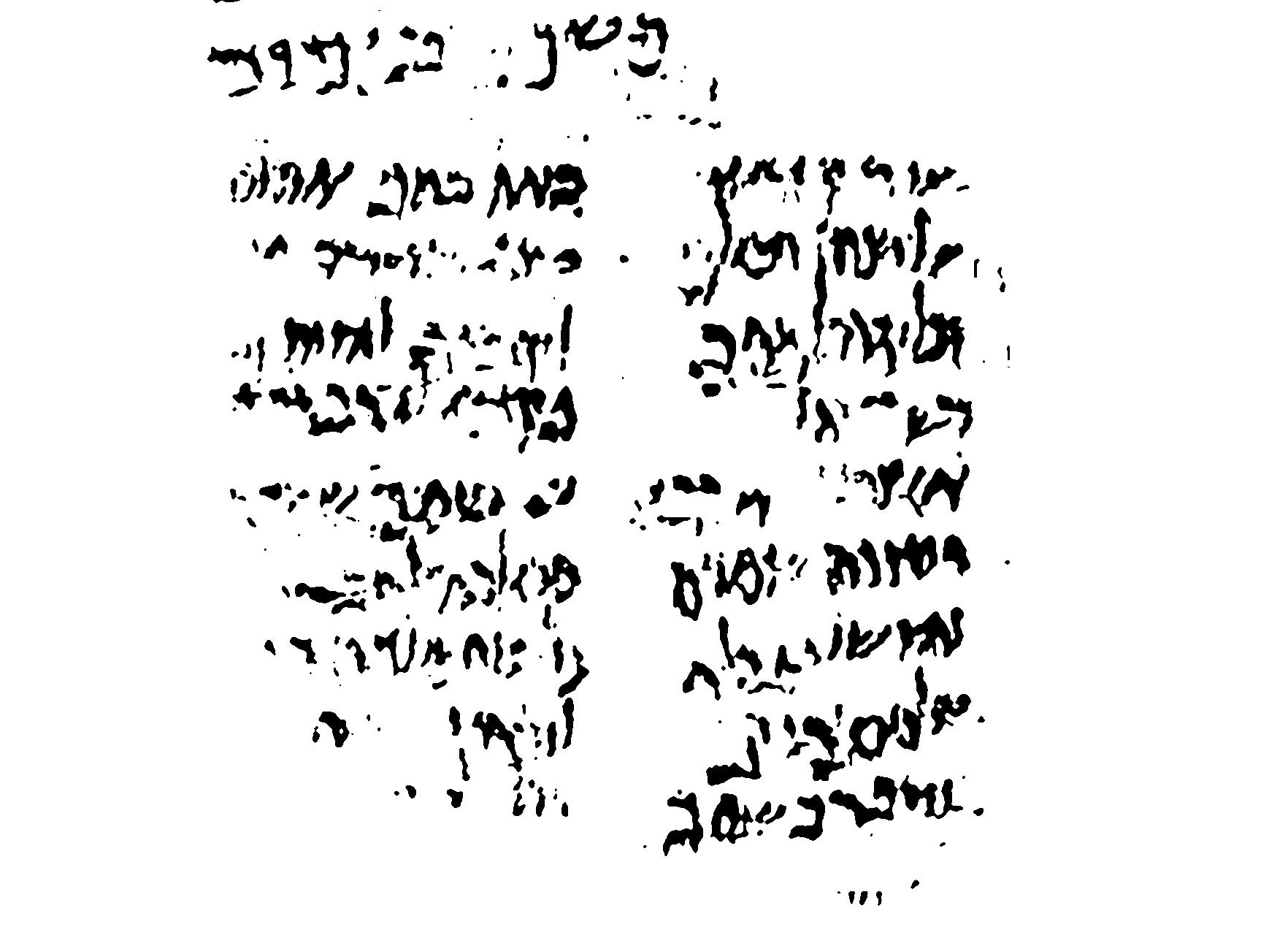}\end{adjustbox}\\ \hline
\begin{adjustbox}{valign=c}\includegraphics[width=0.30\textwidth]{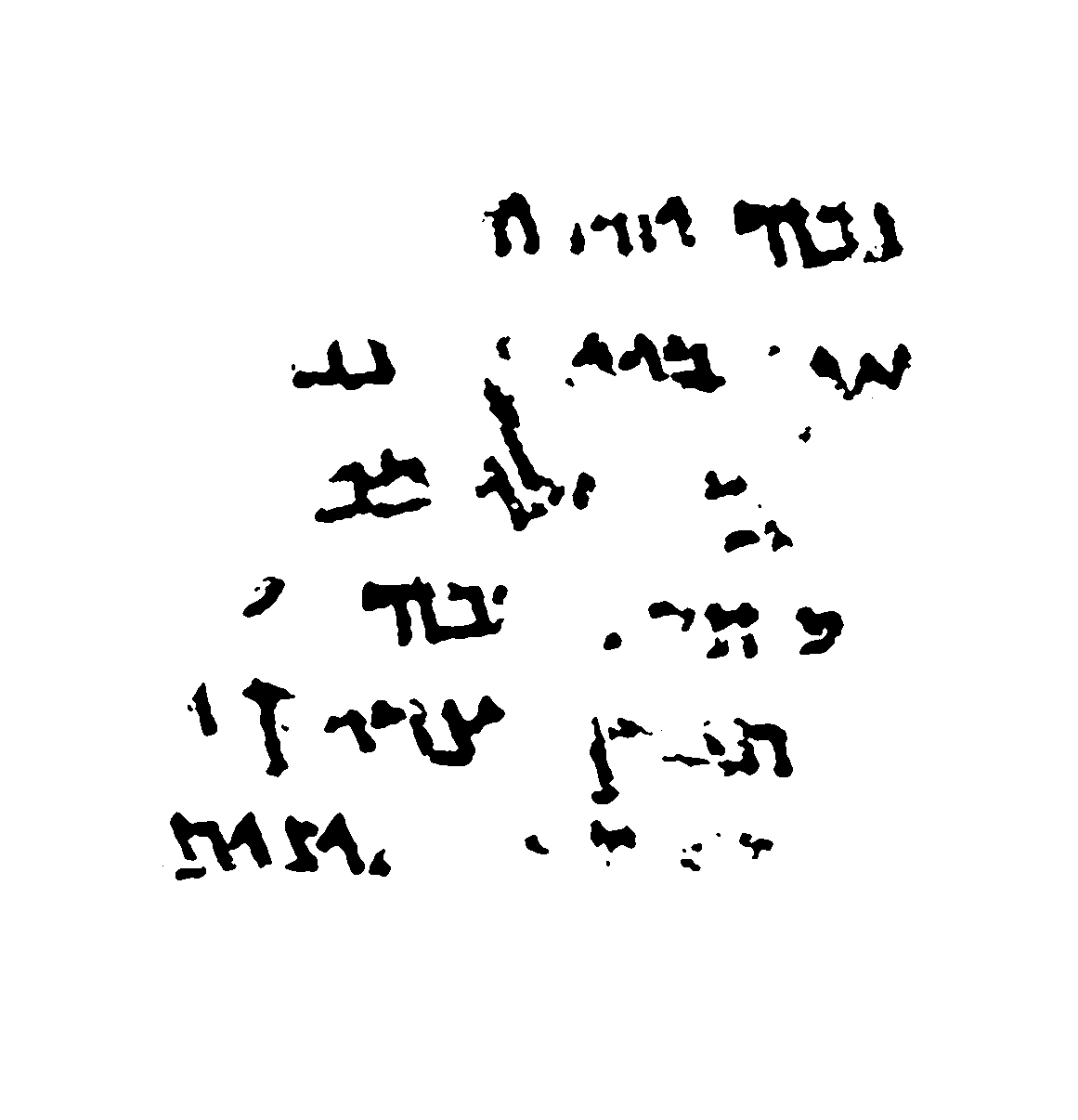}\end{adjustbox}  &  \begin{adjustbox}{valign=c}\includegraphics[width=0.30\textwidth]{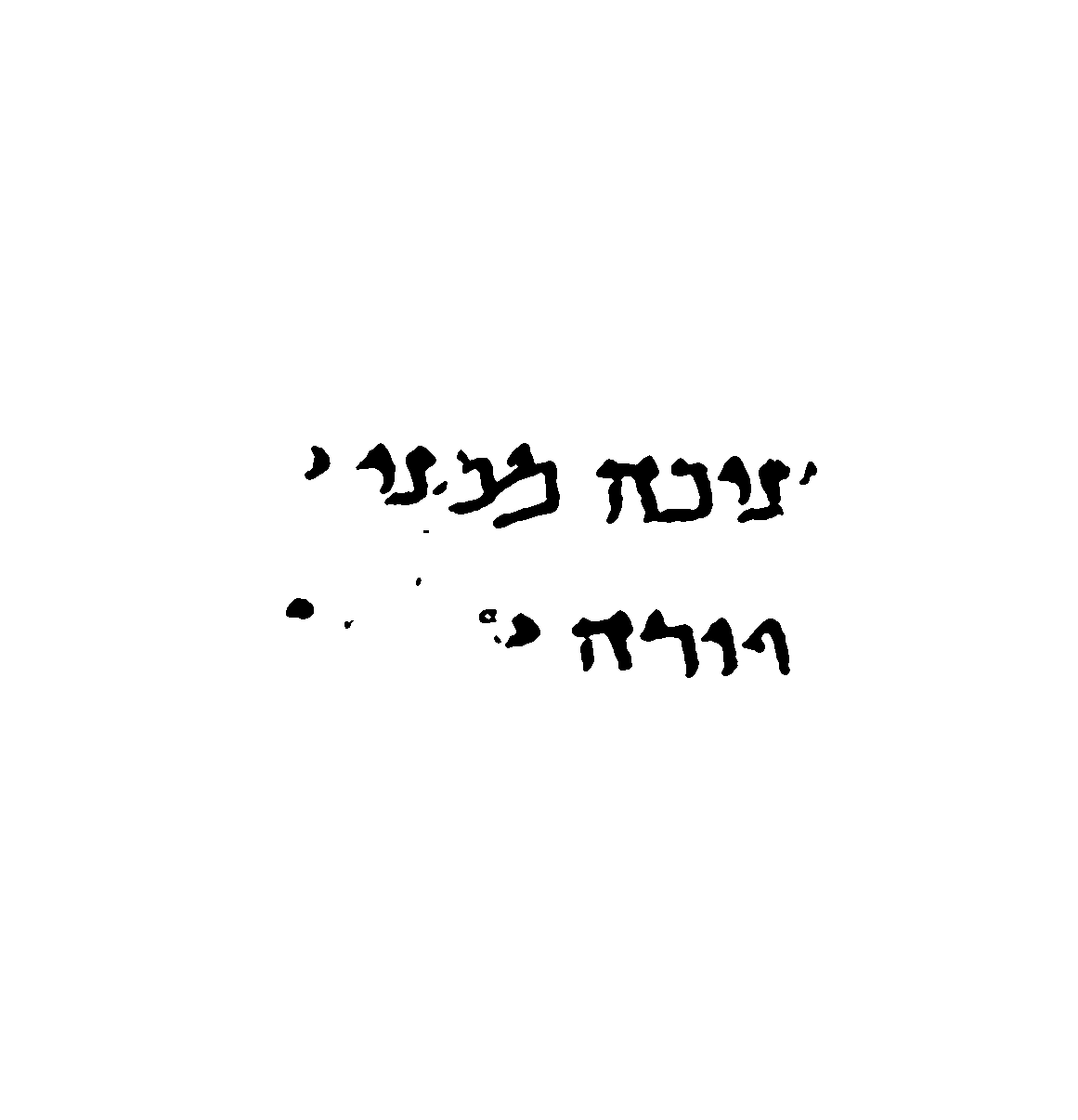}\end{adjustbox}\\ \hline
\begin{adjustbox}{valign=c}\includegraphics[width=0.30\textwidth]{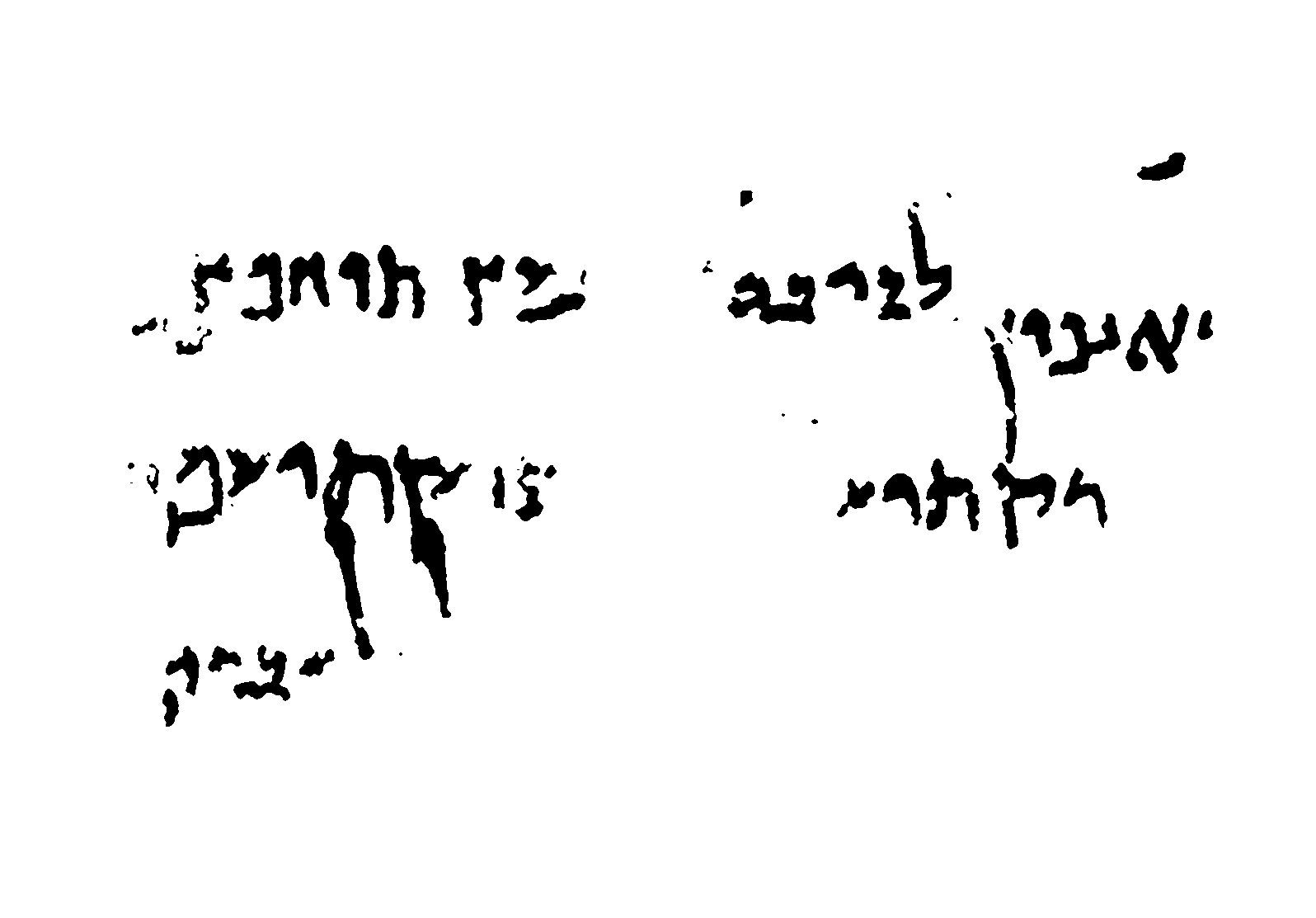}\end{adjustbox}  &  \begin{adjustbox}{valign=c}\includegraphics[width=0.25\textwidth]{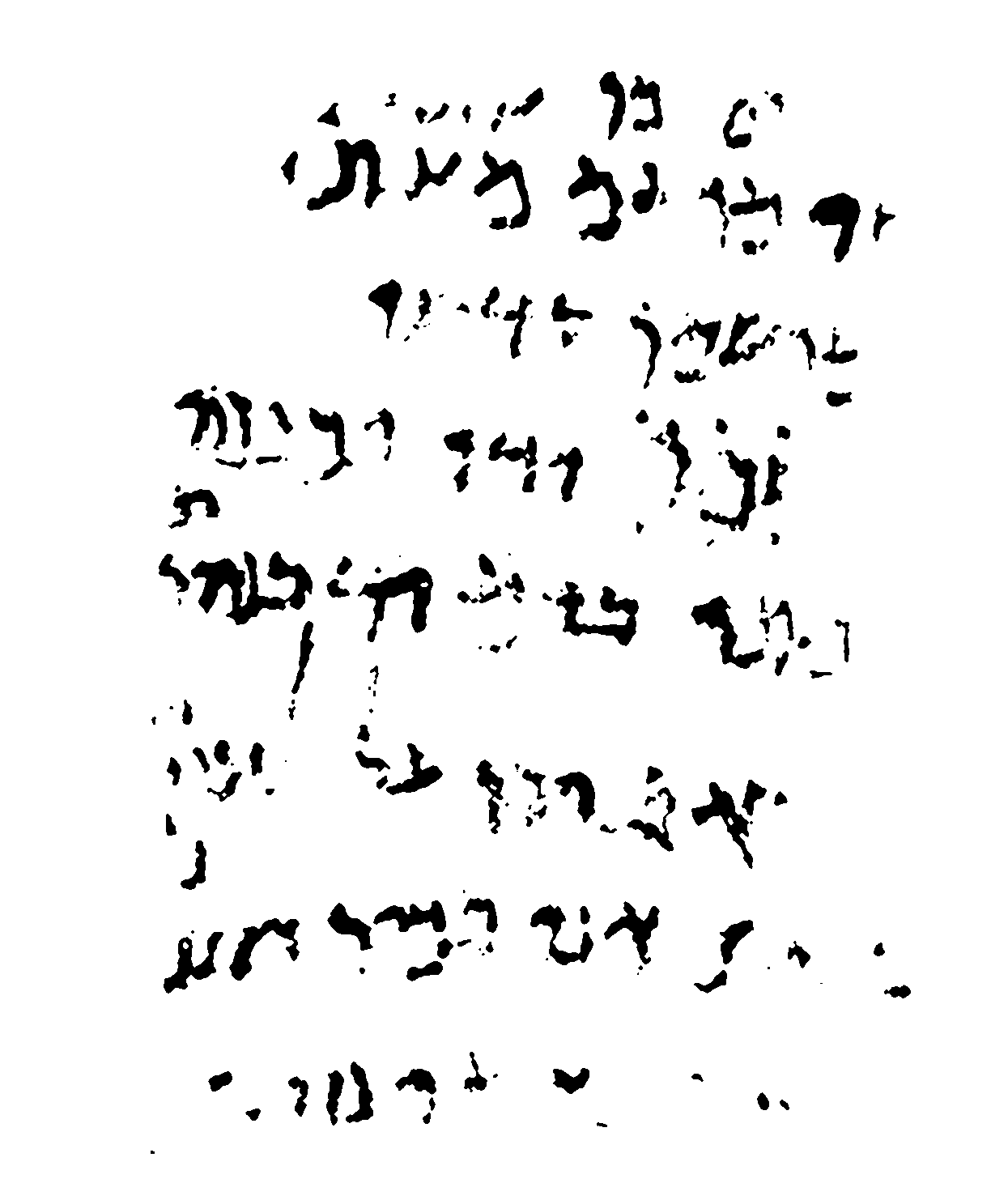}\end{adjustbox}\\ \hline
\begin{adjustbox}{valign=c}\includegraphics[width=0.30\textwidth]{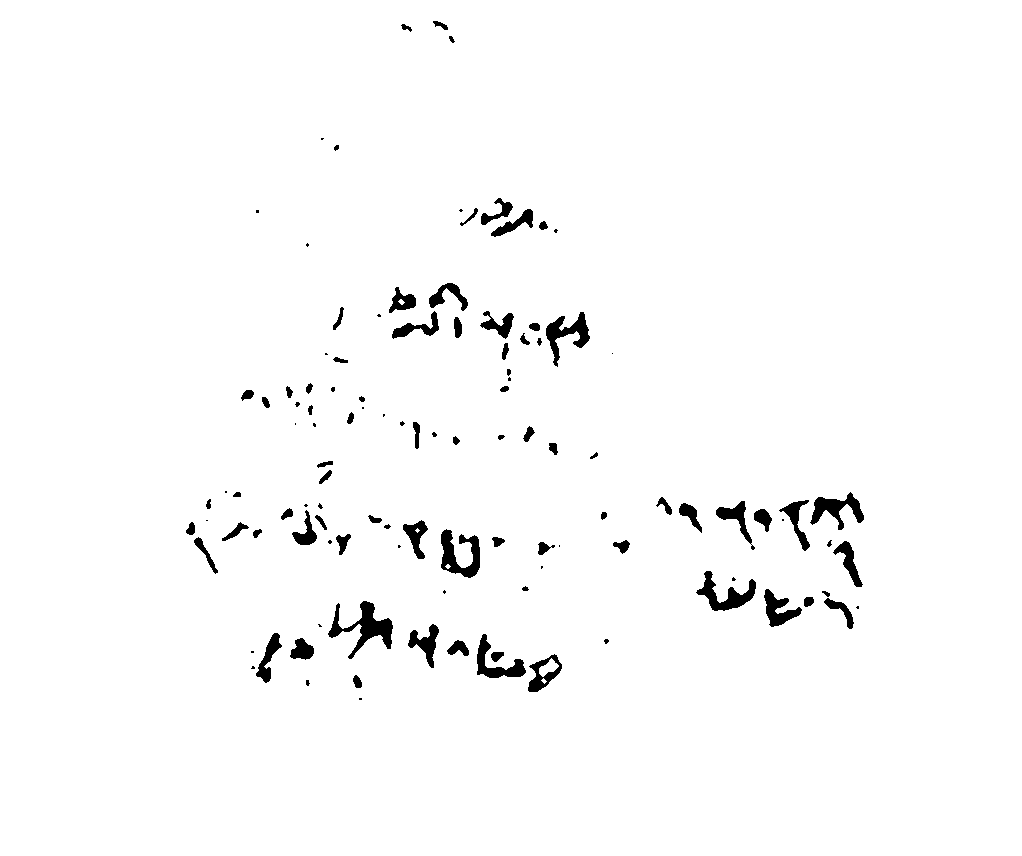}\end{adjustbox}  &  \begin{adjustbox}{valign=c}\includegraphics[width=0.30\textwidth]{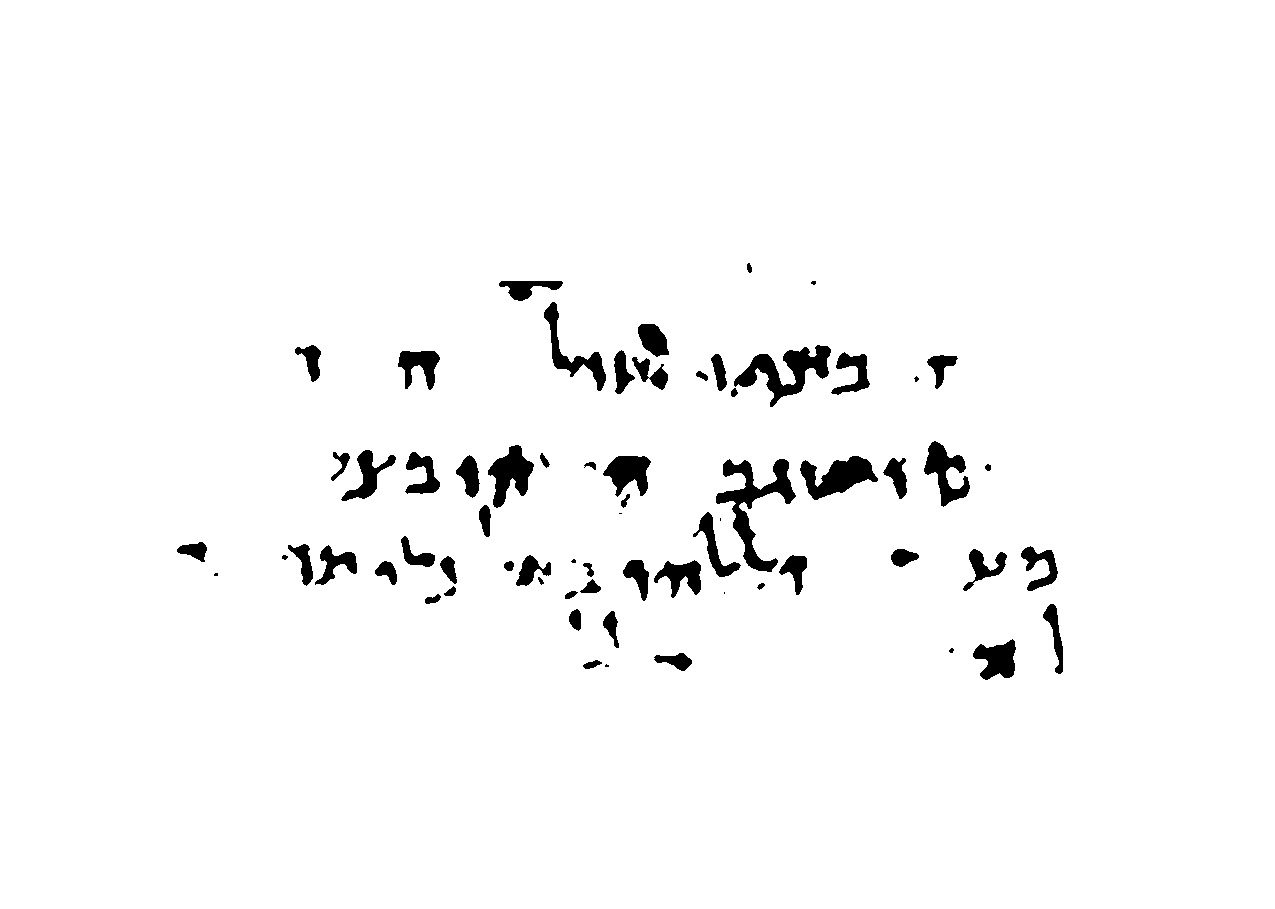}\end{adjustbox}\\ \hline
\end{longtable}

%% file: tex-sections/H-combineplots.tex
\section{Comparative plots for different information sources}\label{appen:E-plots}

\begin{figure}[!ht]
    \centering
    \includegraphics[width=0.92\textwidth]{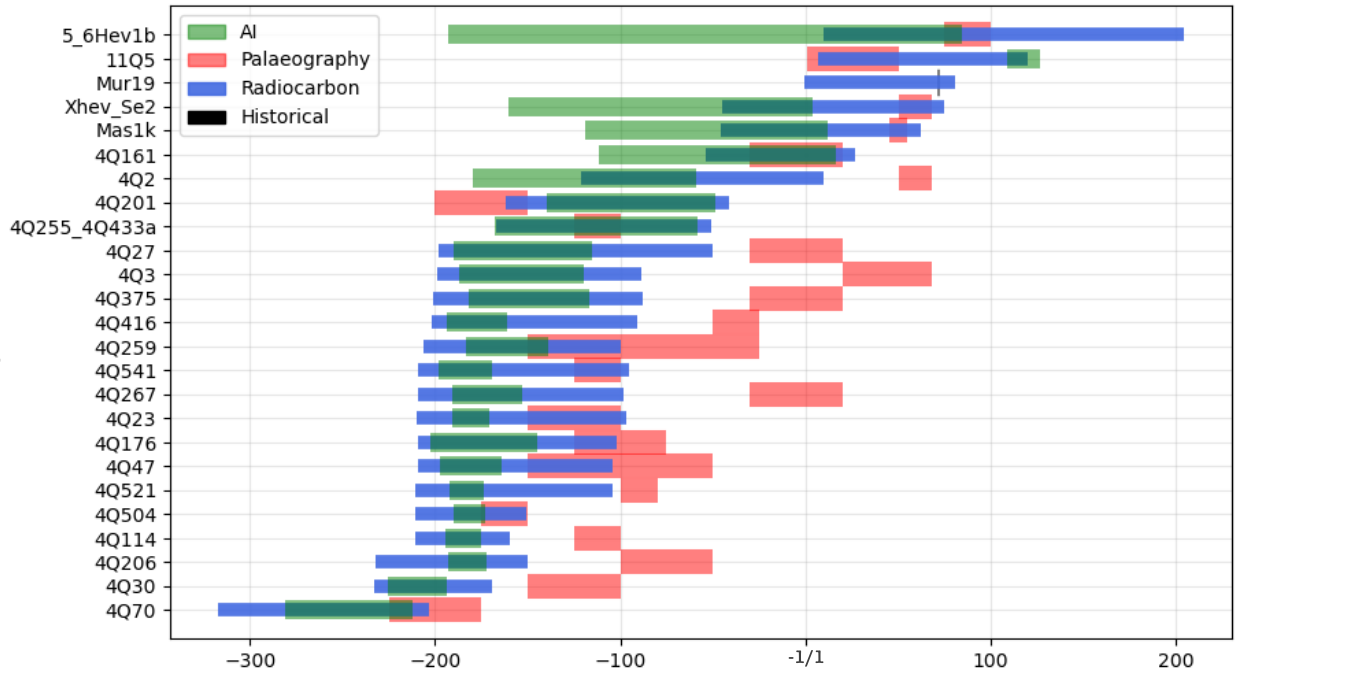}
    \caption{Overview of date estimations by three information sources and a calendar date: accepted 2$\sigma$ calibrated ranges \textsuperscript{14}C without minor peaks (blue), Enoch (green), palaeography (red), and historical (black). The vertical axis contains the manuscript numbers, and the horizontal axis contains dates: BCE in negative and CE in positive.}
    \label{fig:AIvsC14PAL-wo-minorpeaks}
\end{figure}

\begin{figure}[!ht]
    \centering
    \includegraphics[width=0.92\textwidth]{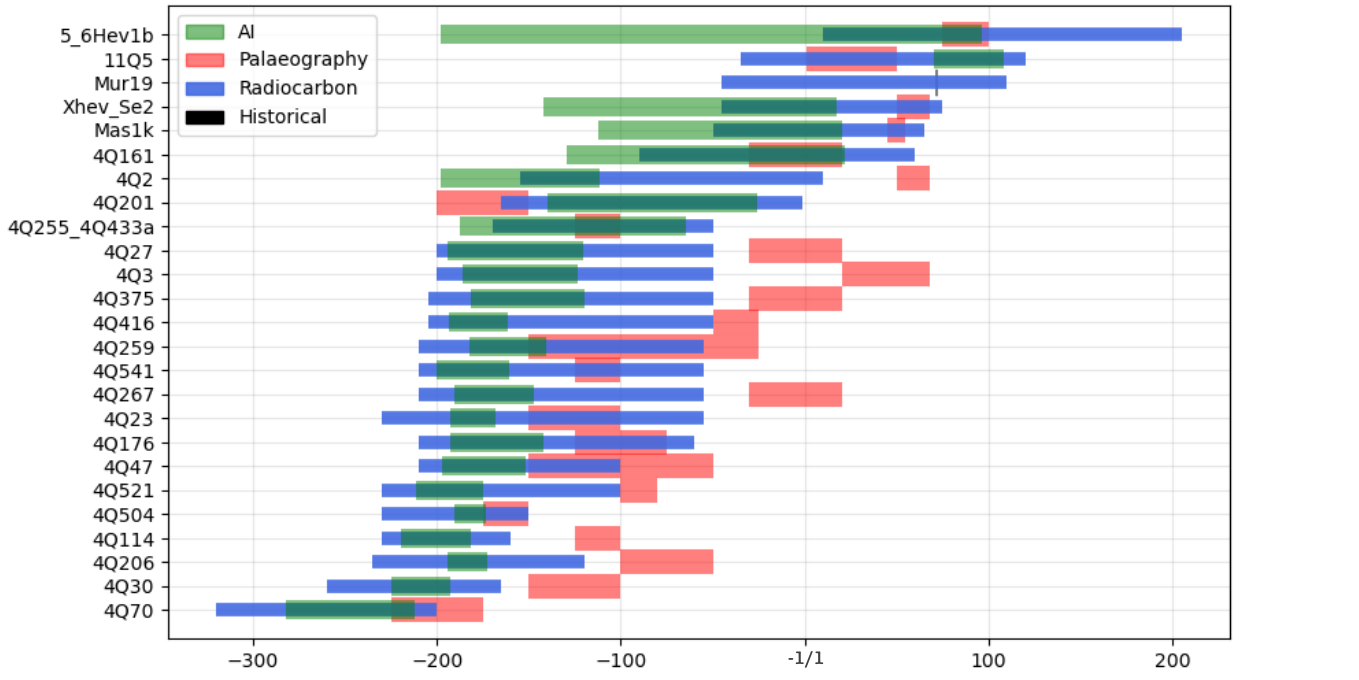}
    \captionsetup{format=hang}
    \caption{Overview of date estimations by three information sources and a calendar date: accepted 2$\sigma$ calibrated ranges \textsuperscript{14}C with minor peaks (blue), Enoch (green), palaeography (red), and historical (black). Please note that this is the same as in Figure~\ref{fig:AIvsC14PAL} in the main article, except here, the minor peaks are taken as a continuous range within the 2$\sigma$ calibrated range.} 
    \label{fig:AIvsC14PAL-continuous}
\end{figure}

\clearpage

\begin{figure}[!ht]
    \centering
    \includegraphics[width=\textwidth]{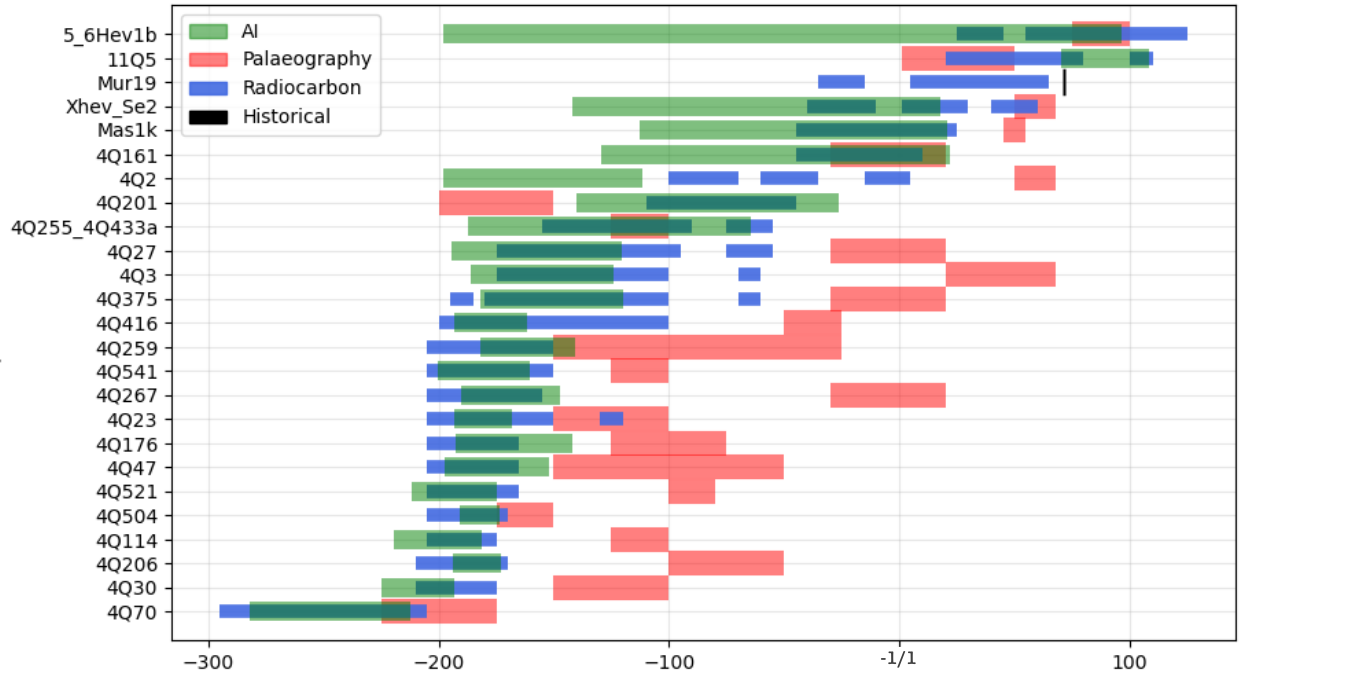}
    \captionsetup{format=hang}
    \caption{Overview of date estimations by three information sources and a calendar date: accepted 1$\sigma$ calibrated ranges \textsuperscript{14}C (blue), Enoch (green), palaeography (red), and historical (black).} 
    \label{fig:AIvsC14PAL-1sigma}
\end{figure}

~\\
~\\
\vfill
\clearpage

%% file: tex-sections/I-imagelist.tex
\section{List of images for different tests}\label{appen:D}
{\tabcolsep=2pt
\begin{table}[!ht]
\centering
\caption{Complete list of 64 training images (including 4Q52; for the date prediction model) from the radiocarbon-dated manuscripts.}
\resizebox{\textwidth}{!}{%
\begin{tabular}{llll|llll}
\hline \hline
\textbf{Q-numbers} & \textbf{Name}     & \textbf{Plate} & \textbf{Fragment}        & \textbf{Q-numbers}  & \textbf{Name}             & \textbf{Plate} & \textbf{Fragment}           \\ \hline
4Q2      & 4Q2\_1   & 215   & 1,4         & 4Q255     & 4Q255\_4Q433a\_1 & 177   & 3R             \\
         & 4Q2\_2   & 215   & 3           &           & 4Q255\_4Q433a\_2 & 177   & 4R             \\ \hline
4Q3      & 4Q3\_1   & 393   & 5           & 4Q259     & 4Q259\_1         & 683   & 6              \\ \cline{1-4}
4Q23     & 4Q23\_1  & 271   & 1,4         &           & 4Q259\_2         & 683   & 7              \\ \cline{1-4}
         & 4Q23\_2  & 272   & 5, 6, 18, 7 &           & 4Q259\_3         & 695   & 1,3            \\
         & 4Q23\_3  & 272   & 19          &           &                  & 696   & 6              \\
         &          & 401   & 3, 4, 5     &           & 4Q259\_4         & 810   & 3R, 5R, 7R, 8R \\
\hline \hline
\end{tabular}}
\end{table}}

\begin{table}[!ht]
\centering
\caption{List of 135 manuscripts used for making date predictions. Please note that one manuscript may contain several images in the test dataset.}
\label{tab:allTest135}
\begin{tabular}{l|l|l|l}
\hline \hline
\textbf{Q-numbers} & \textbf{Q-numbers} & \textbf{Q-numbers} & \textbf{Q-numbers} \\ \hline
1QapGen            & 4Q98               & 4Q301              & 4Q508              \\ \hline
1QH\textsuperscript{a}                & 4Q98a              & 4Q303              & 4Q511              \\ \hline
1QIsa\textsuperscript{a}  & 4Q98b              & 4Q319              & 4Q522              \\ \hline
1QpHab             & 4Q98c              & 4Q325              & 4Q525              \\ \hline
1QS                & 4Q98f              & 4Q373              & 4Q530              \\ \hline
2Q3                & 4Q98g              & 4Q375              & 4Q531              \\ \hline
2Q14               & 4Q109              & 4Q379              & 4Q540              \\
\hline \hline
\end{tabular}
\end{table}

\begin{table}[!ht]
\centering
\caption{List of all 13 images that are split from training manuscripts and added to test.}
\label{tab:trainingc14-in-testing}
\begin{tabular}{l|l}
\hline \hline
\textbf{Q-number} & \textbf{Number of images} \\ \hline
4Q27              & 2                         \\
4Q161             & 2                         \\
4Q267             & 2                         \\
4Q375             & 1                         \\
5-6Hev1b          & 1                         \\
11Q5              & 5                         \\ \hline \hline
\end{tabular}
\end{table}

{\tabcolsep=2pt
\begin{table}[!ht]
\centering
\caption{Complete list of 23 training images from a selection of previously \textsuperscript{14}C-tested manuscripts~\cite{Bonani1992, Jull1995}.}
\label{old-c14-list}
\begin{tabular}{l|c|l}
\hline \hline
\textbf{Manuscript} & \textbf{Radiocarbon (BP)} & \textbf{Image IDs}                             \\ \hline
Mas1l             & 2086,28                                 & MasJosh.png                                    \\ \hline
1QIsa\textsuperscript{a}              & 2141,32                                 & 1QIsaa\_col01\_cleaned.png                     \\ \hline
                    &                                         & 1QIsaa\_col02\_cleaned.png                     \\ \hline
                    &                                         & 1QIsaa\_col03\_cleaned.png                     \\ \hline
                    &                                         & 1QIsaa\_col34\_cleaned.png                     \\ \hline
                    &                                         & 1QIsaa\_col35\_cleaned.png                     \\ \hline
1QpHab              & 2054,22                                 & QIrug-Qumran\_extr09\_2305\_1QpHab\_crpcln.png \\
\hline \hline
\end{tabular}
\end{table}}

\begin{table}[!ht]
\centering
\caption{Complete list of 30 images for date-bearing manuscripts from the fifth–fourth centuries BCE and the second century CE. }
\label{tab:internally-dated}
\begin{tabular}{l|l|l|l}
\hline \hline
\textbf{Manuscript} & \textbf{Date} & \textbf{Manuscript}      & \textbf{Date} \\ \hline
A6\_11R    & -411 & IA06            & -353 \\ \hline
A6\_12R    & -411 & IA17            & -324 \\ \hline
A6\_13R    & -411 & IA21            & -330 \\ \hline
A6\_14     & -411 & MareshaOstracon & -176 \\ \hline
A6\_15     & -411 & Mur24\_1        & 133  \\ \hline
A6\_16     & -411 & Mur24\_2        & 133  \\ \hline
A6\_3      & -411 & NS\_A1r         & -353 \\ \hline
A6\_4      & -411 & NS\_A2r         & -351 \\ \hline
A6\_5      & -411 & NS\_A4r         & -348 \\ \hline
A6\_7      & -411 & NS\_A5r         & -348 \\ \hline
A6\_8      & -411 & NS\_A6r         & -349 \\ \hline
B3\_1      & -456 & NS\_C1r         & -330 \\ \hline
IA01       & -348 & NS\_C4r         & -324 \\ \hline
IA03       & -348 & WDSP1\_1        & -335 \\ \hline
IA04       & -351 & WDSP2           & -352 \\ \hline \hline
\end{tabular}
\end{table}

~\\
\vfill
\clearpage

%% file: tex-sections/J-sample-locations.tex
\section{Radiocarbon sample information}\label{appen:F-sample}

{\tabcolsep=3pt
\footnotesize
\begin{longtable}[c]{lp{1cm}p{2cm}lp{4cm}}
\caption{Information of physical samples for radiocarbon dating}
\label{appne:tableE1:sampleinfo}
\\
\hline \hline
\multicolumn{1}{|l|}{\textbf{IAA plate}}  & \multicolumn{1}{l|}{\textbf{Sample}}       & \multicolumn{1}{l|}{\textbf{Work}}                                & \multicolumn{1}{l|}{\textbf{Plate-fragment}}   & \multicolumn{1}{p{4cm}|}{\textbf{Info from IAA on DJD references, places in fragments where samples were taken, previous treatments from the 1950s onward, and sample weights}}                                                                                                                                                                                       \\ \hline
\multicolumn{1}{|l|}{206}   & \multicolumn{1}{l|}{4Q52}         & \multicolumn{1}{l|}{4QSam\textsuperscript{b}}                             & \multicolumn{1}{l|}{{\href{https://www.deadseascrolls.org.il/explore-the-archive/image/B-371248}{\href{http://www.deadseascrolls.org.il/explore-the-archive/image/B-371248}{P206-Fr003}}}}       & \multicolumn{1}{p{4cm}|}{DJD 17: pl XXIV, fr 7; Bottom left; Maybe glues? Japanese Tissue Paper + Methylcellulose glue (2001)}                                                                                                                                                                                \\ \hline
\multicolumn{1}{|l|}{421}   & \multicolumn{1}{l|}{4Q504}        & \multicolumn{1}{l|}{4QDibHam\textsuperscript{a}}                          & \multicolumn{1}{l|}{\href{https://www.deadseascrolls.org.il/explore-the-archive/image/B-298264}{P421-Fr004}}       & \multicolumn{1}{p{4cm}|}{DJD 7: pl LII, fr 7; Bottom; Maybe castor oil?}                                                                                                                                                                                                                                   \\
\hline\hline
\end{longtable}}

%% file: tex-sections/K-datasheetrun.tex
\section{Data-sheet radiocarbon runs}\label{appen:G}

The samples were dated by two different AMS machines characterised by codes GrA and GrM. For the GrA machine the $\delta$\textsuperscript{13}C values of the IRMS are shown in the table; for GrM these are AMS values.

\footnotesize
\begin{landscape}
{\tabcolsep=0.7pt  
\begin{longtable}{l|l|l|l|l|l|l|l|l|l|l|l|l|l|l|l|l}
\caption{Data-sheet \textsuperscript{14}C runs}
\\
\hline \hline
\textbf{fragment} & \textbf{KLR} & \textbf{samplenr} & \textbf{GrA} & \textbf{GrM} & \textbf{\textsuperscript{14}a \%} & \textbf{$\sigma$} & \textbf{$\delta$\textsuperscript{13}C ‰} & \textbf{C\%} & \textbf{\textless{}\textsuperscript{14}a\textgreater{}} & \textbf{\textless{}$\sigma$\textgreater{}} &\textbf{age BP} & \textbf{$\sigma$} & \textbf{remarks} & \textbf{used} & \textbf{calibrated 1$\sigma$} & \textbf{calibrated 2$\sigma$} \\ \hline
\endhead
\hline
\endfoot
\endlastfoot
P206-Fr003 & 11089 & 65369 & 69793 &  & 75.69 & 0.39 & -21.22 & 39.5 &  &  &  &  &  &  &  &  \\
 &  & 67017 &  & 10677 & 74.38 & 0.42 & -21.20 &  &  &  &  &  &  &  &  &  \\
 &  & 67017 &  & 10678 & 75.05 & 0.43 & -21.40 &  &  &  &  &  &  &  &  &  \\
 &  &  &  &  &  &  &  &  & 75.69 & 0.39 & 2237 & 41 & 1 GrA only &  &  &  \\
 &  &  &  &  &  &  &  &  & 74.71 & 0.30 & 2342 & 33 & 2 GrM averaged &  &  &  \\
 &  &  &  &  &  &  &  &  & 75.07 & 0.24 & 2303 & 26 & 3 averaged & X & 401–369 BCE & \begin{tabular}[c]{@{}l@{}}407–356,\\ 281–232 BCE\end{tabular} \\
 &  &  &  &  &  &  &  &  &  &  &  &  &  &  &  &  \\ \hline
P421-Fr004 & 11090 & 65370 & 68446 &  & 76.20 & 0.32 & -17.85 & 48.8 &  &  &  &  &  &  &  &  \\
\hline
\end{longtable}}
\end{landscape}
\normalsize

%% file: tex-sections/L-worksheet-comparison.tex
\section{\texorpdfstring{Worksheet of comparative data for 2$\sigma$ calibrated ranges and
traditional palaeographic estimates}{}}\label{appen:I}

Whole or partial overlap between 2$\sigma$ calibrated ranges and palaeographic estimates in 17 of the 26 accepted samples:
4Q23, 4Q47, 4Q52, 4Q70, 4Q161, 4Q176, 4Q201/4Q338, 4Q255/4Q433a, 4Q259, 4Q504, 4Q521, 4Q541, 11Q5, Mas1k, Mur19, 5/6Hev1b, XHev/Se2 (see Appendix~\ref{appen:B8.1:overlap}).

\begin{enumerate}
    \item \textbf{4Q23 (4QLevNum\textsuperscript{a})}
    \item[$\bullet$] 355–285 BCE (29.8\%), 230–220 BCE (0.8\%), 210–95 BCE (62.8\%), 75–55 BCE (2.1\%)
    \item[$\bullet$] DJD 12:154 (Ulrich): early Hasmonaean formal script, dating from approximately the middle or latter half of the second century BCE (150–100 BCE).
    \item[]

    \item \textbf{4Q47 (4QJosh\textsuperscript{a})}
    \item[$\bullet$] 355–290 BCE (33.8\%), 210–100 BCE (61.6\%)
    \item[$\bullet$] DJD 14:143 (Ulrich): referring to Cross Hasmonaean formal bookhand, second half of the second century or the first half of the first century BCE (150-50 BCE).
    \item[$\bullet$] Puech, Revue Biblique 122/4 (2015), 482: hasmonéenne au mieux dans la première moitié du 1\textsuperscript{er} s. avant J.-C. (100–50 BCE).
    \item[]

    \item \textbf{4Q52 (4QSam\textsuperscript{b})}
    \item[$\bullet$] 410–355 BCE (78.9\%), 285–230 BCE (16.6\%)
    \item[$\bullet$] DJD 17:220 (Cross, Parry, and Saley) (ca. 250 BCE).
    \item[]

    \item \textbf{4Q70 (4QJer\textsuperscript{a}))}
    \item[$\bullet$] 375–345 BCE (16.3\%), 320–200 BCE (79.2\%)
    \item[$\bullet$] DJD 15:150 (Tov): quoting Yardeni 1990 and Cross 1985 (Cross shifting between earlier and later dates to settle on an earlier date), the late third or early second century BCE (225–175 BCE).
    \item[]

    \item \textbf{4Q161 (4QpIsa\textsuperscript{a})}
    \item[$\bullet$] 90–80 BCE (1.7\%), 55 BCE–30 CE (92.1\%), 45–60 CE (1.7\%)
    \item[$\bullet$] Strugnell 1970 groups this manuscript with other manuscripts such as 4Q166 and 4Q171 and gives a general indication of the script as developed rustic semiformal Herodian (see also DJD 19:112). Yardeni 2007 also lists this manuscript as part of those copied by the prolific scribe she identified and dates it to the late first century BCE to the beginning of the first century CE (30 BCE–20 CE).
    \item[]

    \item \textbf{4Q176 (4QTanh)}
    \item[$\bullet$] 355–300 BCE (30.5\%), 210–100 BCE (64.2\%), 70–60 BCE (0.7\%)
    \item[$\bullet$] Strugnell 1970:229 and Tigchelaar RevQ 2019; “middle Hasmonaean” (ca. 125–75 BCE).
    \item[]

    \item \textbf{4Q201/4Q338 (4QEn\textsuperscript{a} ar/4QGenealogical List)}
    \item[$\bullet$] 165–40 BCE (93.6\%), 10–1 BCE (1.9\%)
    \item[$\bullet$] Milik 1976:140: first half of the second century BCE. Mixed evidence: archaic and connections with semicursive scripts of third and second centuries BCE, perhaps dependent upon the Aramaic scripts and scribal customs of northern Syria or Mesopotamia.
    \item[$\bullet$] Puech 2017:99: ca. 200 BCE.
    \item[$\bullet$] Langlois Le premier manuscrit du Livre d’Hénoch, 62–68: ca. 150 BCE.
    \item[$\bullet$] 200–150 BCE
    \item[]

    \item \textbf{4Q255/4Q433a (4QpapS\textsuperscript{a}/4QpapHodayot-like Text B)}
    \item[$\bullet$] 170-50 BCE (95.4\%)
    \item[$\bullet$] DJD 26:8, 20, 24, 29 (Alexander/Vermes, following Cross): 125–100 BCE.
    \item[]

    \item \textbf{4Q259 (4QS\textsuperscript{e})}
    \item[$\bullet$] 350–310 BCE (24.3\%), 210–100 BCE (69.7\%), 70–55 BCE (1.4\%)
    \item[$\bullet$] DJD 26:8, 20, 24, 133 (Alexander and Vermes, also referring to Cross): 50–25 BCE. Late Hasmonaean/Early Herodian semicursive, with mixed semicursive and semiformal features. But 4Q259 is difficult to date palaeographically. Suggestions range from 50–25 BCE (Cross), second half second century BCE, 150–100 BCE (Milik), to first half first century BCE, preferably shortly after 100 BCE, 100–75 BCE (Puech).
    \item[]

    \item \textbf{4Q504 (4QDibHam\textsuperscript{a})}
    \item[$\bullet$] 355–285 BCE (45.4\%), 230–150 BCE (50.1\%)
    \item[$\bullet$] DJD 7:137 (Baillet): “L’écriture est une calligraphie asmonéenne, qui peut dater des environs de 150 avant J.-C.” Cross: 175–150 BCE.
    \item[]

    \item \textbf{4Q521 (4QMessianic Apocalypse)}
    \item[$\bullet$] 355–285 BCE (38.0\%), 230–100 BCE (57.5\%)
    \item[$\bullet$] DJD 25:3–5 (Puech): formal Hasmonaean script, following Cross; first quarter of the
first century BCE (100–80 BCE).
    \item[]

    \item \textbf{4Q541 (4QapocrLevi\textsuperscript{b} ar)}
    \item[$\bullet$] 355–300 BCE (24.6\%), 210–95 BCE (68.2\%), 75–55 BCE (2.7\%)
    \item[$\bullet$] DJD 31:227 (Puech): Hasmonaean, to the end of the second century BCE, ca. 100 BCE; the writing is of the type of 1QS, 1QIsa\textsuperscript{a}, 4Q175, but posterior to 4Q504 (125–100 BCE).
    \item[]

    \item \textbf{11Q5 (11QPs\textsuperscript{a})}
    \item[$\bullet$] 35–15 BCE (3.3\%), 5–120 CE (92.2\%)
    \item[$\bullet$] DJD 4:6–9 (Sanders): first half of the first century CE (1–50 CE).
    \item[]

    \item \textbf{Mas1k (ShirShabb)}
    \item[$\bullet$] 50 BCE–65 CE (95.4\%)
    \item[$\bullet$] Masada 6:120 (Newsom and Yadin; Newsom HSS 27:168): developed Herodian formal hand, late Herodian formal hand, ca. 50 CE. Also: DJD 11:239.
    \item[]

    \item \textbf{Mur19 (pap WrDiv)}
    \item[$\bullet$] 45 BCE–85 CE (91.5\%), 95–110 CE (3.9\%)
    \item[$\bullet$] Cursive script with internal date of 71/72 CE validates radiocarbon date. The text refers to “year 6 of Masada”. See Section~\ref{appen:B4:amsdating} in Appendix~\ref{appen:B}.
    \item[]

    \item \textbf{5/6Hev1b (Ps)}
    \item[$\bullet$] 10–205 CE (95.4\%)
    \item[$\bullet$] DJD 38:143: late Herodian, understood as 50-68 CE. Cross: 75–100 CE.
    \item[]

    \item \textbf{XHev/Se2 (XHev/Se Num\textsuperscript{a})}
    \item[$\bullet$] 45 BCE–75 CE (95.4\%)
    \item[$\bullet$] DJD 38:174 (Flint): late Herodian, 50–68 CE.
    \item[]

\end{enumerate}

Older 2$\sigma$ calibrated ranges in 9 of the 26 accepted samples: 4Q2, 4Q3, 4Q27, 4Q30, 4Q114, 4Q206, 4Q267, 4Q375, 4Q416 (see Appendix~\ref{appen:B8.2:nooverlap}).

\begin{enumerate}
    \item \textbf{4Q2 (Gen\textsuperscript{b})}
    \item[$\bullet$] 155–130 BCE (5.2\%), 125 BCE–10 CE (90.3\%)
    \item[$\bullet$] DJD 12:31 (Davila): late Herodian or even post-Herodian formal hand, ca. 50–68+ CE.
    \item[]

    \item \textbf{4Q3 (4QGen\textsuperscript{c})}
    \item[$\bullet$] 340–325 BCE (3.5\%), 200–50 BCE (92.0\%)
    \item[$\bullet$] DJD 12:39 (Davila): Herodian formal hand, dating from the middle to the end of that period, ca. 20–68 CE.
    \item[]

    \item \textbf{4Q27 (4QNum\textsuperscript{b})}
    \item[$\bullet$] 340–330 BCE (1.3\%), 200–50 BCE (94.2\%)
    \item[$\bullet$] DJD 12:211 (Jastram): following Cross, early Herodian semiformal, 30 BCE–20 CE, earlier in that range.
    \item[]

    \item \textbf{4Q30 (4QDeut\textsuperscript{c})}
    \item[$\bullet$] 360–275 BCE (57.4\%), 260–245 BCE (1.4\%), 235–165 BCE (36.7\%)
    \item[$\bullet$] DJD 14:15 (White Crawford): following Cross, typical Hasmonaean book hand, 150–100 BCE. But Cross 2003 gives a more narrow date of 125–100 BCE.
    \item[]

    \item \textbf{4Q114 (4QDan\textsuperscript{c})}
    \item[$\bullet$] 355–285 BCE (49.5\%), 230–160 BCE (45.9\%)
    \item[$\bullet$] DJD 16:270 (Ulrich, following Cross): late second century BCE, no more than about a half century younger than the autograph, 125–100 BCE.
    \item[]

    \item \textbf{4Q206 (4QEn\textsuperscript{e} ar)}
    \item[$\bullet$] 360–280 BCE (48.6\%), 235–145 BCE (45.8\%), 135–120 BCE (1.1\%)
    \item[$\bullet$] Milik 1976:225: Hasmonaean, probably first half first century BCE, also referring to Cross 1961: p. 138, fig. 2, lines 2 (4Q30) and 3 (4Q51) and p. 149, fig. 4, lines 2 (4Q114) and 4 (4Q398), 100–50 BCE.
    \item[]

    \item \textbf{4Q267 (4QDamascus\textsuperscript{b})}
    \item[$\bullet$] 355–290 BCE (28.6\%), 210–95 BCE (65.3\%), 70–55 BCE (1.6\%)
    \item[$\bullet$] DJD 18:1, 96 (Yardeni): formal early Herodian, Cross’s round semiformal; connects this manuscript with 4Q397 as possibly same scribe, 30 BCE–20 CE.
    \item[]

    \item \textbf{4Q375 (4QapocrMoses\textsuperscript{a})}
    \item[$\bullet$] 345–320 BCE (6.0\%), 205–50 BCE (89.5\%)
    \item[$\bullet$] DJD 19:112 (Strugnell): early Herodian, rustic semiformal, 30 BCE–20 CE. Compare with 4Q27, 4Q161, both radiocarbon and palaeography.
    \item[]

    \item \textbf{4Q416 (4QInstruction\textsuperscript{b})}
    \item[$\bullet$] 345–320 BCE (8.0\%), 205–90 BCE (78.1\%), 80–50 BCE (9.4\%)
    \item[$\bullet$] DJD 34:74–76 (Strugnell and Harrington): Herodian, between 4Q51 and 1QM, hence “in a date transitional between the late Hasmonaean and the earliest Herodian hands”, and Strugnell judged the hand of 4Q416 to be earlier than the hands of 4Q415, 4Q417, and 4Q418 by some twenty-five years (76), 50–25 BCE.
    \item[]
    
\end{enumerate}